%% file: main.tex
\documentclass[a4paper, 11pt]{book}

\pdfoutput=1

\usepackage[T1]{fontenc}
\usepackage[utf8]{inputenc}
\usepackage{doi}
\usepackage[autostyle]{csquotes}
\usepackage[
    backend=biber,
    style=numeric-comp,
    date=year,
    giveninits=true,
    sorting=none,
    url=false, 
    doi=true,
    eprint=true,
]{biblatex}
\usepackage{biblatex}
\usepackage[french,german,english]{babel}

\usepackage{lmodern} 

\usepackage{fourier} 
\usepackage{setspace} 

\usepackage{graphicx,xcolor}
\graphicspath{{images/}}

\usepackage{subfig}
\usepackage{booktabs}
\usepackage{lipsum}
\usepackage[final]{microtype}
\usepackage{url}

\usepackage{fancyhdr}

\fancypagestyle{plain}{
	\fancyhf{}

	\fancyfoot[EL,OR]{\thepage}}
\fancypagestyle{addpagenumbersforpdfimports}{
	\fancyhead{}
	
	\fancyfoot{}
	\fancyfoot[RO,LE]{\thepage}
}

\usepackage{listings}
\usepackage{hyperref}
\hypersetup{pdfborder={0 0 0},
	colorlinks=true,
	linkcolor=black,
	citecolor=black,
	urlcolor=black}
\ifpdf
\usepackage[final]{pdfpages}
\else
\usepackage{calc}
\usepackage{breakurl}
\usepackage[nlwarning=false]{hypdvips}
\usepackage{backref}
\renewcommand*{\bacref}[1]{}
\fi
\usepackage{bookmark}

\makeatletter
\renewcommand\@pnumwidth{20pt}
\makeatother

\makeatletter
\def\cleardoublepage{\clearpage\if@twoside \ifodd\c@page\else
    \hbox{}
    \thispagestyle{empty}
    \newpage
    \if@twocolumn\hbox{}\newpage\fi\fi\fi}
\makeatother \clearpage{\pagestyle{plain}\cleardoublepage}

\usepackage{color}
\usepackage{tikz}
\usepackage[explicit]{titlesec}
\newcommand*\chapterlabel{}
\titleformat{\chapter}[display]  
	{\normalfont\bfseries\Huge} 
	{\gdef\chapterlabel{\thechapter\ }}     
 	{0pt} 
 	  {\begin{tikzpicture}[remember picture,overlay]
    \node[yshift=-8cm] at (current page.north west)
      {\begin{tikzpicture}[remember picture, overlay]
        \draw[fill=black] (0,0) rectangle(35.5mm,15mm);
        \node[anchor=north east,yshift=-7.2cm,xshift=34mm,minimum height=30mm,inner sep=0mm] at (current page.north west)
        {\parbox[top][30mm][t]{15mm}{\raggedleft \rule{0cm}{0.6cm}\color{white}\chapterlabel}};  
        \node[anchor=north west,yshift=-7.2cm,xshift=37mm,text width=\textwidth,minimum height=30mm,inner sep=0mm] at (current page.north west)
              {\parbox[top][30mm][t]{\textwidth}{\rule{0cm}{0.6cm}\color{black}#1}};
       \end{tikzpicture}
      };
   \end{tikzpicture}
   \gdef\chapterlabel{}
  } 
\titlespacing*{name=\chapter,numberless}{-3.7cm}{83.2pt-\parskip}{-3.2pt+\parskip}
\titlespacing*{\chapter}{-3.7cm}{50pt-\parskip-\parskip}{30pt+\parskip+\parskip}
\titlespacing*{\section}{0pt}{13.2pt}{1em-\parskip}  
\titlespacing*{\subsection}{0pt}{13.2pt}{1em-\parskip}
\titlespacing*{\subsubsection}{0pt}{13.2pt}{1em-\parskip}
\titlespacing*{\paragraph}{0pt}{13.2pt}{1em-\parskip}

\newcounter{myparts}
\newcommand*\partlabel{}
\titleformat{\part}[display]  
	{\normalfont\bfseries\Huge} 
	{\gdef\partlabel{\thepart\ }}     
 	{0pt} 
 	  {\ifpdf\setlength{\unitlength}{20mm}\else\setlength{\unitlength}{0mm}\fi
	  \addtocounter{myparts}{1}
	  \begin{tikzpicture}[remember picture,overlay]
    \node[anchor=north west,xshift=-65mm,yshift=-6.9cm-\value{myparts}*20mm] at (current page.north east) 
      {\begin{tikzpicture}[remember picture, overlay]
        \draw[fill=black] (0,0) rectangle(62mm,20mm);   
        \node[anchor=north west,yshift=-6.1cm-\value{myparts}*\unitlength,xshift=-60.5mm,minimum height=30mm,inner sep=0mm] at (current page.north east)
        {\parbox[top][30mm][t]{55mm}{\raggedright \color{white}Part \partlabel \rule{0cm}{0.6cm}}};  
        \node[anchor=north east,yshift=-6.1cm-\value{myparts}*\unitlength,xshift=-63.5mm,text width=\textwidth,minimum height=30mm,inner sep=0mm] at (current page.north east)
              {\parbox[top][30mm][t]{\textwidth}{\raggedleft \rule{0cm}{0.6cm}\color{black}#1}};
       \end{tikzpicture}
      };
   \end{tikzpicture}
   \gdef\partlabel{}
  } 
\titlespacing*{\part}{11.06cm}{26.4pt-\parskip-\parskip}{0pt}

\usepackage{amsmath}
\usepackage{amsfonts}
\usepackage{amssymb}
\usepackage{mathtools}
\makeatletter
\def\resetMathstrut@{%
  \setbox\z@\hbox{%
    \mathchardef\@tempa\mathcode`\(\relax
      \def\@tempb##1"##2##3{\the\textfont"##3\char"}%
      \expandafter\@tempb\meaning\@tempa \relax
  }%
  \ht\Mathstrutbox@1.2\ht\z@ \dp\Mathstrutbox@1.2\dp\z@
}
\makeatother

\usepackage{braket}
\usepackage{bm}
\usepackage{qcircuit}
\usepackage{algorithm}
\usepackage{algorithmic}
\usepackage{pdflscape}
\usepackage[framemethod=TikZ]{mdframed}
\usepackage{makecell}

\DeclareFieldFormat
  [article,inbook,incollection,inproceedings,patent,thesis,
   unpublished,techreport,misc,book]
  {title}{\mkbibquote{#1}}

\mdfdefinestyle{summarybox}{
    roundcorner=10pt,
    innertopmargin=\topskip,
    frametitle={Summary},
    frametitlerule=true,
    frametitlebackgroundcolor=gray!20,
}
\mdfdefinestyle{paperbox}{
    roundcorner=10pt,
    innertopmargin=\topskip,
    frametitle={Note},
    frametitlerule=true,
    frametitlebackgroundcolor=gray!20,
}

\newcommand{\summary}[1]{
\begin{mdframed}[style=summarybox]
#1
\end{mdframed}
}
\newcommand{\paper}[1]{
\begin{mdframed}[style=paperbox]
#1
\end{mdframed}
}
\newcommand{\comment}[1]{}
\newcommand{\printpublication}[1]{\AtNextCite{\defcounter{maxnames}{99}}\fullcite{#1}}

\newcommand{\qed}{\hfill\square}
\newcommand{\argmin}{\mathrm{argmin}}

\newcommand{\tr}{\mathrm{Tr}}
\newcommand{\iso}{\cong}
\newcommand{\pihalf}{\frac{\pi}{2}}

\renewcommand{\vec}[1]{\bm{#1}}

\DeclareMathAlphabet{\mathcal}{OMS}{cmsy}{m}{n}

\DeclareSourcemap{
  \maps[datatype=bibtex]{
     \map{
        \step[fieldsource=doi,final]
        \step[fieldset=isbn,null]
        }
      }
}
\DeclareSourcemap{
  \maps[datatype=bibtex]{
     \map{
        \step[fieldsource=doi,final]
        \step[fieldset=issn,null]
        }
      }
}
\DeclareSourcemap{
  \maps[datatype=bibtex]{
    \map{
      \pertype{incollection}
      \step[fieldset=address, null]
      \step[fieldset=publisher, null]
    }
  }
}
\DeclareSourcemap{
  \maps[datatype=bibtex]{
    \map{
      \pertype{inproceedings}
      \step[fieldset=address, null]
      \step[fieldset=publisher, null]
      \step[fieldset=location, null]
      \step[fieldset=series, null]
    }
  }
}

\AtEveryBibitem{\clearfield{pages}} 

\begin{document}
\frontmatter
\input{head/titlepage.tex}

\include{head/dedication}

\setcounter{page}{0}
\include{head/abstracts}
\include{head/publications}
\include{head/acknowledgements}

\cleardoublepage
\pdfbookmark{\contentsname}{toc}
\tableofcontents

\cleardoublepage
\phantomsection
\addcontentsline{toc}{chapter}{List of Figures} 
\listoffigures

\cleardoublepage
\phantomsection
\addcontentsline{toc}{chapter}{List of Tables} 
\listoftables

\mainmatter
\include{main/ch0_introduction}
\include{main/ch1_basics}
\include{main/ch2_quantum_hardware}

\include{main/ch3_vqa}
\include{main/ch4_qnspsa}
\include{main/ch5_saqite}

\include{main/ch6_dualqte}
\include{main/ch7_conclusion}

\addtocontents{toc}{\vspace{\normalbaselineskip}}
\cleardoublepage
\bookmarksetup{startatroot}

\include{tail/appendix}
\backmatter
\include{tail/biblio}

\end{document}

%% file: head/titlepage.tex
\begin{titlepage}
\begin{otherlanguage}{french}

\sffamily

\begin{flushleft}
\parbox{0.3\textwidth}{\includegraphics[width=4cm]{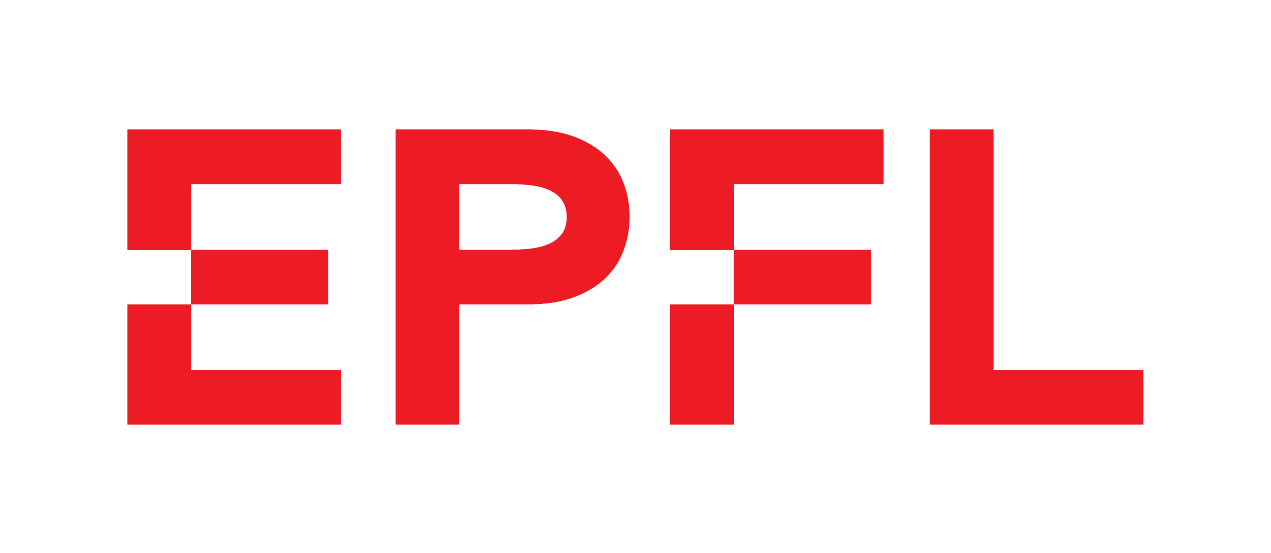}}
\end{flushleft}

\begin{flushright}
Thèse n.~11~132
\end{flushright}

\null\vspace{2cm}

\begin{minipage}{4cm}
\end{minipage}
    \hfill
\begin{minipage}{11cm}
{\Large Scalable Quantum Algorithms for Noisy Quantum \\ [8pt] Computers} \\

\vspace{2cm}

\small
Présentée le 29 février 2024\\[8pt]
Faculté des sciences de base\\
Laboratoire de sciences quantiques numériques \\
Programme doctoral en physique\\

pour l'obtention du grade de Docteur ès Sciences\\[8pt]
par\\ [12pt]
{\Large \textbf{Julien Sebastian GACON}}\\[9pt]

Acceptée sur proposition du jury\\[4pt]
    Prof. F. Mila, président du jury\\
    Prof. G. Carleo, Dr S. Woerner, directeurs de thèse\\
    Prof. G. Mazzola, rapporteur\\
    Prof. J. Goold, rapporteur\\
    Prof. Z. Holmes, rapporteuse\\
\end{minipage}
\vspace{2cm}
\begin{flushright}
    2024
\end{flushright}

\end{otherlanguage}
\end{titlepage}

%% file: head/dedication.tex
\cleardoublepage
\thispagestyle{empty}

\vspace*{3cm}

\begin{raggedleft}
    Humanity must seek what is  
    \emph{not} simple and obvious \\
    using the simple and obvious.\\
     --- Musonius Rufus\\
\end{raggedleft}

\vspace{4cm}

\begin{center}
    \textit{To my parents.}
\end{center}

%% file: head/abstracts.tex
\cleardoublepage
\chapter*{Abstract}
\markboth{Abstract}{Abstract}
\addcontentsline{toc}{chapter}{Abstract (English/Français)} 

Quantum computing not only holds the potential to solve long-standing problems in quantum physics, but also to offer speed-ups across a broad spectrum of other fields.
Access to a computational space that incorporates quantum effects, such as superposition and entanglement, enables the derivation of promising quantum algorithms for important tasks, including preparing the ground state of a quantum system or predicting its evolution over time.
Successfully tackling these tasks promises insights into significant theoretical and technological questions, such as superconductivity and the design of new materials.

The aim of quantum algorithms is to use a series of quantum operations, organized in a quantum circuit, to solve a problem beyond the reach of classical computers.
However, the noise and limited scale of current quantum computers restricts these circuits to moderate sizes and depths. As a result, many prominent algorithms are currently infeasible to run for problem sizes of practical interest.
In response, recent research focused on variational quantum algorithms, which allow the selection of circuits that act within a quantum device's capabilities.
Yet, these algorithms can require the execution of a large number of circuits, leading to prohibitively long computation times. 

This doctoral thesis develops two main techniques to reduce these quantum computational resource requirements, with the goal of scaling up application sizes on current quantum processors.
The first approach is based on stochastic approximations of computationally costly quantities, such as quantum circuit gradients or the quantum geometric tensor (QGT).
The second method takes a different perspective on the QGT, leading to a potentially more efficient description of time evolution on current quantum computers.
Both techniques rely on maintaining available information and only computing necessary corrections, instead of re-computing possibly redundant data. 
The main focus of application for our algorithms is the simulation of quantum systems, broadly defined as including the preparation of ground and thermal states, and the real- and imaginary-time propagation of a system.
The developed subroutines, however, can further be utilized in the fields of optimization or machine learning.
Our algorithms are benchmarked on a range of representative models, such as Ising or Heisenberg spin models, both in numerical simulations and experiments on the hardware. In combination with error mitigation techniques, the latter is scaled up to 27 qubits; into a regime that variational quantum algorithms are challenging to scale to on noisy quantum computers without our algorithms.

\begin{otherlanguage}{french}
\cleardoublepage
\chapter*{Résumé}
\markboth{Résumé}{Résumé}

L’informatique quantique propose de résoudre des grands problèmes de la physique contemporaine et d’autres domaines en utilisant les lois de la mécanique quantique. En utilisant des effets quantiques tels que la possibilité de superposer et d’intriquer les configurations d’un système, de nouveaux types d’algorithmes peuvent être proposés pour des problèmes essentiels comme, par exemple, préparer l’état d’énergie minimale d’un système quantique ainsi que son évolution temporelle. La réalisation de ces objectifs complexes aurait de nombreuses implications théoriques et technologiques, notamment en matière de supraconductivité et de conception de nouveaux matériaux. 

Le principe des algorithmes quantiques est d’utiliser une série d’opérations quantiques dans un circuit informatique pour résoudre des problèmes dépassant les limites des ordinateurs classiques. Cependant, les technologies actuelles des ordinateurs quantiques restreignent ces circuits à des dimension modérées. En conséquence, des nombreux algorithmes sont actuellement impossibles à exécuter pour des applications d’intérêt pratique. Pour cette raison, la recherche dans ce domaine s’est concentrée sur une approche variationnelle des algorithmes quantiques, qui permet de sélectionner des circuits opérant dans les limites des capacités des processeurs quantiques. Malgré cela, ces méthodes peuvent nécessiter un grand nombre de circuits et, en conséquence, un temps de calcul prohibitif. 

Cette thèse de doctorat développe deux techniques principales pour réduire les ressources de calcul quantique, avec l’objectif d’augmenter la dimension des applications avec les processeurs quantiques actuels. La première méthode est basée sur des approximations stochastiques de quantités coûteuses en calculs, telles que les gradients des circuits ou le tenseur géométrique quantique (QGT en anglais). La seconde technique adopte un point de vue différent sur le QGT, pouvant permettre une description plus efficace de l’évolution temporelle. Nous présentons nos algorithmes pour une variété de problèmes, incluant la préparation des états d’énergie minimale, des états thermiques, et la propagation en temps réel et imaginaire d’un système. Les algorithmes développés trouvent d’autres applications dans les domaines de l’optimisation ou de l’intelligence artificielle. Nos algorithmes sont évalués sur des modèles représentatifs, notamment le modèle d’Ising et d’Heisenberg, par des simulations numériques et directement sur des processeurs quantiques. En combinaison avec des techniques d’atténuation des erreurs, ce dernier modèle est étudié avec jusqu’à 27 qubits dans un régime qui présente des obstacles conséquents sans nos algorithmes.

\end{otherlanguage}

%% file: head/publications.tex
\cleardoublepage
\chapter*{Publications}
\markboth{Publications}{Publications}
\addcontentsline{toc}{chapter}{Publications} 

The following publications are covered in this thesis:
\begin{itemize}
    \item \printpublication{gacon_qnspsa_2021} 
    \item \printpublication{gacon_saqite_2023} \textit{Best paper award (2nd place)} 
    \item \printpublication{gacon_dual_2023} 
    \item \printpublication{zoufal_blackbox_2023} 
    \item \printpublication{jones_efficient_2020} 
\end{itemize}

The following research has been conducted during the thesis, but is not explicitly covered:
\begin{itemize}
    \item \fullcite{weidenfeller_scaling_2022}
    \item \fullcite{abbas_optimization_2023}
\end{itemize}

%% file: head/acknowledgements.tex
\chapter*{Acknowledgements}
\markboth{Acknowledgements}{Acknowledgements}
\addcontentsline{toc}{chapter}{Acknowledgements}

This thesis completes a collective journey—a path made possible by, and filled with good memories thanks to, the support and contributions of many.

Stefan Woerner, my supervisor at IBM Quantum, has been instrumental to both my academic and personal growth over the past years. Thank you for always finding time for discussions, for supporting my individual path, and for your clear scientific insights.
Giuseppe Carleo, my professor at EPFL, enabled this thesis by guiding a non-physicist through the world of quantum physics. I've learned a lot from your creative ideas and way of thinking (and I'm grateful for your patience with my extensive use of Overleaf).

This work would not have been possible without my colleagues at EPFL and IBM.
I owe particular thanks to Christa Zoufal who has always been there to provide invaluable help with scientific, administrative, and lifestyle questions.
My appreciation also goes to Almudena Carrera Vazquez whose expertise in mathematics and mountaineering improved both the quality of this thesis and my mountain expeditions.
The past years would not have been half as much fun without traveling with Abby Mitchell, Max Rossmannek, Elena Peña Tapia, Amira Abbas, and David Sutter—thank you!
Furthermore, I'm particularly thankful to Ali Javadi-Abhari for initiating this thesis with his collaboration and welcoming me onto the Qiskit team.
What started with fixing small bugs led to an incredibly enriching learning experience and becoming a maintainer, allowing me to give back to the open-source community. Special thanks to Kevin Krsulich, Matthew Treinish, and Jake Lishman for their open and enthusiastic manner, and their patience in answering code-related questions. 
These acknowledgments could go on and on, but, as I have to cover the printing costs of this thesis, I’ll come to an end. 
A big thank you for scientific discussions and reviewing this thesis go to
Jannes Nys, Stefano Barison, Daniel Egger, Francesco Tacchino, Alexander Miessen, Matthis Lehmkuehler, Laurin Fischer, Marc Drudis, Anthony Gandon, Samuele Piccinelli, Julian Schuhmacher, Riccardo Rossi and Alberto Baiardi. 

Finally, I would like to thank my parents, who gave me the confidence to tackle challenges of academic and adventurous kind and fully supported me along the way. The same is true for my partner, Anna Rind, who encouraged and motivated me with her overwhelming, positive attitude. 
I'm also thankful for my friends, including Philip Verwegen, Jonas Kunath, Andreas Mono, Jasper Jonker, Simon Galli, Tabitha Knoor, Christian Digel, and Max Deutsch (and to whom I forgot: contact me for an apology cake).

\hfill Thank you all! 

\hfill Julien

%% file: main/ch0_introduction.tex
\chapter{Introduction}

Computing power is a catalyst in advancing humankind's capabilities.
From science and medicine to engineering and economics, modern society depends on the growing capacity of our computing systems.
For example, the ability to process large data sets is the foundation of machine learning, which is applied in virtually every technological field.
Solving large systems of equations, on the other hand, enables the study of materials and the design of new structures.

In the past, a reliable strategy to boost the computing power has been to increase the number of transistors on a microchip. 
The growth of this number is famously summarized by Moore's law, stating that it doubles approximately every two years.
So far, this is mainly achieved by reducing the size of a single transistor. 
This trend, however, is slowly grinding to a halt as transistor sizes reach physical limits where transistors are made up of just a few individual atoms and quantum mechanical effects become significant.
We inevitably encounter the question:
If transistors cannot be further miniaturized, how can we keep on solving harder problems?
But tackling more complex problems by increasing computational power represents only one possible, brute-force, approach---the development of faster algorithms and more efficient computing platforms plays an equally crucial role.
In machine learning, for example, the implementation of the backpropagation algorithm~\cite{baydin_autodiff_2017} to train neural networks has been critical to scale up to the large models we see today.
Specialized hardware based on application-specific integrated circuits (ASICs) or field programmable gate arrays (FPGAs), on the other hand, promises an increase in speed and reduction of energy consumption~\cite{machupalli_review_2022, ringlein_fgpa_2023}.

However, even if we were able to further scale up computational capabilities at the current rate, some problems remain notoriously difficult.
Quantum effects are not only limiting transistor sizes, but they can also be enormously challenging to simulate on a conventional computer.
The state of a quantum system is described by a wave function, which captures non-classical effects of entanglement and superposition, and scales exponentially in the system size.
Storing a single full wave function of a system of only 60 particles would require more than 9 exabytes (EB) memory,\footnote{The wave function of 60 spin-1/2 particles contains up to $2^{60}$ complex amplitudes. Using single-point precision, this equals $2^{60} \times 2 \times 32$ bits $\approx 9.22$ EB.} much more than available even on the largest supercomputers today~\cite{frontier}.
In response to this prohibitive scaling on current computers, an entirely new computing paradigm has been put forward by Feynman in the early 1980s.
Instead of attempting to mitigate the effects of quantum mechanics, it \emph{leverages} them to build a computer playing by its rules: a quantum computer~\cite{feynman_theboss_1982}.

A quantum computer is built of quantum bits (qubits), which are fundamental units of information that obey the laws of quantum mechanics. 
As such, storing the state of 60 particles only requires a linear amount of qubits and enables an efficient representation of the wave function.
In addition to an efficient representation of the wave function, quantum computers have access to a new operational space, which allows the development of novel algorithms for applications beyond quantum mechanics.
Among the most famous examples is Shor's algorithm~\cite{shor_factoring_1994} for prime number factorization, which achieves an exponential speedup compared to known algorithms on classical computers. 
Other examples include Grover's algorithm~\cite{grover_search_1996}, which could provide a quadratic speedup in an unstructured search, or the HHL algorithm~\cite{harrow_hhl_2009}, which may solve specific sparse linear systems in exponentially less time than classical algorithms.

Due to these promising theoretical results, recent years have seen tremendous technological advances in the physical realization of quantum computers.
In the late 90s, the first quantum algorithms were demonstrated, such as a 2-qubit Grover search~\cite{chuang_experimental_1998} and a 3-qubit quantum Fourier transform~\cite{weinstein_qft_2001}, implemented using nuclear magnetic resonance performed on chloroform isotopes and alanine molecules, respectively.
By now, quantum computers with several 100 qubits are commonly available~\cite{ibm_quantum} with up to 1000-qubit chips being released~\cite{ibm_condor_2023}.
These sizes would suffice to solve a wide range of open problems in physics, chemistry, material sciences and optimization~\cite{reiher_elucidating_2017, daley_practical_2022, abbas_optimization_2023}---if the quantum computers were perfectly error-free.
However, current devices are subject to several sources of noise and only a limited number of operations can be implemented to a high fidelity.
This restricts the problems that today's quantum computers can solve to relatively small sizes and, so far, no advantage in a practically relevant application has been demonstrated.

The goal of this thesis is to develop algorithms that enable the scaling of problems to larger sizes on current, noisy quantum computers.
This is achieved by understanding and circumventing the limitations of the algorithms at hand, or deriving novel approaches.
We are mainly focusing on simulating quantum systems, such as approximating their real- and imaginary-time dynamics and preparing their ground state, which is among the most natural applications on quantum computers.

\section{Why quantum computing?}

In this section we provide a more quantitative motivation for quantum computing.
We begin by discussing the previously outlined problem of quantum simulation in more detail. 
But, since quantum computing sits at the interface of computer science and physics, we additionally motivate quantum computing from a purely computer-scientific perspective.

\subsection{From a physics perspective}

The change of a physical system is modeled by differential equations.
The ability to solve these equations to find the dynamics of a system is core to modern technologies.
For example, humans venture into space on trajectories derived by Euler-Lagrange equations~\cite{troutman_eulerlagrange_1996} in spacecrafts that obey the rocket equation~\cite{tsiolkovsky_rocket_1954, moore_rocket_1810} and are designed using Navier-Stokes equations~\cite{fefferman_navierstokes_2015}.
Finally, the majority of these equations are solved on computers based on electric circuits which are again described by differential equations~\cite{lewis_diffeq_2017}.
Reducing the scale even further to the atomic and subatomic level, we reach the realm we are interested in in this thesis: quantum mechanics.
There, the (non-relativistic) dynamics can be described by the Schrödinger equation, given by 
\begin{equation}
    i \hbar \frac{\partial}{\partial t} \ket{\psi} = H \ket{\psi},
\end{equation}
where $\hbar = h / (2\pi)$ with the Plank constant $h$, $\ket{\psi}$ is the wave function encoding the current configuration of the system, and $H$ is the Hamiltonian describing the energy of the system~\cite{guertin_relativistic_1974}. The objects $\ket{\psi}$ and $H$ will be formally introduced in Chapter~\ref{ch:basics}. Further, we set $\hbar \equiv 1$ in the remainder of this thesis.

The Schrödinger equation is a first-order, linear differential equation. Depending on $H$, this type of equation is not particularly difficult to solve. For example, if $H$ is diagonal or tridiagonal we can solve for the derivative of $\ket{\psi}$ with a number of operations scaling linearly with the dimension of $\ket{\psi}$. 
The challenge of simulating quantum mechanics, instead, lies in the sheer size of the equation, which scales exponentially with the number of particles.
Assume we would like to simulate a spin-1/2 system, common in a variety of models, such as the Hubbard and Ising models which are important to study, e.g., superconductivity and ferromagnetism~\cite{thijssen_comphys_2007}.
Being a quantum mechanical system, each spin can be described as a superposition of ``up'' state, $\ket{\uparrow}$, and ``down'' state, $\ket{\downarrow}$, as
\begin{equation}
    \ket{\psi} = c_{\uparrow} \ket{\uparrow} + c_\downarrow \ket{\downarrow},
\end{equation}
for complex coefficients $c_\uparrow$ and $c_\downarrow$. Note, that these coefficients have to fulfill a normalization condition which we will also discuss in Chapter~\ref{ch:basics}.
Since particles obeying the laws of quantum mechanics can be in arbitrary superpositions of their basis states, a system of $n$ particles is described a wave function with all possible $2^n$ combinations of $\ket{\uparrow}$ and $\ket{\downarrow}$, that is
\begin{equation}
    \ket{\psi} = c_{\uparrow \cdots \uparrow \uparrow} \ket{\uparrow \cdots \uparrow \uparrow}
               + c_{\uparrow \cdots \uparrow \downarrow} \ket{\uparrow \cdots \uparrow \downarrow}
               + \cdots
               + c_{\downarrow \cdots \downarrow \downarrow} \ket{\downarrow \cdots \downarrow \downarrow}.
\end{equation}
It will be useful to use a more concise notation for this state, given by
\begin{equation}
    \ket{\psi} = \sum_{k=0}^{2^n-1} c_k \ket{k},
\end{equation}
where $\ket{k}$ represents a spin basis state by associating $0$ bits in the binary representation of $k$ with a spin up, $\ket{\uparrow}$, and the $1$ bits with spin down, $\ket{\downarrow}$.
For example, the value $k=6$ represents the following spin state, 
\begin{equation}
     \ket{6} = \ket{0 \cdot 2^3 + 1 \cdot 2^2 + 1 \cdot 2^1 + 0 \cdot 2^0} \equiv \ket{0110} \equiv \ket{\uparrow \downarrow \downarrow \uparrow}.
\end{equation}
The basis states are time-independent and inserting this representation into the Schrödinger equation results in a linear system of an exponential number of equations for the time-dependent coefficients $c_k = c_k(t)$, given by
\begin{equation}
    \forall j \in \{0, ..., 2^n - 1\}:~ \frac{\partial c_j}{\partial t} = -i \sum_{k=0}^{2^n-1} c_k \braket{j|H|k}.
\end{equation}
The time-evolved state is then obtained by integrating the coefficients $c_k(t)$ in time, where the initial value is determined by the initial state of the system $\ket{\psi_0}$ as $c_k(0) = \braket{k|\psi_0}$.

In a physical system, interactions between spins are typically local, i.e., the number of partners a single particle interacts with does not scale with system size.
Though this leads to a sparse Hamiltonian $H$ with a polynomial number of non-zero elements per row, there are still $\mathcal{O}(2^n \text{poly}(n))$ entries due to the exponential number of rows.
Directly solving the Schrödinger equation by exactly diagonalizing the Hamiltonian is therefore only possible for small systems or if only a few basis states are relevant.

Instead of directly solving for the coefficients of the wave function, recent years have seen a surge of promising computational techniques to approximate solutions in specific cases. 
Broadly, these can be categorized into stochastic and compression methods, which are combined with a variational principle to determine the system dynamics.
Stochastic approaches, such as quantum Monte Carlo, rely on sampling relevant configurations of the wave function~\cite{carleo_tvmc_2012, carleo_tvmc_2017, sinibaldi_ptvmc_2023}.
These, however, struggle from a sign problem which can, for example, occur in systems of strongly interacting fermions~\cite{troyer_sign_2005}.
Compression techniques, on the other hand, attempt to provide a memory-efficient representation of the wave function using tensor networks~\cite{verstraete_mps_2008, czarnik_tn_2018} or artificial neural networks~\cite{carleo_nqs_2017}.
While these computational techniques excel at specific tasks, their general applicability is eventually limited by the exponentially large Hilbert space the wave function lives in and there is a palette of important tasks that remain unsolved, such as simulating frustrated models~\cite{wu_variational_2023} or the precise phase diagram of the Fermi-Hubbard model~\cite{qin_hubbard_2022}.

A solution that is as radically different as it is intuitive, was proposed by Richard Feynman~\cite{feynman_theboss_1982}: 
Instead of imposing the rules of quantum mechanics on a classical computer, could we build a computer that inherently implements these properties? Could we build a computer of particles that exhibit the quantum mechanical effects we would like to model?
Such a machine would be able to store an exponential number of complex-valued coefficients $c_k$ of $\ket{\psi}$ in only $n$ particles.\footnote{assuming spin-1/2 particles}

Several years later, Seth Lloyd showed such a quantum computer would be able to solve the Schrödinger equation efficiently for Hamiltonians $H$ that act only locally~\cite{lloyd_universal_1996}. 
Instead of solving a differential equation for the coefficients, we can now directly implement approximations of the time-evolution operator
\begin{equation}
    U(t) = e^{-iHt},
\end{equation}
to evolve an initial state $\ket{\psi_0}$ to $\ket{\psi(t)} = U(t)\ket{\psi_0}$, and solve the Schrödinger equation.
In Section~\ref{sec:simulating} we discuss this decomposition in detail.

By leveraging efficient representations, quantum algorithms have been proposed for a plethora of hard tasks in quantum physics.
Beyond implementing the real-time evolution of a system, these also include determining the ground state, preparing thermal states or performing imaginary-time evolution, to name a few. 
As such, quantum computing holds enormous potential to solve hard tasks in quantum physics. 

\subsection{From a computer science perspective}

Though originally proposed for the simulation of quantum physics, quantum computers present an impactful computational paradigm able to formulate new algorithms for typically classical tasks, such as period-finding or database search. To understand the difference to a classical computer, let us focus on what novel \emph{operations} a quantum computer has access to.
On a fundamental level, a classical computer represents data as binary information, or bits, which can be in either the state $0$ or $1$. A program takes a set of such bits as input, acts on them with logical operations, and returns their final state.
A quantum computer may act in a similar fashion. The qubits can initially be in either the $0$ or $1$ state and at the end of the quantum computation, measuring yields classical bit values. 
The key difference is what operations a quantum and classical computer have access to \emph{during} the computation.

\begin{figure}[thbp]
    \centering
\includegraphics[width=0.9\textwidth]{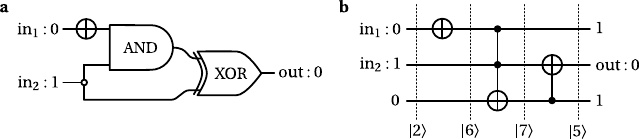}
    \caption[A classical computer program]{A classical program. 
    (a) Irreversible circuit. (b) Reversible version with an additional bit as scratch space. The basis state representation $\ket{k}$ of the bits is shown in each step of the reversible computation. See Appendix~\ref{app:gates} for a definition of the logic gates.}
    \label{fig:classical_program}
\end{figure}

At each point of the classical computer program the bits are in a deterministic state $b_n b_{n-1} \cdots b_1$, with $b_j \in \{0, 1\}$.
To illustrate the difference to a quantum computer, we assume a reversible computational model, as shown in Fig.~\ref{fig:classical_program}. This is not a restriction, as any irreversible logic circuit can be translated into a reversible network~\cite{bennett_reversible_1973}.
Since the number of bits remains constant at each step, we can write the binary state in an integer format, as
\begin{equation}
    k = \sum_{j=1}^{n} 2^{j-1} b_j. 
\end{equation}
Each state of the program can therefore be represented by a single integer with values in $\{0, ..., 2^{n}-1\}$ or, equivalently, as a $2^n$ dimensional unit vector, where exactly one entry has value $1$ and the remaining are 0.
This allows to see the operations a classical computer has access to as a permutation matrix $P \in \{0, 1\}^{2^n \times 2^n}$, where each row and column have exactly one entry that is $1$. 
For example, the initial state of the computation in Fig.~\ref{fig:classical_program} is $b_2 b_1 b_0 = 010$, which is
\begin{equation}
    \ket{k} = \ket{0 \cdot 2^0 + 1 \cdot 2^1 + 0 \cdot 2^2} = \ket{2} \equiv 
    {\small
    \begin{pmatrix}
        0 \\ 0 \\ 1 \\ 0 \\ 0 \\ 0 \\ 0 \\ 0     
    \end{pmatrix}},
    {\small
    \begin{matrix}
         \phantom{0} \\ \phantom{0} \\ \leftarrow $2$\text{nd entry} \\ \phantom{0}  \\ \phantom{0} \\ \phantom{0} \\ \phantom{0} \\ \phantom{0}     
    \end{matrix}
    }
\end{equation}
where we counted indices starting from $0$.
The NOT operation acts as bitflip on bit $b_2$ and as identity $I$ on the others, therefore the classical circuit performs the operation
\begin{equation}
    \underbrace{{\small \begin{pmatrix}
        0 & 0 & 0 & 0 & 1 & 0 & 0 & 0 \\
        0 & 0 & 0 & 0 & 0 & 1 & 0 & 0 \\
        0 & 0 & 0 & 0 & 0 & 0 & 1 & 0 \\
        0 & 0 & 0 & 0 & 0 & 0 & 0 & 1 \\
        1 & 0 & 0 & 0 & 0 & 0 & 0 & 0 \\
        0 & 1 & 0 & 0 & 0 & 0 & 0 & 0 \\
        0 & 0 & 1 & 0 & 0 & 0 & 0 & 0 \\
        0 & 0 & 0 & 1 & 0 & 0 & 0 & 0 \\
    \end{pmatrix}}}_{\equiv \text{NOT} \otimes I \otimes I}
    {\small
    \begin{pmatrix}
        0 \\ 0 \\ 1 \\ 0 \\ 0 \\ 0 \\ 0 \\ 0     
    \end{pmatrix} 
    }
    =
    {\small
    \begin{pmatrix}
        0 \\ 0 \\ 0 \\ 0 \\ 0 \\ 0 \\ 1 \\ 0     
    \end{pmatrix}
    }.
\end{equation}

Clearly, this is not an efficient manner to represent a computation. However, it allows to outline the difference between a classical and quantum computation in a clear fashion: a classical computer can only act with permutations on the bit-states, whereas a quantum computer can perform any unitary operation $U$ on the qubits. A unitary operation is represented by a matrix $U \in \mathbb{C}^{2^n \times 2^n}$, with $U^\dagger U = \mathbb{1}$ being the identity matrix of suitable dimension. 
Permutations, since $P^\dagger P = P^2 = \mathbb{1}$, are a subset of the available operations on a quantum computer which can implement a far larger class of operations.
In fact, instead of having access to a single unit vector $\ket{k}$ a quantum program can prepare, and perform computations on, a linear combination (superposition) of \emph{all} $2^n$ states, 
\begin{equation}
    U\ket{\psi} = \sum_{k=0}^{2^n-1} c_k U\ket{k}.
\end{equation}
Such an operation is implemented with a single quantum operation $U$ and acts on all states simultaneously. 
Upon measurement, however, the quantum state collapses to a single basis state, which is further discussed in Chapter~\ref{ch:basics}. By repeating the operations and measurements multiple times, the superpositions can be probed, but, to remain efficient, the number of measurements cannot scale exponentially in the number of qubits. This measurement problem requires an reduction of the quantum state at the end of a computation to an efficiently measurable state. 
For example, the quantum Fourier transform (QFT) of a $2^n$-dimensional vector can be implemented with only $O(n^2)$ gate operations. The Fast Fourier Transform (FFT), on the other hand, requires exponentially more classical operations, namely $\mathcal{O}(n2^n)$~\cite{cleve_qft_2000}. 
While the results of the QFT cannot be directly read-out, it is the center piece of a variety of quantum algorithms that promise a speed-up over classical algorithms~\cite{shor_factoring_1994, brassard_quantum_2002}.

\begin{figure}[thbp]
    \centering
    \includegraphics[width=0.65\textwidth]{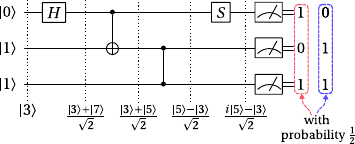}
    \caption[A quantum computer program]{A quantum program. As the classical program, a quantum program can have a classical input state and a classical output. In contrast to classical bits, however, which are represented by a single unit vector, qubits can be in complex superpositions of basis states. See Appendix~\ref{app:gates} for a definition of the gates.}
    \label{fig:quantum_program}
\end{figure}

In Fig.~\ref{fig:quantum_program} we show an example of a quantum program. The states $\ket{\psi}$ at each point in the computation can be understood as the likelihood of observing a sequence of bits if we were to measure the qubits at this point. In classical computations, each state is a unit vector and therefore the probability of observing the corresponding bits is 1. 
A quantum state has varying probabilities for each basis states and repeating the measurements yields different outcomes according to the distribution of the coefficients $c_k$.
This is explained further in Section~\ref{sec:quantum_computing}.

From a computational perspective, quantum computing is a new paradigm with access to a previously inaccessible operational space of exponential dimension $2^n \times 2^n$.
However, since both in- and output are still $n$ bits, developing algorithms that outperform classical algorithms is a difficult endeavor.
Still, the potential advantage of this paradigm is demonstrated by famous algorithms like Shor's algorithm~\cite{shor_factoring_1994}, which performs a prime factorization with exponentially less quantum operations than classical operations, or Grover's algorithm~\cite{grover_search_1996}, which promises a  quadratic speedup for unstructured search problems.  

\section{How to quantum compute?}

The fundamental unit of information in a quantum computer is a qubit, which can have can have two states labeled $\ket{0}$ and $\ket{1}$---analogous to a classical bit taking on values $0$ or $1$.
This qubit must be realized by a quantum mechanical system since our quantum computer should be able to leverage quantum mechanical effects, such as entanglement and superposition of qubit states.
In addition to defining a qubit state, we also need to be able to perform computations on it and read out its state.
These conditions, along with other requirements that a universal quantum computer must meet, are summarized in the DiVincenzo criteria~\cite{divincenzo_criteria_2000}, introduced in Chapter~\ref{ch:basics}.

\begin{figure}[th]
    \centering
    \includegraphics[width=\textwidth]{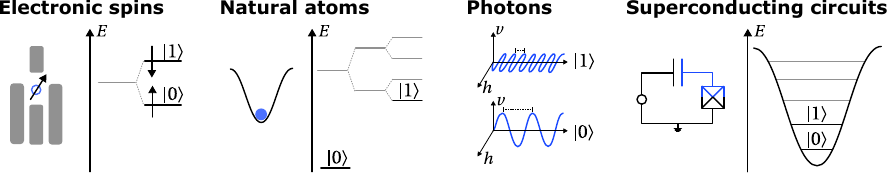}
    \caption[Qubit platforms]{Examples of qubit encodings using different platforms. Electronic spins in quantum dots use spin energy levels, which are split using the Zeeman effect~\cite{burkard_spinqubits_2023}. Natural atoms in optical lattices can use two stable energy levels, such as the singlet ground state and the lowest hyperfine state~\cite{henriet_neutralatoms_2020}.
    Photons can be used as qubit in numerous ways, e.g., by using different polarizations (horizontal vs. vertical), different frequencies, or combinations thereof~\cite{moody_integratedphotonics_2022}.
    Superconducting qubits use energy levels of the non-linear resonant circuit~\cite{kjaergaard_superconducting_2020}.
    }
    \label{fig:qubits}
\end{figure}

What, then, would be a suitable platform for a qubit? 
The state of a single nuclear spin can be isolated and maintained for long times~\cite{zhong_optically_2015}. Plus, it certainly is a quantum mechanical system that exhibits superposition and entanglement.
However, this isolation also makes it hard to measure the spin's state~\cite{nielsen_chuang_2010}.
On the other end of the spectrum: a coin has two states, which are easily measured by looking at it. But, this macroscopic system cannot maintain superposition for a useful amount of time.
Indeed, qubit requirements present an inherent trade-off, as we desire the system to be both well-isolated, to store the quantum information, and controllable, to perform computation. 
The question for the wide range of proposed quantum computing platforms is therefore not whether they fulfill the DiVincenzo criteria, but \emph{how well} they do. We we list a selection of common platforms in Fig.~\ref{fig:qubits} and discuss the state of the art in Section~\ref{sec:stateofart}.

A qubit type we will be focusing on lot a during this thesis is the superconducting qubit, whose physical implementation is discussed in detail in Chapter~\ref{chap:hardware}.
This qubit is based on an anharmonic oscillator at sufficiently low temperatures and large enough dissipation such that the energy levels become quantized.
The lowest two energy levels $E_0$ and $E_1$ define a qubit and by driving the oscillator with an electrical pulse at frequency $\omega = |E_0 - E_1|$, we can coherently change between the states $\ket{0}$ and $\ket{1}$. To create a non-classical superposition state, the drive can be applied for a time that does not fully flip the state.
A strong advantage of superconducting qubits is their strong coupling to the external electrical field, which
allows for very fast gate operations and readout.
However, they have short lifetimes in comparison to other architectures like trapped ions systems---which, conversely, have slower operations.
Developing and executing quantum computations therefore requires careful considerations of which qubit type is used.

\section{State of the art}\label{sec:stateofart}

Due to the promises quantum computers hold in theory, tremendous effort has been undertaken to build such devices.
Since the first experimental demonstration of few-qubit quantum algorithms, academic institutions and computing-centric companies have proposed and realized increasingly larger quantum computers.
Among the largest digital quantum computers are, at time of writing this thesis, superconducting qubits systems that count 1121 qubits~\cite{ibm_condor_2023}.
Ideal, noise-free quantum computers of this size would, in theory, already be able to outperform classical computers on tasks such as quantum simulation~\cite{daley_practical_2022} or preparing molecular ground states~\cite{reiher_elucidating_2017}. 
Current devices, however, are subject to a wide range of noise sources which restrict the number of operations that can be applied in a circuit and reduce the solution quality.
In practice, demonstrations of practical quantum algorithms have therefore only reached limited problem sizes, which are still in reach of classical algorithms.

\subsection{Limitations}

While classical computers are able to protect the computation to virtually error-free levels~\cite{lin_error_2004}, building a scalable, fault-tolerant quantum computer (FTQC) able to correct for errors occurring to during the computation is more challenging due to the sensitivity of quantum information with respect to its environment. 
In addition, quantum information cannot simply be copied, therefore protection against errors cannot be achieved by performing operations on redundant copies of the quantum state.
Instead, quantum error correction uses an array of $n$ noisy, physical qubit to encode $m (\leq n)$ logical qubits and performs intermediate stabilizer measurements to detect if an error occurred.
First implementations of the so-called surface code have been demonstrated on superconducting hardware~\cite{krinner_ec_2022, acharya_surfacecode_2023} and natural atoms~\cite{bluvstein_logical_2023}, however, these do not yet preserve the qubit states for long enough to perform complex operations. 
This holds especially true, as a FTQC is only able to implement a specific set of discrete basis gates in a fault-tolerant fashion and compiling a target operation can result in a high gate count~\cite{dawson_sk_2005, ross_optimal_2016}.
Another disadvantage of the surface code is a low rate $r = m/n$ of logical to physical qubits, which scales as $n^{-1}$ 
as the code quality increases and, therefore, becomes inefficient as an increasing number of errors must be corrected~\cite{bravyi_ldpc_2023}.
Recently, focus therefore shifted to a different family of low-density parity check (LDPC) codes, which provide a constant rate $r$, but introduce non-local interactions~\cite{bravyi_ldpc_2023}.
This overhead could be mitigated by building code-specific hardware.

Until error-correcting codes can be reliably implemented, the complexity of quantum circuits that can be executed on hardware is, thus, limited.
To quantify the capability of a given device,
several holistic metrics have been introduced.
Quality measures include volumetric benchmarks~\cite{blume_volumetric_2020}, such as the Quantum Volume (QV)~\cite{jurcevic_qv_2021}, which measure the largest rectangular circuit of a certain type of operations that can be reliably executed, or the Layer Fidelity (LF)~\cite{mckay_lf_2023}, which characterizes the fidelity of layers of gate operations.
In addition to the quality and scale of a device, the execution speed plays a crucial role in the practicality of quantum algorithms. 
This is captured by the Circuit Layer Operations Per Second (CLOPS) metric~\cite{wack_clops_2021}, a quantum analog to the classical Floating-point Operations Per Second (FLOPS), which is a measure for the number of operations executed in a given time frame. 
Importantly, this includes overhead due to classical pre- and post-processing to provide a practical estimate of an algorithm runtime.

As each qubit platform has its individual advantages and disadvantages, the performance metrics can vary significantly in-between them.
Superconducting qubit systems, for example, scale to over 1000 qubits~\cite{ibm_condor_2023} and have fast gate times, leading to a large CLOPS. However the short qubit lifetime limits the number of applicable operations.
In contrast, a long qubit lifetime is an advantage of trapped ion systems, which can demonstrate large QV~\cite{moses_racetrack_2023}, but these devices are currently not as scalable.
To address the shortcomings of different qubit types, there are proposals for hybrid schemes that aim to combine the advantages of different architectures, e.g., using solid-state qubits for processing and phonons for state storage~\cite{hann_hybrid_2019} or photons for transmission~\cite{togan_hybrid_2010}.

Current error rates put quantum algorithms with provable advantage, such as Suzuki-Trotter, Shor or Quantum Phase Estimation, out of reach of noisy quantum processors.
For instance, implementing the dynamics of a practically relevant system for times that are challenging on classical computers with a Suzuki-Trotter expansion yields estimates of 200 qubits and $2.5 \cdot 10^5$ CX gates~\cite{daley_practical_2022} or 121 qubits and $10^7$ CX gates~\cite{miessen_performance_2021}. With CX error rates of 99.9\% achieved on superconducting qubits~\cite{kjaergaard_superconducting_2020} or ions~\cite{bruzewicz_trapped-ion_2019}, the fidelity of the final state is virtually 0.
As a result, research in recent years has focused on a variational algorithm paradigm~\cite{cerezo_variational_2021}, which is discussed in-depth in Chapter~\ref{ch:vqa}.
In a nutshell, this family of algorithms tunes free parameters in an ansatz quantum circuit to solve a given task.
The ability to select quantum circuits that operate within the capability of a given device, makes variational algorithms a promising candidate for noisy quantum computers.
As we will see in Chapters~\ref{chap:qnspsa} to~\ref{chap:dual}, however, these approaches introduce new bottlenecks that we aim at circumventing in this thesis.

\subsection{Demonstrations}

Though aforementioned metrics and device characteristics provide estimations for the problem dimensions a quantum computer could solve, only a practical implementation can truly show the capability of a device. 
Such a demonstration typically focuses on a problem that is particularly simple to implement on a given hardware but, ideally, is hard to solve classically.
In this context, "simple" refers to a problem that can be solved with shallow quantum circuits on the specific device at hand, e.g., by using native gates or qubit connections that reflect the device's topology.

For the simulation of quantum dynamics, analog quantum computers (discussed in Chapter~\ref{ch:basics}) go a step further and recreate the system's Hamiltonian in the lab. 
Using ultracold Rydberg atoms in an optical lattice, this approach currently allows the largest demonstrations of quantum dynamics for a transverse-field Ising model with 196 to 256 spins~\cite{scholl_simulation_2021, ebadi_256_2021}.
Analog quantum computers, however, are typically restricted to systems available in the lab and realizing general systems involves significant overhead.
Digital quantum computers are able to solve paradigmatic model systems up to 127 qubits by leveraging an array of error mitigation techniques~\cite{kim_utility_2023, farrell_schwinger_2023, shtanko_manybody_2023, rosenberg_dynamics_2023}.
Practically-relevant application models are typically restricted to smaller sizes. 
Ground-state calculations of the standard Fermi-Hubbard model, to the best of our knowledge, have been demonstrated\footnote{we only consider algorithms where the complete optimization/calculation has been performed on a quantum computer} for up to 8 spins (16 qubits) on a $2 \times 4$ lattice~\cite{stanisic_observing_2022}.

A related field concerned with classical Ising models, i.e., diagonal in the computational basis, is quantum optimization. 
Here, a quantum computer is not advantageous in representing the solution of the system---which, per definition, is a single bitstring---but rather in how the optimization is performed. Algorithms such as the annealing-based quantum approximation optimization algorithm (QAOA)~\cite{farhi_quantum_2014} or quantum imaginary time evolution~\cite{motta_determining_2020, mcardle_varqite_2019} may provide a better approach to finding solutions that classical heuristics.
Such problems can also be a useful test bed for quantum dynamics algorithms, since the Hamiltonian is closely related to a quantum Ising model but is classically verifiable to multiple 100s of variables (a.k.a. spins).
Demonstrations of QAOA range from 40-qubits on 3-regular graphs~\cite{sack_3r3_2023}, over 72 qubits in a Sherrington-Kirkpatrick model~\cite{dupont_gqaoa_2023}, to up to 434 qubits on hardware-native graphs~\cite{pelofske_quantum_2023, pelofkse_scaling_2023} on superconducting devices.

\section{Outline}

This thesis is written to be mostly self-contained for readers with a backgrounds in physics or computer science.
In Chapters~\ref{ch:basics}-\ref{ch:vqa}, we discuss the foundations of quantum computing and algorithms, with a particular focus on current, noisy quantum computers. Readers already familiar with the material may want to skip these chapters, or only skim over topics they are interested in.
The main contributions of this thesis are presented in Chapters~\ref{chap:qnspsa}-\ref{chap:dual}, which are based on Refs.~\cite{gacon_qnspsa_2021, gacon_saqite_2023, gacon_dual_2023}, respectively.
We summarize the main results in Chapter~\ref{ch:conclusion} and provide an outlook on the field of simulating quantum systems.
In the following we provide a brief overview of each Chapter:
\begin{itemize}
    \item In Chapter~\ref{ch:basics} we introduce the basic concepts of quantum computation. 
    This includes qubits, the fundamental unit of information in this paradigm, and how computations are described in form of a quantum circuit.
    These, together with measurements and the evaluation of expectation values, allow to formulate quantum algorithms.
    As example of a quantum algorithm, we again take up the problem of simulating quantum dynamics and formally introduce the Suzuki-Trotter decomposition for time evolution.
    The chapter is concluded with a first experiment on quantum hardware.
    \item Chapter~\ref{chap:hardware} complements the quantum computing theory by discussing practical considerations for realizing a quantum computing system.
    We focus on a superconducting architecture and explain the basic principles of how qubits, and operations thereon, are implemented.
    This understanding reveals the noise sources in today's hardware, and we demonstrate how these can be reduced through error mitigation techniques.
    Finally, every part of the computation is set into context by discussing how the quantum computing stack transforms the input problem to executable instructions and extracts the results.
    \item In Chapter~\ref{ch:vqa}, we introduce on the variational algorithm paradigm, ubiquitous approach for on noisy quantum computers.
    In this context, we discuss approaches for ground-state preparation and for simulating the real- and imaginary-time evolution of a quantum system.
    These algorithms rely on computing the quantum geometric tensor (QGT) and quantum circuit gradients, for which we review a wide spectrum of available techniques.
    \item Chapter~\ref{chap:qnspsa} (based on Ref.~\cite{gacon_qnspsa_2021}) focuses on reducing the computational cost of quantum natural gradient descent, which is a promising algorithm to find the ground state of a Hamiltonian.
    This information-geometric approach to optimization leverages the QGT, which becomes computationally expensive to evaluate on quantum hardware as the number of parameters in the variational circuit model increases.
    To circumvent this scaling issue, we propose a routine to compute unbiased samples of the QGT to construct a stochastic approximation at a constant cost.
    Since a large number of iterations is performed in the optimization, only a few samples per step are sufficient and we can significantly reduce the resource costs of the algorithm.
    We apply the resulting algorithm for ground-state preparations for applications in physics, optimization and machine learning.
    \item In Chapter~\ref{chap:saqite} (based on Ref.~\cite{gacon_saqite_2023}) we build upon the previously introduced stochastic approximation of the QGT and extend it for quantum time evolution. 
    The proposed improvements also allow to speed up the convergence for optimization problems in cases where the initial state is efficiently simulable classically, for example, as Clifford circuit.
    We apply the algorithm for a classical optimization problem and perform an experiment on quantum hardware to compute the imaginary-time evolution of a 27-qubit Ising model. 
    \item Chapter~\ref{chap:dual} (based on Ref.~\cite{gacon_dual_2023}) follows a different approach to variational time evolution. 
    Instead of sampling the QGT, we introduce a dual formulation that relies on evaluating the infidelity of the variational ansatz for two sets of parameters.
    This leads to an optimization problem that must be solved in each iteration. 
    Leveraging available information from previous timesteps, we show that the dual formulation promises an asymptotic speedup for real- and imaginary-time evolution.
    To strengthen this statement we derive bounds on the sample complexity of our algorithm and the standard variational approach.
    We apply the algorithm for imaginary- and real-time evolution of the Heisenberg model and to compute thermal averages using the quantum minimally entangled typical thermal states algorithm.
    \item This thesis concludes in Chapter~\ref{ch:conclusion}, where we review the novel results and reflect their potential to scale up algorithms on noisy quantum computers.
    Finally, we discuss further research directions and open questions related to the introduced methods.
\end{itemize}

%% file: main/ch1_basics.tex
\chapter{Quantum computing basics}\label{ch:basics}

\summary{In this chapter we review the basic principles of digital quantum computers and quantum simulation.
We introduce qubits, gates and measurements, and how they are composed to describe an algorithm in form of a quantum circuits. 
Then, we explain how physical systems are mapped to qubits, discuss product formulas for time evolution and apply them to simulate an Ising model on a superconducting quantum processor.
}

\noindent
Computers come in two flavors: analog and digital.
An analog computer performs a computation by continuously modeling the physical target process. 
A famous example is Michelson's harmonic analyzer~\cite{hammack_fourier_2014}, which computes the Fourier coefficients of a function by using wheels of different circumference.
A digital computer, on the other hand, abstracts away its internal mechanics and provides a set of discrete, logical operations.
While both can, in theory, perform universal computations analog computers are prone to the accumulations of errors due to their continuous state and typically less flexible than digital computers. 
Digital computers are also subject to errors, but due to their discrete states, these can be detected and corrected more easily.

The same concepts apply to quantum computers.
An analog quantum computer simulates a quantum system by physically realizing a target environment in the lab, whereas a digital quantum computer provides a set of basis instructions to perform universal quantum computation. 
A strength of analog quantum computers is that they have very little overhead for the simulation of systems they are modeling and currently outperform their digital counterpart for these systems~\cite{ebadi_256_2021}. 
Albeit analog platforms are universal~\cite{aharonov_adiabatic_2007, lloyd_qaoa_2018}, digital quantum computers are typically more flexible to program.
In addition, digital processors are compatible with quantum error correction, which would allow to control error rates as devices scale up~\cite{nielsen_chuang_2010}.
Since we are interested in performing a variety of tasks, including non-physical processes such as imaginary-time evolution, this thesis is focusing on digital quantum computers.
In the remainder of this work we are thus dropping the specifier "digital" and explicitly specify if a quantum computer is analog.

\section{Quantum computing}\label{sec:quantum_computing}

In the introduction we have discussed that a quantum computer holds the potential to efficiently implement certain operations that are challenging on a classical machine.
For example, to perform the time evolution of a quantum system a quantum computer could implement the unitary operation on an $n$-particle system, given by
\begin{equation}
    U(t) = e^{-iHt},
\end{equation}
where $H$ is the Hamiltonian of the system and $t$ the simulation time, using a number of operations scaling only polynomially in $n$---even though, for spin-$1/2$ particles, $U(t)$ can be thought of a unitary matrix in $\mathbb{C}^{2^n \times 2^n}$.
In this section, we discuss the building blocks required to realize such an operation and then demonstrate how it can be used to simulate quantum mechanical systems in Section~\ref{sec:simulating}.

The fundamental unit of information in a quantum computer is a quantum bit, or qubit, which we explain in detail in Section~\ref{sec:qubit}.
In addition, we require the ability to apply operations and read out information. 
The conditions for universal quantum computation are summarized in the DiVincenzo criteria~\cite{divincenzo_criteria_2000}, which require a quantum computer to have
\begin{enumerate}
    \item a scalable, well-defined qubit,
    \item the ability to initialize the qubits in a desired (possibly simple) state,
    \item gate times much shorter than the qubit lifetime,
    \item a universal set of gates, and
    \item the ability of measuring the qubits.
\end{enumerate}
In Chapter~\ref{chap:hardware} we will see how these criteria can be realized for superconducting qubits.
In the remainder of this section, we discuss the building blocks for defining programs on quantum computer, namely \emph{qubits}, \emph{gates} and \emph{measurements}. Together, they form a quantum circuit.
This is a central concept for quantum computing, as quantum algorithms attempt to outperform classical algorithm by making use of quantum circuits that are hard to simulate efficiently classically.

\subsection{Quantum bits}\label{sec:qubit}

A qubit is a quantum mechanical two-level system. 
The two levels are the two computational basis states $\ket{0}$ and $\ket{1}$ which span all possible qubit states. 
A classical bit also has access to binary states, but the fact that a qubit is a \emph{quantum mechanical} two level system means that its state is described by a superposition, given by
\begin{equation}
    \ket{q} = c_0 \ket{0} + c_1\ket{1},
\end{equation}
with $c_0, c_1 \in \mathbb{C}$ and $|c_0|^2 + |c_1|^2 = 1$, whereas a classical state would always be strictly in one of the two basis states.
The qubit state $\ket{q}$ lives in a 2-dimensional complex vector space, therefore we can express it as vector in $\mathbb{C}^2$ by choosing vector representations of the basis states.
Using 2-dimensional unit vectors as basis provides a natural representation, which allows to read off the coefficients from the vector representation.
We have
\begin{equation}\label{eq:unitvector}
    \ket{0} \equiv \begin{pmatrix}
        1 \\ 0
    \end{pmatrix},~~
    \ket{1} \equiv \begin{pmatrix}
        0 \\ 1
    \end{pmatrix},~~
    \ket{q} \equiv \begin{pmatrix}
        c_0 \\ c_1
    \end{pmatrix}.
\end{equation}
States of multiple qubits are described using the tensor product of single qubit states, as
\begin{equation}
    \ket{\psi} = \ket{q_n} \otimes \ket{q_{n-1}} \otimes \cdots \otimes \ket{q_1},
\end{equation}
where we typically drop the tensor product for brevity and write $\ket{\psi} = \ket{q_n \cdots q_1}$.
Since each qubit belong to a 2-dimensional vector space, $\ket{\psi}$ is a $2^n$-dimensional object which can be written as
\begin{equation}
    \ket{\psi} = \sum_{k=0}^{2^n - 1} c_k \ket{k}.  
\end{equation}
The vector representation of Eq.~\eqref{eq:unitvector} defines a vector for $\ket{k}$ by writing the integer $k$ in binary format with $n$ bits. We have
\begin{equation}
    \ket{k} = \bigotimes_{m=1}^n \ket{\text{bin}(k)_m},
\end{equation}
where $\text{bin}(k)_m$ is the $m$th bit in the binary representation of $k$. Via the Kronecker product~\cite{nicholson_kronecker_2001}, $\ket{k}$ can then be identified with the $k$th unit vector of dimension $2^n$, as shown in an explicit example in Appendix~\ref{app:tensoryoga}.

A more general description of quantum states, which we seldom need in this thesis but introduce for completeness, is given by density matrices.
The density matrix $\rho$ of a state $\ket{\psi}$ is defined as 
\begin{equation}
    \rho = \ket{\psi}\bra{\psi}.
\end{equation}
If a given $\rho$ can be written in such a way, it is called a pure state.
In addition, density matrices allow the description of so-called mixed states, which are a classical mixture of pure states, given by
\begin{equation}
    \rho = \sum_k p_k \ket{\psi_k}\bra{\psi_k},
\end{equation}
where $p_k \geq 0$ and $\sum_k p_k = 1$.
A density matrix can be associated with a positive semi-definite, Hermitian matrix in $\mathbb{C}^{2^n \times 2^n}$ and has a trace of 1.
We call a state maximally mixed if $\rho \equiv \mathbb{1}/2^n$, which is an important state that plays a role, e.g., in Gibbs state preparation in Section~\ref{sec:ite} or the trainability of variational quantum algorithms in Section~\ref{sec:training_vqa}.
The vector space quantum states belong to can be equipped with the Hilbert-Schmidt inner product,
\begin{equation}
    (\rho_1, \rho_2) \mapsto \tr(\rho_1^\dagger \rho_2) = \tr(\rho_1 \rho_2),
\end{equation}
to form a Hilbert space~\cite{nielsen_chuang_2010}. In the case of pure states, the inner product becomes 
\begin{equation}
    (\ket{\psi_1}, \ket{\psi_2}) \mapsto \tr\big(\ket{\psi_1}\bra{\psi_1} \ket{\psi_2}\bra{\psi_2}\big) = |\braket{\psi_1|\psi_2}|^2,
\end{equation}
which is called the fidelity between the states $\ket{\psi_1}$ and $\ket{\psi_2}$.

To measure properties of a quantum state we are frequently interested in evaluating expectation values of an observable $O$.
Observables are Hermitian operators, $O = O^\dagger$, which represent measurable quantities of a quantum system.
The expectation value of an observable its average value over each possible quantum state, weighted by the probability of the state to occur. It is given by
\begin{equation}
    \mathbb{E}_\rho[O] = \tr(O\rho),
\end{equation}
for mixed states, which simplifies to 
\begin{equation}
    \mathbb{E}_{\psi}[O] = \tr(O\ket{\psi}\bra{\psi}) = \braket{\psi|O|\psi},
\end{equation}
for pure states.
Note, that multiplying a pure state with a modulo 1 coefficient (i.e., of the form $\exp(i\varphi)$, $\varphi \in \mathbb{R}$) does not impact the expectation value. Such a factor is called a global phase and typically dropped from a quantum state as it cannot be measured.

The Pauli operators are a useful tool to understand the space of single-qubit states. 
They are defined as 
\begin{equation}
    \begin{aligned}
    X &= \ket{0}\bra{1} + \ket{1}\bra{0} \equiv \begin{pmatrix} 0 & 1 \\ 1 & 0 \end{pmatrix}, \\
    Y &= i\ket{1}\bra{0} - i\ket{0}\bra{1} \equiv \begin{pmatrix} 0 & -i \\ i & 0 \end{pmatrix}, \\
    Z &= \ket{0}\bra{0} - \ket{1}\bra{1} \equiv \begin{pmatrix} 1 & 0 \\ 0 & -1 \end{pmatrix},
    \end{aligned}
\end{equation}
and together with the identity matrix they form an orthonormal basis of the real vector space of Hermitian $2\times 2$ matrices\footnote{complex linear combinations of Pauli matrices form a basis for $\mathbb{C}^{2\times 2}$} and a Hilbert space with the Hilbert-Schmidt inner product.
Since a qubit state $\rho$ has trace 1, it can be fully specified by computing the projection onto the Pauli matrices. 
This defines a vector $\vec v \in \mathbb{R}^3$, as
\begin{equation}
    \vec v = \begin{pmatrix}
        \tr(X\rho) \\
        \tr(Y\rho) \\ 
        \tr(Z\rho)
    \end{pmatrix}.
\end{equation}
Since both $\rho$ and the Pauli matrices are Hermitian we can drop the complex conjugate in the Hilbert-Schmidt inner product and the expectation values are ensured to be real numbers.
For pure states $\vec v$ is normalized, while mixed states have $\|\vec v\|_2 < 1$.

\begin{figure}
    \centering
    \includegraphics[width=\textwidth]{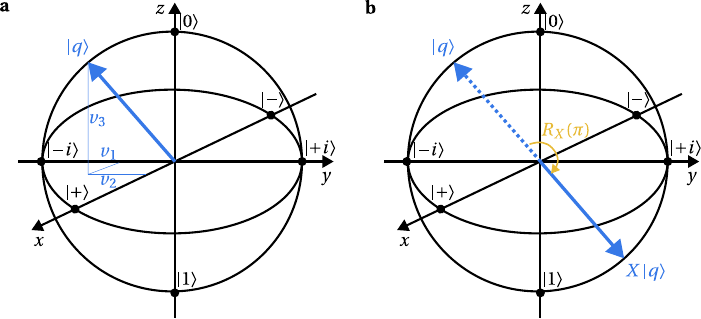}
    \caption[Qubit states on the Bloch sphere]{(a) A pure qubit state $\ket{q}$ on the Bloch sphere. Its position is determined by the vector $\vec{v} = (v_1~v_2~v_3)^\top$.
    (b) A bitflip operation is implemented as $\pi$-rotation around the $x$-axis.
    }
    \label{fig:bloch}
\end{figure}

In Fig.~\ref{fig:bloch}(a) we show a diagram of the possible single qubit states, which is also referred to as the Bloch sphere.
Instead of specifying $\ket{q}$ using the computational basis states $\ket{0}$ and $\ket{1}$, which are the eigenstates of the Pauli-$Z$ operator, any set of orthogonal states can be used. 
Frequently used basis sets include the eigenstates of the Pauli-$X$ operator
\begin{equation}
    \ket{+} = \frac{\ket{0} + \ket{1}}{\sqrt{2}},~~ \ket{-} = \frac{\ket{0} - \ket{1}}{\sqrt{2}},
\end{equation}
and of the Pauli-$Y$ operator
\begin{equation}
    \ket{+i} = \frac{\ket{0} + i\ket{1}}{\sqrt{2}},~~ \ket{-i} = \frac{\ket{0} - i\ket{1}}{\sqrt{2}}.
\end{equation}
These states are indicated on the Bloch sphere in Fig.~\ref{fig:bloch}(a).
Note that orthogonal basis states are antiparallel on the Bloch sphere.

\subsection{Quantum circuits}

\begin{figure}[th]
    \centering
    \includegraphics[width=0.6\textwidth]{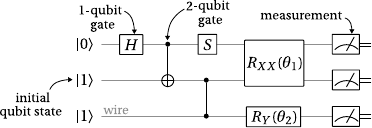}
    \caption[Description of a quantum circuit]{A quantum circuit. Qubits, represented as wires, start in an initial state and are acted on by gates. Gates can depend on parameters, such as the Pauli rotations $R_X$ and $R_{ZZ}$. To read out information, a qubit is measured and the doubled line represents a classical bit. See Appendix~\ref{app:gates} for a definition of each gate.}
    \label{fig:circuit_explained}
\end{figure}

In a quantum computation, we apply operations onto a quantum state to construct a final target state. 
Apart from measurements, which we will discuss in the following section, and resets, which reset any qubit state to $\ket{0}$, we are mostly concerned with operations $U$ that are unitary, i.e., $UU^\dagger = \mathbb{1}$.
These act on quantum states as
\begin{equation}
    \begin{aligned}
        \ket{\psi} &\mapsto U\ket{\psi}, \\
        \rho &\mapsto U\rho U^\dagger,
    \end{aligned}
\end{equation}
depending on how the state is represented.
A collection of qubits and operations acting thereon are organized in a quantum circuit, such as shown in Fig.~\ref{fig:circuit_explained}. 
Visually, each qubit is represented by a wire on which gates and measurements are applied from left to right.

Unitary operations can be written as exponentials of a Hermitian generator $G$, as $e^{-i\theta G}$ for some $\theta \in \mathbb{R}$.
An important type of such operations are Pauli rotations, defined as
\begin{equation}
    R_P(\theta) = e^{-i\theta/ 2 P},
\end{equation}
for $P \in \{X, Y, Z\}^{\otimes n}$ and with the convention of an additional factor 1/2.
A bitflip operation $\ket{0} \leftrightarrow \ket{1}$, for example, represented by the Pauli-$X$ operation can be implemented up to global phase, by
\begin{equation}
   R_X(\pi) = e^{-i\pi/2} X,
\end{equation}
which is visualized in Fig.~\ref{fig:bloch}(b).
A series of three Pauli rotations allows to implement a general single-qubit transformation, as  
\begin{equation}
    U(\phi, \theta, \lambda) = e^{i(\phi + \lambda)/2} R_Z\left(\phi - \frac{\pi}{2}\right) R_Y(\theta) R_Z\left(\lambda + \frac{\pi}{2}\right),
\end{equation}
though this decomposition is not unique.
The matrix representation of these gates (and the following ones) is shown in Appendix~\ref{app:gates}.

Another important family of gates is the $n$-qubit Clifford group $\mathcal{C}$, defined by the property that they map the group of $n$-qubit Pauli operators $\mathcal{P} = \{I, X, Y, Z\}^{\otimes n}$ onto itself. More formally, 
\begin{equation}
    \forall C \in \mathcal{C}~\forall P \in \mathcal{P}: CPC^\dagger \in \mathcal{P}.
\end{equation}
Frequently used elements of the single-qubit Clifford group in this thesis are the Hadamard gate $H$, defined by the action
\begin{equation}
    H: \begin{cases}
        \ket{0} \mapsto \ket{+}, \\
        \ket{1} \mapsto \ket{-},
    \end{cases}
\end{equation}
the phase gate $S = \sqrt{Z}$ and the $\sqrt{X}$ gate.
These can be used to transform between different Pauli basis states, e.g.,
\begin{equation}
    SH\ket{0} = S\ket{+} = \ket{+i}.
\end{equation}

In addition to single qubit gates, operations can act on multiple qubits simultaneously, such as the controlled-NOT (CX) gate, which is defined as 
\begin{equation}
    \begin{array}{c}
        \Qcircuit @R=2em @C=.7em {
            & \ctrl{1} & \qw \\
            & \targ & \qw
        }
    \end{array}:~~
    \text{CX} = \ket{0}\bra{0} \otimes I + \ket{1}\bra{1} \otimes X.
\end{equation}
In the circuit gate representation the state of the top qubit controls whether the $X$ gate is applied on the bottom qubit.
An interesting property of the CX gate is that it is also Clifford and, together with $H$ and $S$, generates any $n$-qubit Clifford group~\cite{gottesman_clifford_1998}.
Some two-qubit gates are symmetric, i.e., the action does not change if top and bottom qubit are exchanged, which is reflected in the circuit symbol. The controlled-$Z$ (CZ) gate, for example, is
\begin{equation}
    \begin{array}{c}
        \Qcircuit @R=2em @C=.7em {
            & \ctrl{1} & \qw \\
            & \control\qw & \qw
        }
    \end{array}:~~
    \text{CZ} = \ket{0}\bra{0} \otimes I + \ket{1}\bra{1} \otimes Z.
\end{equation}

A universal quantum computer requires a set of basis gates which allows to implement an arbitrary unitary operation.
For example, the general-single qubit rotation $U$ together with the CX gate is universal~\cite{williams_basisgates_2011}, or the commonly used set $\{R_Z(\theta), \sqrt{X}, X, \text{CX}\}$ on IBM's superconducting hardware~\cite{ibm_quantum}.
A basis can also consist of discrete basis gates, where no gate is parameterized, such as Clifford gates in combination with the $T$ gate.
Such a decomposition is, in general, only approximate and of importance mainly for fault-tolerant quantum computing~\cite{dawson_sk_2005, ross_optimal_2016}, not current quantum computers.
The process of decomposing the operations to the supported basis gates of a quantum computer is called compilation, which is discussed in detail in Section~\ref{sec:stack}.

\subsection{Measurements}

Retrieving information of a quantum state requires measurements. 
These can generally be defined using a set of positive operator-valued measurements (POVM) $\{(j, M_j)\}_j$ where the measurement is represented by a positive semi-definite matrix $M_j \in \mathbb{C}^{2^n \times 2^n}$ and associated with the outcome labeled as $j$. 
The likelihood of observing outcome $j$ on a state described by a density matrix $\rho \in \mathbb{C}^{2^n \times 2^n}$ is given by taking the expectation value of the measurement operator $p_j = \tr(M_j \rho)$.
The measurements further satisfy a completeness relation, defined by
\begin{equation}
    \sum_j M_j = \mathbb{1},
\end{equation}
The state after the measurement depends on the physical realization of the measurement process. 

\begin{figure}
    \centering
    \includegraphics[width=\textwidth]{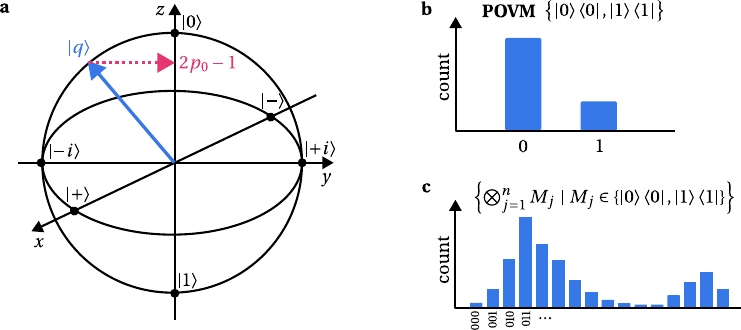}
    \caption[Qubit measurements]{(a) Projection of the qubit state $\ket{q}$ onto the $z$-axis, corresponding to the POVM $\{\ket{0}\bra{0}, \ket{1}\bra{1}\}$. The likelihood of observing the $\ket{0}$ state is given by $p_0$, which is determined by the location of the projection on the $z$-axis.
    (b) To estimate the likelihood of observing a state in an experiment, the measurement process is repeated to obtain sampling statistics.
    (c) A collection of $n$ single-qubit measurements has $2^n$ possible outcomes, corresponding to a POVM with tensored single-qubit projectors. Here, we show an example distribution for $n=3$.
    }
    \label{fig:bloch_collapse}
\end{figure}

A special class of POVMs we are focusing on are projective, single-qubit Pauli measurements. For a Pauli operator $P$ with eigenstates $\{\ket{\lambda_j}\}_{j=1,2}$, the POVM matrices are given by $M_j = \ket{\lambda_j}\bra{\lambda_j}$ and the post-measurement state is $\ket{\lambda_j}$.
For example, measuring a single qubit state $\ket{q}$ in Pauli-$Z$ basis, also called the computational basis, yields either
of the eigenvalues $\pm 1$ of the Pauli-$Z$ operator, according to the overlap with its eigenstates $\ket{0}$ or $\ket{1}$. Explicitly,
\begin{equation}
    \ket{q} \stackrel{\text{measure}}{\xrightarrow{\hspace{1cm}}} \begin{cases}
        \ket{0} \text{ (outcome $+1$), with probability } p_0 = |\braket{q|0}|^2, \\
        \ket{1} \text{ (outcome $-1$), with probability } p_1 = |\braket{q|1}|^2.
    \end{cases}
\end{equation}
This process has an intuitive visualization in the Bloch sphere, see Fig.~\ref{fig:bloch_collapse}(a).
Though for specific tasks general POVMs can be more efficient than projective Pauli measurements~\cite{garcia_measure_2021, fischer_povm_2022}, the latter are commonly supported natively by quantum hardware and sufficient for the purposes of this thesis.

Per measurement only a single outcome is recorded. If the measured qubit is in an eigenstate of a POVM operator, one measurement is sufficient to determine its state. Otherwise, the circuit with its measurement is repeated multiple times to obtain sampling statistics of the observed states as shown in Fig.~\ref{fig:bloch_collapse}(b-c). 
The probability $p_k$ of measuring state $\ket{k}$ is then estimated by 
\begin{equation}
    p_k \approx \hat p_k = \frac{\text{count}(k)}{N},
\end{equation}
where $N$ is the total number of times the circuit has been sampled.

\subsubsection{Expectation values}

With access to a sampling procedure of a quantum states, we can also evaluate expectation values of an observable, $\braket{\psi|O|\psi}$.
As quantum computing platforms typically only support measurements in a specific Pauli basis we first have to decompose the observable $O$ into a suitable representation before measuring it.
If we assume we are able to perform measurements in the Pauli-$Z$ basis, we have to write $O$ as a sum of terms that can be diagonalized in this basis.
Since the observable is Hermitian, it can be expressed in the basis of $n$-qubit Pauli operators,
\begin{equation}
    O = \sum_{j} \alpha_j P_j,
\end{equation}
for $\alpha_j \in \mathbb{R}$.
We then regard the expectation value per Pauli term,
\begin{equation}
    \braket{\psi|O|\psi} = \sum_{j} \alpha_j \braket{\psi|P_j|\psi}.
\end{equation}
Each of the Pauli operators can then be diagonalized using Clifford gates, as
\begin{equation}\label{eq:basistrafo}
    \begin{aligned}
        X &\rightarrow Z = H X H \\
        Y &\rightarrow Z = H S^\dagger Y S H,
    \end{aligned}
\end{equation}
where $I$ and $Z$ are diagonal in the $Z$-basis.
If $B \in \{I, H, H S^\dagger\}^{\otimes n}$ describes the basis transformation on all $n$ qubits, the Pauli $P$ is diagonalized as 
\begin{equation}
    B P B^\dagger = \Lambda \in \{I, Z\}^{\otimes n}.
\end{equation}
This allows to reformulate the expectation value of each Pauli as expectation value of a diagonal observable, i.e.,
\begin{equation}
    \begin{aligned}
        \braket{\psi|P|\psi} &= \braket{\psi|B^\dagger \Lambda B |\psi} \\
        &= \sum_{k, k'=0}^{2^n - 1} c^*_k c_{k'} \braket{k|\Lambda|k'} \\
        &= \sum_{k=0}^{2^n - 1} |c_{k}|^2 \lambda(k),
    \end{aligned}
\end{equation}
where we expanded the basis-transformed state in the computational basis, $B \ket{\psi} = \sum_k c_k \ket{k}$ and $\Lambda\ket{k} = \lambda(k)\ket{k}$.

The probabilities $p_k = |c_k|^2$ are obtained by sampling the state $B\ket{\psi}$ and 
the eigenvalues $\lambda(k)$ can be efficiently evaluated classically using the binary representation $\text{bin}(k)$ of $k$. 
We have
\begin{equation}
    \begin{aligned}
        \lambda(k) &= \braket{k|\Lambda_n \otimes \cdots \otimes \Lambda_1|k} \\
                     &= \prod_{m=1}^{n} \braket{\text{bin}(k)_m|\Lambda_m|\text{bin}(k)_m} \\
                     &= \prod_{m=1}^n \begin{cases}
                         (-1)^{\text{bin}(k)_m}, \text{ if } \Lambda_m = Z, \\
                         1, \text{ otherwise}.
                        \end{cases}
    \end{aligned}
\end{equation}
In this simple protocol, the state $\ket{\psi}$ must be measured with a new basis transformation $B$ for each Pauli term in the Hamiltonian. This can be drastically reduced by simultaneously diagonalizing commuting operators~\cite{bravyi_tapering_2017}.
However, in quantum chemistry applications Hamiltonians can consist of a large number of non-commuting Pauli terms, other POVMs than Pauli terms might offer more efficient fashion to compute expectation values~\cite{fischer_povm_2022}.

\section{Simulating quantum systems}\label{sec:simulating}

The study of a quantum system's time evolution provides insights into important quantum mechanical processes, including the behaviour of a system when it is driven out of its equilibrium state and how it behaves in the long-time limit.
For example, in a system accurately described by classical mechanics we expect the ergodicity axiom to hold. 
This states that an initial configuration entirely explores the available phase space and the long-term time average of a local observable matches both the microcanoncial and Gibbs ensembles, i.e., it thermalizes~\cite{polkovnikov_nonequilibrium_2011, deutsch_eth_2018}.
While there exist exceptions in classical systems, such as spin glasses~\cite{mezard_spin_1986}, quantum effects might further prevent ergodicity, e.g., by many-body localizations of the system~\cite{nandkishore_mbl_2015}. 

\begin{figure}[th]
    \centering
    \includegraphics[width=\textwidth]{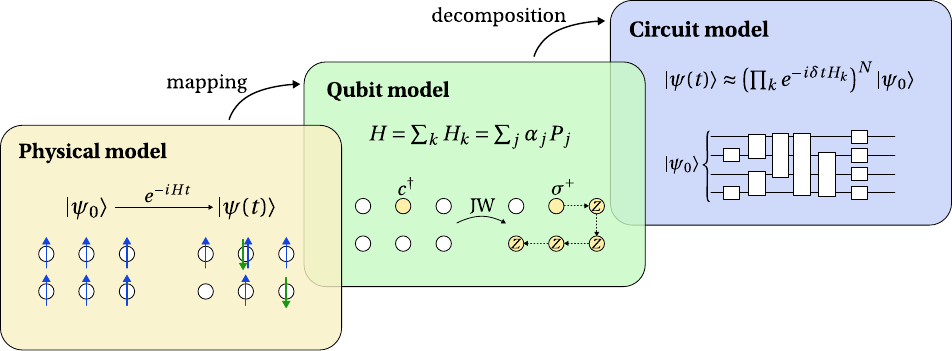}
    \caption[Mapping of a quantum system to a quantum computer]{Steps involved in the simulation of a physical system on a quantum computer. 
    The Hamiltonian is mapped to Pauli operators using, e.g., a Jordan-Wigner (JW)~\cite{jordan_wigner_1928} transformation, followed by a decomposition using a product formula to map the evolution onto a quantum circuit. With permission, this figure is based on Ref.~\cite{tacchino_simulators_2020}.}
    \label{fig:mapping}
\end{figure}

The dynamics of a quantum system are described by the Schrödinger equation,
\begin{equation}
    \frac{\partial}{\partial t}\ket{\psi(t)} = -iH\ket{\psi(t)},
\end{equation}
for a quantum state $\ket{\psi(t)}$, describing the  configuration of the system at time $t$, and the Hamiltonian $H$, which is a Hermitian operator describing the total energy of the system.
For a static Hamiltonian, $H \neq H(t)$, this equation is solved by
\begin{equation}
    \ket{\psi(t)} = U(t)\ket{\psi_0} = e^{-iHt}\ket{\psi_0},
\end{equation}
for an initial state $\ket{\psi_0}$ and evolution time $t$.
In this section, we review the implementation of the time evolution operator $U(t)$ on a quantum computer using product formulas.
The is performed in two steps, by
\begin{enumerate}
    \item mapping the physical model onto a qubit model and Pauli algebra,
    \item decomposing the Pauli Hamiltonian into circuit operations,
\end{enumerate}
as outlined in Fig.~\ref{fig:mapping}.

\subsection{Mapping quantum systems to qubits}\label{sec:mapping}

To simulate a quantum mechanical system on a quantum computer its state must be expressed using a set of qubits, say $n$, and the system Hamiltonian mapped to the in Pauli basis, as
\begin{equation}
    H = \sum_j \alpha_j P_j,
\end{equation}
where $P_j \in \{I, X, Y, Z\}^{\otimes n}, \alpha_j \in \mathbb{R}$.
Systems of spin-1/2 particles are natural candidates for such a mapping, as their spin algebra is already described by Pauli matrices and they have two basis states which can directly be associated with qubit states, i.e.,
\begin{equation}
    \begin{aligned}
    \ket{\uparrow} &\equiv \ket{0}, \\
    \ket{\downarrow} &\equiv \ket{1}.
    \end{aligned}
\end{equation}
For example, the transverse-field Ising model~\cite{thijssen_comphys_2007} on $n$ particles in both the spin-1/2 and qubit representation reads
\begin{equation}\label{eq:tfim}
    H = J \sum_{\braket{jk}} Z_j Z_k + h \sum_{j=1}^n X_j,
\end{equation}
where $J$ is the interaction strength, $h$ the field strength, $\braket{jk}$ sums over interacting particles $j$ and $k$ and the subscript on the Paulis denotes which particle is acted on, meaning that
\begin{equation}
    Z_j = \underbrace{I \otimes \cdots \otimes I}_{\times (j - 1)} \otimes Z \otimes \underbrace{I \otimes \cdots \otimes I}_{\times (n - j)}.
\end{equation}
Due to this natural mapping, we usually focus on spin-1/2 models in this thesis. However, for completeness we discuss how to map more general systems to qubits.

A variety of encodings exist for particles with spins $S\geq 1/2$, bosons and fermions. 
These schemes typically trade off the number of qubits required for the encoding with the number of Pauli terms and their locality. 
A single spin-$S$ particle can, for example, be represented using $\lceil\log_2(2S + 1)\rceil$~\cite{mathis_toward_2020} qubits, which we have also used for the $S=1/2$ case so far.
For $S>1/2$, a linear encoding using $2S+1$ qubits can be beneficial, as it generates less dense Pauli operators~\cite{mathis_toward_2020}.
Here,e formulate a qubit basis of the spin-$S$ space by mapping the eigenstates of the spin $S_Z$ operator, $\{\ket{m_z} : m_z \in \{-S, ..., S\}\}$, to a computational basis state given by
\begin{equation}\label{eq:spinz}
    \ket{m_z = S - k} \mapsto \ket{2^k} = \ket{\underbrace{0 \cdots 0}_{\times (2S -k)}1~\underbrace{0 \cdots 0}_{\times k}~},
\end{equation}
where the qubit at index $k+1$ is in the state $\ket{1}$.
This allows to write the $S_Z$ operator in qubit form as 
\begin{equation}
    S_Z = \sum_{m_z = -S}^S m_z \frac{(Z + I)_{S + m_z + 1}}{2} = \sum_{m_z=-S}^{S} \frac{m_z}{2} Z_{S + m_z + 1},
\end{equation}
where the identities cancel and it is easily verifiable that $S_Z\ket{m_z} = m_z \ket{m_z}$. 
For concrete forms of other spin operators we refer to Ref.~\cite{mathis_toward_2020}.
Bosons can be encoded in a similar fashion to spin-$S$ particles. Using a cutoff of $n_\text{max}$ bosons, the number of particles per site can be counted using $n_\text{max} + 1$ qubits as in Eq.~\eqref{eq:spinz} and the number operator takes on a form similar to the $S_Z$ operator~\cite{somma_simulation_2003}.

Fermions are less straightforward to encode, as the operators must obey the fermionic anti-commutation rules,
\begin{equation}
    \begin{aligned}
        &\{c^\dagger_{j}, c^\dagger_{j^\prime}\} = \{c_{j}, c_{j^\prime}\} = 0, \\
        &\{c_{j}, c^\dagger_{j^\prime}\} = \delta_{jj^\prime},
    \end{aligned}
\end{equation}
where $c_{j}$ is the annihilation operator on site $j$ and $c_j^\dagger$ the creation operator.
These are not intrinsically reflected by the corresponding qubit operators, which instead obey the commutation rules
\begin{equation}
    \begin{aligned}
        &[\sigma^+_{j}, \sigma^+_{j^\prime}] = [\sigma^-_{j}, \sigma^-_{j^\prime}] = 0 \\
        &[\sigma^-_{j}, \sigma^+_{j^\prime}] = \delta_{jj^\prime} Z_j,
    \end{aligned}
\end{equation}
where $\sigma^+ = \ket{1}\bra{0} = (X + iY) / 2$ and $\sigma^- = \ket{0}\bra{1} = (X - iY)/2$ create or annihilate a qubit excitation, respectively.
Numerous techniques exist to enforce the anticommutation rules on qubit operators, the most famous arguably being the Jordan-Wigner mapping~\cite{jordan_wigner_1928}, which has already been introduced in the late 1920s.
Here, each fermionic operator is mapped onto the corresponding qubit operator, plus a series of Pauli-$Z$ operations to ensure the anticommutativity. The mapping reads
\begin{equation}
    \begin{aligned}
        c_j &= \sigma^-_j \otimes \left( \bigotimes_{k=0}^{j-1} Z_k \right), \\
        c^\dagger_j &= \sigma^+_j \otimes \left( \bigotimes_{k=0}^{j-1} Z_k \right),
    \end{aligned}
\end{equation}
and requires picking a line (or ``snaking'') on the spin topology, as indicated in the central panel of Fig.~\ref{fig:mapping}.
This mapping requires $n$ qubits for $n$ spinless fermions. To include spin, the number of qubits can be increased such that each qubit excitation indicates whether a particle of spin $\sigma$ is present at site $j$.
For example, the Fermi-Hubbard model~\cite{thijssen_comphys_2007, qin_hubbard_2022} on two sites for spin-1/2 particles is 
\begin{equation}
    H = -J \sum_{\sigma\in\{\uparrow, \downarrow\}} \left(c^\dagger_{1,\sigma}c_{2,\sigma}  + c_{2,\sigma}^\dagger c_{1,\sigma}\right) + U \sum_{k=1}^{2} \left(\prod_{\sigma \in \{\uparrow, \downarrow\}} c^\dagger_{k,\sigma} c_{k, \sigma} \right),
\end{equation}
where $c^{(\dagger)}_{k, \sigma}$ annihilates (creates) a spin $\sigma$ at site $k$, $J$ is the hopping rate and $U$ describes the Coulomb interaction. 
Introducing a single index capturing both variables, we can map the Hamiltonian as 
\begin{equation}
    H = \frac{J}{2} \left(X_1 X_2 + Y_1 Y_2 + X_3 X_4 + Y_3 Y_4 \right) + \frac{U}{4} \left(Z_1 + Z_2 + Z_3 + Z_4 + Z_1 Z_4 + Z_2 Z_3\right),
\end{equation}
where qubit at index $j$ is $\ket{1}$ if a fermion with spin $\sigma$ is present at site $k$, i.e., the index mapping is $(k, \sigma) \mapsto j = k + 2 \delta_{\sigma \uparrow}$~\cite{tacchino_simulators_2020}.

The Jordan-Wigner mapping has the disadvantage that it may encode a local fermionic interactions into non-local qubit operations and a range of results attempts to decrease this effect~\cite{bravyi_tapering_2017, setia_bksf_2018, jiang_jkmn_2020, li_mapping_2022}.
However, since the Hamiltonian is mapped onto physical qubits with a specific connectivity determined by the quantum chip, interactions that are local on paper might eventually result in non-local operations.
Recent mapping algorithms therefore minimize the non-locality with the hardware connectivity in mind~\cite{miller_bonsai_2023}.

\subsection{Product formulas}\label{sec:trotter}

With a qubit-representation of the quantum system at hand, we now seek an efficient implementation of quantum time evolution on a quantum computer.
As outlined in the introduction, the time evolution of a quantum state under a Hamiltonian $H$ is determined by the Schrödinger equation, that is
\begin{equation}
    \frac{\partial}{\partial t}\ket{\psi(t)} = -i H\ket{\psi(t)},
\end{equation}
and for an initial state $\ket{\psi_0} = \ket{\psi(0)}$ the time-evolved state at time $t$ is given by 
\begin{equation}
    \ket{\psi(t)} = U(t)\ket{\psi_0} = e^{-iHt} \ket{\psi_0}.
\end{equation}
We now discuss how to find an efficient implementation of the time-evolution operator $U(t)$.

To find such a method, we assume the Hamiltonian can be written as finite sum of interactions, i.e.,
\begin{equation}
    H = \sum_{k=1}^K H_k.
\end{equation}
To decompose the exponential of this sum we leverage a product formula, such as the Suzuki-Trotter expansion~\cite{lie_theorie_1970, trotter_product_1959, suzuki_trotter_1976}. 
The first-order formula, also referred to as the Lie-Trotter expansion, for two bounded operators $A$ and $B$ reads
\begin{equation}
   e^{A + B} = \lim_{N \rightarrow \infty} \left(e^{A/N} e^{B/N}\right)^N.
\end{equation}
Clearly, if $A$ and $B$ commute it is already exact at $N=1$.
By applying the formula to the Hamiltonian for a finite $N$, we obtain the first-order approximation
\begin{equation}
    U(t) = \left(\prod_{k=1}^{K} e^{-i \delta t H_k}\right)^N + \mathcal{O}(\delta t^2),
\end{equation}
where $\delta t = t/N$ and we assume the norm of every commutator $[H_k, H_{k^\prime}]$ is bounded. 
This bound can be improved by, e.g., taking into account the values of the commutators, but for our purposes the scaling in $\delta t$ is sufficient to assess the convergence~\cite{berry_trotter_2007, low_trotter_2019, childs_trotter_2021}.
Higher, order-$p$, expansions allow to reduce the error to $\mathcal{O}(\delta t^p)$ by repeating the number of products for effectively smaller timesteps in a symmetrized manner~\cite{suzuki_trotter_1976}.
To increase accuracy, so-called multi-product formulas combine several expansions of the same order but at different timesteps $\delta t$~\cite{carreravazquez_mpf_2023}.
Alternatively, probabilistic versions of product formulas reduce the terms in the expansion by selecting Pauli terms according to their coefficient~\cite{campbell_qdrift_2019}.

\begin{figure}[th]
    \centering
    \includegraphics[width=\textwidth]{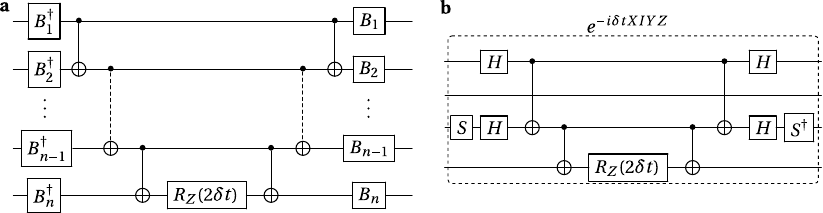}
    \caption[Decomposition of an arbitrary Pauli rotation]{(a) Decomposition of Pauli exponentials $e^{-i\delta t P}$ for Paulis $P \in \{X, Y, Z\}^{\otimes n}$. This operation can be implemented by a chain of CX gates on the qubits in combination with a basis transformation to the $Z$ basis, as shown in Eq.~\eqref{eq:basistrafo}. The dashed line in the CX gate denotes that the pairwise CX connections continue. Note that identities in the exponential are excluded.
    Other decompositions with different CX layouts are possible and can be leveraged to optimize the circuit for different hardware topologies.
    (b) An example decomposition for $e^{-i\delta t XIYZ}$.
    }
    \label{fig:pauliexp}
\end{figure}

We consider the expansion of $U(t)$ to be efficient if the number of qubit operations scales at most polynomially in system size $n$. 
This is achieved if the number of Hamiltonian terms $K$ does not scale exponentially and each term $\exp(-i\delta t H_k)$ must is implementable efficiently. 
One important case where both conditions are typically satisfied is in physical systems where particles can be modeled with local interactions, which implies a polynomial number of Hamiltonian terms.
An $n$-particle Ising model as in Eq.~\eqref{eq:tfim} on a square lattice, for example, has $K=3n - 2\sqrt{n}$ Pauli terms. 
Each Pauli exponential on $n>1$ qubits can then be implemented using $2n$ CX gates, as shown in Fig.~\ref{fig:pauliexp}. 
Also see the following subsection for a concrete example on the Suzuki-Trotter expansion of this model.

The Suzuki-Trotter expansion allows a rigorous error analysis of the decomposition. However, the operations required to be executed on the quantum computer are determined by the qubit Hamiltonian. Since for important physical systems, such as the Fermi-Hubbard model, the qubit interactions are non-local, realizing even a single Suzuki-Trotter step can become a challenge.
In Chapter~\ref{ch:vqa} we therefore follow a different approach to time evolution, based on projecting the time evolution to variational parameters in a fixed-structure circuit.

\subsection{Implementing the Ising model}

As an example for the Suzuki-Trotter formula, we consider the Ising model on a periodic chain in a tilted, transversal field, given by
\begin{equation}
    H = \underbrace{J\sum_{i=1}^{n} Z_i Z_{i+1~\text{mod}~n} + h_Z \sum_{i=1}^{n} Z_i}_{=H_{Z}} + \underbrace{h_X \sum_{i=1}^{n} X_i}_{=H_X},
\end{equation}
with $J=1$, $h_X = (\sqrt{5} + 5)/8$ and $h_Z = (\sqrt{5} + 1)/4$. 
These settings are chosen such that the system is robustly non-integrable~\cite{kim_ballistic_2013}.
The initial state is the ground state of the non-interacting model ($J=0$), which is the product state
\begin{equation}
    \ket{\psi_0} \propto \big((\sin(\theta) - 1) \ket{\uparrow} + \cos(\theta) \ket{\downarrow} \big)^{\otimes n},
\end{equation}
where $\theta = \tan^{-1}(h_Z/h_X)$ and the state is normalized.
Since we are dealing with spin-1/2 particles, this Hamiltonian is already in a suitable qubit formulation and we map the spin state to a qubit state by identifying $\ket{\uparrow} \equiv \ket{0}$ and $\ket{\downarrow} \equiv \ket{1}$.

\subsubsection{Decomposing the evolution operator}

To implement the evolution operator $U(t) = e^{-itH}$ on a quantum computer we are looking for a decomposition in exponentials of Pauli terms only, $e^{-itP}$ for $P \in \{I, X, Y, Z\}^{\otimes n}$, which we can implement efficiently using the circuit of Fig.~\ref{fig:pauliexp}.
We observe that the Hamiltonian can be split into two contributions, $H_Z$ and $H_X$, which both consist of commuting terms only.
Each of these can therefore be directly decomposed into Pauli evolutions without any error, e.g.,
\begin{equation}
    e^{-it H_X} = e^{-it h_X \sum_{j=1}^n X_j} = \prod_{j=1}^n e^{-it h_X X_j}.
\end{equation}
To find an approximation of the non-commuting terms $H_X$ and $H_Z$ we can leverage a Suzuki-Trotter expansion.

\begin{figure}[thbp]
    \centering
    \includegraphics[width=0.8\textwidth]{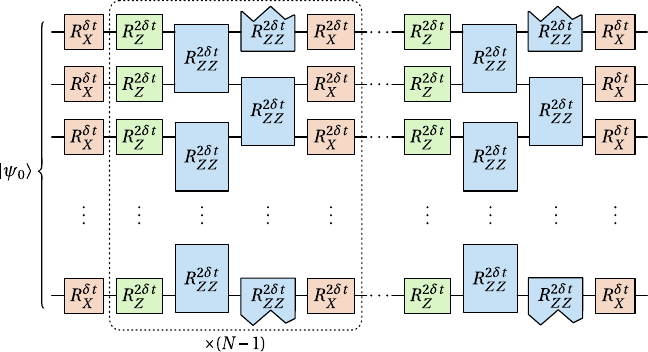}
    \caption[Suzuki-Trotter expansion for an Ising model]{The Suzuki-Trotter expansion of Eq.~\eqref{eq:trotter_example}. 
    For conciseness, the rotation angle is shown as superscript and the coefficients of the Ising model are not shown.
    For example, the red $R_X^{\alpha}$ gates represent a rotations of $R_X(h_X \alpha)$, whereas the green $R_Z$ rotations have an additional factor $h_Z$ and the $R_{ZZ}$ rotation a factor $J$.}
    \label{fig:trotter_example}
\end{figure}

The first-order expansion of the time-evolved state $\ket{\psi(t)}$ at time $t$ for $N$ timesteps would read
\begin{equation}
    \ket{\psi(t)} = \left(e^{-i\delta t H_X} e^{-i\delta t H_Z}\right)^N\ket{\psi_0} + \mathcal{O}(\delta t^2).
\end{equation}
However, we can also employ a second-order expansion, which, for two bounded, complex-valued matrices $A$ and $B$, is given by
\begin{equation}
    \begin{aligned}
        e^{A + B} &= \lim_{N\rightarrow\infty} \left(e^{B/(2N)} e^{A/N} e^{B/(2N)}\right)^N \\
                  &= \lim_{N\rightarrow\infty} e^{B/(2N)} \left(e^{A/N} e^{B/N}\right)^{N-1} e^{A/N} e^{B/(2N)}.
    \end{aligned}
\end{equation}
Applied to the time evolution operator we obtain
\begin{equation}\label{eq:trotter_example}
        e^{-itH} = e^{-\frac{i\delta t}{2} H_X} \left(e^{-i\delta t H_Z} e^{-i\delta t H_X}\right)^{N-1}
       e^{-i\delta t H_Z} e^{-\frac{i\delta t}{2} H_X} + \mathcal{O}(\delta t^3).
\end{equation}
Compared to the first-order expansion, the number of times the exponential of $H_Z$ is implemented is still $N$ but we have to implement an additional of the $H_X$ exponential.
However, this exponential is realized as layer of only single-qubit rotation gates which come at an negligible overhead compared to the two-qubit evolutions in $H_Z$. 
Hence, with the second-order expansion we gain an order of magnitude improvement of the error rate at at close to zero additional cost.
In Fig.~\ref{fig:trotter_example} we show the circuit decomposition of this expansion. 

\subsubsection{Implementation on quantum hardware}

To demonstrate the behavior of the error, we integrate the Hamiltonian up to time $T=2$ for different timesteps $\delta t$ and $n=12$ spins. The smaller the timestep, the smaller the error, but at the cost of more Trotter steps $N = \lceil T/\delta t\rceil$.
These additional Trotter steps do not pose an error if the circuit could be executed without noise, however, on real hardware, the additional operations increase the error.
We therefore expect a trade-off between circuit error and Trotter error, which leads to an optimal timestep $\delta t^*$.

We implement the circuits on \texttt{ibm\_auckland}, which is an IBM Quantum Falcon processor~\cite{ibm_quantum}, and provides a 12-qubit circle that fits the spin topology. 
First, the circuits are executed without any optimization or error mitigation.
Then, to show the potential of error mitigation, we use a pulse-efficient decomposition of the $R_{ZZ}$ evolutions for superconducting hardware~\cite{earnest_pulseefficient_2021} to reduce the length of the circuit pulse sequence, combined with Pauli twirling~\cite{carreravazquez_mpf_2023} over 8 randomly selected twirling sequences, see Section~\ref{sec:em} for more detail.
Each expectation value is averaged over $4000$ measurements and we show mean and standard deviation over 5 independent experiments.
To quantify the error in the evolution we track the value of the $X_0$ observable throughout the evolution.
Due to the translational invariance of the model it does not matter which spin we pick.

\begin{figure}[thbp]
    \centering
    \includegraphics[width=0.6\textwidth]{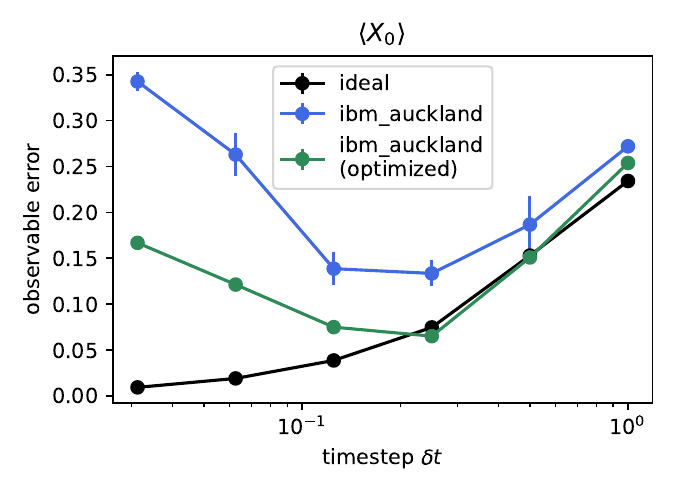}
    \caption[Hardware experiment results of a Suzuki-Trotter expansion]{Absolute errors in the observable $\braket{X_0}$ for a Suzuki-Trotter decomposition. We compare noise-free simulations without sampling noise with hardware results on $\texttt{ibm\_auckland}$.
    }
    \label{fig:trotter_errors}
\end{figure}

In Fig.~\ref{fig:trotter_errors} we show the results of the noise-free simulation and the hardware execution. We find that in the simulation the error decreases as $\delta t$ becomes smaller, but on hardware $\delta t^* \approx 0.3$ yields the smallest error. 
In addition we find that the optimized and error-mitigated circuits have a smaller error and standard deviation than the original circuits.
The fact that the circuit error increases for more timesteps $N$ is a significant limitation of the Suzuki-Trotter expansion on current, noisy devices.

\section{Conclusion}

In this chapter, we reviewed the fundamental principles of a digital quantum computer and defined central vocabulary for the rest of this thesis, such as qubits, quantum circuits, measurements and expectation values.
We have seen how physical systems, described in terms of spin-$S$ particles or fermions, can be represented on a quantum computer in terms of qubits and Pauli operators.
This allows to use product formulas, such as the Suzuki-Trotter expansion, to efficiently implement the time evolution operator of a locally interacting system in a quantum circuit.
Using this approach, we demonstrated the time evolution of an Ising model with a tilted external field on 12 spins on the superconducting qubit chip $\texttt{ibm\_auckland}$.

In the following chapter, we will dive into the details of how a specific quantum computer architecture, based on superconducting resonant circuits, is realized. 
This gives insights into the challenges of implementing quantum algorithms on current hardware and explains why current demonstrations of practical algorithms do not yet leverage the full size of quantum chips.
In Chapter~\ref{ch:vqa} we will then discuss a potentially more suitable family of algorithms for noisy quantum computers.

%% file: main/ch2_quantum_hardware.tex
\chapter{Realizing a quantum computer}\label{chap:hardware}

\summary{%
In this chapter we review the superconducting qubit architecture, which is used in the hardware experiments throughout this thesis.
We show how qubits and operations thereon are physically realized, discuss sources of noise and techniques to suppress and mitigate errors.
Finally, we show how quantum computers are accessed in practice and describe the quantum computing stack from the input problem to obtaining the final result.
}

\noindent
While a fault-tolerant quantum computer is hardware agnostic and allows to work with qubits as purely logical entities, the performance and limitations of near-term quantum computers are closely tied to the underlying physical implementation.
For example, the duration and fidelity of gate operations varies significantly between different qubit architectures, or even on the same quantum chip~\cite{nation_mapomatic_2023}.
Different quantum processors might also allow for hardware-specific optimization, which could allow to represent operations more efficiently than on other hardware~\cite{earnest_pulseefficient_2021}.
Scaling up demonstration therefore requires a detailed knowledge of the hardware and in this chapter we outline the fundamental principles of superconducting qubits.

\section{Superconducting qubits}\label{sec:sc}

\begin{figure}[t]
    \centering
    \includegraphics[width=1\textwidth]{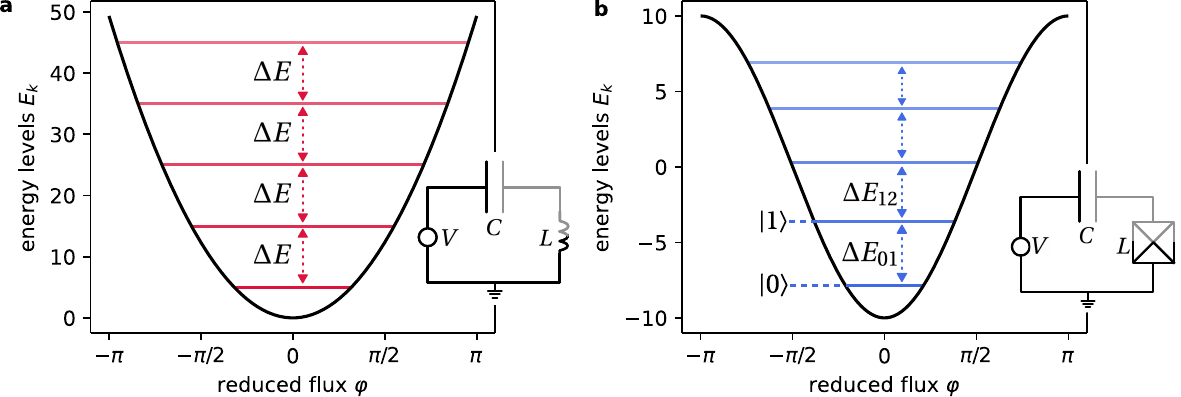}
    \caption[Harmonic and anharmonic oscillators with their energy levels]{Energy levels of harmonic and anharmonic resonant circuits with their potential energy as function of the reduced flux, see Appendix~\ref{app:sc}. The structure of the circuits is shown as inset, where the charge islands are marked grey, showing the voltage (V), capacitance (C) and inductance (L). 
    (a) Quantum harmonic oscillator with equidistant energy levels $\Delta E$.
    (b) Basic setup of a superconducting qubit using a Josephson junction as inductance. The resulting energy levels are not equidistant and suitable to implement a qubit.
    }
    \label{fig:qho}
\end{figure}

A qubit is a quantum mechanical system with two stable and uniquely addressable energy states, which can be identified with the computational states $\ket{0}$ and $\ket{1}$.
A possible candidate to realize such a two-level system is a quantum harmonic oscillator like, for example, a resonator circuit at sufficiently low temperature and dissipation such that the energy levels discretize~\cite{kjaergaard_superconducting_2020}.
Since the oscillator is harmonic the resulting energy level spacings $\Delta E$ are equidistant, as shown in Fig.~\ref{fig:qho}(a).
This implies that inducing a state transition of the oscillator at energy $\Delta E$ could transfer between any two energy levels. To implement a qubit, however, we require two levels with \emph{unique} transitions, in order to knowingly change between them.
To create a superconducting (SC) qubit with unique energy level spacings, we can create an anharmonic potential by replacing the linear inductance with a nonlinear inductance, such as a Josephson junction. 
We refer to Appendix~\ref{app:sc} to guide the reader from a harmonic to an anharmonic potential.
The resulting system is shown in Fig.~\ref{fig:qho}(b), where the lowest two energy levels are typically used as computational states $\ket{0}$ and $\ket{1}$.

The SC qubit setup in Fig.~\ref{fig:qho}(b) is referred to as a charge qubit since the energy levels correspond to the presence of electron pairs on the ``charge island'' in between the capacitance and inductance~\cite{nakamura_cooper_1999, lapierre_superconducting_2021}. 
From an experimental perspective, this is a very fundamental setup and there exist numerous variations that address shortcomings of this original architecture, like the transmon~\cite{koch_transmon_2007}, Xmon~\cite{barends_xmon_2013} or fluxonium~\cite{manucharyan_fluxonium_2009}.
These evolutions allowed to improve the qubit lifetime from the order of nanoseconds in the initial designs to more than milliseconds~\cite{kjaergaard_superconducting_2020}.

\begin{figure}
    \centering
    \includegraphics[width=0.52\textwidth]{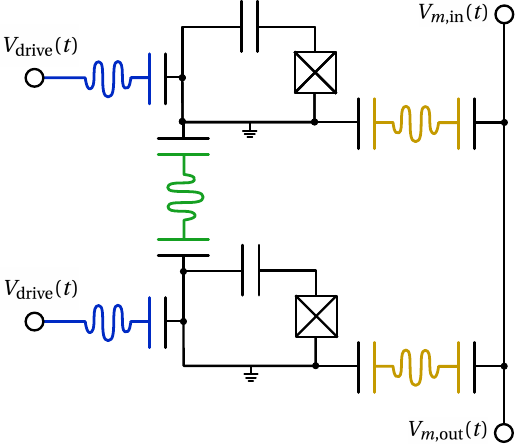}
    \caption[Basic setup of two superconducting charge qubits]{
    Basic setup of two charge qubits coupled by a resonator (green).
    The drive lines (blue) modulate the qubit states and are used for both single and two-qubit gates.
    For readout of the qubit states, each charge qubit is coupled to a readout line via a resonator (yellow).
    Each squiggly line represents a resonator cavity, i.e., a quantum harmonic oscillator with circuit diagram as in Fig.~\ref{fig:qho}(a).
    }
    \label{fig:sc_setup}
\end{figure}

With a qubit at hand, we now discuss how operations on the computational states $\ket{0}$ and $\ket{1}$ can be realized.
To implement any unitary operation on a set of $n$ qubits it suffices to have access to a universal basis gate set, such as arbitrary single qubit rotations $U(\phi, \theta, \lambda)$ and the CX gate~\cite{williams_basisgates_2011}.
These gates, as well as the readout of the state, can physically be realized by driving the qubit with microwave pulses sent via lines coupled to the qubit via a capacitance.
We show the general setup in Fig.~\ref{fig:sc_setup} for two charge qubits, each with its individual drive line used to implement single- and two-qubit gates, a coupling resonator and a readout line. In the following section we discuss each of these building blocks.

\subsubsection{Single qubit gates}

The action of a microwave pulse on a qubit at its resonance frequency $\omega_q = \Delta E_{01}$ (remember we are using units with $\hbar\equiv 1$) can be described by the effective Hamiltonian
\begin{equation}
    H_\text{drive}(t) = -\frac{\Omega_R s(t)}{2} \left(\cos(\phi) X + \sin(\phi) Y\right),
\end{equation}
where $\Omega_R$ is called the Rabi frequency, $s(t)$ is the envelope of the pulse, $\phi$ its phase, and $X$ and $Y$ are Pauli operators~\cite{kjaergaard_superconducting_2020}.
The envelope describes the amplitude of the pulse at each point in time and can have different shapes, such as a Gaussian or Gaussian square pulse shown in Appendix~\ref{app:compile_trotter}.
The Rabi frequency describes how strongly the qubit couples to the microwave drive and determines how fast the qubit state can be changed.
For SC qubits, as in natural atoms, this coupling strength is determined by the electric dipole moment.
A key difference to natural atoms, however, is that SC qubits are built in a lab and their dipole moment can be engineered to much larger values than occurring in nature~\cite{wallraff_strong_2004}.
This allows for very fast qubit operations and is one of the strengths of the SC architecture. 

\begin{figure}
    \centering
    \includegraphics[width=0.75\textwidth]{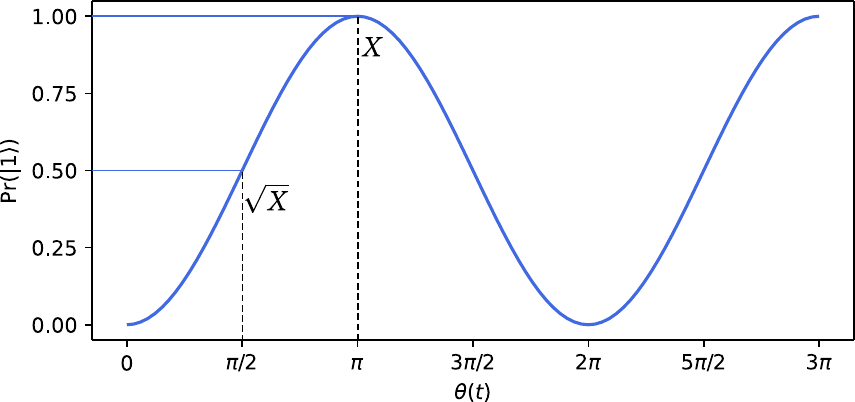}
    \caption[Rabi oscillations]{
    Rabi oscillations starting from the state $\ket{0}$ as function of rotation angle $\theta(t)$.
    After a $\pi/2$-pulse the likelihood of the qubit have transitioned to $\ket{1}$ is 50\%, and after a complete $\pi$-pulse we have applied the action of an $X$ gate.
    }
    \label{fig:rabi}
\end{figure}

Evolving the qubit under the drive Hamiltonian $H_\text{drive}$ allows to rotate the qubit around an axis in the $x-y$ plane determined by the phase $\phi$ about the angle
\begin{equation}
    \theta(t) = -\Omega_R \int_0^t s(t')\mathrm{d}t'.
\end{equation}
This includes the $R_X$ rotation for $\phi=0$ and $R_Y$ rotation for $\phi=\pi$.
The in-phase pulse ($\phi=0$) causes the qubit to oscillate between the states $\ket{0}$ and $\ket{1}$, which is known as Rabi oscillations and shown in Fig.~\ref{fig:rabi}. 
For $\theta(t) = \pi$, which is called a $\pi$-pulse, the operation acting on the qubit is given by 
\begin{equation}
    R_X(\theta(t)) = \exp\left(-i\int_0^t H^{(\phi=0)}_\text{drive}(t') \mathrm{d}t'\right) = e^{-i \theta(t) / 2 X} = e^{-i\pi /2} X,
\end{equation}
i.e., a $\pi$-pulse equals an $X$ gate up to a global phase.
By applying a shorter pulse, we can implement fractional $X$ gates and construct superpositions of the qubit states. 
For example, a $\pi/2$-pulse implements a $\sqrt{X}$ gate, since 
\begin{equation}
    e^{-i\pi/4 X} = R_X\left(\frac{\pi}{2}\right) = e^{-i\pi/4} \sqrt{X}.
\end{equation}
This gate can be used to create an equal superposition state,
\begin{equation}
    \sqrt{X} \ket{0} = \frac{(1 + i)\ket{0} + (1 - i)\ket{1}}{2},
\end{equation}
and is equivalent to the Hadamard up to $R_Z$ rotations. 
That a $\pi/2$-pulse creates an equal superposition state can be intuitively understood: a $\pi$-pulse performs a complete bitflip, but after a $\pi/2$-pulse the qubit only has a 50\% likelihood for having changed the state. This is also schematically shown in Fig.~\ref{fig:rabi}.

While rotating around arbitrary axes in the $x-y$ plane is already sufficient to implement any single-qubit unitary~\cite{nielsen_chuang_2010}, we also discuss how to implement a rotation around the $z$-axis.
This $R_Z$ operation can be practical, since it can be implemented without sending a physical pulse to the qubit and can be convenient to define a basis gate set with minimal gate calibrations.
Being a diagonal gate in the computational basis, the $R_Z$ gate only affects the relative phases of the qubit basis states. 
This leads to the special property that its effect can only be measured if it is used in combination with other gates.
We can leverage this property to implicitly implement the $R_Z$ gate by shifting phases of the subsequent pulses---which is also known as a virtual gate~\cite{mckay_virtualz_2017}.
This can be illustrated in a simple example where we apply an $R_Z$ rotation followed by an $R_X$ gate, which can be written as
\begin{equation}
    \begin{aligned}
    R_X(\theta) R_Z(\phi) &\equiv R_Z(-\phi) R_X(\theta) R_Z(\phi) \\
    &= e^{i\frac{\phi}{2} Z} e^{-i\frac{\theta}{2} X} e^{-i\frac{\phi}{2} Z} \\
    &= e^{-i\frac{\theta}{2} \left(\cos(\phi) X + \sin(\phi) Y\right)} \\
    &= R_{\phi}(\theta),
    \end{aligned}
\end{equation}
where on the first line we inserted an additional $R_Z(-\phi)$ gate which does not affect the final measurement and $R_{\phi}(\theta)$ describes a rotation on the Bloch sphere around the axis $(\cos(\phi),~\sin(\phi),~0)^\top$ about the angle $\theta$.
More generally, if we implement an $R_Z(\phi)$ rotation in between a set of pulses with rotation angles $\theta_j$ around axes defined by the phases $\phi_j$, the resulting operation can be decomposed as
\begin{equation}\label{eq:virtualz}
    \begin{aligned}
    &R_{\phi_d}(\theta_d) \cdots R_{\phi_j}(\theta_j) R_Z(\phi) R_{\phi_{j-1}}(\theta_{j-1}) \cdots R_{\phi_1}(\theta_1) \\
    &= 
    R_Z(-\phi) R_{\phi_d + \phi}(\theta_d) \cdots R_{\phi_j + \phi}(\theta_j) R_{\phi_{j-1}}(\theta_{j-1}) \cdots R_{\phi_1}(\theta_1).
    \end{aligned}
\end{equation}
See Appendix~\ref{app:virtualz} for the derivation.

The $\sqrt{X}$ and virtual $R_Z$ gate together allow to implement arbitrary single qubit unitaries, since a general unitary can be written as~\cite{mckay_virtualz_2017}
\begin{equation}
    U(\theta, \phi, \lambda) = 
    R_Z\left(\phi - \frac{\pi}{2}\right) \sqrt{X} R_Z(\pi - \theta) \sqrt{X} R_Z\left(\lambda - \frac{\pi}{2}\right).
\end{equation}
These two gates form a convenient basis gate set as it only requires calibrating the $\pi/2$-pulse on the device, which comes at much smaller experimental overhead compared to providing reliable pulse shapes for any angle.
Indeed, $\{\sqrt{X}, X, R_Z\}$ is a commonly used basis for single qubit gates on SC qubits currently, for example, provided by IBM Quantum~\cite{ibm_quantum}.

\subsubsection{Two qubit gates}

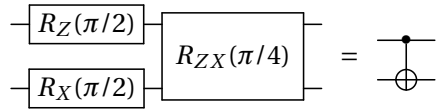
\begin{figure}[t]
    \centering
    \[
    \begin{array}{c}
        \Qcircuit @R=1em @C=.7em {
            & \gate{R_Z(\pi/2)} & \multigate{1}{R_{ZX}(\pi/4)} & \qw \\
            & \gate{R_X(\pi/2)} & \ghost{R_{ZX}(\pi/4)} & \qw
        }
    \end{array}
        =
    \begin{array}{c}
        \Qcircuit @R=1em @C=.7em {
            & \ctrl{1} & \qw \\
            & \targ & \qw 
        }
    \end{array}
    \]
    \caption[Decomposition of an $R_{ZX}$ rotation]{The CX gate can be decomposed as an $R_{ZX}(\pi/4)$ rotation and single qubit gates, up to a global phase.}
    \label{fig:cx_as_cr}
\end{figure}

For a universal basis gate set we now lack a two-qubit gate.
Depending on the specific qubit implementation different types of two-qubit gates can be realized~\cite{kjaergaard_superconducting_2020}.
Here, we consider the implementation of a CX gate by using a cross-resonance (CR) gate.
A CR gate applies a microwave pulse to the control qubit, but with the resonance frequency of the target qubit and a phase $\phi = \pi$. 
For a detailed discussion of this interaction, we refer to Refs.~\cite{malekakhlagh_cr_2020, magesan_cr_2020}, as for our purposes the simplified, effective Hamiltonian suffices. 
It is given by
\begin{equation}
    H_\text{CR} = \frac{\omega_{IX}}{2} I \otimes X + \frac{\omega_{ZI}}{2} Z \otimes I + \frac{\omega_{IZ}}{2} I \otimes Z + \frac{\omega_{ZZ}}{2} Z \otimes Z + \frac{\omega_{ZX}}{2} Z \otimes X
\end{equation}
which acts on control and target qubits as $\exp(-it H_\text{CR}) \ket{q_\text{control}} \otimes \ket{q_\text{target}}$~\cite{malekakhlagh_cr_2020, magesan_cr_2020}, where $t$ is the time of interaction.
The coefficients $\omega$ depend, among other factors, on the coupling strength of the two qubits and the difference in resonance frequencies (called detuning).
To synthesize a CX gate from the CR interaction, we would like to isolate only the ZX interaction, since it can be used to create a CX as shown in Fig.~\ref{fig:cx_as_cr}.

One approach to suppress the undesired terms, in particular the ZZ interactions which cannot be countered with single qubit gates, is to tune the coefficients $\omega$ of undesired terms to be close to zero. 
Another approach is to echo the CR gate with a second CR pulse with flipped amplitude and single-qubit rotations in between. This allows to cancel the ZZ interaction at the expense of promoting other single qubit errors. The effective Hamiltonian of the echoed CR (ECR) gate is given by~\cite{malekakhlagh_cr_2020}
\begin{equation}
    H_\text{ECR} = \frac{\omega_{II}}{2} I \otimes I + \frac{\omega_{IY}}{2} I \otimes Y + \frac{\omega_{IZ}}{2} I \otimes Z + \frac{\omega_{ZX}}{2} Z \otimes X.
\end{equation}

\begin{figure}[t]
    \centering
    \includegraphics[width=0.6\textwidth]{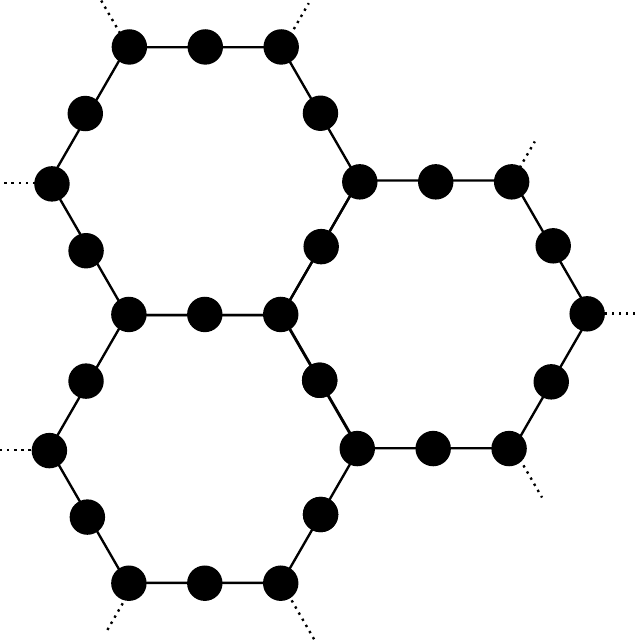}
    \caption[Heavy-hex qubit topology]{The heavy-hex structure used in IBM Quantum devices. Each filled, black circle represents a qubit and two-qubit gates are enabled between connected circles.}
    \label{fig:heavyhex}
\end{figure}

The main sources of errors for this two-qubit interaction are undesired terms in the CR and ECR Hamiltonians and resonances of the microwave pulse with other qubit transition frequencies~\cite{malekakhlagh_cr_2020}. 
To avoid such ``cross-talk'' it is crucial to tune the qubits to have distinct transition frequencies, throughout as many energy levels as possible.
This frequency splitting problem becomes more problematic as the number of qubits on a single chip increases and they become more interconnected.
In frequency-tunable architectures, where the resonance frequency of qubits can be tuned via an external magnetic field, this problem can, to some extent, be mitigated~\cite{chavez_fluxtunable_2022}. 
Another option is building chips with sparse qubit connectivity~\cite{hertzberg_sctopology_2021}, such as IBM's heavy hex topology~\cite{ibm_heavyhex}, shown in Fig.~\ref{fig:heavyhex}.
A drawback of sparse connectivities, however, is that implementing operations between qubits that are not natively connected requires additional Swap gates, which increases the overall gate count and circuit error. This is also discussed further in Section~\ref{sec:stack}.

\subsubsection{Qubit readout}

Beyond the ability to implement single- and two-qubit gates, a quantum computation requires reliable and, ideally, fast measurements of the qubit states. 
It is often desirable to implement a qubit readout as quantum non-demolition (QND) measurement, where the qubit state is projected to the computational basis but not destroyed and the qubits can further used in the collapsed states~\cite{magnard_reset_2018}.

\begin{figure}
    \centering
    \includegraphics[width=\textwidth]{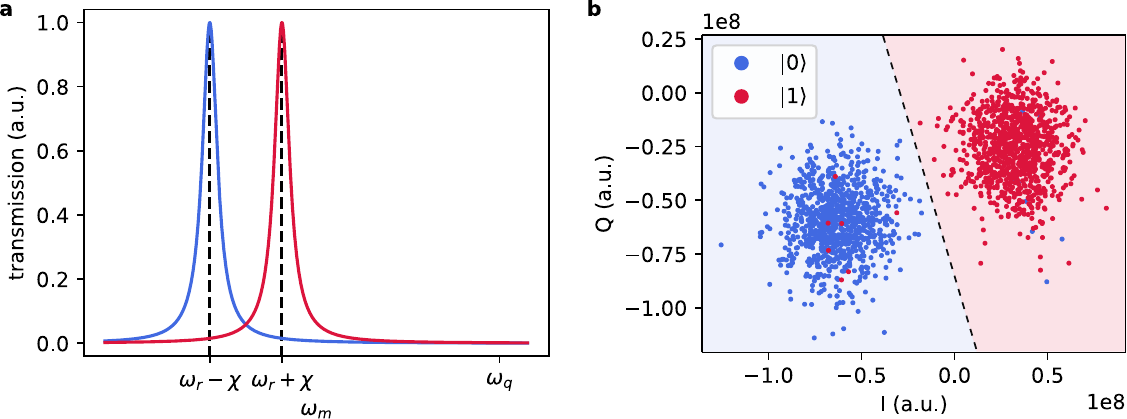}
    \caption[IQ measurement of a qubit state]{
    (a) Transmission as a function of probing frequency $\omega_m$. The qubit frequency is far detuned from the resonator frequency $\omega_r$, which is shifted by $\chi$. The width of the Lorenzians is given by the resonator linewidth.
    (b) ($I$, $Q$) data for prepared states $\ket{0}$ and $\ket{1}$ with a linear separator. New measurements in the blue zone are classified as $\ket{0}$ and in the red zone as $\ket{1}$. The plot shows 1000 single shot readouts on \texttt{ibm\_canberra}~\cite{ibm_quantum}, analyzed using Qiskit Experiments~\cite{kanazawa_qiskitexperiments_2023}.}
    \label{fig:readout}
\end{figure}

Superconducting qubits can implement QND measurements by dispersively coupling to a resonator circuit,
i.e., the detuning between the qubit and resonator frequencies $\Delta = \omega_q - \omega_r$ is much larger than their coupling $g$ and they are only weakly entangled.
In this dispersive coupling limit with $|g / \Delta| \ll 1$, the resonance frequency of the readout resonator is shifted depending on the state of the qubit~\cite{blais_cqed_2021}, as 
\begin{equation}
    \tilde\omega_r = 
    \begin{cases}
        \omega_r + \chi, \text{ if } \ket{q} = \ket{0}, \\
        \omega_r - \chi, \text{ if } \ket{q} = \ket{1}, \\
    \end{cases}
\end{equation}
where $\chi = g^2/\Delta$ and $\ket{q}$ denotes the qubit state. This is schematically shown in Fig.~\ref{fig:readout}(a).

To probe the resonance frequency of the readout resonator we send a microwave pulse through the readout line,
where it will be transmitted or reflected.
After interaction with the resonator, the envelope of the signal is given by
\begin{equation}
    s(t) = A_m \cos(\omega_m t + \phi_m),
\end{equation}
where $A_m$, $\omega_m$ and $\phi_m$ describe the amplitude, frequency and phase of the measurement signal, respectively.
To determine the quantum state from this measurement, we write the signal in a complex-valued representation as
\begin{equation}
   s(t) = \mathrm{Re}\big((I + iQ)e^{i\omega_m t}\big),
\end{equation}
where $I = A_m\cos(\phi_m)$ and $Q = A_m\sin(\phi_m)$~\cite{kjaergaard_superconducting_2020}, plot the $(I, Q)$ values on a plane and perform a classification, as shown in 
Fig.~\ref{fig:readout}(b). To ensure a reliable labelling of the qubit states they should be maximally separated, which can be achieved by tuning the measurement frequency $\omega_m$~\cite{kjaergaard_superconducting_2020}.

For a high readout fidelity, fast probing of the resonator is crucial to ensure the qubit does not decay during the process.
This is particularly important since coupling the qubit to the readout resonator can decrease the coherence time, which is known as the Purcell effect~\cite{houck_purcell_2008}. This issue can be mitigated by an intermediate Purcell filter in between the qubit and the resonator to suppress transitions at the qubit frequency $\omega_q$~\cite{reed_purcellfilter_2010, bronn_purcellfilter_2015}.
In some architectures the readout line connects to readout resonators of multiple qubits which allows for joint readout or the states~\cite{filipp_2qmeas_2009}.

\subsubsection{Comparison to other qubit architectures}

The fabrication of quantum dot spin qubits and SC qubits is building upon the highly-developed semiconductor industry, potentially allowing to scale up to large devices quickly.
Indeed, SC qubit architectures are among the largest, currently available devices with up to 1121 qubits on a single chip~\cite{ibm_condor_2023}. Spin qubits, however, are in practice more challenging to fabricate due to their significantly smaller size of $\sim 100$nm~\cite{burkard_spinqubits_2023}.
Quantum computers based on trapped ions are reaching sizes up to 30 operational qubits~\cite{chen_ionq_2023} and 49 for natural atoms~\cite{chen_ionq_2023}. 
Natural atoms have also been used as analog quantum computers for experiments of up to 256 spins~\cite{ebadi_256_2021} with up to 324 being reported as trapped in optical tweezers~\cite{schymik_300atoms_2022}.

The gate fidelities of all aforementioned platforms can reach similar values. Single-qubit gate fidelities are reliably around 99.9\%, two-qubit fidelities are typically averaged around 99\%, as are measurement errors~\cite{ibm_quantum, chen_ionq_2023, graham_atoms_2022, mills_spin2q_2022, mills_spinmeas_2022}.
For two-qubit and readout errors, specifically optimized, small devices have been able to demonstrate fidelities up to 99.9\%~\cite{gaebler_ions_2016, ding_sc_2023}, but larger devices suffer from more error sources.
The two-qubit error is the bottleneck of each architecture since usually a large number of two-qubit gates are applied in circuits.
An important difference concerning two-qubit gates across architectures is the available qubit connectivity. 
While spin and SC qubits typically have a sparse, fixed connectivity, trapped ions support interactions between arbitrary qubit pairs and natural atoms are able to change the connectivity by physically moving the atoms in the optical lattice.

Devices also differ on the time scales they operate on.
The coherence time describes how long a qubit remains in a prepared basis states. It is measured in the typical decay time $T_1$, which describes the probability of remaining at the prepared state after a time $t$ has passed as $1 - e^{-t/T_1}$.
SC qubits have $T_1$ times on the order of up to milliseconds~\cite{ding_sc_2023}, whereas spin states can be preserved about 10 times longer~\cite{mills_spinmeas_2022}.
This stands in contrast to natural atoms and trapped ions, which reliably exhibit coherence times of a few seconds~\cite{gaebler_ions_2016, graham_atoms_2022} or even minutes~\cite{chen_ionq_2023}.
However, the duration of gates and readout are can be two to three magnitudes larger than on semiconductor-based devices~\cite{ding_sc_2023, gaebler_ions_2016, graham_atoms_2022}, reducing the number of operations that can be applied at a high fidelity.

As already discussed in the introduction, every quantum computing architecture has individual trade-offs.
Selecting the optimal qubit platform to run a particular application, thus, depends on many factors, such as the required number of qubits, the circuit depth, the connectivity, and the total number of circuits, to name a few. 
In the following chapters we will see that this thesis is mainly concerned with a large number of circuits in an iterative workflow.
SC qubits are, thus, a suitable candidate, as they have fast single- and two-qubit gates with high fidelities.
The execution speed is also beneficial for error mitigation techniques, which we discuss in the following section.
One drawback of this choice is that the connectivity of SC devices is often limited in order to reduce cross-talk, which requires care in selecting a problem that can be mapped efficiently to the device.
An additional advantage from a more pragmatic perspective is that SC hardware is currently among most readily available architectures.
For IBM Quantum devices, which we use in this thesis, typical operation fidelities are over $99.9\%$ single-qubit gate fidelities, over $99\%$ two-qubit gate fidelities and approximately $99\%$ measurement accuracy. 
The CX gate duration is on the order of $300$ns and qubit readout varies between $700$ns and 5$\mu$s~\cite{ibm_quantum}. Qubits are arranged in the sparse heavy-hex structure, shown in Fig.~\ref{fig:heavyhex}, and devices with up to 1121 qubits~\cite{ibm_condor_2023} exist. 

\section{Error suppression \& mitigation}\label{sec:em}

In an ideal scenario, we would have access to a fault-tolerant quantum computer (FTQC) with the capability to correct any errors that occur during the circuit execution.
This would allow us to employ canonical algorithms with deep circuits, as for example required for product formulas for time evolution or quantum phase estimation for ground state preparation, which have established error bounds.
However, FTQCs are built upon quantum error correcting codes that necessitate a large number of qubits and low physical gate errors to a degree that is typically not achieved by current devices~\cite{wang_surfacecode_2010, krinner_ec_2022, bravyi_ldpc_2023}.
Therefore, instead of fully correcting errors, recent research is focusing on reducing errors with pre- and post-processing of the circuits~\cite{ezzell_dd_2023, nation_m3_2021, bravyi_mitigating_2021, berg_trex_2022, kim_scalable_2023, berg_pec_2023, kim_utility_2023, giurgica_zne_2020, czarnik_cdr_2021, berg_checks_2023, mcclean_decoding_2020}.

We differentiate into two types of schemes: error suppression (ES) modifies the circuit to counter the accumulation of errors during the circuit execution, while error mitigation (EM) attempts to reduce the occurred errors by combining the outcomes of circuit ensembles.
While certain EM schemes are, in principle, able to fully remove the effect of errors, this comes at an exponential classical cost~\cite{berg_pec_2023, quek_exponential_2023, takagi_limits_2022}.
Nevertheless, ES and EM schemes allow to extend the reach of near-term quantum computers and enable demonstrations on the order of 100 qubits~\cite{kim_utility_2023}. 
While these experiments do not present a quantum advantage~\cite{begusic_classical_2023, tindall_classical_2023}, but show that noisy quantum computers are able to access regimes that are not trivially simulable.
In the following sections, we review common ES and EM techniques for noisy quantum computers, focusing mainly on methods we apply in the hardware experiments in this thesis.

\subsection{Dynamical decoupling}

Since qubits are not perfectly isolated, they accumulate errors from interactions with their environment even while they are idle and not used in a computation.
A well-known error source, for example, is dephasing, where a qubit is prepared in a superposition $\ket{0} + e^{i\alpha}\ket{1}$ but the phase information $\alpha$ is lost over time.
Dynamical decoupling (DD) is a technique to reduce such an accumulation of errors during idle times of the qubits by decoupling the qubit from its environment.
This method was among the first proposed ES techniques for quantum computing systems and has since been demonstrated to extend the decoherence times of qubits on a variety of platforms~\cite{biercuk_dddemo_2009, alvarez_dddemo_2010, wang_dddemo_2012, pokharel_dddemo_2018}.
A key advantage of DD is that it decouples the qubit by applying additional operations only during idle times of the qubits and can be used without increasing the circuit depth.

From a more quantitative point of view, consider a qubit interacting with an environment described by the following Hamiltonian,
\begin{equation}
    H = H_Q + H_E + H_\text{QE},
\end{equation}
where $H_Q$ is the error-free qubit Hamiltonian, $H_E$ describes the environment and $H_\text{QE}$ induces errors due to unwanted interactions of the qubit and the environment.
In an ideal setting, DD stroboscopically decouples the qubit and the environment by adding a sequence of instantaneous pulses,
\begin{equation}
    H_{DD}(t) \propto \sum_{k=1}^{m} \delta(t - t_k) H_{D_k},
\end{equation}
where $\delta$ is the Dirac-Delta function, $H_{D_k}$ is the Hamiltonian of the decoupling pulse and $t_k$ are the times at which the pulses are applied.
The evolution under $H + H_{DD}$ for time $T = \sum_{k=1}^m \Delta t_k$ is given by
\begin{equation}
    \mathcal{U}(T) = U(\Delta t_{m}) D_m U(\Delta t_{m-1}) D_{m-1} \cdots U(\Delta t_1) D_1,
\end{equation}
where $U(\Delta t_k) = \exp(-i\Delta t_k H)$ evolves under $H$ for the duration $\Delta t_k = t_k - t_{k-1}$ (with $t_0 = 0$) and $D_k = \exp(-i\pi/2 H_{D_k})$ is the evolution under the decoupling pulse.
It can be shown that this evolution can be written as
\begin{equation}
    \mathcal{U}(T) = \left(e^{-iT H_Q} \otimes \mathbb{1}_E \right) e^{-i T (\mathbb{1}_Q \otimes H_E + H_\text{error})} 
    \approx e^{-iTH_Q} \otimes e^{-iTH_E}.
\end{equation}
By designing specific DD sequences, we can then try to minimize the error term, which leads to an approximate separate evolution of the qubit and its environment.

\begin{figure}[th]
    \centering
    \includegraphics[width=0.9\textwidth]{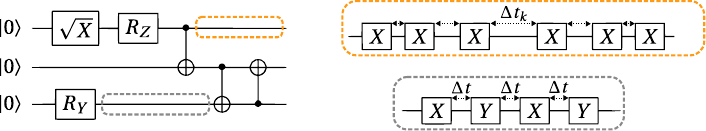}
    \caption[Dynamical decoupling sequences]{A possible selection of DD sequences in a circuit.
    The top, orange sequence uses a spacing of $m=6$ $X$-pulses according to Eq.~\eqref{eq:dd_uhrig}, as the qubit is known to be in the $x-y$ plane.
    The bottom, grey sequence applies the XY4 sequence, which protects arbitrary states.
    }
    \label{fig:dd}
\end{figure}

For states close to the $x-y$ plane, which can be described as $\ket{0} + e^{i\alpha}\ket{1}$, the qubit can be optimally decoupled using a sequence of $m$ $X$-pulses up to an error scaling as $\mathcal{O}(\Delta t^m)$~\cite{ezzell_dd_2023}. 
The times of the pulses are distributed according to a sine-squared, as
\begin{equation}\label{eq:dd_uhrig}
    t_k = T\sin^2\left( \frac{k\pi}{2(m + 1)} \right),
\end{equation}
for $k \in \{1, ..., m\}$.
To account for arbitrary states, however, only $X$-pulses are not sufficient and they must be further combined with rotations around different axes.
The simplest scheme protecting any states up to order $\mathcal{O}(\Delta t)$ is the XY4 scheme, consisting of four equidistant, alternating $X$ and $Y$ pulses.
More elaborate, higher-order schemes, nest $X$ and $Y$ sequences distributed as sine~\cite{wang_nudd_2011} or leverage rotations around arbitrary axes~\cite{genov_robustdd_2017}.
In Fig.~\ref{fig:dd} we show an illustrative example of selected DD sequences. 

\subsection{Measurement error mitigation}\label{sec:mem}

On near-term quantum computers, errors in the qubit measurement operation contribute significantly to the overall noise in the calculation. As previously discussed in Section~\ref{sec:sc}, readout fidelities are typically on the order of 99\% on current SC hardware, which is below the single- and two-qubit gate fidelities.
The task of measurement EM is to reduce these readout errors.

A direct approach to correcting errors in qubit measurements is to learn how errors transform a given input state and then invert the process.
For $n$ qubits, the measurement process can be described by a transfer matrix $A \in \mathbb{R}^{2^n \times 2^n}$ with entries
\begin{equation}
    A_{jk} = \mathrm{Pr}(\ket{k} \rightarrow \ket{j})
\end{equation}
that describe the probability of preparing state $\ket{k}$ and measuring $\ket{j}$ for $j,k \in \{0, ..., 2^{n} - 1\}$.
The transfer matrix is a left-stochastic matrix, i.e., the entries in each column sum up to 1 and each element is non-negative, $A_{jk} \geq 0$.
For example, in a single qubit experiment we could prepare the states $\ket{0}$ and $\ket{1}$ 100 times each and record a transfer given by
\begin{equation}
    A = 
    \begin{pmatrix}
        0.99 & 0.02 \\
        0.01 & 0.98
    \end{pmatrix}.
\end{equation}
In this experiment, the state $\ket{0}$ had a 99\% likelihood of being measured correctly, but in 1\% of the cases the wrong state $\ket{1}$ was measured. 
Conversely, the $\ket{1}$ state had a 98\% chance of being classified correctly and a 2\% chance of being measured as $\ket{0}$.

Using the transfer matrix, this measurement process can be described as
\begin{equation}
    \tilde{\vec p} = A \vec p,
\end{equation}
where $\vec p \in \mathbb{R}^{2^n}$ describes the noise-free probability distribution of the qubit states and $\tilde{\vec p} \in \mathbb{R}^{2^n}$ are the noisy measurements.
To revert the measurement errors, we can then invert the transfer matrix, which is possible under the assumption of sufficiently low noise as $A$ becomes strictly diagonally dominant.
If we apply the inversion to noisy probabilities $\tilde{\vec p}$ the mitigated probabilities are given by
\begin{equation}
    \vec q = A^{-1} \tilde{\vec p}.
\end{equation}
where $\vec q \in \mathbb{R}^{2^n}$ is a quasi-probability vector, since its entries sum up to 1 but are not necessarily positive, as the inverse of a left-stochastic matrix may have negative entries but the columns still sum up to 1~\cite{maciejewski_mem_2020}.
Note that $A$ is constructed once using calibration measurements and then applied to new, unseen measurements, for which the inversion will not perfectly correct for errors.
For example, inverting above $A$ gives
\begin{equation}
    A^{-1} \approx 
        \begin{pmatrix}
        1.01 & -0.02 \\
        -0.01 & 1.02
    \end{pmatrix},
\end{equation}
and mitigating the measurements $\tilde{\vec p} = (1, 0)^\top$ yields $\vec q \approx (1.01, -0.01)^\top$.

\subsubsection{Tensored measurement mitigation}

Since constructing the full transfer matrix $A$ requires evaluating an exponential number of states, it can only be computed for limited system sizes. 
One possible approach to improve the scalability is to only calibrate transfer matrices on single qubits and construct $A$ as tensor product. 
That is, $A = A_n \otimes \cdots \otimes A_1$ with single-qubit transfer matrices given by
\begin{equation}
    A_j = 
    \begin{pmatrix}
        \mathrm{Pr}(\ket{0}_j \rightarrow \ket{0}_j) &  \mathrm{Pr}(\ket{1}_j \rightarrow \ket{0}_j) \\
        \mathrm{Pr}(\ket{0}_j \rightarrow \ket{1}_j) &  \mathrm{Pr}(\ket{1}_j \rightarrow \ket{1}_j)
    \end{pmatrix},
\end{equation}
where $\ket{0}_j$ ($\ket{1}_j$) denotes state $\ket{0}$ ($\ket{1}$) on qubit $j$. 
If the target observable is also a tensor product of single-qubit operators, i.e., $O = O_n \otimes \cdots \otimes O_1$, the EM can be performed using only $\mathcal{O}(n)$ operations and is efficiently implementable~\cite{bravyi_mitigating_2021}. See Appendix~\ref{app:mem} for the derivation of this result.

The tensored model, however, disregards correlations between errors and can become imprecise in practice~\cite{chen_correlated_2019}. 
Several methods have been developed to circumvent these truncations by including correlations~\cite{bravyi_mitigating_2021} or considering corrections in a local subspace~\cite{nation_m3_2021}.
Another family of measurement EM techniques avoids assuming a model on the occurred errors and attempts to remove the introduced bias through readout error by Pauli twirling~\cite{berg_trex_2022}, a technique we discuss below.

\subsubsection{Matrix-free measurement mitigation}

Instead of constructing the full transfer matrix or assuming uncorrelated readout errors, the matrix-free measurement mitigation (M3)~\cite{nation_m3_2021} method constructs a truncated matrix $A'$ on only those states that have been observed in the measurement $\tilde{\vec p}$.
This approach is scalable in the sense that the computational overhead scales with the number of measurements $N$ instead of the dimension of the state space of the qubits.
In addition, only corrections within states that are close in Hamming distance $h_D$ can be considered. This distance measures the number of bits that differ in two states, i.e.
\begin{equation}
    h_D(j, k) = \sum_{b=1}^n \begin{cases}
        1, \text{ if } \text{bin}(j)_b \neq \text{bin}(k)_b, \\
        0, \text{ otherwise},
    \end{cases}
\end{equation}
where $j, k \in \{0, ..., 2^n - 1\}$.
In terms of the full matrix $A$, the entries of the M3 matrix are then defined as
\begin{equation}
    A'_{jk} \propto 
    \begin{cases}
        A_{jk}, \text{ if } h_D(j, k) \leq D, \\
        0, \text{ otherwise},
    \end{cases}
\end{equation}
where the indices run only over non-zero entries in $\tilde{\vec p}$, or, more formally, $j, k \in \{m:~ \tilde p_m > 0\}$.
After the construction, the columns of $A'$ must be normalized to ensure the matrix is still left-stochastic.

The dimension of $A'$ is limited by the number of measurements $N$ and the selected Hamming distance $D$.
Depending on the value of these parameters, $A'$ might already be small enough to invert it directly.
However, to find the error-mitigated probabilities more efficiently we can employ matrix-free, iterative algorithms, such as bi-conjugate gradient descent~\cite{vorst_bicgstab_1992}.
These methods only require a subroutine to compute $A'\tilde{\vec p}$, which can be evaluated without explicitly constructing the complete matrix and only accessing the required indices directly.

To calibrate M3, a set of circuits with known states must be measured to obtain. Crucially, we cannot iterate over $2^n$ circuits to prepare all possible initial basis states.
Instead, the authors propose different techniques, such as calibrating only on $\ket{0}^{\otimes n}$ and $\ket{1}^{\otimes n}$ or on a balanced set of states, where each qubit is set to $\ket{1}$ $n$ times.
In practice, and where possible, it might be beneficial to construct the transfer matrix for circuits close to the circuit of interest that have known computational basis states as output.

\subsection{Pauli twirling}

\begin{figure}[thbp]
    \centering
    \includegraphics[width=\textwidth]{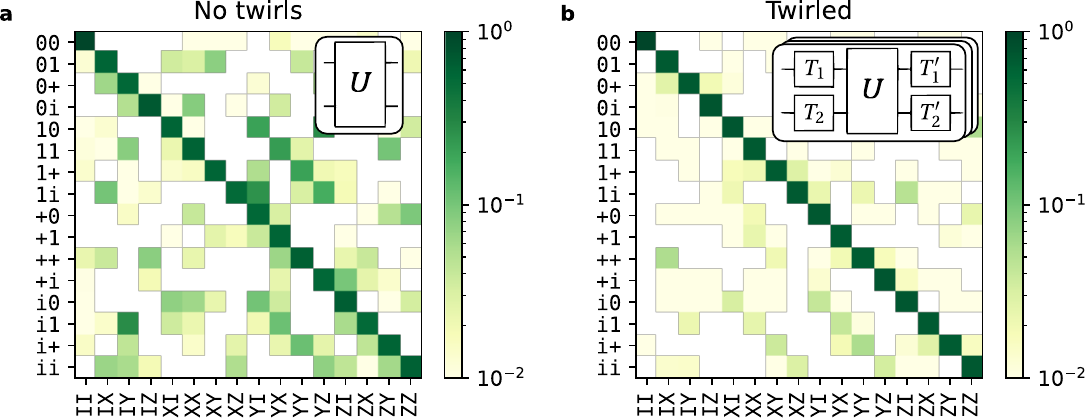}
    \caption[Effect of Pauli twirling]{The effect of twirling on two-qubit noise on \texttt{ibm\_cairo}. (a) The process tomography without PT, performed with 1000 shots per circuit. (b) The same process but each $R_{ZZ}$ gate is twirled 10 times.}
    \label{fig:twirling}
\end{figure}

Pauli twirling (PT) allows to convert noise channels into stochastic Pauli errors by suppressing coherent error contributions~\cite{wallman_twirling_2016}. In terms of the Pauli transfer matrix, PT suppresses the off-diagonal terms of the noise channel.
While reducing coherent errors in itself can already improve the result of quantum computations~\cite{kim_scalable_2023}, PT is especially powerful in combination with EM techniques which benefit from errors being incoherent, such as zero-noise extrapolation~\cite{kandala_zne_2019, kim_scalable_2023} discussed in Section~\ref{sec:zne}.

To apply PT to a circuit, we independently twirl the operations $U$ in the circuit by sandwiching it between random single qubit Pauli rotations. Importantly, the overall action of the twirled operation remains the same, i.e., PT is a unitary-preserving operation.
For $n$-qubit Clifford operations $C$, we can choose twirls $T, T' \in \{I, X, Y, Z\}^{\otimes n}$ such that~\cite{wallman_twirling_2016}
\begin{equation}
    C = T C T'.
\end{equation}
However, twirling can be generalized to non-Clifford operations by allowing to change the sandwiched operation~\cite{kim_scalable_2023}. 
For example, twirling the $R_{ZZ}$ gate with $T = T' = X \otimes Z$ requires flipping its sign, as
\begin{equation}
    R_{ZZ}(\theta) = (X \otimes Z) R_{ZZ}(-\theta) (X\otimes Z),
\end{equation}
which we discuss is further discussed in Appendix~\ref{app:pt}.

To build an intuitive understand of PT, we demonstrate its effect on the two qubit circuit
\begin{equation}
    U = \left(R_{ZZ}\left(\frac{\pi}{2}\right)R_{ZZ}\left(-\frac{\pi}{2}\right)\right)^{r},
\end{equation}
for $r=10$ executed on \texttt{ibm\_cairo}. 
Since the angles are chosen to have opposite signs per $R_{ZZ}$ pair, the overall action is the identity, and a noise-free Pauli process tomography~\cite{nielsen_tomography_2021} would result in the identity matrix.
In the presence of noise, the diagonal entries are damped and off-diagonal terms appear, which is shown in Fig.~\ref{fig:twirling}(a).
By twirling each $R_{ZZ}$ gate individually, as described in detail in Appendix~\ref{app:pt}, and averaging over 10 independent executions the off-diagonal terms in the transfer matrix are reduced, see Fig.~\ref{fig:twirling}(b).

\subsection{Probabilistic error cancellation}\label{sec:pec}

\begin{figure}
    \centering
    \includegraphics[width=\textwidth]{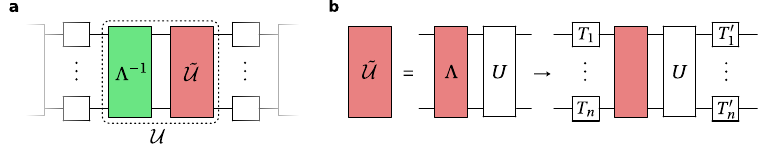}
    \caption[Probabilistic error cancellation]{Conceptual representation of probabilistic error cancellation. Note, that some operations are implemented as averaging multiple circuits, such as the quasi-probability decomposition of $\Lambda^{-1}$ or the Pauli twirling.
    (a) The inverted noise model aims at canceling the noise in a specific circuit layer $\tilde{\mathcal{U}}$. 
    (b) The noise is assumed to be Pauli noise, which can be ensured by averaging over a Pauli-twirled noisy operations, as shown here for the case that $U$ is Clifford.
    }
    \label{fig:pec}
\end{figure}

Probabilistic error cancellation (PEC) is among the earliest proposed methods to reduce errors in shallow quantum circuits~\cite{temme_shallow-em_2017}.
The underlying idea is intuitive; if we can learn a model that describes the noise acting in the circuit, we can attempt to add operations to invert the process in the circuit.
More formally, if the ideal, noise-free operation acts on a state described by a density matrix $\rho$ as
\begin{equation}
    \mathcal{U}(\rho) = U\rho U^\dagger,
\end{equation}
the goal is to learn a noise channel $\Lambda$ that describes the observed, noisy operation 
\begin{equation}
    \tilde{\mathcal{U}}(\rho) = \mathcal{U}(\Lambda(\rho)),
\end{equation}
to then implement $\Lambda^{-1}$ and reconstruct $\mathcal{U}(\rho) = \tilde{\mathcal{U}}(\Lambda^{-1}(\rho))$. 
Importantly, the inverse noise model $\Lambda^{-1}$ is generally not a unitary operation, but can be implemented as a quasi-probabilistic mixture of unitaries.
This workflow is visualized in Fig.~\ref{fig:pec}(a) for a single layer of operations $U$ and which must, in practice, be performed for each unique layer.

In Ref.~\cite{berg_pec_2023}, the error is assumed to be Pauli noise, which can be ensured by PT, as shown schematically in Fig.~\ref{fig:pec}(b).
Such a noise process can be modeled with a Pauli-Lindblad channel given by $\Lambda(\rho) = \exp[\mathcal{L}](\rho)$ with the Lindblad generator
\begin{equation}
    \mathcal{L}(\rho) = \sum_{P \in \mathcal P} \lambda_P (P\rho P^\dagger - \rho),
\end{equation}
where $\mathcal{P}$ is a subset of $n$-qubit Pauli operators and $\lambda_P \in \mathbb{R}_{\geq 0}$ describes the contribution of the Pauli $P$ to the noise channel.
For an efficient decomposition, $\mathcal{P}$ is assumed to have $\text{poly}(n)$ elements of the $4^n$ possible Paulis.
The coefficients $\lambda_P$ can be learned by measuring the expectation value of Paulis at after repeatedly applying the noisy operation $\tilde{\mathcal{U}}$~\cite{berg_pec_2023, berg_learning_2023, chen_learnability_2023}.
The inverse channel, $\Lambda^{-1}(\rho) = \exp[-\mathcal{L}](\rho)$, is then implemented through a quasi-probabilistic sampling. 

Under the assumption the learned noise model correctly represents the device, PEC allows to obtain an unbiased estimator of an observable, but at the cost of a sampling overhead.
Implementing $\Lambda^{-1}$ increases the variance of observable estimation by a factor $\gamma^2$, where
\begin{equation}
    \gamma = e^{2\sum_{P \in \mathcal{P}} \lambda_P} \geq 1
\end{equation}
depends on the strength of the noise.
Note that this factor is computed per mitigated layer and the overall circuit overhead is obtained by multiplying all $\gamma$ factors.
While this sampling overhead can be small for few-qubit circuits~\cite{berg_pec_2023}, it can reach impractical values for large circuits, e.g., $\gamma^2$ is estimated to be of order $10^{128}$ for the 127-qubit circuit of CX depth 60 investigated in Ref.~\cite{kim_utility_2023}.
This leads to the pursuit of another strategy: instead of canceling the noise, probabilistic error amplification (PEA) implements $\exp[\alpha\mathcal{L}]$ to \emph{amplify} the noise by a controlled factor $\alpha \geq 0$.
In combination with zero-noise extrapolation, which we discuss in the next section, this technique allows to construct accurate expectation values without the $\gamma^2$ overhead but at the cost of a bias.

\subsection{Zero-noise extrapolation}\label{sec:zne}

In zero-noise extrapolation (ZNE) the error in a quantum computation is systematically increased and then extrapolated in the opposite direction; to the zero-noise limit~\cite{temme_shallow-em_2017, li_zne_2017}.
In contrast to PEC, ZNE produces a biased estimator.
This is, for example, due to imperfect noise amplifications~\cite{kim_scalable_2023} or the extrapolation technique, which rely on heuristic models~\cite{giurgica_zne_2020} or Richardson extrapolation~\cite{tran_znebias_2023}.
Despite this, ZNE is a frequently used tool on near-term quantum computers due its simplicity and low overhead~\cite{carreravazquez_mpf_2023, rossmannek_embedding_2023, gacon_saqite_2023, kim_utility_2023}. 
However, it is important to note that it is a sensitive tool, which is substantially affected by the chosen extrapolation method and how errors are amplified.

\begin{figure}[th]
    \centering
    \includegraphics[width=\textwidth]{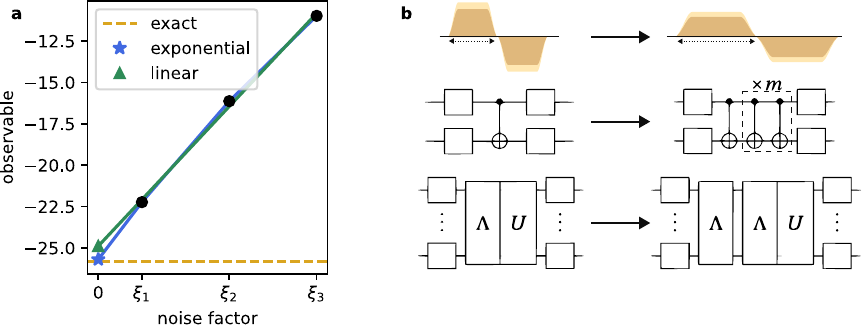}
    \caption[Zero-noise extrapolation models]{
    (a) ZNE extrapolation for an exponential and linear model along with the exact target value. Black circles show the noise-amplified observable evaluations for noise factors $\{\zeta_1, \zeta_2, \zeta_3\}$ and the extrapolation to $0$ noise.
    (b) Noise amplification techniques for ZNE. Top to bottom shows pulse stretching~\cite{kandala_zne_2019}, repeating CX gate in a redundant fashion~\cite{giurgica_zne_2020}, and amplifying the learned noise~\cite{kim_scalable_2023}.
    }
    \label{fig:zne}
\end{figure}

To systematically amplify the noise in a quantum computation, a variety of techniques can be employed.
Due to short decoherence times, one option is to slow down gate operations by stretching the pulse sequences~\cite{kandala_vqe_2017}, if such low-level access to the hardware is available.
A more generally applicable technique is based on inserting redundant gates that act as the identity~\cite{he_zne_2020, carreravazquez_mpf_2023}.
Typically, gates that already occur in the circuit are repeated. 
For example, to increase the error of a CX gate it can be replaced by a sequence of $2m + 1 (m\in\mathbb{N})$ CX gates, or a two qubit Pauli gate $R_{P_1 P_2},~ P_1, P_2 \in \{X, Y, Z\}$ can be followed by canceling rotations, $R_{P_1 P_2}(\theta) R_{P_1 P_2}(-\theta)$.
A drawback of increasing noise with these techniques is that they work independently of the underlying error processes and it is a-priori unknown how the noise is actually amplified.
PEA remedies this issue and allows to amplify the noise in a high-controlled manner by learning the noise model~\cite{kim_scalable_2023}.
In Fig.~\ref{fig:zne} we summarize different noise amplification techniques and present a ZNE example for linear and exponential models, with experimental data from Chapter~\ref{chap:saqite} using \texttt{ibm\_peekskill} and amplifying noise by introducing redundant CX gates.

\section{Quantum computing stack}\label{sec:stack}

In the previous sections we discussed how a set of few basis gates on connected qubits can be implemented on SC hardware using microwave pulses, and how errors in the physical device can be suppressed and mitigated. 
While it can be beneficial to design quantum algorithms with a particular physical qubit implementation in mind, we often work on a higher, device-agnostic level of abstraction.
This allows to formulate algorithms independently of their physical realization, verify their correctness in a simpler mathematical framework, and perform high-level optimizations that might not be evident on lower levels.
Working in different levels of abstraction leads to a quantum computing stack that is closely related to classical computers: each level has its own representation and we move in between levels using a set of a translation rules, often contained in the form of a compiler~\cite{jones_qstack_2012, maronese_compiling_2022}.

\begin{figure}[th]
    \centering
    \includegraphics[width=\textwidth]{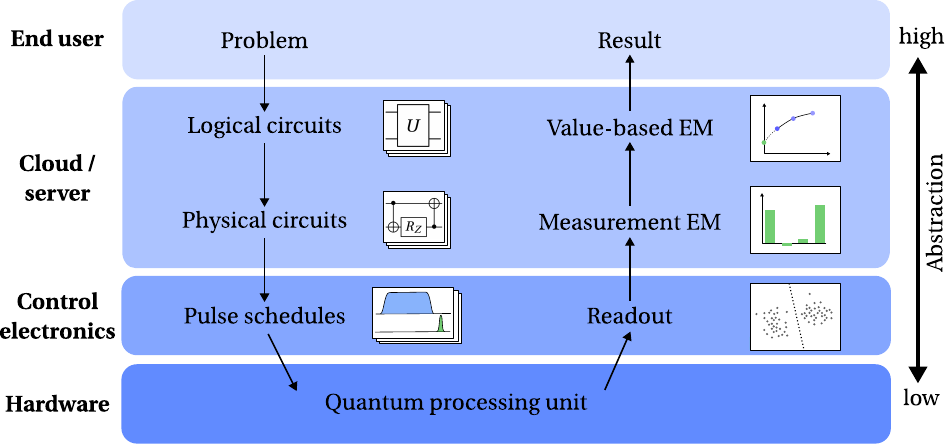}
    \caption[Quantum computing stack]{An implementation of a quantum computing stack.}
    \label{fig:stack}
\end{figure}

In Fig.~\ref{fig:stack}, we present the flowchart of a current quantum computing stack:
\begin{enumerate}
    \item Initially, we select a quantum algorithm to solve our problem. The algorithm requires executing a set of logical circuits, formulated using a device-agnostic set of operations.
    \item These logical circuits are then compiled to the target device by translating and optimizing the operations to use a minimal number of supported basis gates.
    \item To obtain the physical device instructions, the basis gates are transformed into pulse schedules according to the device's calibrations.
    \item Finally, the pulse schedule is executed on the device, once per shot. The measurements and expectation values can then be error mitigated before the final result is returned to the user.
\end{enumerate}
The following sections discuss this workflow in more detail.

\subsubsection{Problem} 

At the highest level of abstraction, quantum computers are one possible subroutine to solve a given problem.
This level is not only device-agnostic but also does not consider which circuits are being generated or which quantum error correction or mitigation is employed.
Instead, users interact with the quantum computer via an interface or software package~\cite{steiger_projectq_2018, Qiskit, bergholm_pennylane_2022, cirq, jones_quest_2019, luo_yaojl_2020} that provides a set of available operations to implement a quantum algorithm.
On this level, classical subroutines in established workflows can easily be replaced with quantum computing methods without taking into account the physical implementation on the device. 

To facilitate the following discussion about the quantum computing stack, we can consider the concrete problem of implementing the real-time evolution of a pair of Bell states $\ket{\psi_0} = (\ket{00} + \ket{11})^{\otimes 2}/2$ under the Hamiltonian
\begin{equation}
    H = Z_3 Z_4 + Z_1 Z_3 + Z_1 Z_2 + X_2 X_3 + Y_1 Y_3,
\end{equation}
by means of a first-order Lie-Trotter product formula, which we discussed in Section~\ref{sec:trotter}.

\subsubsection{Logical circuits}

At this level, we are constructing quantum circuits using gates as mathematical operations---irrespective of whether they are natively supported on the quantum computer we are later running them on. This enables us to design quantum algorithms and investigate their theoretical properties in an ideal, noise-free environment. 

\begin{figure}[th]
    \centering
    \includegraphics[width=0.9\textwidth]{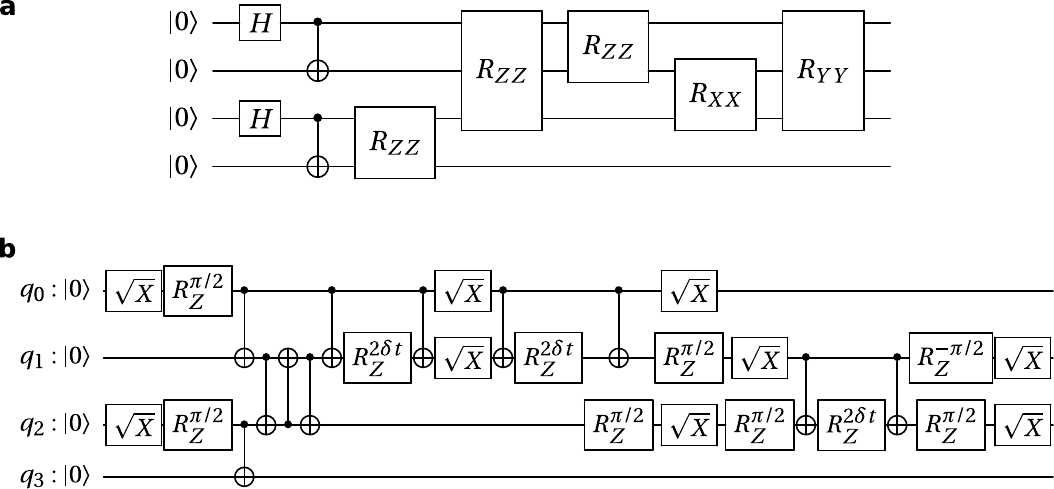}
    \caption[Compiling a product formula circuit]{
    (a) A logical circuit implementing a first-order Suzuki-Trotter step on a pair of Bell states. The wires have no labels $q_j~:~\ket{0}$ as they have not been mapped to physical qubits yet.
    (b) The compiled circuit for a basis gate set of $\{\sqrt{X}, X, R_Z, \text{CX}\}$ on a linear connectivity. The qubit labels $q_j~:~\ket{0}$ denote which physical qubit is acted on. Here, $R_Z^\alpha$ is a shorthand notation for $R_Z(\alpha)$.
    We allow algorithmic knowledge, e.g., re-ordering exponentials in the first-order Lie-Trotter expansion. See Appendix~\ref{app:compile_trotter} for details on the involved compilation steps.
    }
    \label{fig:trotter_logical}
\end{figure}

For example, the first-order Trotter step of the time evolution Hamiltonian above is given by
\begin{equation}
    e^{-i\delta t H} \approx e^{-i\delta t Z_3 Z_4} e^{-i\delta t Z_1 Z_3} e^{-i\delta t Z_1 Z_2} e^{-i\delta t X_2 X_3} e^{-i\delta t Y_1 Y_3},
\end{equation}
for a small $\delta t \in \mathbb{R}_{>0}$. 
These operations can be written as a logical circuit using two-qubit Pauli rotations $R_{P_1, P_2}$, $P_{1,2} \in \{X, Y, Z\}$ as shown in Fig.~\ref{fig:trotter_logical}(a). 
Here, we assume that gates can act on arbitrary pairs of qubits, even though on most quantum computers this is not the case. 

\subsubsection{Physical circuits \& pulse schedules}

The physical circuit level is device-dependent and only allows to use operations natively supported by the specific qubit implementation at hand. 
To transform the logical, device-agnostic circuit to this representation we use a compiler, which maps the instructions to physical qubits and translates logical operations to a set of basis operations. These are then implemented as a pulse schedule for qubit drive lines on the device according to the available gate calibrations.
The device might not natively support all required connections of the logical circuit, especially if the qubit connections are sparse, such as in Fig.~\ref{fig:heavyhex}.
In this case, additional Swap gates can be inserted into the circuit to enable new connections in a process called routing. 
While there are provably optimal routing strategies for specific settings, such as creating an all-to-all connectivity on a line topology~\cite{weidenfeller_scaling_2022}, finding the optimal solution is NP-hard~\cite{nanninci_routing_2023}. In practice, heuristic algorithms are therefore often used~\cite{li_sabre_2019}.

In addition to phrasing the logical circuit using physical gates, a compiler typically performs a series of optimizations to reduce the number of operations applied in the circuit. 
These optimization can be performed before the circuit is mapped to basis instructions, e.g., by leveraging high-level description of the gates to cancel or commute operations, or after the mapping, which often enables optimizations by accumulating or canceling a series of basis gates.
If the hardware does not only expose a set of supported basis gates, but also the physical implementation of how the pulses are applied, some gates can be mapped directly to a pulse instead of a basis gate.
For example, if a SC device is able to perform a CR pulse, which drives a $R_{ZX}$ rotation, then two-qubit Pauli rotation can be expressed as a single CR interaction~\cite{earnest_pulseefficient_2021}.
This is more efficient than the standard decomposition using two CX gates, which are implemented with two CR or ECR pulses in total.

Some EM techniques require a pre-processing of circuits, which can take place at different levels of the circuit description.
Typically, most changes are performed on the physical circuits after redundant gates have been removed and the final structure is fixed.
At this level, the remaining two-qubit gates can be twirled or categorized into layers for noise learning. 
The idle times of all qubits are also determined and DD sequences can be inserted.
At which level the circuit has to be modified for ZNE depends on the noise amplification technique. 
If, for example, noise is amplified by inserting redundant two-qubit gates, the additional circuit can be generated at the physical circuit level, but stretching the microwave pulses happens in the pulse schedules.

In Fig.~\ref{fig:trotter_logical}(b) we show the Trotter circuit compiled to a basis gate set of $\{\sqrt{X}, X, R_Z, \text{CX}\}$, which is common on IBM Quantum devices~\cite{ibm_quantum} and a linear connectivity, i.e., CX gates are only supported in-between neighboring qubits. Here, we do not include any EM preparation or pulse-level optimization. Appendix~\ref{app:compile_trotter} provides a detailed description of the performed compilation steps, including the final pulse schedule.

\subsubsection{Post-processing}

At the lowest level of the stack, the pulse schedules are executed on the quantum processor and the readout signal is mapped to a computational basis state. Note that this is repeated once per shot, per measurement basis. 
The resulting probability distribution can then be measurement-error-mitigated before computing the desired expectation values and applying further EM techniques, such as ZNE. Finally, the error-mitigated results are combined to obtain the problem solution.

\section{Conclusion}

This chapter reviewed the foundations of the superconducting qubit architecture, as well as important practical considerations in performing quantum computations.
Superconducting qubits are based on superconducting resonant circuits and their states are manipulated using microwave pulses.
A range of undesired effects, such as spontaneous exchanges with the environment, thermal excitations or cross-talk, lead to imperfect gate operations and readouts.
On current IBM Quantum devices used in this thesis, typical gate fidelities are around 99.9\% for single-qubit operations and 99\% for two-qubit gates and measurements.
Error suppression and mitigation techniques attempt to reduce these error rates by modifying the quantum circuits and combining the output of multiple copies.
While fully eliminating the errors with such methods is exponentially expensive in general~\cite{takagi_limits_2022, quek_exponential_2023}, error mitigation can still extend the reach of current, noisy quantum computers.
To close the gap to the high-level description of quantum algorithms, such as the Suzuki-Trotter expansion in the previous chapter, we have seen how abstract quantum circuits are compiled to machine-level instructions.

The execution of quantum algorithms at scale on noisy quantum computers is tightly coupled to the physical realization the qubits.
As quantum processors advance, we must carefully consider which platform to use depending on the requirements of the algorithm, such as the type of operations and the number of circuit executions.
In addition, the compilation becomes an increasingly important step as circuit sizes grow and manual optimization becomes challenging. 
Recent techniques focus e.g. on taking into account individual gate error rates~\cite{nation_mapomatic_2023} or reducing circuit depths by approximate compiling, which trades off gate errors and errors due to incorrect compilation of the target unitary~\cite{khatri_quantum-assisted_2019, madden_aqc_2022}.

In the following chapter, we will focus on a prominent algorithm paradigm for current quantum computers: variational quantum algorithms.
These replace the deterministic construction of a wave function with the optimization of parameters in a pre-defined circuit.
Crucially, this allows the manual selection of quantum circuits that are within the capabilities of the device, but trades off guarantees on the solution quality.

%% file: main/ch3_vqa.tex
\chapter{Variational quantum algorithms}\label{ch:vqa}

\summary{%
This chapter introduces the concept of variational quantum algorithms. 
These are a ubiquitous family of algorithms on near-term quantum computers, since they allow to select circuits that are within a device's capabilities.
In particular, we discuss approaches for ground-state preparation and for simulating real- and imaginary-time evolution of a quantum system.
These algorithms rely on computing quantum circuit gradients and the quantum geometric tensor, for which we introduce a wide spectrum of available techniques.
}

Canonical quantum algorithms hold the promise of polynomial~\cite{grover_search_1996} or, in some cases, even exponential speed-ups~\cite{shor_factoring_1994, lloyd_universal_1996, harrow_hhl_2009} over classical methods.
These algorithms follow deterministic circuit-construction rules, which are agnostic to the device they are executed on, to create a target state that encodes the problem solution.
As we have seen in the previous chapters, these circuits quickly become 
infeasible to run on near-term quantum computers, since, for practical problem sizes, they typically become too deep and have non-local interactions.
To leverage the full potential of devices available today, another promising strategy emerged over the recent years: variational quantum algorithms (VQAs).

VQAs are a family of algorithms that use parameterized quantum states $\ket{\phi(\vec\theta)}$ with tunable parameters $\vec\theta \in \mathbb{R}^d$ to encode the solution of a problem.
In contrast to canonical quantum algorithms, VQAs are restricted to the available solution space provided by $\ket{\phi(\vec\theta)}$ and attempt to find the optimal parameters $\vec\theta^*$, such that $\ket{\phi(\vec\theta^*)}$ is the best possible approximation of the target state. 
Crucially, the parameterized quantum state $\ket{\phi(\vec\theta)}$, called an ``ansatz'', can be chosen to operate within the limited capabilities of near-term quantum computers.
The proposed ansatz for a VQA generally depends on the problem, though adaptive and problem-agnostic models exist, which we discuss further in Section~\ref{sec:ansatz}.
The VQA workflow was first proposed to approximate ground states of quantum systems, dubbed the variational quantum eigensolver (VQE)~\cite{peruzzo_vqe_2014}. 
Since then, however, VQAs have been developed for virtually any problem class that promises a potential quantum advantage, ranging from simulating quantum dynamics~\cite{yuan_varqte_2019, cirstoiu_vff_2020,barison_pvqd_2021, benedetti_timeevo_2021,lin_simulation_2021, dborin_simulating_2022, slattery_simulationnitary_2022, gacon_dual_2023, gacon_saqite_2023, miessen_quantum_2023} over combinatorial optimization problems~\cite{farhi_quantum_2014, hegade_portfolio_2021, barkoutsos_cvar_2020, chai_sqaoa_2021} and quantum machine learning~\cite{havlicek_supervised_2019, zoufal_qbm_2021, abbas_power_2021, tacchino_qml_2021} to quantum circuit compilation~\cite{heya_vqgo_2018, khatri_quantum-assisted_2019, jones_compilation_2022}. 
For a detailed review see Ref.~\cite{cerezo_variational_2021}.

Typically, the optimal parameters of a VQA are defined as the minimum of a loss function $\mathcal{L}: \mathbb{R}^d \rightarrow \mathbb{R}$, that is
\begin{equation}
    \vec\theta^* = \argmin_{\vec\theta \in \mathbb{R}^d}\mathcal{L}(\vec\theta).
\end{equation}
This loss function can, in most cases, be written as
\begin{equation}\label{eq:vqa_loss}
    \mathcal{L}(\vec \theta) = \sum_k f_k(\braket{\phi_k(\vec\theta) | O_k | \phi_k(\vec\theta)}),
\end{equation}
where $O_k$ are observables, $\ket{\phi_k(\vec\theta)}$ are the ansatz states (which may differ per observable) and $f_k: \mathbb{R} \rightarrow \mathbb{R}$ are post-processing functions of the expectation values~\cite{cerezo_variational_2021}. 
A quantum computer is then used for the efficient evaluation of the expectation values $\braket{\phi_k(\vec\theta)|O_k|\phi_k(\vec\theta)}$---a task which can be expensive on classical computers---while a classical subroutine optimizes the parameters using e.g. gradient-free techniques, such as simplex methods~\cite{spendley_simplex_1962, nelder_simplex_1965, powell_cobyla_1994}, pattern-search methods~\cite{powell_search_1973} or evolutionary algorithms~\cite{beyer_evolution_2002, hansen_cmaes_2006}, or with gradient-based methods~\cite{cauchy_gd_1857, nesterov_gd_1983, spall_spsa_1988, fletcher_optimization_2000, kingma_adam_2017}. 

Even though VQAs have shown a lot of advances in the recent years, they still face formidable challenges.
One of the main open issues is the definition of suitable ansatz states, $\ket{\phi(\vec\theta)}$, which is further discussed in Section~\ref{sec:ansatz}, or providing convergence guarantees.
Both of these are no issues for canonical quantum algorithms, which provide a recipe for constructing the required circuits and, in many cases, have a well-understood error scaling~\cite{brassard_quantum_2002, harrow_hhl_2009, childs_trotter_2021}.
The error of a first-order Trotter expansion, for example, scales as $\mathcal{O}(T^2/N)$, where $T$ is the simulation time and $N$ the total number of timesteps~\cite{motta_emerging_2022}. 
Variational quantum time evolution, which we introduce in Section~\ref{sec:varqte}, on the other hand currently only allows for a-posteriori error bounds, which can be computationally costly to evaluate~\cite{zoufal_errorbounds_2021, endo_varqte_2020, gacon_dual_2023}.

\subsubsection{Applications of VQAs}

In this thesis, we are focusing on finding ground states of quantum systems and evolving quantum states in time. 
These, however, are merely two tasks for which VQAs have been proposed. Before diving into these topics, we provide a wider view of available VQA flavors and list a few examples of different loss functions:

\begin{itemize}
    \item \textit{Ground state search.} To find the ground state of a system described by the Hamiltonian $H$, we can minimize its energy, i.e.,
    \begin{equation}
        \mathcal{L}(\vec\theta) = \braket{\phi(\vec\theta) | H | \phi(\vec\theta)}.   
    \end{equation}
    This is discussed in detail in Section~\ref{sec:groundstate}.
    \item \textit{Combinatorial optimization.}
    Certain problems that seek to optimize a discrete, classical cost function can be formulated as ground-state search of a specifically-constructed Hamiltonian, which is diagonal in the computational basis~\cite{farhi_quantum_2014, abbas_optimization_2023}.
    For example, a quadratic unconstrained binary optimization (QUBO) problem corresponds to an Ising model with pairwise interactions.
    \item \textit{Black box optimization.} The optimization of a black box binary function $f: \{0,1\}^n \rightarrow \mathbb{R}$ can be implemented using 
    \begin{equation}
        \mathcal{L}(\vec\theta) = \sum_{k=0}^{2^{n - 1}} p_k(\vec\theta) f(\text{bin}(k)),
    \end{equation}
    where $p_k(\vec\theta) = \braket{\phi(\vec\theta)|k}\braket{k|\phi(\vec\theta)}$ is the probability to measure state $\ket{k}$ and $\text{bin}(k)$ is the binary representation of $k$~\cite{zoufal_blackbox_2023}. The function $f$, thus, implicitly defines a diagonal Hamiltonian, see Section~\ref{sec:blackbox} for more details.
    \item \textit{Time evolution.}
    To reduce the circuit depth of a Suzuki-Trotter expansion, each individual timestep can be projected onto a variational circuit by optimizing
    \begin{equation}
        \mathcal{L}(\vec\theta) = \left|\Braket{\phi(\vec\theta)|e^{-i\delta t H}|\phi\left(\vec\theta^{(t)}\right)}\right|^2,
    \end{equation}
    where $\vec\theta^{(t)}$ are the parameters at the previous timestep~\cite{barison_pvqd_2021}.
    \item \textit{Classification.} The variational quantum classifier (VQC) is a supervised learning algorithm for classification tasks~\cite{havlicek_supervised_2019}. 
    Training this quantum machine learning model on a dataset $D = \{(\vec x_k, y_k)\}_k$ with features $\vec x_k$ and labels $y_k$ can be implemented with the loss function
        \begin{equation}
            \mathcal{L}(\vec\theta) = \sum_{(\vec x, y) \in D} |y - \braket{\psi_{\vec x}|U^\dagger(\vec\theta) O U(\vec\theta)|\psi_{\vec x}}|,
        \end{equation}
    where $\ket{\psi_{\vec x}}$ is an initial state encoding the feature, $U(\vec\theta)$ is a trainable unitary, and $O$ maps the state to the label. For example, in binary classification, $O$ could be chosen as the projector to states with odd or even parity. 
\end{itemize}

\section{Variational ground-state preparation}\label{sec:groundstate}

The ground state is the lowest energy eigenstate of a quantum system. 
Studying these states provides insights into fundamental properties of the system, including the occurrence of quantum phase transitions or the system's stability.
Since it is the state with minimal possible energy, the following relation holds for the ground state $\ket{\psi_0}$ and any quantum state $\ket{\psi}$ in the system,
\begin{equation}
    \braket{\psi|H|\psi} \geq \braket{\psi_0|H|\psi_0} = E_0,
\end{equation}
where $H$ is the system's Hamiltonian and $E_0$ is the energy of the ground state.
We can leverage this relation to approximate the ground state with a variational quantum state $\ket{\phi(\vec\theta)}$ by tuning the parameters $\vec\theta$ to minimize the energy, that is
\begin{equation}
    \vec\theta^* = \argmin_{\vec\theta} E(\vec\theta),
\end{equation}
where $E(\vec\theta) = \braket{\phi(\vec\theta)|H|\phi(\vec\theta)}$. 
This approach is known as the Rayleigh-Ritz variational principle~\cite{ritz_variational_1909}.

Because the problem dimensionality grows exponentially with the size of the physical system, defining a suitable ansatz $\ket{\phi(\vec\theta)}$ and evaluating the variational energy $E(\vec\theta)$ on classical computers can pose a significant challenge---tasks that a quantum computer could potentially perform efficiently.
This gives rise to an iterative scheme called the variational quantum eigensolver (VQE)~\cite{peruzzo_vqe_2014, mcclean_vqe_2016, tilly_vqe_2022}, where the energies are evaluated on a quantum computer and the parameters updated on a classical machine.
The VQE is, thus, an archetype of VQAs where the loss function is given by the system energy, $\mathcal{L}(\vec\theta) = E(\vec\theta)$, which has been used to find the ground state of molecules~\cite{omalley_uccsd_2016, kandala_vqe_2017, hempel_ion_2018, nam_h2o_2020, google_hartree-fock_2020, motta_vqe-ef_2023, rossmannek_embedding_2023}, condensed matter systems, like the Fermi-Hubbard model~\cite{stanisic_observing_2022}, and statistical physics models~\cite{obrien_gaudin_2022}.

A common approach to optimizing the loss function is the family of gradient descent (GD) algorithms, which have shown tremendous success in training large machine learning models~\cite{alzubaidi_review_2021} and have provably better convergence properties than gradient-free methods in certain settings~\cite{harrow_gd_2021}.
Given an initial point $\vec\theta^{(0)} \in \mathbb{R}^d$, GD iteratively moves into the direction of the steepest descent in the loss landscape, that is
\begin{equation}\label{eq:gd}
    \vec\theta^{(k+1)} = \vec\theta^{(k)} - \eta_k \vec\nabla \mathcal{L}\left(\vec\theta^{(k)}\right),
\end{equation}
where $\eta_k \in \mathbb{R}_{>0}$ is a small learning rate. 
The parameter updates are then performed until a maximal allowed number of iterations is reached, or until a convergence criterion, such as a minimal gradient norm, is satisfied.
Several modifications of this update rule exist, which attempt to improve the convergence. These include momentum terms~\cite{qian_momentum_1999, nesterov_gd_1983}, adaptive learning rates~\cite{kingma_adam_2017}, or preconditioning~\cite{amari_natural_1998}, to name a few. 
In this thesis, we focus on two GD algorithms which solve two different fundamental shortcomings and are of particular interest in context of near-term quantum computing: simultaneous perturbation stochastic approximation (SPSA) and quantum natural gradients (QNG).

\subsection{SPSA}\label{sec:spsa_descent}

Evaluating circuits on quantum computers entails a substantial overhead. As previously discussed in Chapter~\ref{chap:hardware}, circuits must be compiled for the hardware, sent to the quantum processor, and returned to the end-user. Noisy quantum computers may also require error mitigation, which can further increase the number of circuits and required post-processing of the results.
In combination with a moderate number of available quantum computers and slower processing speeds than classical computers, these factors collectively lead to a limitation of the currently available quantum computing resources.
Consequently, minimizing the number of circuits is essential for enhancing variational quantum algorithms. 
Within the VQE optimization, the computation of energy gradients is often the limiting factor, which require the evaluation $\mathcal{O}(d)$ circuits, as discussed in detail in Section~\ref{sec:gradients}.
Furthermore, while these methods can provide the analytic gradient values, if the circuit are evaluated exactly, in practice, the results are subject to noise from a finite number of measurements and the device itself.

Constructing stochastic approximations of the loss function gradients, instead, can offer a resource-efficient alternative.
The simultaneous perturbation stochastic approximation (SPSA) algorithm~\cite{spall_spsa_1988,spall_overview_1998} is a stochastic formulation of GD, which has originally been developed for stochastic optimization problems and is, thus, inherently compatible with a noisy loss function.
Its update rule is
\begin{equation}
    \vec\theta^{(k+1)} = \vec\theta^{(k)} - \eta_k \widehat{\vec\nabla\mathcal{L}}^{(k)},
\end{equation}
with learning rate $\eta_k > 0$. 
Instead of the exact gradient, SPSA uses a gradient sample 
\begin{equation}
    \widehat{\vec\nabla\mathcal{L}}^{(k)} = \frac{\mathcal{L}(\vec\theta^{(k)} + \epsilon_k \vec\Delta) - \mathcal{L}(\vec\theta^{(k)} - \epsilon_k \vec\Delta)}{2\epsilon_k}\vec\Delta^{-1},
\end{equation}
where $\vec\Delta \in \mathbb{R}^d$ is a randomly sampled perturbation direction, $\vec\Delta^{-1}$ is an element-wise inverse, and $\epsilon_k$ is the perturbation magnitude.

The idea of SPSA is that, since every gradient sample is unbiased (up to $\mathcal{O}(\epsilon_k^2)$), it, on average, has the same performance as GD.
In fact, under certain conditions on the hyperparameters detailed in Refs.~\cite{spall_spsa_1988, spall_implementation_1998}, it can be shown that the iteration converges to the optimal solution $\vec\theta^*$.
Commonly used settings fulfilling these conditions are $\vec\Delta \in \mathcal{U}(\{1, -1\}^d)$, i.e., each entry is Bernoulli distributed over $\{1, -1\}$, and exponential decays for learning rate and perturbation, that is
\begin{equation}
    \begin{aligned}
        \eta_k &= A(B + k)^{-\alpha},  \\
        \epsilon_k &= Ck^{-\gamma},
    \end{aligned}
\end{equation}
with hyperparameters $A, B, C \in \mathbb{R}_{>0}$ and decay rates $\alpha, \gamma \in \mathbb{R}_{>0}$. 
Asymptotically optimal values for the decay rates are $\alpha = 1$ and $\gamma = 1/6$~\cite{spall_implementation_1998}, however, if a lot of noise is present in the loss function
it is typically better to choose smaller values. Ref.~\cite{spall_implementation_1998} suggests $\alpha = 0.602$ and $\gamma = 0.101$, 
which still satisfy the conditions.
The remaining hyperparameters can be calibrated according to the loss function as, for example, done in Ref.~\cite{kandala_vqe_2017}. See also Eq.~\eqref{eq:spsacal} for more detail.

\subsubsection{Example}

\begin{figure}[thp]
    \centering
    \includegraphics[width=0.49\textwidth]{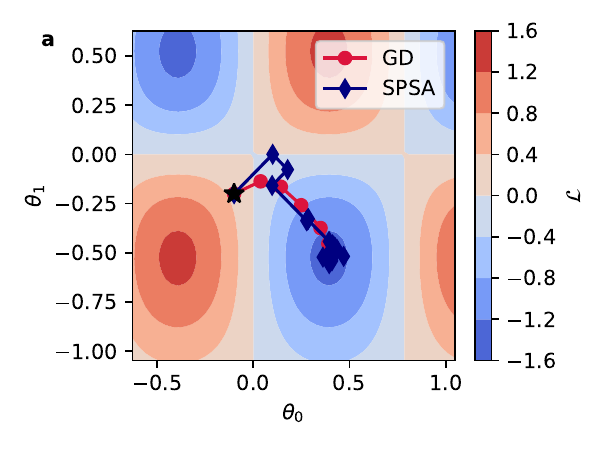}
    \includegraphics[width=0.49\textwidth]{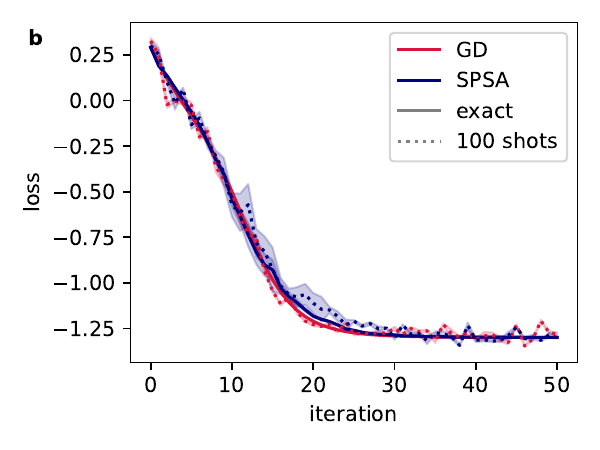}
    \caption[Comparison of gradient descent and SPSA]{Comparison of GD and SPSA, which uses $\vec\Delta \in \mathcal{U}(\{1, -1\}^d)$. (a) The path of 10 iterations at learning rate $\eta_k = 0.05$ and perturbation $\epsilon_k = 0.1$ in a noise-free (exact) setting. (b) The loss for 50 iterations and $\eta_k = 0.01$ using exact evaluations or shot noise with 100 measurements per expectation value. For SPSA, we show mean and standard deviation of 10 independent runs.}
    \label{fig:spsa_path}
\end{figure}

To obtain an intuition for SPSA we compare it against GD to find the ground state of $H = Z_1 Z_2 + Z_1 Z_3$.
Being a diagonal Hamiltonian in the computational basis it can be related to a combinatorial optimization problem and we therefore use the ansatz of the Quantum Approximate Optimization Algorithm (QAOA)~\cite{farhi_quantum_2014}.
This ansatz is defined as 
\begin{equation}\label{eq:qaoa_simple}
    \ket{\phi(\vec\theta)} = e^{-i\theta_1 M} e^{-i\theta_0 H}  \ket{+}^{\otimes 2},
\end{equation}
where $M = X_1 + X_2$.
If the initial parameter values of the optimization are set to $\vec 0$, the initial state is an equal superposition over all computational basis states and is, thus, guaranteed to already have an overlap with the ground state.
However, to avoid a saddle point at the beginning of the optimization we use a slightly perturbed initial point $\vec\theta^{(0)} \approx \vec 0$.
The optimizations with SPSA and GD are conducted twice: once for an ideal, noise-free evaluation of the system energy, and a second time that include statistical noise from a finite number of circuit measurements. The results are shown in Fig.~\ref{fig:spsa_path}.

We observe that GD moves into the direction of steepest descent, whereas SPSA ``jumps'' around the GD trajectory according to the random direction $\vec\Delta$.
On average, however, SPSA converges with the same number of iterations as GD while using only $\mathcal{O}(1)$ function evaluations, in contrast to $\mathcal{O}(d)$ for GD.
Unless loss function evaluations can be heavily parallelized, SPSA typically converges more quickly, as each iteration is cheaper to implement~\cite{spall_multivariate_1992, gacon_qnspsa_2021}. This property also holds if only noisy evaluations of the loss functions are available~\cite{chin_comparative_1994}.
An additional advantage of SPSA is that the random perturbation directions can help to escape saddle points or local minima, which is shown in Chapter~\ref{chap:qnspsa}.
These properties, in combination with its simplicity, make SPSA a prominent optimization scheme on near-term quantum computers.

\subsection{Quantum natural gradients}\label{sec:qng}

\begin{figure}[th]
    \centering
    \includegraphics[width=0.49\textwidth]{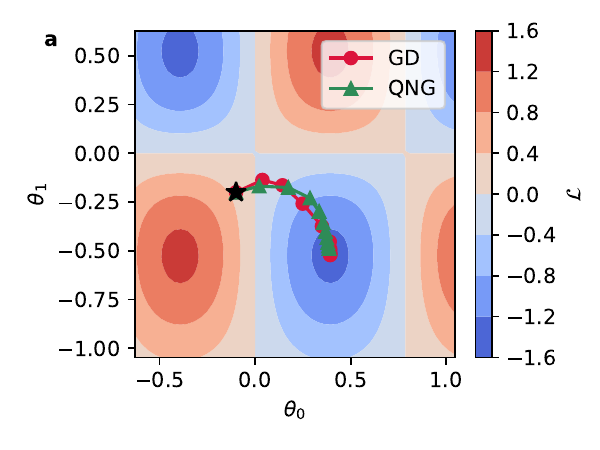}
    \includegraphics[width=0.49\textwidth]{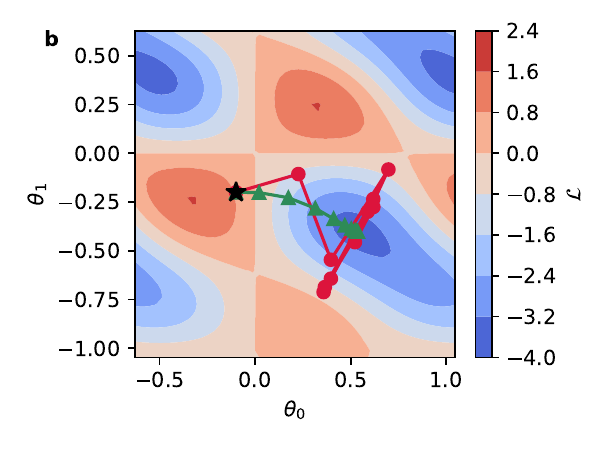}
    \caption[Comparison of gradient and quantum natural gradient descent]{Optimization of two Hamiltonians $H_a$ and $H_b$ on 3 qubits with the ansatz of Eq.~\eqref{eq:qaoa_simple}. (a) GD and QNG in a regular setting, $H_a = Z_0 Z_1 + Z_0 Z_2$. (b) GD fails to converge due to difference in parameter sensitivities, under the modified Hamiltonian $H_b = H_a - 2 Z_1 Z_2$.
    }
    \label{fig:qng-vs-gd}
\end{figure}

Taking on a geometric perspective, the direction of the GD update step is given by the gradient and it's magnitude is limited by the $\ell_2$ norm. 
This becomes clear upon rewriting the GD iteration of Eq.~\eqref{eq:gd} in the following, equivalent, form
\begin{equation}
    \vec\theta^{(k+1)} = \argmin_{\vec\theta} \Braket{\vec\theta - \vec\theta^{(k)}, \vec\nabla\mathcal{L}\left(\vec\theta^{(k)}\right)} + \frac{1}{2\eta_k} \left\|\vec\theta - \vec\theta^{(k)}\right\|^2_2.
\end{equation}
The restriction of the Euclidean distance of the parameters reveals a crucial drawback of GD: It does not consider the sensitivity of the underlying model $\ket{\phi(\vec\theta)}$ with respect to the parameters. 
If, for example, the loss function changes more significantly upon changing a subset of parameters, but is insensitive to others, GD will require a small learning rate for stable convergence and will suffer from a slow convergence. 
To demonstrate this behavior we investigate two experiments in Fig.~\ref{fig:qng-vs-gd}. In the first setting, the model has similar sensitivities in each parameter dimension and GD converges as expected. In the second however, the loss function is more challenging and GD oscillates around the solution.

The quantum natural gradient (QNG) resolves these shortcomings of GD by limiting the size of the update step by the change induced in the model~\cite{amari_natural_1998, stokes_qng_2020}.
This change is quantified by the Fubini-Study metric $d_\text{FS}$, such that the update rule becomes
\begin{equation}\label{eq:qng_argmin}
    \vec\theta^{(k+1)} = \argmin_{\vec\theta} \Braket{\vec\theta - \vec\theta^{(k)}, \vec\nabla\mathcal{L}\left(\vec\theta^{(k)}\right)} + \frac{1}{2\eta_k} d^2_\text{FS}\left(\vec\theta, \vec\theta^{(k)}\right),
\end{equation}
with 
\begin{equation}
    d_\text{FS}(\vec\theta, \vec\theta') = \arccos\left(\left|\braket{\phi(\vec\theta)|\phi(\vec\theta')}\right|\right).
\end{equation}
In principle, the QNG update step can be computed in this representation, which is the core idea of Chapter~\ref{chap:dual}. 

In practice, however, a more direct formulation is common, which allows to view the QNG as a pre-conditioned gradient.
This derivation requires the assumption that the QNG update step,
\begin{equation}
    \vec{\delta\theta}^{(k)} = \vec\theta^{(k+1)} - \vec\theta^{(k)},
\end{equation}
is small, which can be imposed, for example, with a small learning rate $\eta_k$.
The Fubini-Study metric can, then, be expanded as 
\begin{equation}\label{eq:fs_approximation}
    \begin{aligned}
    d^2_\text{FS}(\vec\theta, \vec\theta + \vec{\delta\theta}) 
    &= \arccos^2\left(\left|\braket{\phi(\vec\theta)|\phi(\vec\theta + \vec{\delta\theta})}\right|\right) \\
    &= 1 - \left|\braket{\phi(\vec\theta)|\phi(\vec\theta + \vec{\delta\theta})}\right|^2 + \mathcal{O}\left(\|\vec{\delta\theta}\|_2^4\right) \\
    &= \vec{\delta\theta}^\top g(\vec\theta) \vec{\delta\theta} + \mathcal{O}\left(\|\vec{\delta\theta}\|_2^3\right),
    \end{aligned}
\end{equation}
where $g(\vec\theta) = \mathrm{Re}(G(\vec\theta))$ is the real part of the quantum geometric tensor (QGT)~\cite{stokes_qng_2020, gacon_dual_2023}.
The QGT encodes the sensitivity of a parameterized quantum state $\ket{\phi(\vec\theta)}$ with respect to perturbation of its parameters. 
It is defined as 
\begin{equation}\label{eq:qgt}
    G_{jk}(\vec\theta) = \braket{\partial_j\phi(\vec\theta) | \partial_k\phi(\vec\theta)} - \braket{\partial_j \phi(\vec\theta)|\phi(\vec\theta)}\braket{\phi(\vec\theta)|\partial_k\phi(\vec\theta)},
\end{equation}
where we introduce the notation $\partial_j \equiv \partial/\partial \theta_j$.
The QGT is an object that plays a central role in this thesis, as it also determines the dynamics in a variational approach to quantum time evolution.
It is interesting to note that, despite its name and being self-adjoint, the QGT does not implement a metric on the space of quantum states since $G$ is only positive \emph{semi}-definite.
Furthermore, the QGT is closely connected to the quantum Fisher information matrix (QFI), which is given by $4g(\vec\theta)$.
Thereby, the QGT is related to further applications e.g. in quantum sensing or analyzing quantum machine learning models~\cite{meyer_qfi_2021}.
The imaginary part of the QGT, on the other hand, is related to the Berry curvature~\cite{graf_berry_2021}.

Plugging the QGT formulation into Eq.~\eqref{eq:qng_argmin} and solving the minimization we obtain the common QNG representation
\begin{equation}\label{eq:qng}
    \vec\theta^{(k+1)} = \vec\theta^{(k)} - \eta_k g^{-1}\big(\vec\theta^{(k)}\big) \vec\nabla\mathcal{L}\big(\vec\theta^{(k)}\big).
\end{equation}
While the QNG is often directly defined in this form in literature, we point out that this equation only follows the true natural gradient dynamics up to an error of order $\|\vec{\delta\theta}^{(k)}\|_2^3$.
In a practical implementation, where the evaluation of the gradient and QGT are subject to shot and device noise, however, this approximation error might be negligible.

If we only consider the performance per update step, QNG is the superior choice to GD, which becomes even clearer with its connection to imaginary-time evolution discussed in the following section.
A significant drawback, however, and a reason it is not frequently used in practice, is that the QNG requires the calculation of the QGT in each update step, which comes at a cost of evaluation $\mathcal{O}(d^2)$ circuits, compared to $\mathcal{O}(d)$ for GD and $\mathcal{O}(1)$ for SPSA.
This issue, and possible solutions, are the topic of Chapter~\ref{chap:qnspsa}.

\section{Variational quantum time evolution}\label{sec:varqte}

\begin{figure}[thbp]
    \centering
    \includegraphics[width=\textwidth]{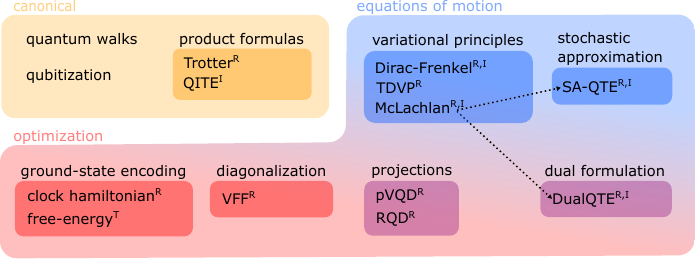}
    \caption[Overview of quantum time evolution methods]{Overview of time evolution algorithms.
    Superscripts indicate supported applications: (R)eal time evolution, (I)maginary time evolution or (T)hermal state preparation. Note, that the ability to perform imaginary time evolution implies thermal state preparation.
    Variational principles are discussed in this section, and the quantum walks~\cite{childs_quantumwalk_2010}, qubitization~\cite{low_qubitization_2019} and QITE~\cite{motta_determining_2020} are not explicitly mentioned, but only set in the context for the reader familiar with the topic.
    Arrows indicate algorithms that are part of this thesis, which are derived from McLachlan's variational principle.
    This map is not exhaustive, see also Ref.~\cite{miessen_quantum_2023} for a review.
    }
    \label{fig:time-map}
\end{figure}

Variational approaches to quantum time evolution attempt to represent the time-evolved state $\ket{\psi(t)}$ by a parameterized ansatz $\ket{\phi(\vec\theta(t))}$ with time-dependent parameters $\vec\theta(t)$.
A fundamental technique we discuss in detail in the remainder of this section is based on variational principles, that project the state dynamics $\partial \ket{\psi(t)} / \partial t$ onto equations of motion for the parameter dynamics $\partial\vec\theta / \partial t$.
However, a wide variety of other algorithms have been put forward, which we briefly discuss and put into context here.
We provide an overview in Fig.~\ref{fig:time-map}, including contributions presented in this thesis: stochastic approximation to time evolution (Chapter~\ref{chap:saqite}), and a dual formulation (Chapter~\ref{chap:dual}).

Instead of deriving equations for the parameter dynamics, the Suzuki-Trotter step can be directly projected onto the ansatz
~\cite{otten_rqd_2019, barison_pvqd_2021, benedetti_timeevo_2021, lau_nisq_2021}.
For instance, in real-time evolution, the restarted quantum dynamics (RQD)~\cite{otten_rqd_2019} or the projected variational quantum dynamics (p-VQD) algorithm~\cite{barison_pvqd_2021} might offer a scalable option for near-term devices, provided a Trotter step can be efficiently executed.
However, even a single step can become a bottleneck if the Hamiltonian features long-distance interactions or a high number of Pauli terms, as seen e.g. in molecular dynamics. 

Other strategies focused on real-time evolution attempt to learn the unitary of the Suzuki-Trotter expansion.
Variational Fast Forwarding (VFF) methods~\cite{cirstoiu_vff_2020, commeau_vff_2020, gibbs_vff_2022}, for example, aim to diagonalize the evolution operator using a variational ansatz.
Access to the diagonalization would allow arbitrary long time evolution, however, determining the diagonalizing unitary is a challenging task, which currently restricts its application to a limited number of qubits, and may not be possible efficiently for some systems~\cite{atia_fastforwarding_2017}.
Classical pre-processing techniques, on the other hand, variationally learn the evolution operator on a small, classically simulable system. The solution is then systematically bootstrapped to larger system sizes by relying on constraints such as translational invariance~\cite{mansuroglu_classical_2021} or low entanglement~\cite{dborin_simulating_2022}.
While these methods can be scaled up to larger systems within these specific contexts, they do not support general quantum time evolution.

Instead of modeling the system dynamics, ground-state encodings prepare an artificial system whose ground state allows to read out the evolved state. 
For real-time evolution, the clock Hamiltonian~\cite{mcclean_clock_2013, barison_clock_2022} uses auxiliary qubits to encode the time evolved states at discrete target times.
If we are instead interested in performing imaginary time evolution to prepare thermal states (see Section~\ref{sec:ite}), the time evolution could be avoided by directly optimizing the Helmholtz free energy of the system, which is minimized by the system's thermal state~\cite{wang_gibbsprep_2021, sbahi_mirror_2022, consiglio_gibbs_2023}.
Finding the ground state of a system allows to use different techniques than the direct time evolution, but is generally not an easier task.

\subsection{Real-time evolution}

The dynamics of a state $\ket{\psi(t)}$ under a Hamiltonian $H$ are determined by the Schrödinger equation,
\begin{equation}
    \frac{\partial}{\partial t} \ket{\psi(t)} = -i H\ket{\psi(t)}.
\end{equation}
If the initial state of the system is $\ket{\psi_0}$, the time-evolved state after time $t$ is given by 
\begin{equation}
    \ket{\psi(t)} = e^{-iHt}\ket{\psi_0}.
\end{equation}
Note that we assume a static Hamiltonian, but the following techniques can equally be applied to time-dependent Hamiltonians.
As previously discussed in Section~\ref{sec:trotter}, the time evolution can be approximated on a quantum computer using a Suzuki-Trotter expansion of the evolution operator. This, however, may lead to deep and complex circuits, as the gates in the circuits reflect the locality of the Hamiltonian and the depth increases with simulation time and desired accuracy.

In this section we follow a different approach: instead of constructing a circuit to implement the evolution operator, we select a variational quantum state $\ket{\phi(\vec\theta)}$ and attempt to find the parameters $\vec\theta = \vec\theta(t) \in \mathbb{R}^d$ that optimally approximate $\ket{\psi(t)}$ at any time $t$.
The goal of variational (real) quantum time evolution (VarQRTE) is thus to solve the following equation as precisely as possible
\begin{equation}\label{eq:variational_schroedinger}
    \frac{\partial}{\partial t}\ket{\phi(\vec\theta)} \approx -iH\ket{\phi(\vec\theta)},
\end{equation}
where the initial parameters $\vec\theta^{(0)}$ are determined by the condition $\ket{\phi(\vec\theta^{(0)})} = \ket{\psi_0}$. This approach can be thought of mapping the dynamics of the \emph{state} onto \emph{parameters} of the ansatz state. Here, and in the following, we drop the explicit time-dependency of $\vec\theta$.

\subsubsection{Variational projection}

\begin{figure}
    \centering
\begin{tikzpicture}
\node at (0,0) {\includegraphics[width=8cm]{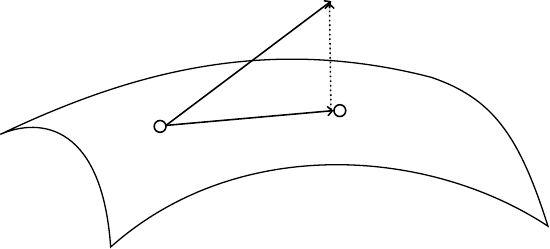}};
\node at (3.3, 0.5) {$\mathcal{M}$};
\node at (1.3, 1.3) {$\ket{r(\vec\theta)}$};
\node at (-1, 1.3) {$-iH\ket{\phi(\vec\theta)}$};
\node at (-2.1, -0.3) {$\ket{\phi(\vec\theta)}$};
\node at (0, -0.2) {$\frac{\partial}{\partial t}\ket{\phi(\vec\theta)}$};
\end{tikzpicture}
\caption[McLachlan's variational principle]{Projection of the time evolution onto the variational manifold $\mathcal{M} = \{\ket{\phi(\vec\theta)}~|~\vec\theta \in \mathbb{R}^d\}$. McLachlan's VP minimizes the residual norm $\sqrt{\braket{r(\vec\theta)|r(\vec\theta)}}$.
}
\label{fig:varqte_manifold}
\end{figure}
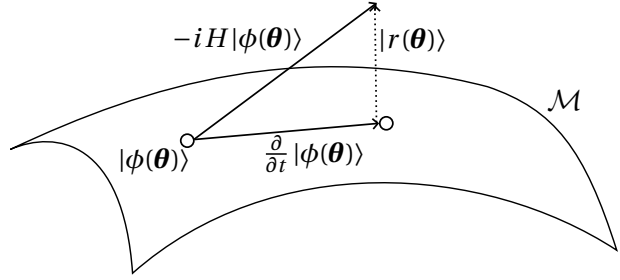

Solving Schrödinger's equation for a variational state $\ket{\phi(\vec\theta)}$
requires projecting the time evolution from the complete Hilbert space onto the variational manifold\footnote{The state $\ket{\phi(\vec\theta)}$ could, in theory, be engineered to not fulfill all conditions for a manifold using, e.g., specifically constructed parameterizations or gates. Here we assume a setting where the ansatz forms a manifold.} $\mathcal{M} = \{\ket{\phi(\vec\theta)} | \vec\theta \in \mathbb{R}^d\}$, which is schematically shown in Fig.~\ref{fig:varqte_manifold}.
The projection can be determined with variational principles (VPs), such as the Dirac-Frenkel~\cite{frenkel_wave_1933}, McLachlan~\cite{mclachlan_variational_1964}, or time-dependent variational principle (TDVP)~\cite{kramer_tdvp_1981}.
These principles differ in how the residual, 
\begin{equation}
    \ket{r(\vec\theta)} = \frac{\partial}{\partial t}\ket{\phi(\vec\theta)} - \left(-iH\ket{\phi(\vec\theta)}\right) 
    = \left(\frac{\partial}{\partial t} + iH\right)\ket{\phi(\vec\theta)},
\end{equation}
is minimized. A good overview is given in Refs.~\cite{broeckhove_equivalence_1988,yuan_varqte_2019}.

In this thesis we focus McLachlan's formulation~\cite{mclachlan_variational_1964}, which minimizes the $\ell_2$-norm of the residual, i.e., $\|\ket{r(\vec\theta)}\|_2 = \sqrt{\braket{r(\vec\theta)|r(\vec\theta)}}$.
The other VPs are discussed and compared in Appendix~\ref{app:vp}.
By variation of the parameter derivatives $\dot{\vec\theta} = \partial\vec\theta / \partial t$ we then obtain the rule
\begin{equation}\label{eq:varqte_lse}
    g(\vec\theta) \dot{\vec\theta} = \vec b^{(R)}(\vec\theta),
\end{equation}
where $g(\vec\theta) = \mathrm{Re}(G(\vec\theta)) \in \mathbb{R}^{d \times d}$ is the real part of the QGT, see Eq.~\eqref{eq:qgt}, and we call $\vec b^{(R)}(\vec\theta)$ the evolution gradient for real-time evolution. It is given by 
\begin{equation}
    b_k^{(R)}(\vec\theta) = \mathrm{Im}\left(\braket{\partial_k\phi(\vec\theta) | H|\phi(\vec\theta)} - \braket{\partial_k\phi(\vec\theta)|\phi(\vec\theta)} E(\vec\theta)\right),
\end{equation}
where $E(\vec\theta) = \braket{\phi(\vec\theta)|H|\phi(\vec\theta)}$ is the system energy.
The interested reader can find the explicit derivation of McLachlan's parameter dynamics in Appendix~\ref{app:mclachlan_proof}.

\subsection{Imaginary-time evolution}\label{sec:ite}

In the imaginary time-evolution we consider the Schrödinger equation under the variable transformation $\tau = it$ given by
\begin{equation}\label{eq:imag_schroedinger}
    \frac{\partial}{\partial \tau} \ket{\psi(\tau)} = (E(\tau) -H)\ket{\psi(\tau)},
\end{equation}
where $E(\tau) = \braket{\psi(\tau)|H|\psi(\tau)}$.
Though it is not a physical process, the imaginary-time evolution is an important tool to understand quantum mechanical systems, as it allows to prepare ground states or thermal states.
Thermal states allow the calculation of thermodynamic observables at finite temperature, used in the study of condensed matter systems~\cite{kliesch_thermalstates_2018}, but also find applications in machine learning, e.g., as a subroutine in quantum Boltzmann machines~\cite{amin_qbm_2018, zoufal_qbm_2021}.
Ground-state preparation is an even more ubiquitous task used to determine material properties~\cite{skriver_groundstate_1984, lehtovaara_ite_2007, shi_variational_2018}, solve optimization problems~\cite{abbas_optimization_2023} or even simulate real-time dynamics~\cite{mcclean_clock_2013, barison_clock_2022}. 

For an initial state $\ket{\psi_0}$, the imaginary-time evolved state is defined as
\begin{equation}
    \ket{\psi(\tau)} = \frac{e^{-\tau H} \ket{\psi_0}}{\sqrt{\braket{\psi_0|e^{-2\tau H}|\psi_0}}},
\end{equation}
where the normalization is required as the $\exp(-\tau H)$ is not unitary. 
Written in the eigenbasis $\{\ket{\lambda_n}\}_{n\geq 0}$ of the Hamiltonian, the evolution is, up to normalization, given by
\begin{equation}\label{eq:itevolved}
    \ket{\psi(\tau)} \propto c_0\ket{\lambda_0} + \sum_{n>0} c_n e^{-(E_n - E_0)\tau} \ket{\lambda_n},
\end{equation}
where $E_n$ are the ordered eigenstate energies, $H\ket{\lambda_n} = E_n \ket{\lambda_n}$ with $E_n \leq E_{n+1}$, $c_n = \braket{\lambda_n|\psi_0}$ and we assume $c_0 \neq 0$.
Hence, whereas in real-time evolution the phase of each eigenstate is oscillating with the frequency $E_n$, in the imaginary-time evolution each eigenstate is damped exponentially according to its energy difference with the ground-state energy.
Given that the initial state has an overlap with the ground state, its imaginary time evolution converges to the ground state with an exponential rate given by the spectral gap $E_1 - E_0$ (or to a superposition of lowest energy states if the ground state is degenerate). See Appendix~\ref{app:ite} for the derivation and details on the convergence.

Another application area of imaginary-time evolution is the preparation of thermal states of a Hamiltonian $H$ at temperature $T > 0$. 
The thermal state, also known as Gibbs-state, is a mixed state defined as,
\begin{equation}
    \rho_G(\beta) = \frac{e^{-\beta H}}{Z(\beta)} = \frac{1}{Z(\beta)} \sum_{n\geq 0} e^{-\beta E_n} \ket{\lambda_n}\bra{\lambda_n},
\end{equation}
where $\beta = (k_\mathrm{B} T)^{-1}$ is called the inverse temperature, $k_\mathrm{B}$ is the Boltzmann constant and $Z(\beta) = \mathrm{Tr}(\exp(-\beta H)) = \sum_n \exp(-\beta E_n)$ is the partition function. 

\begin{figure}[t]
    \centering
    \[
    \begin{array}{c}
    \Qcircuit @C=1em @R=.7em {
        \lstick{\ket{0_A}} & \gate{H} & \ctrl{1} & \multigate{5}{e^{-\frac{\beta}{2} (H_A \otimes \mathbb{I}_B)}} & \qw & \qw & \\
        \lstick{\ket{0_B}} & \qw & \targ & \ghost{e^{-\frac{\beta}{2} (H_A \otimes \mathbb{I}_B)}} & \meter & & \dstick{~~~~~~\rho_G} \\
        \lstick{\vdots} & & & & & &  \\
         & & & & & & \\
        \lstick{\ket{0_A}} & \gate{H} & \ctrl{1} & \ghost{e^{-\frac{\beta}{2} (H_A \otimes \mathbb{I}_B)}} & \qw & \qw & \\
        \lstick{\ket{0_B}} & \qw & \targ & \ghost{e^{-\frac{\beta}{2} (H_A \otimes \mathbb{I}_B)}} & \meter \gategroup{1}{6}{5}{6}{1em}{\}} & & 
    }
    \end{array}
    \]
    \caption[Thermal state preparation]{Schematics of thermal state preparation with imaginary-time evolution. Pairs of qubits in system $A$ and $B$ are initially prepared in the Bell state $(\ket{0_A 0_B} + \ket{1_A 1_B}) / \sqrt{2}$. The imaginary time evolution is performed on the extended system for time $\beta/ 2$, such that after tracing out system $B$, system $A$ is in state $\rho_G$. 
    Note that the time evolution operation here includes normalization.}
    \label{fig:gibbs-state-prep}
\end{figure}
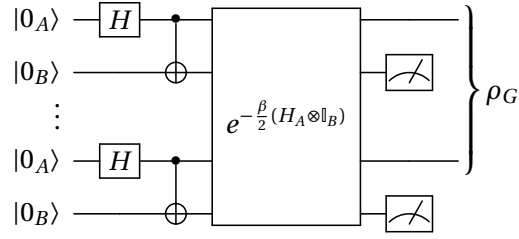

The thermal state can be explicitly prepared with the setup presented in Fig.~\ref{fig:gibbs-state-prep} by leveraging a second, auxiliary system~\cite{zoufal_qbm_2021}, as shown explicitly in Appendix~\ref{app:gibbsprep_proof}.
If, however, only thermal averages of the form
\begin{equation}
    \braket{A}_\mathrm{th} = \mathrm{Tr}\left(\rho_G A\right),
\end{equation}
are required, the secondary system can be avoided using techniques such as QMETTS~\cite{motta_determining_2020}. 
There, imaginary-time evolution is used as subroutine which is explained in detail in Chapter~\ref{chap:dual}.

\subsubsection{Variational projection}

Equations of motion for the parameter dynamics in the imaginary-time evolution can be derived analogously to the real-time case.
McLachlan's VP for variational quantum imaginary-time evolution (VarQITE) yields an update rule given by
\begin{equation}
    g(\vec\theta) \dot{\vec\theta} = \vec b^{(I)}(\vec\theta),
\end{equation}
where 
\begin{equation}
    b^{(I)}_k(\vec\theta) = -\mathrm{Re}\left(\braket{\partial_k \phi(\vec\theta)|H|\phi(\vec\theta)}\right),
\end{equation}
which is related to the energy gradient as $\vec b^{(I)}(\vec\theta) = -\vec\nabla E(\vec\theta) / 2$. Changing in-between real- and imaginary-time evolution, thus, only requires computing a different right-hand side in the equations of motion.
The Dirac-Frenkel VP yields a similar update rule shown in Appendix~\ref{app:vp}.
The TDVP forms an exception as, for circuits with only real parameters, it leads to purely imaginary parameter derivatives and does not apply to imaginary-time evolution~\cite{yuan_varqte_2019}.

\subsubsection{Relation to quantum natural gradients}

McLachlan's VP for imaginary-time evolution is closely related to the QNG.
Integrated with a Forward-Euler method, the parameters in the imaginary-time evolution are updated as
\begin{equation}
    \begin{aligned}
    \vec\theta(\tau + \Delta_t) &= \vec\theta(\tau) + \Delta_t \dot{\vec\theta}(\tau) \\
    &= \vec\theta(\tau) - \frac{\Delta_t}{2} g^{-1}(\vec\theta(\tau)) \vec\nabla E(\vec\theta(\tau)),
    \end{aligned}
\end{equation}
where we assume that $g^{-1}$ exists---otherwise the inversion is to be understood as solving a regularized system of equations with the right-hand side $\vec\nabla E$.
This update rule coincides with the QNG iteration in Eq.~\eqref{eq:qng},
\begin{equation*}
    \vec\theta^{(k+1)} = \vec\theta^{(k)} - \eta_k g^{-1}\left(\vec\theta^{(k)}\right) \vec\nabla \mathcal{L}\left(\vec\theta^{(k)}\right),
\end{equation*}
if the loss function is the system energy, $\mathcal{L}\equiv E$, the learning rate is fixed at $\eta_k = \Delta_t/2$, and by identifying $\vec\theta^{(k)} = \vec\theta(k\Delta_t)$.

Since the imaginary-time evolution is guaranteed to converge to the ground state of a system, given that the initial state has sufficient overlap with it, the connection to VarQITE is a strong motivation for the convergence of QNG. 
It is important to highlight, however, that the convergence is not guaranteed. Instead, it hinges on a sufficiently small timestep and a suitable ansatz, such that the imaginary-time dynamics are correctly captured with McLachlan's VP. As, in practice, these conditions are typically not certain, the connection to imaginary-time evolution generally only provides a motivation, rather than a convergence guarantee.

\section{Ansatz selection}\label{sec:ansatz}

The ansatz is a key component impacting the solution quality and efficiency of a VQA.
A suitable ansatz for a given problem should be
\begin{itemize}
    \item device compatible: this usually implies shallow circuits with local operations,
    \item sufficiently expressive: the ansatz should accurately approximate the target state, 
    \item trainable: finding the optimal parameters is achievable in a reasonable amount of time. 
\end{itemize}
These conditions present multiple trade-offs to navigate. For example, a circuit preparing a highly expressive ansatz, which has a high chance to represent the target state, is often not near-term compatible as it typically has a large gate count. In addition, a highly expressive ansatz with a large number of parameters can be difficult to train, due to accessing sizable parts of the exponentially large Hilbert space~\cite{mcclean_barren_2018, cerezo_cost-induced_2021, holmes_expressibility_2022, ragone_unified_2023}. However, if the circuit becomes too shallow it might not require a quantum computer to train it~\cite{tindall_classical_2023, begusic_classical_2023, cerezo_simulable_2023}.

Finding compact, problem-specific, and trainable ansatz states is an active field of research.
We can broadly categorize a spectrum of states ranging from \emph{physically-motivated}, which are built on insights into the physical model, to \emph{hardware-efficient}, which are aim at fully leveraging the available device.
In principle, the first category is desirable as it may allow for theoretical guarantees and might be easier to train, however, these circuits can be costly to implement on current devices to a high fidelity.
In addition to the ansatz types discussed below, recent works reduce the requirements of a single quantum processor by embedding a quantum circuit model into a classical model~\cite{rossmannek_embedding_2023, barison_embedding_2023} or by re-combining the result of multiple smaller quantum circuits to model a larger system~\cite{motta_vqe-ef_2023, gentinetta_cutting_2023}.

\subsection{Ansatz types}

We now review different families of ansatz circuits, ranging from physically-motivated to hardware-efficient, and discuss their trade-offs.

\subsubsection{Symmetry-preserving}

Symmetries of a physical system describe its invariance under certain transformations.
They are integral tools to understanding a system, as symmetries give rise to conserved quantities in both the real- and imaginary-time evolution of the system.
More formally, if an observable $C$ commutes with the Hamiltonian, $[H, C] = 0$, it describes a conservation law.
This can be seen by considering an eigenstate $\ket{\lambda}$ of $C$ with eigenvalue $\lambda$, which is preserved under real-time evolution,
\begin{equation}
    \braket{\lambda(t)|C|\lambda(t)} = \braket{\lambda|e^{iHt}Ce^{-iHt}|\lambda} = \braket{\lambda|C|\lambda} = \lambda,
\end{equation}
and imaginary-time evolution
\begin{equation}
     \braket{\lambda(\tau)|C|\lambda(\tau)} =\frac{\braket{\lambda|e^{-H\tau}Ce^{-H\tau}|\lambda}}{\braket{\lambda|e^{-2H\tau}|\lambda}} =  \frac{\braket{\lambda|e^{-2H\tau}C|\lambda}}{\braket{\lambda|e^{-2H\tau}|\lambda}} =\frac{\braket{\lambda|e^{-2H\tau}|\lambda} \lambda}{\braket{\lambda|e^{-2H\tau}|\lambda}} = \lambda.
\end{equation}
This also holds if $\ket{\lambda}$ is a superposition of degenerate eigenstates, that share the same eigenvalue. 

We would now like to construct an ansatz that, like the time-evolution, preserves the expectation value. 
To this end, we define a Hermitian operator $G$ which generates states that obey the conservation law, i.e.,
\begin{equation}
    \forall \theta \in \mathbb{R}: \braket{\lambda|e^{i\theta G} C e^{-i\theta G}|\lambda} = \lambda.
\end{equation}
This is always satisfied for generators that commute with the conserved observable, $[C, G] = 0$. 
An ansatz consisting of such building blocks allows to explore the space spanned by the state $\ket{\lambda}$, while preserving the eigenvalue $\lambda$. The energy, however, is not generally preserved, which makes these ansatz suitable for ground-state preparation. 

\begin{figure}[thbp]
    \centering
    \includegraphics[width=0.6\textwidth]{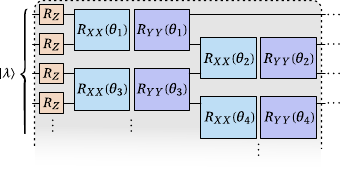}
    \caption[Symmetry-preserving ansatz]{An ansatz preserving the number of $\ket{1}$ in the initial state $\ket{\lambda}$. The $R_Z$ gates are freely parameterized, but adjacent $R_{XX}R_{YY}$ pairs are coupled with the same parameter, as indicated. As long as they are applied in pairs, the $R_{XX}R_{YY}$ gates can be applied on any qubit pair. The dashed box can be repeated several times.}
    \label{fig:symmpreserving}
\end{figure}

We illustrate this concept for the conservation of the particle number which is, for example, given in the Heisenberg model.
The Hamiltonian, given by
\begin{equation}
    H = \sum_{\braket{jk}} (X_j X_k + Y_j Y_k + Z_j Z_k) + \sum_{j} Z_j,
\end{equation}
commutes with the number of spin (or qubit) excitations, defined by
\begin{equation}
    N = \sum_{j} \frac{1 + Z_j}{2}.
\end{equation}
The eigenstates of this operator are the computational basis states, where the eigenvalue equals the number of qubits in the $\ket{1}$ state.
The initial state $\ket{\lambda}$ could, thus, be chosen as superposition of all basis states that have the desired number of qubit excitations.
A possible generator for this symmetry is the Swap operation between any two qubits, which can be written as
\begin{equation}
    \text{Swap}_{jk} = \frac{X_j X_k + Y_j Y_k}{2},
\end{equation}
and we have $[N, \text{Swap}_{jk}] = 0$.
We can now write a particle-conserving ansatz by using interactions of the form
\begin{equation}
    e^{-i\theta \text{Swap}} = R_{XX}(\theta) R_{YY}(\theta) \equiv 
    \begin{pmatrix}
        1 & 0 & 0 & 0 \\
        0 & \cos(\theta) & -i\sin(\theta) & 0 \\
        0 & -i\sin(\theta) & \cos(\theta) & 0 \\
        0 & 0 & 0 & 1
    \end{pmatrix},
\end{equation}
which coherently change amplitudes of $\ket{01} \leftrightarrow \ket{10}$. See Fig.~\ref{fig:symmpreserving} for an example structure, which additionally uses $R_Z$ rotations that do not change the distribution of $\ket{0}$ and $\ket{1}$ states.
Circuits with these building blocks have been used to find the ground state of the J1-J2 model~\cite{feulner_j1j2_2022} or of chemistry Hamiltonians in the particle-hole picture~\cite{barkoutsos_ph_2018}, which are both particle-preserving.
Another notable example is the unitary coupled-cluster (UCC) ansatz, which has successfully been employed in a variety of molecular ground state calculations. See Appendix~\ref{app:uccsd} for more detail.

Implementing a state that obeys the system's symmetries is a powerful ability and provides an intuition about the optimization process.
This approach is used, for example, to model molecular~\cite{gard_symmetry_2020, setia_symmetry_2020, barison_embedding_2023}, fermionic~\cite{anselmetti_symmetry_2021} or strongly correlated systems~\cite{sun_ansatz_2023}, but also finds applications in quantum machine learning to construct equivariant models~\cite{meyer_symmetries_2023, le_invariant_2023}.
However, since these models are not developed with the hardware constraints in mind, they typically lead to deep circuits once compiled and may be infeasible to execute on current devices. 
Hence, we now turn to models which allow for shallower circuits.

\subsubsection{Hamiltonian variations}

Motivated by the UCC ansatz, which leverages exponentials of terms in the Hamiltonian, the Hamiltonian variational ansatz (HVA) has been proposed~\cite{wecker_hva_2015}.
This model attempts to address shortcomings of the UCC approach, such as the large circuit depth, and in particular focuses on modelling physical systems, where interactions are local.
For a Hamiltonian decomposed in non-commuting summands, given by
\begin{equation}
    H = \sum_{k=1}^K H_k,
\end{equation}
with $[H_k, H_{k'}] \neq 0$ if $k \neq k'$, the HVA is defined as 
\begin{equation}
    \ket{\phi(\vec\theta)} = \left(\prod_{j=1}^r \prod_{k=1}^K e^{-i\theta_{k + jK} H_k} \right)\ket{\psi_0},
\end{equation}
where $\ket{\psi_0}$ is an un-parameterized initial state and we call $r$ the number of repetitions.

This circuit model is commonly used to prepare ground states of spin models, such as the Ising~\cite{schindler_ansatz_2022}, Heisenberg~\cite{ho_hva_2019, bosse_hvakagome_2021, kattemolle_hvakagome_2022} or Hubbard~\cite{wecker_hva_2015, stanisic_observing_2022} model.
To reduce the circuit depth, adaptive models have been proposed~\cite{gomes_adaptvarqite_2021}, analogously to the adaptive UCC methods.
Other variations include term-wise diagonalizations of Hamiltonian terms via Fourier transformations to reduce the circuit depth~\cite{babbush_lowdepth_2018, choquette_qoca_2021} or purposefully breaking symmetries of the system~\cite{vogt_symmetry_2020, choquette_qoca_2021}.

The HVA is a general model closely connected to different ansatz classes.
Connecting to the previous symmetry-preserving circuits, the HVA also respects a symmetry if it is satisfied by each individual term $H_k$. 
By choosing such terms $H_k$ and preparing an initial state $\ket{\psi_0}$ with a desired quantum number, the HVA provides a flexible, symmetry-preserving circuit.

\subsubsection{Annealing}

Another strategy for the HVA initialization is based on quantum adiabatic annealing (QAA).
In QAA, we first prepare the ground state $\ket{\psi_0}$ of a simple Hamiltonian $H_0$ and then slowly transition to a target Hamiltonian $H_1$, whose ground state we are interested in.
This is done by evolving under the time-dependent Hamiltonian
\begin{equation}
    H(t) = \left(1 - \alpha\left(\frac{t}{T}\right)\right) H_0 + \alpha\left(\frac{t}{T}\right) H_1,
\end{equation}
with a bijective function $\alpha: [0, 1] \rightarrow [0, 1]$ with boundary conditions $\alpha(0) = 0$ and $\alpha(1) = 1$.
The adiabatic theorem states that for a large enough annealing times $T$, which depends on the eigenvalue spectrum of $H(t)$, and an overlap of the ground states of $H_0$ and $H_1$, the system will remain in the ground state of $H(t)$ at all times. Thus, the evolution will end up in the ground state of $H_1$ at $t=T$~\cite{born_adiabatic_1928}.

The time evolution under $H(t)$ could be implemented using a time-series expansion
\begin{equation}
    \ket{\psi(T)} = \lim_{N \rightarrow \infty} \left(\prod_{j=1}^{N} e^{-i \frac{T}{N} H\left(\frac{jT}{N}\right)} \right)\ket{\psi_0},
\end{equation}
where each exponential is again decomposed using a Suzuki-Trotter expansion.
While explicitly performing the time evolution of $H(t)$ can be costly, we can leverage the structure of this Hamiltonian to define an annealing ansatz as
\begin{equation}\label{eq:qaoa}
    \ket{\phi(\vec\beta, \vec\gamma)} = \left( \prod_{j=1}^r e^{-i\beta_j H_0} e^{-i\gamma_j H_1} \right) \ket{\psi_0},
\end{equation}
for parameters $\vec{\beta}, \vec{\gamma} \in \mathbb{R}^r$.
For finite $r < \infty$ this implements a first-order approximation to the annealing process, but this allows to control the circuit depth and find a heuristic annealing schedule $(\vec\beta, \vec\gamma)$.
The annealing ansatz is related to HVA, as it equals the HVA of the Hamiltonian $H = H_0 + H_1$ with the initial state $\ket{\psi_0}$. 

This ansatz has initially been introduced for combinatorial optimization as quantum approximation optimization algorithm (QAOA)~\cite{farhi_quantum_2014}.
A common choice for the initial Hamiltonian, also referred to as ``mixer'', is $H_0 = -\sum_{j=1}^n X_j$, whose ground state is the product state $\ket{+}^{\otimes n}$ which is easily prepared on a quantum computer using a layer of Hadamard gates. 
Since the Hamiltonian in a combinatorial optimization is diagonal in the computational basis, its ground state is a single basis state and is guaranteed to have an (exponentially small) overlap with $\ket{\psi_0}$.
Several improvements to this initial version exist, including increasing the initial overlap and to speed up the convergence~\cite{egger_warm-starting_2021}, providing annealing-based initial values for the parameter values $(\vec\beta, \vec\gamma)$~\cite{sack_quantum_2021} or introducing additional Hamiltonian terms to allow for faster annealing~\cite{chai_sqaoa_2021, wurtz_counterdiabatic_2022}.

\subsubsection{Hardware-efficient}

\begin{figure}[th]
    \centering
    \includegraphics[width=\textwidth]{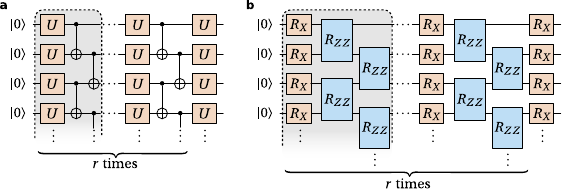}
    \caption[Hardware-efficient ansatze]{Hardware-efficient ansatze, where the layers (marked grey) are repeated $r$ times, plus a final set of rotations. (a) An ansatz based on parameterized single qubit unitaries $U$ and shallow, unparameterized entangling layers. The unitary $U$ can contain multiple parameters, e.g. could be chosen as $U = R_Y(\theta_j)$ or $U = R_Y(\theta_j) R_Z(\theta_{j+1})$. (b) A brickwall based on $R_X$ and $R_{ZZ}$ rotations, e.g. for a transverse-field Ising model. Up to the last $R_X$ layer, this circuit is equivalent to the HVA if the Ising model topology is a line.}
    \label{fig:hw_efficient}
\end{figure}

\noindent
Instead of constructing a problem-dependent ansatz, hardware-efficient ansatz circuits aim at fully leveraging the available operational space of a given quantum chip.
These circuits are typically characterized by a large number of parameters in a densely packed set of gates that are natively implemented in the hardware.
A commonly used, generic structure is shown in Fig.~\ref{fig:hw_efficient}(a), where layers of parameterized, single-qubit rotation gates alternate with layers of local entangling gates.
This ansatz has been employed for a variety of applications, ranging from molecular ground state search~\cite{kandala_vqe_2017, shee_qubit-efficient_2022, barison_embedding_2023} to quantum machine learning~\cite{havlicek_supervised_2019}.

It might still be advantageous to endow some intuition about the problem onto such a generic circuit. 
For instance, if the wave function is known to only have real amplitudes, this property can be imposed on the hardware efficient circuit by allowing only gates which are represented by real matrices, such as $R_Y$ rotation gates and CX or CZ entangling gates.
Another common approach for quantum dynamics, dubbed the ``brickwall'' ansatz and shown in Fig.~\ref{fig:hw_efficient}(b), is building the circuit out of gates appearing in a Suzuki-Trotter expansion~\cite{berthusen_rqd_2022, benedetti_timeevo_2021}.
Note, however, that, unlike the time evolution, this ansatz does generally preserve the energy.
In contrast to physically-motivated ansatze, hardware-efficient circuits are not classified by problem-specific properties but are rather categorized by  generic properties, such as their expressibility or entangling capacity~\cite{sim_expressibility_2019}.

\subsection{Trainability}\label{sec:training_vqa}

In the previous subsection we discussed the ability of circuits to represent the solution and the gate-cost of implementing them on a quantum computer. 
The third crucial property of an ansatz is whether it can be efficiently optimized, meaning that we can find optimal parameters $\vec\theta^*$ of the model within a reasonable amount of time.
Note, that we here do not focus on the absolute quality of the final solution, but only care about the best possible approximation within the ansatz space.
Intuitively, the more information about a problem is available to build a specially tailored circuit that covers only a small region of the Hilbert space containing the solution, the easier it should be to optimize. There is evidence supporting this intuition in some cases for, e.g., HVA~\cite{wiersema_hva_2020} or annealing-based circuits~\cite{zhou_qaoa_2020, mele_barren_2022}.
For problem-agnostic circuits, such as hardware-efficient ansatze, on the other hand, finding the optimum can be significantly more challenging~\cite{larocca_diagnosing_2022}.

While classical neural networks can be engineered to perform well, even if little to no structure of the problem is known, these techniques do not necessarily transfer to the optimization of quantum circuits.
In fact, a range of results show that, under certain conditions, training loss functions of randomly initialized circuits can become exponentially difficult due to their gradients vanishing exponentially in system size---a phenomenon dubbed barren plateaus~\cite{mcclean_barren_2018, ortiz_entanglement-induced_2021, cerezo_cost-induced_2021, wang_noise-induced_2021}.
For example, $n$-qubit circuits consisting of $\mathcal{O}(\text{poly}(n))$ layers of random 2-qubit gates form a 2-design, i.e., they match Haar-random unitaries up to the second moment.
They, therefore, prepare uniformly random states on the $(2^{n}-1)$-dimensional unit surface and both variance and the values of their gradients decay exponentially in $n$~\cite{mcclean_barren_2018}.
For expectation values of global observables this property can, in fact, already be shown for circuits of constant depth~\cite{cerezo_cost-induced_2021}.
Another cause of barren plateaus are random states that exhibit a large amount of entanglement across the qubits, followed by tracing out the majority of them. In the worst case, this leads to the observation of a maximally mixed state, which does not contain sufficient information for efficient training.
Beyond the choice of the circuit, the presence of certain types of noise, such a depolarizing channels, on the quantum device can cause the gradients of a loss function to vanish~\cite{wang_noise-induced_2021}. 
Non-unital noise, such as amplitude damping, which drives the qubit to the $\ket{0}$ state, on the other hand, may help to avoid this effect~\cite{fefferman_nonunital_2023}.
Strategies to mitigate barren plateaus induced by circuit properties attempt to violate conditions like the random initialization or deep circuits.
Techniques that initialize the circuit as identity~\cite{grant_bp-initialization_2019}, perform a layer-wise training or prune operations~\cite{sim_pruning_2021}, or split deep circuits into a combination shallower copies~\cite{tuysuz_splitting_2023} can be shown to avoid vanishing gradients.
When using such approaches, however, it is crucial to assess whether the resulting model still requires a quantum computer to simulate. By sufficient restriction, in particular by limiting the number of operations, the ansatz might become classically simulable~\cite{kechedzhi_effective_2023, cerezo_simulable_2023}.

\section{Quantum circuit gradients}\label{sec:gradients}

As we have seen in the previous sections, evaluating gradients of quantum circuits is a core subroutine in VQA. Beyond enabling gradient-based optimizations of the loss function, it is also a required component of variational algorithms for time evolution.
In this section we introduce a variety of methods to compute quantum circuit gradients and the QGT.

The gradient elements of the VQA loss function of Eq.~\eqref{eq:vqa_loss} are
\begin{equation}
    \partial_j \mathcal{L}(\vec\theta) = \sum_k f_k'(\braket{\phi_k(\vec\theta)| O_k |\phi_k(\vec\theta)}) \partial_j \braket{\phi_k(\vec\theta)|O_k|\phi_k(\vec\theta)}.
\end{equation}
Since the derivative $f_k'$ of the post-processing function $f_k$ is classical and efficiently computable, e.g., by means of automatic differentiation, we here focus on the evaluation of the non-classical part of the loss function: the gradient of the expectation values of form
\begin{equation}
    \ell(\vec\theta) = \braket{\phi(\vec\theta)|O|\phi(\vec\theta)},
\end{equation}
where we drop the subscript $k$ of observable and state for convenience.

The gradient of an expectation value can also be written as
\begin{equation}
    \partial_j\ell(\vec\theta) = 
    \braket{\partial_j \phi(\vec\theta)|O|\phi(\vec\theta)}
    + \braket{\phi(\vec\theta)|O|\partial_j\phi(\vec\theta)}
    = 2 \mathrm{Re}\big(\braket{\partial_j \phi(\vec\theta)|O|\phi(\vec\theta)}\big),
\end{equation}
hence, for optimization of a VQA loss function, we are mainly concerned with the \emph{real} part of the expectation gradient. 
However, other applications, such as variational real-time evolution, also require the \emph{imaginary} part,
$\mathrm{Im}(\braket{\partial_j \phi(\vec\theta)|O|\phi(\vec\theta)})$.

\subsubsection{Product and chain rule}

Some gradient methods in this chapter, such as the parameter-shift rule in Section~\ref{sec:paramshift} or the linear combination of unitaries in Section~\ref{sec:lcu}, assume only unique parameters in the quantum circuit gates.
In these cases, the product and chain rules can be used to reduce the circuits to unique, single-parameter expressions.

Assume a gate $U$ acting on an input state $\ket{\psi}$, where the gate angle $f(\theta)$ is a function of the target parameter.
Then, by the chain rule, the derivative is
\begin{equation}
    \frac{\partial}{\partial\theta} U(f(\theta))\ket{\psi} = f'(\theta) \frac{\partial}{\partial\omega} U(\omega) \ket{\psi},
\end{equation}
with the substituted variable $\omega = f(\theta)$.
If a parameter impacts two distinct gates $U_1$ and $U_2$, the derivative is given by the product rule as
\begin{equation}
    \frac{\partial}{\partial\theta} \big(U_1(\theta)U_2(\theta)\big)\ket{\psi} = 
    \left(\frac{\partial}{\partial\theta} U_1(\theta)\right)U_2(\theta)\ket{\psi} +
    U_1(\theta) \left(\frac{\partial}{\partial\theta} U_2(\theta)\right)\ket{\psi}.
\end{equation}

\subsubsection{Illustrative example}

Assume the parameterized quantum state
\begin{equation}
    \ket{\phi(\theta)} = R_X(\theta) \ket{0} \equiv 
    \begin{pmatrix}
        \cos(\theta/2) \\
        -i\sin(\theta/2)
    \end{pmatrix}.
\end{equation}
Its derivative is
\begin{equation}
        \Ket{\frac{\partial}{\partial\theta} \phi(\theta)} = \frac{-i}{2} R_X(\theta) X \ket{0} 
        \equiv \begin{pmatrix}
            \frac{\partial}{\partial \theta} \cos(\theta/2) \\ 
            -i \frac{\partial}{\partial \theta} \sin(\theta/2)
        \end{pmatrix}
        = \frac{-i}{2} \begin{pmatrix}
            -i\sin(\theta/2) \\ 
            \cos(\theta/2)
        \end{pmatrix}.
\end{equation}
Note, that the derivative is not a normalized state anymore.
The expectation value gradient for the observable $Z$ is, then,
\begin{equation}
    \begin{aligned}
    \frac{\partial}{\partial\theta} \braket{\phi(\theta)|Z|\phi(\theta)} 
    &= 2 \mathrm{Re}\left( \Braket{\frac{\partial}{\partial\theta} \phi(\theta) \Big| Z \Big| \phi(\theta)} \right) \\
    &= -2\cos\left(\frac{\theta}{2}\right)\sin\left(\frac{\theta}{2}\right).
    \end{aligned}
\end{equation}

\subsection{Finite differences}\label{sec:findiff}

Finite differences (FD) approximate the gradient of the loss function with a difference quotient over a perturbation $\epsilon \in \mathbb{R}_{>0}$.
By Taylor expansion of $\ell(\vec\theta)$, the central FD quotient is given by
\begin{equation}
    \partial_j \ell(\vec\theta) = \frac{\ell(\vec\theta + \epsilon \vec{e}_j) - \ell(\vec\theta - \epsilon \vec{e}_j)}{2\epsilon} + \mathcal{O}(\epsilon^2),
\end{equation}
with the $j$th unit vector $\vec{e}_j \in \mathbb{R}^d$ and where the quadratic error scaling holds if $\ell$ is three times differentiable. Forward or backward difference quotients only perturb one term of the difference and have a larger error scaling of $\mathcal{O}(\epsilon)$. However, if the unperturbed point
$\ell(\vec\theta)$ can be re-used, these formulas might be advantageous despite the increased error.

A considerable benefit of FD gradients, in comparison to analytic methods described below, is that they are agnostic to the inner structure of the function.
No product or chain rules have to be taken into account and no derivatives of the post-processing functions $f'_k$ must be computed. 
Only evaluations of the loss function are required, which makes this method simple to implement.
This simplicity, however, comes at the cost of a bias in terms of the perturbation $\epsilon$.
Another limitation of FD gradients is that no decompositions are known to compute the imaginary part of an expectation value gradient, $\mathrm{Im}(\braket{\partial_j\phi(\vec\theta)|O|\phi(\vec\theta)})$.

\subsubsection{Higher-order derivatives}

Higher-order derivatives are obtained by nesting the finite difference shifts. Let $\{j_1, \dots, j_m\} \subset \mathbb{N}$ be an index set, then the finite difference approximation of the $m$th derivative of the loss function is
\begin{equation}\label{eq:fd_higher_order_derivative}
    \begin{aligned}
    \left(\prod_{k = 1}^{m}\partial_{j_k}\right) \ell(\vec\theta) &\approx 
    \left(\prod_{k = 2}^{m} \partial_{j_k}\right) \frac{\ell(\vec\theta + \epsilon \vec{e}_{j_1}) - \ell(\vec\theta - \epsilon \vec{e}_{j_1})}{2\epsilon}  \\
    &\approx ... \\
    &\approx \frac{1}{(2\epsilon)^m}
    \sum_{\vec\alpha\in\{1, -1\}^m} (-1)^{\#\vec\alpha=-1} \ell\left(\vec\theta + \epsilon\sum_{k=1}^{m} \alpha_k \vec e_{j_k}\right),
    \end{aligned}
\end{equation}
where $\#\vec\alpha = -1$ counts the occurrences of $-1$ in the vector $\vec\alpha$.
Each element of an order-$m$ derivative tensor requires $2^m$ loss function evaluations. Since derivative tensors are symmetric, the total number of evaluations is~\cite{mari_higherorder_2021}
\begin{equation}
    2^m \binom{m + d - 1}{m}.
\end{equation}

\subsubsection{Quantum geometric tensor}

Evaluating the QGT with a FD method requires to formulate it as derivative of a function.
This can be achieved for the real part of the QGT, which is the Hessian of the fidelity, i.e.,
\begin{equation}
    g_{jk}(\vec\theta) = -\frac{1}{2} \partial_j\partial_k F(\vec\theta', \vec\theta)\big\vert_{\vec\theta' = \vec\theta},
\end{equation}
with $F(\vec\theta', \vec\theta) = |\braket{\phi(\vec\theta') | \phi(\vec\theta)}|^2$~\cite{gacon_qnspsa_2021}.
Fixing $\vec\theta'$ for the differentiation and then plugging in the value $\vec\theta' = \vec\theta$, we obtain
\begin{equation}
    \begin{aligned}
    g_{jk}(\vec\theta) \approx -\frac{1}{8\epsilon^2} \Big(
        &F(\vec\theta, \vec\theta + \epsilon(\vec e_j + \vec e_k))
        - F(\vec\theta, \vec\theta + \epsilon(\vec e_j - \vec e_k)) \\
        &- F(\vec\theta, \vec\theta + \epsilon(\vec e_k - \vec e_j))
        + F(\vec\theta, \vec\theta - \epsilon(\vec e_j + \vec e_k))
    \Big).
    \end{aligned}
\end{equation}
Each off-diagonal element must be evaluated with all four summands, but the diagonal requires only two fidelity evaluations since we can leverage $F(\vec\theta, \vec\theta + \epsilon(\vec e_j - \vec e_j)) = F(\vec\theta, \vec\theta) = 1$.
Taking into account that $g$ is symmetric, a finite difference approximation requires a total of 
\begin{equation}
    4\frac{d(d - 1)}{2} + 2d = 2d^2,
\end{equation}
fidelity evaluations.
As for the expectation value gradient, there are no known forms to compute the imaginary part of the QGT with FD.

\subsection{Parameter-shift rule}\label{sec:paramshift}

The parameter-shift rule (PSR) is an analytic gradient formula, which means that in the absence of measurement and device noise provides the exact gradient values.
This technique requires the same number of expectation value evaluations as FD, that is $2d$, but does not have a bias. 
However, analytic formulas require knowledge of the internal structure of the loss function and, possibly, the implementation of chain and product rules to correctly compute the gradients.

Deriving analytic formulas requires a unitary representation where each parameter acts on a single unitary only, i.e.,
\begin{equation}\label{eq:ansatz_state}
    \ket{\phi(\vec\theta)} = V_d U_d(\theta_d) \cdots V_1 U_1(\theta_1) V_0\ket{0},
\end{equation}
where $U_j$ is a unitary depending on a single parameter $\theta_j$ and $V_j$ are non-parameterized operations, all acting on $n$ qubits.
More generic cases, where parameters are repeated or are wrapped in a function, can be reduced to this form by means
of the product and chain rules for differentiation. 
Since the $U_j$ are unitary, they can be written in terms of a Hermitian matrix $G_j \in \mathbb{C}^{2^n \times 2^n}$, called the generator, such that $U_j(\theta_j) = \exp(-i\theta_j G_j)$. 

For generators with two distinct eigenvalues $\pm \lambda$, $\lambda \in \mathbb{R}$ the parameter-shift rule 
states that the gradients of the loss function are
\begin{equation}\label{eq:paramshift}
    \partial_j \ell(\vec\theta) = \frac{\lambda}{2} \left(\ell(\vec\theta + s\vec e_j) - \ell(\vec\theta - s\vec e_j)\right),
\end{equation}
with the shift $s = \pi / (4\lambda)$~\cite{schuld_evaluating_2019}. The proof of this equation is given in Appendix~\ref{app:paramshift_proof}.
For generators that satisfy $G_j^2 = \mathbb{I}$, such as the Pauli operators, the parameter shift rule can also be evaluated as 
\begin{equation}
    \partial_j\ell(\vec\theta) = \frac{\ell(\vec\theta + s\vec e_j) - \ell(\vec\theta - s\vec e_j)}{\sin(s)},
\end{equation}
where the shift $s$ can be chosen to be any $s\neq k\pi,~k \in \mathbb{Z}$~\cite{mari_higherorder_2021}.
Furthermore, the parameter-shift rule can be generalized to unitaries whose generator has $R$ equidistantly spaced eigenvalues, which requires evaluating $2R$ shifted circuits~\cite{wierichs_general-paramshift_2022}.
In practice, however, the condition on the generator spectrum is usually not a restriction.
General unitaries can then be decomposed into basis gate sets where only single-qubit gates are parameterized, such as $\{R_Z(\theta), \sqrt{X}, X, \text{CX}\}$. The only parameterized gate in this basis, $R_Z$, has two distinct eigenvalues $\pm 1$, which allows to directly apply Eq.~\eqref{eq:paramshift}.

\subsubsection{Higher-order derivatives}

Higher-order derivatives are obtained by nesting the parameter shifts, analogous to the FD case in Eq.~\eqref{eq:fd_higher_order_derivative}.

\subsubsection{Quantum geometric tensor}

Since the QGT is the Hessian of the Fubini-Study metric, it can be computed with a second-order derivative formula. 
The result is similar to the FD equations, but with a different shift and global coefficient.
We have
\begin{equation}
    \begin{aligned}
    g_{jk}(\vec\theta) = -\frac{\lambda^2}{8} \Big(
        &F(\vec\theta, \vec\theta + s(\vec e_j + \vec e_k))
        - F(\vec\theta, \vec\theta + s(\vec e_j - \vec e_k)) \\
        &- F(\vec\theta, \vec\theta + s(\vec e_k - \vec e_j))
        + F(\vec\theta, \vec\theta - s(\vec e_j + \vec e_k))
    \Big),
    \end{aligned}
\end{equation}
with the appropriate parameter-shift $s$ and generator eigenvalues $\pm\lambda$.
As in the FD case, the PSR only allows to evaluate the real part of the QGT.

\subsubsection{Examples}
\begin{figure}[htbp]
    \centering
    \[
        \begin{array}{c}
        \Qcircuit @C=1em @R=.7em {
            & \gate{\sqrt{Y}} & \ctrl{1} & \qw & \ctrl{1} & \gate{\sqrt{Y}} & \qw \\ 
            & \gate{S^\dagger} & \targ & \gate{R_Y(\theta/2)} & \targ & \gate{S} & \qw \\
        }
        \end{array}.
    \]
    \caption[$XY$ gate decomposition]{The $XY$ gate can be decomposed to use only a single Pauli rotation gate.}
    \label{fig:decompose_xy}
\end{figure}
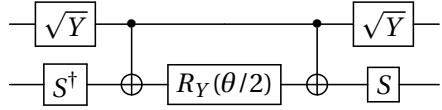

\begin{itemize}
    \item An important family of gates with eigenvalues $\pm \lambda$ are the $n$-qubit Pauli rotations,
        \begin{equation}
            U(\theta) = R_P(\theta), P \in \{X, Y, Z\}^{\otimes n},
        \end{equation}
        which have the generator $G = P / 2$ with eigenvalues $\pm 1/2$ and shift $s = \pi/2$. Note that this does not
        include generators that are \emph{sums} of Paulis.
    \item The controlled phase gate is a 2-qubit, non-Pauli rotation that can be implemented natively on superconducting qubits hardware~\cite{glaser_exp11_2023}.
        It is given by
        \begin{equation}
            \text{CP}(\theta) = e^{-i\theta \ket{11}\bra{11}},
        \end{equation}
        and it's generator $G = \ket{11}\bra{11}$ has eigenvalues $(0, 1)$. 
        By applying a global phase of $\exp(i\theta / 2)$ the generator becomes a diagonal matrix with entries $(1/2, 1/2, 1/2, -1/2)$
        which allows to apply the parameter-shift rule with eigenvalues $\pm 1/2$.
    \item The parameter-shift rule is not directly applicable to the partial swap, or $XY$ gate \cite{schuch_xygate_2003},
        \begin{equation}
            XY(\theta) = e^{i\frac{\theta}{2}(XX + YY)},
        \end{equation}
        whose generator has three distinct eigenvalues, $(2, 0, -2)$.
        This gate can be differentiated either with a generalized parameter-shift rule for arbitrary generators~\cite{wierichs_general-paramshift_2022}
        or by decomposing it to Pauli rotations, such as shown in Fig.~\ref{fig:decompose_xy}.
\end{itemize}

\subsection{Linear combination of unitaries}\label{sec:lcu}

\begin{figure}[htbp]
    \centering
    \[
    \begin{array}{c}
        \Qcircuit @C=1em @R=.7em {
            \lstick{\ket{0}} & \gate{H} & \gate{P(\alpha)} & \ctrlo{1} & \ctrl{1} & \multimeasureD{1}{X \otimes O} \\
            \lstick{\ket{\psi}} & \qw & \qw & \gate{A} & \gate{B} & \ghost{X\otimes O} 
        }
    \end{array}
    \]
    \caption[Linear combination of unitaries]{Circuit to evaluate the real ($\alpha=0$) or imaginary ($\alpha=\pi/2$) parts of $\braket{\psi|A^\dagger O B|\psi}$. The phase gate $P$ acts as $P(\alpha)\ket{1} = e^{i\alpha}\ket{1}$ and leaves the $\ket{0}$ state unchanged.}
    \label{fig:innerprod_lcu_circuit}
\end{figure}
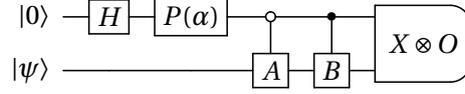

Quantum circuit gradients are terms of the form $\braket{\partial_j \phi(\vec\theta)|O|\phi(\vec\theta)}$, which, if we have access to the unitary $U(\vec\theta)$ preparing the state $\ket{\phi(\vec\theta)} = U(\vec\theta)\ket{0}$, can be directly written as
\begin{equation}\label{eq:unitary_gradient}
    \big\langle0\big|\left(\partial_j U(\vec\theta)\right)^\dagger O  U(\vec\theta)\big|0\big\rangle.
\end{equation}
Instead of decomposing this gradient as sum of expectation values, which we leveraged in the PSR, expressions with different unitaries $A$ and $B$ applied on the left and right of the observable $O$, that is $\braket{\psi|A^\dagger O B|\psi}$, can be evaluated as a conditioned superposition, as shown in Fig.~\ref{fig:innerprod_lcu_circuit}.
Before the measurement, the state prepared in the registers is
\begin{equation}
    \frac{1}{\sqrt{2}} \left( \ket{0} A\ket{\psi} + e^{i\alpha} \ket{1} B\ket{\psi} \right),
\end{equation}
and evaluating the Pauli-$X$ operator on the auxiliary qubit and the target observable on the 
state register yields either the real or imaginary part of $\braket{\psi|A^\dagger O B|\psi}$, depending on the value of the phase shift $\alpha$.

To apply this decomposition for quantum circuit gradient evaluation, the term $\partial_j U$ must be expressed as linear combination of unitaries (LCU), as both $A$ and $B$ must be unitary operators. Since $\partial_j U$ is represented by a complex matrix, it can always be written as
\begin{equation}
    \partial_j U(\vec\theta) = c \left[\left(A_R(\vec\theta) + A_R^\dagger(\vec\theta)\right) + i\left(A_I(\vec\theta) + A_I^\dagger(\vec\theta)\right)\right],
\end{equation}
for unitary matrices $A_R$ and $A_I$, based on the real and imaginary parts of $\partial_j U$, and a coefficient $c \in \mathbb{R}$~\cite{schuld_evaluating_2019}.
Each of the four expectation values can then be derived using the circuit shown in Fig.~\ref{fig:innerprod_lcu_circuit}.
The circuit gradients for general unitaries $U(\vec\theta)$ can thus be computed by plugging the decomposition of the derivative as LCU into Eq.~\eqref{eq:unitary_gradient} and evaluating each of the four expectation values with the circuit of Fig.~\ref{fig:innerprod_lcu_circuit}.

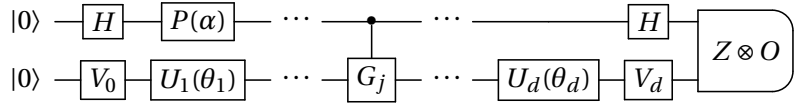
\begin{figure}[t]
    \[
    \begin{array}{c}
    \Qcircuit @C=1em @R=.7em {
        \lstick{\ket{0}} & \gate{H} & \gate{P(\alpha)} & \qw & \cdots & & \ctrl{1} & \qw & \cdots & & \qw & \gate{H} & \multimeasureD{1}{Z \otimes O} \\
        \lstick{\ket{0}} & \gate{V_0} & \gate{U_1(\theta_1)} & \qw & \cdots & & \gate{G_j} & \qw & \cdots & & \gate{U_d(\theta_d)} & \gate{V_d} &  \ghost{Z \otimes O}
    }
    \end{array}
    \]
    \caption[Gradients via a linear combination of unitaries]{The quantum circuit to compute expectation value gradients with a LCU. For $\alpha=0$ the real part of the gradient
             is obtained and for $\alpha=\pi/2$ the imaginary part.
             If $O = I^{\otimes n}$, all gates after the controlled-$G_j$ gate can be omitted.
             }
    \label{fig:lcu_circuit}
\end{figure}

In contrast to the PSR, the LCU method is more generally applicable, as it does not require generators with only two distinct eigenvalues.
In addition, LCU methods enable the evaluation of the \emph{imaginary} part of a gradient, which is a required subroutine for variational real-time evolution.
The LCU circuit for general unitaries needs both open and closed controls of these unitaries, which can be  costly to implement, especially on near-term hardware. 
Luckily, the circuit can be significantly simplified for ansatze of the form in Eq.~\eqref{eq:ansatz_state}.
The simplification is based on the fact that it suffices to control the difference between the original unitary and differentiated unitary, which allows to evaluate the gradients using the circuit in Fig.~\ref{fig:lcu_circuit} with phase shifts $\alpha$ chosen as
\begin{equation}
    \begin{aligned}
        \alpha = 0 &\rightarrow \mathrm{Im}\big(\braket{\partial_j \phi(\vec\theta) | O |\phi(\vec\theta)}\big), \\
        \alpha = \frac{\pi}{2} &\rightarrow -\mathrm{Re}\big(\braket{\partial_j \phi(\vec\theta) | O |\phi(\vec\theta)}\big),
    \end{aligned}
\end{equation}
see Appendix~\ref{app:lcu_proof} for the derivation. If the generator $G_j$ is not unitary, it has to be decomposed into a linear combination of unitaries.
Besides the circuit shown here, there exist different implementations of the LCU circuit. Alternatively, for example, explicitly evaluate the $Z$-expectation value~\cite{schuld_evaluating_2019} or use different auxiliary qubit observables~\cite{zoufal_generative_2021}, but they implement the same operations.

\subsubsection{Higher-order derivatives}

\begin{figure}[htp]
    \[
    \begin{array}{c}
    \Qcircuit @C=1em @R=.7em {
        \lstick{\ket{0}} & \gate{H} & \gate{P(\alpha_1)} & \qw & \cdots & & \ctrl{2} & \qw & \cdots & & \qw & \qw & \cdots & & \qw & \ctrl{1} & \gate{H} & \multimeasureD{2}{Z \otimes Z \otimes O} \\
        \lstick{\ket{0}} & \gate{H} & \gate{P(\alpha_2)} & \qw & \cdots & & \qw & \qw & \cdots & & \ctrl{1} & \qw & \cdots & & \gate{H} & \control\qw & \qw & \ghost{Z \otimes Z \otimes O} \\
        \lstick{\ket{0}} & \gate{V_0} & \gate{U_1(\theta_1)} & \qw & \cdots & & \gate{G_{j_1}} & \qw & \cdots & & \gate{G_{j_2}} & \qw & \cdots &  & \gate{U_d(\theta_d)} & \gate{V_d} & \qw & \ghost{Z \otimes Z \otimes O} 
    }
    \end{array}
    \]
    \caption[Hessians via a linear combination of unitaries]{The quantum circuit to compute second-order derivatives of an expectation value.}
    \label{fig:lcu_hessian_circuit}
\end{figure}
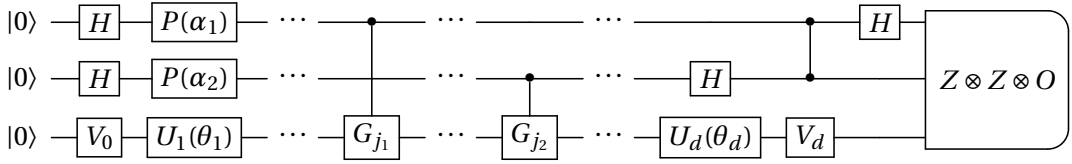

Higher-order derivatives are obtained with a similar LCU technique, which, however, requires a linear number of auxiliary qubits.
For example, the quantum circuit to evaluate second-order derivatives is shown in Fig.~\ref{fig:lcu_hessian_circuit}, where the real part of the Hessian
is obtained for the phase shifts $\alpha_1 = \alpha_2 = 0$. See Refs.~\cite{dallaire-demers_qgan_2018, zoufal_generative_2021} for more details.

\subsubsection{Quantum geometric tensor}

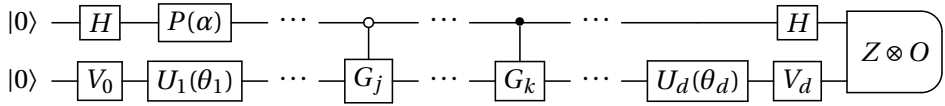
\begin{figure}[htp]
    \[
    \begin{array}{c}
    \Qcircuit @C=1em @R=.7em {
        \lstick{\ket{0}} & \gate{H} & \gate{P(\alpha)} & \qw & \cdots & & \ctrlo{1} & \qw & \cdots & & \ctrl{1} & \qw & \cdots & & \qw & \gate{H} & \multimeasureD{1}{Z \otimes O} \\
        \lstick{\ket{0}} & \gate{V_0} & \gate{U_1(\theta_1)} &\qw & \cdots & & \gate{G_j} & \qw & \cdots & & \gate{G_k} & \qw & \cdots & & \gate{U_d(\theta_d)} & \gate{V_d} &  \ghost{Z \otimes O}
    }
    \end{array}
    \]
    \caption[QGT via a linear combination of unitaries]{The LCU circuit to evaluate real or imaginary parts of $\braket{\partial_j \phi(\vec\theta)|O|\partial_k\phi(\vec\theta)}$, where we assume $j \leq k$. If $O = I^{\otimes n}$, all gates after the controlled-$G_k$ gate are obsolete, as explained in Appendix~\ref{app:lcu_proof}.
    }
    \label{fig:lcu_qgt_circuit}
\end{figure}

With the LCU technique, both the real and imaginary parts of the QGT can be evaluated directly in the form of Eq.~\eqref{eq:qgt}.
The second part of the QGT terms, $\braket{\partial_j\phi(\vec\theta)|\phi(\vec\theta)}\braket{\phi(\vec\theta)|\partial_j\phi(\vec\theta)}$, also referred to as the phase-fix as it arises by taking into account global phase of the variational state, can be evaluated with the circuit in Fig.~\ref{fig:lcu_circuit} and using the identity as observable, $O = I^{\otimes n}$.
Remember that both the real and imaginary part have to be evaluated as 
\begin{equation}
    \braket{\partial_j\phi(\vec\theta)|\phi(\vec\theta)} = \mathrm{Re}(\braket{\partial_j\phi(\vec\theta)|\phi(\vec\theta)}) + i\mathrm{Im}(\braket{\partial_j\phi(\vec\theta)|\phi(\vec\theta)}).
\end{equation}
To obtain the first part of the QGT, the LCU circuit must be extended to prepare a superposition of the two gradient states $\ket{\partial_j \phi(\vec\theta)}$ and $\ket{\partial_k \phi(\vec\theta}$ instead of a gradient state and the unmodified state.
This is achieved by adding an open-control on the generator $G_j$ as shown in Fig.~\ref{fig:lcu_qgt_circuit}.
Leveraging the fact that the QGT is self-adjoint, the evaluation requires $2d$ LCU circuits for the phase fix and $(d + 1) d / 2$ circuits for the first part.

\subsubsection{Examples}

\begin{figure}[htbp]
    \centering
    \[
        \begin{array}{c}
        \Qcircuit @C=1em @R=.7em {
            \lstick{\ket{0}} & \gate{H} & \ctrl{1} & \gate{H} & \multimeasureD{1}{Z \otimes Z} \\
            \lstick{\ket{0}} & \gate{R_X(\theta)} & \gate{X} & \qw & \ghost{Z \otimes Z}
        }
        \end{array}
    \]
    \caption[LCU derivative of the $R_X$ gate]{Derivative of the $R_X$. Note that the order of the $\mathrm{CX}$ and $R_X$ gates does not matter as they commute.}
    \label{fig:rx_derivative}
\end{figure}
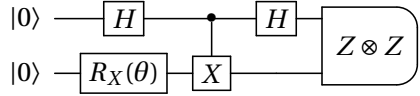

\begin{itemize}
    \item The generator of $R_X(\theta) = e^{-i\theta X/2}$ is $X/2$. The gradient of the $Z$-expectation value can, thus, be written as
    \begin{equation}
        \frac{i}{2} \braket{0|(R_X(\theta) X)^\dagger Z R_X(\theta)|0}.
    \end{equation}
    This allows to use the conditioned superposition of Fig.~\ref{fig:innerprod_lcu_circuit} in it's simplified form, and we show the circuit to evaluate the real part of the gradient in Fig.~\ref{fig:rx_derivative}.
    To obtain the gradient, we have to evaluate the expectation value of the circuit and finally multiply with a coefficient of $i/2$.
    \item The generator of the $XY$ gate is a sum of Paulis, $G = XX + YY$. While this gradient cannot be directly evaluated with the PSR, by linearity of the expectation value,
    the gradient of $\ket{\phi(\theta)} = XY(\theta)\ket{0}$ can be evaluated with the LCU method by expanding the generator into a linear combination of unitaries as 
    \begin{equation}
    \Ket{\frac{\partial}{\partial \theta} \phi(\theta)} = XY(\theta) XX \ket{0} + XY(\theta) YY \ket{0}.
    \end{equation}
    Plugging this into Eq.~\eqref{eq:lcu} yields four expectation values to compute for the gradient.
\end{itemize}

\subsection{Stochastic approximation of gradients}\label{sec:spsa}

All gradient methods discussed so far require the evaluation of $\mathcal{O}(d)$ circuits per gradient evaluation.
For optimization of the loss function $\ell$ on noisy hardware, where circuit evaluations come with a non-negligible overhead and results are subject to sampling error, using
analytic formulas is therefore not necessarily the best option.
Instead, it is often sufficient to construct a stochastic estimator of gradients at a cheaper cost.
The previously introduced SPSA technique~\cite{spall_spsa_1988,spall_overview_1998} allows to draw gradient samples at a $\mathcal{O}(1)$ cost by simultaneously performing a FD estimation on \emph{all} parameter dimensions. 
The sample is defined as
\begin{equation}
    \widehat{\vec\nabla \ell} = \frac{\ell(\vec\theta + \epsilon\vec\Delta) - \ell(\vec\theta - \epsilon\vec\Delta)}{2\epsilon} \vec\Delta^{-1}, 
\end{equation}
where $\epsilon \in \mathbb{R}_{>0}$ is the perturbation magnitude, $\vec\Delta \in \mathbb{R}^d$ is a random perturbation direction and $\vec\Delta^{-1}$ is an element-wise inverse.

Under certain conditions, detailed e.g. in Ref.~\cite{spall_spsa_1988}, it can be shown that this estimator is unbiased up to second order in $\epsilon$, i.e.,
\begin{equation}
    \mathbb{E}\left[ \widehat{\vec\nabla\ell} \right] = \vec\nabla\ell(\vec\theta) + \mathcal{O}(\epsilon^2),
\end{equation}
see Appendix~\ref{app:spsa_unbiased} for the proof.
The conditions include that $\ell$ is differentiable thrice and the elements of $\vec\Delta$ are independent, identically distributed (i.i.d.), symmetrically distributed around 0 and the inverse moments $\mathbb{E}[|\Delta_j|^{-1}]$ exist. A frequently-used distribution fulfilling these criteria is the Bernoulli distribution over the set $\{1, -1\}$, each with probability $1/2$.
Note, that some common continuous distributions, such as the normal distribution or uniform distribution, do not fulfill the existence of inverse moments.

\subsubsection{Higher-order derivatives}

Higher-order gradients can be estimated by nesting the perturbations, similar to the finite difference gradients in Section~\ref{sec:findiff}.
Samples of the Hessian, for example, are obtained as
\begin{equation}\label{eq:spsa_hessian}
    \hat H = \widehat{\vec\nabla\vec\nabla^\top \ell} = \frac{\delta\ell}{4 \epsilon^2} \left(\frac{1}{2 \vec\Delta (\vec\Delta')^\top} + \frac{1}{2 \vec\Delta'\vec\Delta^\top}\right),
\end{equation}
where the matrix in the denominator is understood as element-wise inverse, and
\begin{equation}
\delta\ell = \ell(\vec\theta + \epsilon\vec\Delta + \epsilon\vec\Delta') - \ell(\vec\theta + \epsilon\vec\Delta - \epsilon\vec\Delta') - \ell(\vec\theta - \epsilon\vec\Delta + \epsilon\vec\Delta') + \ell(\vec\theta - \epsilon\vec\Delta - \epsilon\vec\Delta')
\end{equation}
is a measure for the curvature at the point $\vec\theta$~\cite{spall_accelerated_1997}. 
The Hessian estimator reflects the symmetry of the Hessian due to the sum of outer products of the perturbation directions $\vec\Delta$ and $\vec\Delta'$.
This is an unbiased estimator of the Hessian up to first order in $\epsilon$, $\mathbb{E}[\hat H] = H + \mathcal{O}(\epsilon)$, as shown in Appendix~\ref{app:2spsa_unbiased}.

\subsubsection{Quantum geometric tensor}

Since the real part of the QGT is the Hessian of the Fubini-Study metric, it can be approximated with Eq.~\eqref{eq:spsa_hessian}.
Concretely, we have
\begin{equation}
    \hat g = -\frac{\delta F}{8\epsilon^2} \left(\frac{1}{2 \vec\Delta(\vec\Delta')^\top} + \frac{1}{2 \vec\Delta'\vec\Delta^\top}\right),
\end{equation}
with
\begin{equation}
    \delta F = 
F(\vec\theta, \vec\theta + \epsilon\vec\Delta + \epsilon\vec\Delta') - F(\vec\theta, \vec\theta + \epsilon\vec\Delta - \epsilon\vec\Delta') -  F(\vec\theta, \vec\theta - \epsilon\vec\Delta + \epsilon\vec\Delta') + F(\vec\theta, \vec\theta - \epsilon\vec\Delta - \epsilon\vec\Delta'),
\end{equation}
and $F(\vec\theta, \vec\theta') = |\braket{\phi(\vec\theta')|\phi(\vec\theta)}|^2$~\cite{gacon_qnspsa_2021}.
This estimator requires only four evaluations of the fidelity, independent of the number of parameters, and is further discussed in Chapter~\ref{chap:qnspsa}.

\newpage
\subsection{Efficient classical quantum circuit gradients}\label{sec:revgrad}

\paper{This section is based on the co-authored article "Efficient calculation of gradients in classical simulations of variational quantum algorithms" by Jones and Gacon, available on arXiv:2009.02823.}

\noindent
The first step in implementing variational workflows on a quantum computer is typically a small-scale, classical simulation of the algorithm. 
In absence of sampling errors and device noise, the simulation allows the algorithm to be verified for correctness in a controlled environment.
This importance of classical simulation of quantum algorithms is reflected by the growing effort in building high-performance quantum circuit simulators~\cite{kelly_qcgpu_2018, smelyanskiy_qhipster_2016, guerreschi_qhipster_2020, luo_yaojl_2020, jones_quest_2019, cuquantum, Qiskit}.
State-vector simulations allow to directly construct the exponentially large vectors $\ket{\phi(\vec\theta)}$ and $\ket{\partial_j \phi(\vec\theta)}$ needed for the gradient evaluations of form $\braket{\partial_j \phi(\vec\theta)| O |\phi(\vec\theta)}$. For an ansatz of form Eq.~\eqref{eq:ansatz_state} with $d$ parameters, calculating the state vector requires $\mathcal{O}(d)$ gate operations, which equals a total of $\mathcal{O}(d^2)$ operations for the full gradient vector. 
However, quantum circuits are typically constructed of unitary (and thus reversible) operations, which allows for asymptotically faster gradient simulations using only $\mathcal{O}(d)$ gate operations and two state-vectors in memory.

The gradient of the expectation value for states of form Eq.~\eqref{eq:ansatz_state} is
\begin{equation}
        \braket{\partial_j \phi(\vec\theta) | O | \phi(\vec\theta)} 
        = \underbrace{\bra{0} V_0^\dagger U_1(\theta_1) \cdots V_{j-1}^\dagger}_{=\bra{\lambda^{(j)}}}
        iG_j 
        \underbrace{U_j^\dagger(\theta_j) \cdots U_d^\dagger(\theta_d) V_d^\dagger O \ket{\phi(\vec\theta)}}_{=\ket{\zeta^{(j)}}}.
\end{equation}
The two intermediate states $\ket{\lambda^{(j)}}$ and $\ket{\zeta^{(j)}}$ can be evaluated recursively using a constant number of gate operations each
\begin{equation}
    \begin{aligned}
        &\ket{\phi^{(j - 1)}} = U_{j-1}^\dagger(\theta_{j-1}) V_{j-1}^\dagger \ket{\phi^{(j)}}, \\
        &\ket{\zeta^{(j - 1)}} = U_{j-1}^\dagger(\theta_{j-1}) V_{j-1}^\dagger \ket{\zeta^{(j)}}.
    \end{aligned}
\end{equation}
The recursion is initialized with values $\ket{\lambda^{(d)}} = \ket{\phi(\vec\theta)}$ and $\ket{\zeta^{(d)}} = O\ket{\lambda^{(d)}}$ and computes the  gradient entries as $\braket{\partial_j \phi(\vec\theta) | O | \phi(\vec\theta)} = i\braket{\lambda^{(j)}|G_j|\zeta^{(j)}}$, starting from $k=d$ and iterating to $k=1$. 
The steps are explicitly stated in Algorithm~\ref{alg:recursive_gradient}.
The total number of gate applications on the state-vector, counting $U_k, V_k$ and $O$ is $7d + 2$, plus $d$ state-vector inner products.

\begin{algorithm}[thbp]
    \caption{Recursive quantum circuit gradient calculation using $\mathcal{O}(d)$ gate operations.}
    \label{alg:recursive_gradient}
    \begin{algorithmic}
    \REQUIRE Circuit operations $\{V_d, U_d, \dots, V_0\}$, operator $O$, generators $G_j$ of unitary $U_j$
    \ENSURE Gradient $\vec g \in \mathbb{C}^d$ with $g_j = \braket{\partial_j\phi(\vec\theta)| O |\phi(\vec\theta)}$
        \STATE $\ket{\lambda} = V_d U_d(\theta_d) \cdots U_1(\theta_1) V_0 \ket{0}$ \COMMENT{state-vector allocation, $\mathcal{O}(d)$ gates}
        \STATE $\ket{\zeta} = O\ket{\phi}$ \COMMENT{state-vector allocation, operator multiplication}
        \FOR{$k = d ... 1$}
            \STATE $\ket{\lambda} = V_k^\dagger U_k^\dagger(\theta_k)\ket{\lambda}$ \COMMENT{$\mathcal{O}(1)$ gates}
            \STATE $\ket{\zeta} = V_k^\dagger U_k^\dagger(\theta_k)\ket{\zeta}$ \COMMENT{$\mathcal{O}(1)$ gate}
            \STATE $g_k = i\braket{\lambda|G_k|\zeta}$ \COMMENT{operator multiplication, vector-product}
        \ENDFOR
        \RETURN $\vec g$
        \end{algorithmic}
\end{algorithm}

\begin{figure}[thbp]
    \centering
    \includegraphics[width=\textwidth]{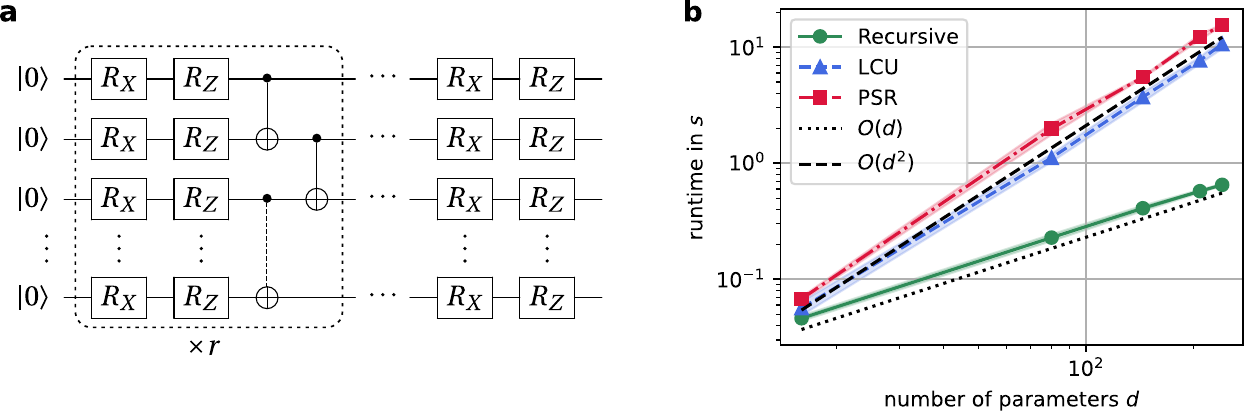}
    \caption[Benchmark circuit for classical gradient evaluation]{(a) Structure of the benchmark circuit. The dotted box is repeated $r$ times and each Pauli gate has a unique parameter,
    leading to a total of $2n(r+1)$ parameters for $n$ qubits.
    (b) Runtimes of the fast, recursive gradient calculation compared to simulations of LCU or PSR gradient methods for $n=8$ qubits and varying repetitions $r$. The observable for the expectation value is $O = \sum_{j} Z_j$.}
    \label{fig:revgrad_benchmark_circuit}
\end{figure}

In Fig.~\ref{fig:revgrad_benchmark_circuit} we show a benchmark for the gradient calculation runtime of a hardware-efficient ansatz circuit as a function of the number of parameters. We compare the reverse mode to direct circuit simulation of the parameter-shift rule and the LCU method. 
The benchmark clearly shows the improved scaling of the reverse gradient calculation, which only applies $\mathcal{O}(d)$ unitaries, whereas the other techniques scale as $\mathcal{O}(d^2)$. 
The parameter-shift rule has an offset from the LCU method, as it requires simulating two circuits per gradient and LCU only a single circuit.

The same asymptotic cost as the algorithm introduced above can be derived via automatic differentiation (AD)~\cite{bartholomew_autodiff_2000, maclaurin_reversible_2015, margossian_autodiff_2019}, which is ubiquitous in classical machine learning frameworks~\cite{baydin_autodiff_2017}.
The ``reverse mode'' of AD allows to evaluate gradients of a function with $M$ computational nodes and $d$ parameters using $\mathcal{O}(d)$ operations and storing $\mathcal{O}(M)$ intermediate states.
If applied to state-vectors or a quantum system each intermediate state corresponds to an exponentially large vector, which makes the standard form of the reverse AD mode unsuitable for quantum circuit gradients.
However, by leveraging the fact that quantum circuit operations are reversible and applying techniques developed for reversible neural networks~\cite{gomez_reversible_2017}, the memory overhead can be reduced to storing only a constant number of state vectors~\cite{luo_yaojl_2020}.
See Appendix~\ref{app:ad} for an introduction to AD and a detailed discussion of the application to quantum circuit gradients.

\subsubsection{Quantum geometric tensor}

The recursive gradient technique can be extended to computing the QGT at an asymptotic cost of $\mathcal{O}(d^2)$ gate operations, instead of $\mathcal{O}(d^3)$~\cite{jones_qgt_2020}.
We consider the QGT, as defined in Eq.~\eqref{eq:qgt}, separately from its phase fix, which allows it to be formulated as
\begin{equation}
    G = S - \vec{p} \vec{p}^\dagger,
\end{equation}
with $S \in \mathbb{C}^{d \times d}$ and $\vec p \in \mathbb{C}^d$, defined as
\begin{equation}
    \begin{aligned}
        S_{j, k}(\vec\theta) &= \braket{\partial_j \phi(\vec\theta)| \partial_k \phi(\vec\theta)}, \\
        p_{j}(\vec\theta) &= \braket{\partial_j \phi(\vec\theta)| \phi(\vec\theta)}.
    \end{aligned}
\end{equation}
The phase fix vector $\vec p$ corresponds to an expectation value gradient with the identity as observable and could be efficiently evaluated with the sweeping gradients of the previous section. 
However, with the identity as observable, gates in the bra- and ket-vectors cancel and we can compute $\vec p$ as a by-product of the symmetric $S$ matrix.

To compute $S$ in $\mathcal{O}(d^2)$ gate operations and with a constant number of state vectors in memory, we expand the definition under the assumption of quantum states in the form of Eq.~\eqref{eq:ansatz_state}.
We obtain
\begin{equation}
    \begin{aligned}
        S_{j, k}(\vec\theta) &= 
        \braket{0|V_0^\dagger \cdots (i G_j U_j(\theta_j))^\dagger V_j^\dagger \cdots U_d^\dagger(\theta_d) V_d^\dagger V_d U_d(\theta_d) \cdots V_k (-i G_k U_k(\theta_k)) \cdots V_0|0} \\
        &= 
        \underbrace{\bra{0} V_0^\dagger \cdots}_{=\bra{\lambda^{(j)}}} G_j^\dagger \underbrace{U_j^\dagger(\theta_j) V_j^\dagger \cdots U_{k+1}^\dagger(\theta_{k+1}) V_{k+1}^\dagger G_k V_{k-1} U_{k-1}(\theta_{k-1}) \cdots V_0 \ket{0}}_{=\ket{\eta^{(k:j)}}},
    \end{aligned}
\end{equation}
and
\begin{equation}
    p_j(\vec\theta) = i\bra{0} V_0^\dagger \cdots U_{j-1}^\dagger(\theta_{j-1}) V_{j-1}^\dagger G_j V_{j-1} U_{j-1}(\theta_{j-1}) \cdots V_0 \ket{0}.
\end{equation}
In each iteration, starting from $j=1$, we begin by computing the diagonal element at $k=j$.
Then, the rows are iterated upwards by transforming
\begin{equation}
    \begin{aligned}
    \ket{\lambda^{(j)}} &\rightarrow \ket{\lambda^{(j-1)}} = V_{j-1}^\dagger U_{j-1}^\dagger(\theta_{j-1}) \ket{\lambda^{(j)}}, \\
    \ket{\eta^{(k:j)}} &\rightarrow \ket{\eta^{(k:j-1)}} = V_{j-1}^\dagger U_{j-1}^\dagger(\theta_{j-1}) \ket{\eta^{(k:j)}},
    \end{aligned}
\end{equation}
and evaluate the inner product $S_{j, k} = \braket{\lambda^{(j)}|G_j^\dagger|\eta^{(k:j)}}$.
At the end of the row iteration we can simply evaluate $p_k = i\braket{\eta^{(k:1)}|\lambda^{(1)}}$. The full algorithm is formulated in Algorithm~\ref{alg:recursive_qng}.

\begin{algorithm}[th]
    \caption{Recursive quantum circuit gradient calculation using $\mathcal{O}(d)$ gate operations.}
    \label{alg:recursive_qng}
    \begin{algorithmic}
    \REQUIRE Circuit operations $\{V_d, U_d, \dots, V_0\}$, generators $G_j$ of unitary $U_j$
    \ENSURE QGT $G \in \mathbb{C}^{d \times d}$ 
        \STATE $\ket{\psi} = V_0\ket{0}$ \COMMENT{Add the initial unitary, next $V_j U_j$ are applied in pairs only}
        \FOR{$j = 1 ... d$}
            \STATE $\ket{\lambda} = \ket{\psi}$  \COMMENT{copy of $\ket{\psi}$ we modify}
            \STATE $\ket{\eta} = G_j \ket{\lambda}$ \COMMENT{note this is not a normalized state as $G_j$ may not be unitary}
            \STATE $S_{j, j} = \braket{\eta|\eta}$  \COMMENT{special case for the diagonal}
            \FOR{$k = j - 1 ... 1$}
                \STATE $\ket{\lambda} = V_k^\dagger U_k^\dagger(\theta_k) \ket{\lambda}$ \COMMENT{shift the unitaries leftwards}
                \STATE $\ket{\lambda} = V_k^\dagger U_k^\dagger(\theta_k) \ket{\lambda}$ 
                \STATE $S_{j, k} = \braket{\lambda|G_k^\dagger|\eta}$
            \ENDFOR
            \STATE $p_j = i\braket{\eta|\lambda}$
            \STATE $\ket{\psi} = V_j U_j(\theta_j)\ket{\psi}$
        \ENDFOR

        \STATE $G = L + L_\text{triu}^H$  \COMMENT{$L_\text{triu}$ is upper triangular, without diagonal}
        \STATE $G = G + \vec p \vec{p}^H$  \COMMENT{outer vector product}
        \RETURN $G$
        \end{algorithmic}
\end{algorithm}

\section{Conclusion}

This chapter reviewed the concept of variational quantum algorithms (VQAs) with focus on ground-state preparation and quantum time evolution.
In the context of preparing ground states we discussed how loss functions of VQAs can be optimized using gradient descent, including some of its shortcomings. 
In response, we reviewed key concepts for the remainder of this thesis, including simultaneous perturbation stochastic approximation (SPSA), which allows to construct gradient samples at a constant cost, independent of the number of parameters in the circuit, and the quantum natural gradient, which adjusts the gradient with respect to the model sensitivity by leveraging the quantum geometric tensor (QGT).
The QGT also plays a central role in the variational formulation of quantum time evolution, which we derived using McLachlan's variational principle.
Finally, we introduced a range of techniques to evaluate the quantum circuit gradients and the QGT, and we provide an overview in Table~\ref{tab:gradients}.

\begin{table}[htp]
    \centering
    \begin{tabular}{|l|c|c|c|c|}
        \hline
        & FD & PSR & LCU & SPSA \\
        \hline\hline
        gradient cost & $\mathcal{O}(d)$ & $\mathcal{O}(d)$ & $\mathcal{O}(d) + 1$ qubit & $2^*$ \\
        QGT cost & $\mathcal{O}(d^2)$ & $\mathcal{O}(d^2)$ & $\mathcal{O}(d^2) + 2$ qubits & $4^*$ \\
        bias & $\mathcal{O}(\epsilon^2)$ & 0 & 0 & $\mathcal{O}(\epsilon^2)$ \\
        complex parts & no & no & yes & no \\
        \hline
    \end{tabular}
    \caption[Comparison of gradient and QGT methods]{Comparison of gradient and QGT calculations via finite difference (FD), parameter-shift rules (PSR), linear combination of unitaries (LCU) and SPSA.
    Among all methods, LCU is the only one which allows to compute both real and imaginary parts of the gradient and QGT. The perturbation for FD and SPSA should be chosen as $\epsilon \geq N^{-1/2}$, where $N$ is the number of measurements to evaluate the expectation value.
    $({}^*)$ Cost per gradient or QGT sample.
    }
    \label{tab:gradients}
\end{table}

An important open question for VQAs is the construction of suitable ansatz circuits, such that the algorithm is able to represent the solution accurately, and the solution can be efficiently determined. 
In contrast to canonical algorithms, such as product formulas for time evolution or quantum phase estimation for ground-state preparation, convergence guarantees are more challenging to provide. For variational ground-state preparation, lower bounds on the minimal energy could provide insights into the solution quality~\cite{westerheim_dualvqe_2023}. The error of a variational time evolution, could be assessed using a-posteriori error bounds~\cite{zoufal_errorbounds_2021} or Hamiltonian learning~\cite{wiebe_learning_2014}.
Another interesting question for variational time evolution is how knowledge of preserved quantities could be integrated into the variational principles, which is e.g. discussed in Ref.~\cite{hackl_varqte_2020}.

With this chapter, the foundations for the novel research in this thesis are laid out. 
The next chapter begins with the first research project with the goal to reduce the computational requirements for quantum natural gradients, by applying SPSA techniques to the calculation of the QGT.

%% file: main/ch4_qnspsa.tex
\chapter{Quantum Natural SPSA}\label{chap:qnspsa}

\summary{This chapter is based on the article "Simultaneous Perturbation Stochastic Approximation of the Quantum Fisher Information" by Gacon et al., published in Quantum \textbf{5} 567 (2021).
To remedy the prohibitive cost of computing the quantum geometric tensor, which scales quadratically with the number of parameters in the circuit, we introduce a constant-cost, unbiased stochastic estimator.
We apply this estimator to quantum natural gradient descent and show that this new method, called QN-SPSA, has the favorable properties of natural gradients while requiring significantly less resources.
}

In the previous chapters we have seen that near-term quantum computers are subject to imperfect quantum operations and short qubit decoherence times. 
Therefore, quantum algorithms require careful selection of the quantum circuits that can be reliably executed, which lead to the emergence of variational quantum algorithms as promising paradigm. 
In this chapter, we focus on the quantum natural gradient (QNG) algorithm~\cite{stokes_qng_2020} for ground-state preparation, which belongs to the family of gradient descent (GD) methods, but takes on an information-geometric point of view.
Compared to standard GD, this change comes with advantages such as invariance under re-parameterization~\cite{amari_natural_1998} and the ability to adjust the learning rate with respect to parameter sensitivity in each dimension.
As we have seen in Section~\ref{sec:varqte}, another interesting property is that QNG is closely connected to variational quantum imaginary time evolution (VarQITE)~\cite{mcardle_varqite_2019, yuan_varqte_2019}.

A significant drawback of both VarQITE and QNG is that they require the evaluation of the real part of the quantum geometric tensor (QGT) in every iteration.
For a variational ansatz $\ket{\phi(\vec\theta)}$ depending on a $d$-dimensional parameter vector $\vec\theta \in \mathbb{R}^d$, this is a real, symmetric, positive semi-definite (PSD) $d\times d$ matrix with entries
\begin{equation}\label{eq:qgt_direct}
    g_{jk}(\vec\theta) = \mathrm{Re}\left\{\braket{\partial_j\phi(\vec\theta) | \partial_k\phi(\vec\theta)} - \braket{\partial_j \phi(\vec\theta)|\phi(\vec\theta)}\braket{\phi(\vec\theta)|\partial_k\phi(\vec\theta)}\right\}.
\end{equation}
Evaluating $g(\vec\theta)$ requires evaluating a number of circuit scaling quadratically in the number of parameters $d$, which quickly becomes expensive for large number of variational parameters.
To reduce this cost, some methods propose to approximate the QGT by a diagonal or block-diagonal matrix~\cite{stokes_qng_2020}, however this approximation misses correlations in-between parameters and might fail in systems where these are strong.

\begin{figure}[thbp]
    \centering
    \includegraphics[width=\textwidth]{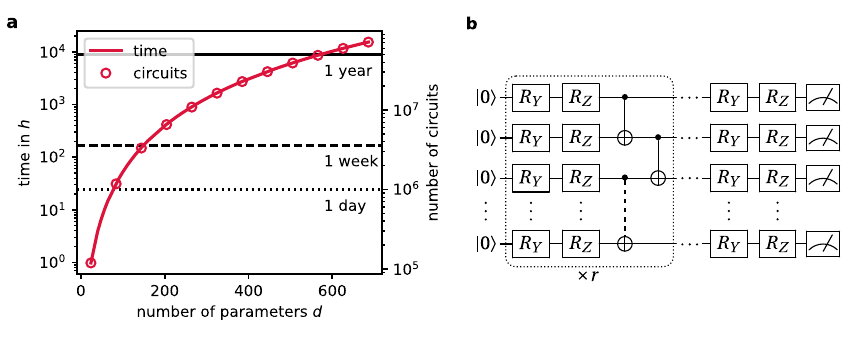}
    \caption[QNG runtime estimates]{(a) Runtime estimates to perform QNG on a superconducting qubit chip as a function of the number of parameters in the circuit. (b) The ansatz circuit used in the benchmark estimates.}
    \label{fig:qng_scaling}
\end{figure}

In Fig.~\ref{fig:qng_scaling}(a) we count the number of circuits required for a full QNG optimization and estimate the runtime on a current superconducting qubit device. 
We assume the near-term-friendly circuit structure in Fig.~\ref{fig:qng_scaling}(b), which only has next-neighbor interactions and $r = \lceil\log_2(n)\rceil$ layers for $n$ qubits. 
For 300 QNG steps, the number of circuits to execute is already of the order of $10^6$ for only 100 parameters. Assuming an optimistic $N=10^3$ measurements per circuit and idealistic execution times for gate operations and qubit reset, which is detailed in Appendix~\ref{app:qng_scaling}, we find that the runtime of the optimization already exceeds 1 week for a mere 200 parameters.
With quantum processors reaching sizes over 1000 qubits~\cite{ibm_condor_2023}, multiple hundreds of parameters are reasonable circuit dimensions but such a a system is prohibitively costly to simulate with QNG or VarQITE. 

In this chapter, we propose a technique to reduce the cost of approximating the QGT to constant in the number of parameters.
The idea is to construct stochastic estimators by first formulating $g$ as Hessian and, then, applying simultaneous perturbation stochastic approximation (SPSA)~\cite{spall_spsa_1988} techniques to draw unbiased samples using only four circuit evaluations, independent of the number of parameters.
The resulting algorithm, quantum natural SPSA (QN-SPSA), is a stochastic approximation of the QNG. We benchmark the algorithm and investigate its efficiency and convergence properties on various model systems. Then, we apply it to more practically-relevant tasks including the ground-state preparation of a molecular system on quantum hardware, a generative learning task, and black-box function optimization.

\section{Stochastic approximation of the QGT}

In Section~\ref{sec:spsa}, we have seen the SPSA algorithm, which is a stochastic approximation of gradient descent, and defines iterations as
\begin{equation}
    \vec\theta^{(k+1)} = \vec\theta^{(k)} - \eta_k \widehat{\vec\nabla\mathcal{L}}^{(k)},
\end{equation}
with learning rate $\eta_k \in \mathbb{R}_{>0}$ and the gradient sample
\begin{equation}
    \widehat{\vec\nabla\mathcal{L}}^{(k)} = \frac{\mathcal{L}\big(\vec\theta^{(k)} + \epsilon\vec\Delta\big) - \mathcal{L}\big(\vec\theta^{(k)} - \epsilon\vec\Delta\big)}{2\epsilon} \vec\Delta^{-1},
\end{equation}
for a small perturbation $\epsilon \in \mathbb{R}_{>0}$
and a random $\vec\Delta \in \mathbb{R}^d$ where each element has zero mean, $\mathbb{E}[\Delta_j] = 0$, and existing inverse moment, $\mathbb{E}[|\Delta_j^{-1}|] \leq \infty$. The inverse $\vec\Delta^{-1}$ is to be understood element-wise.
A commonly used distribution is the Bernoulli distribution over $\{1, -1\}$ for each dimension.

In contrast to a gradient descent iteration, which requires $\mathcal{O}(d)$ circuit evaluations to compute the gradient, SPSA only requires two---independent of the number of parameters $d$.
We can leverage the same technique to reduce the number of evaluations for the QNG, which has the update rule
\begin{equation}
    \vec\theta^{(k+1)} = \vec\theta^{(k)} - \eta_k g^{-1}\big(\vec\theta^{(k)}\big) \vec\nabla\mathcal{L}\big(\vec\theta^{(k)}\big).
\end{equation}
The gradient of the loss function can directly be approximated with the SPSA formula above. To extend the technique to the real part of the QGT, we formulate it as Hessian of the fidelity,
\begin{equation}
    g(\vec\theta) = -\frac{1}{2} \vec\nabla\vec\nabla_{\vec\theta}^\top F(\vec\theta', \vec\theta)\Bigg|_{\vec\theta'=\vec\theta},
\end{equation}
where $F(\vec\theta', \vec\theta) = |\braket{\phi(\vec\theta')|\phi(\vec\theta)}|^2$ is the fidelity of the variational ansatz at two different parameter points, $\vec\theta'$ and $\vec\theta$. 
See Appendix~\ref{app:qgt_formulas} for the equivalence of this QGT formulations with the previously given Eq.~\eqref{eq:qgt_direct}.
Using the formula for stochastic approximations of Hessians, see Section~\ref{sec:spsa}, we then define a sample of $g$ at parameters $\vec\theta$ as 
\begin{equation}\label{eq:qgt_sample}
    \hat g = -\frac{\delta F}{8\epsilon^2} \left(\frac{1}{2 \vec\Delta(\vec\Delta')^\top} + \frac{1}{2 \vec\Delta'\vec\Delta^\top}\right),
\end{equation}
with
\begin{equation}
    \delta F = 
F(\vec\theta, \vec\theta + \epsilon\vec\Delta + \epsilon\vec\Delta') - F(\vec\theta, \vec\theta + \epsilon\vec\Delta - \epsilon\vec\Delta') -  F(\vec\theta, \vec\theta - \epsilon\vec\Delta + \epsilon\vec\Delta') + F(\vec\theta, \vec\theta - \epsilon\vec\Delta - \epsilon\vec\Delta'),
\end{equation}
for two random directions $\vec\Delta, \vec\Delta' \in \mathbb{R}^d$.
This is an unbiased estimator of the QGT, i.e., $\mathbb{E}(\hat g) = g$, since the second-order SPSA approximation produces unbiased Hessian samples (up to $\mathcal{O}(\epsilon)$).

\begin{figure}[htp]
    \centering
    \includegraphics[width=0.5\textwidth]{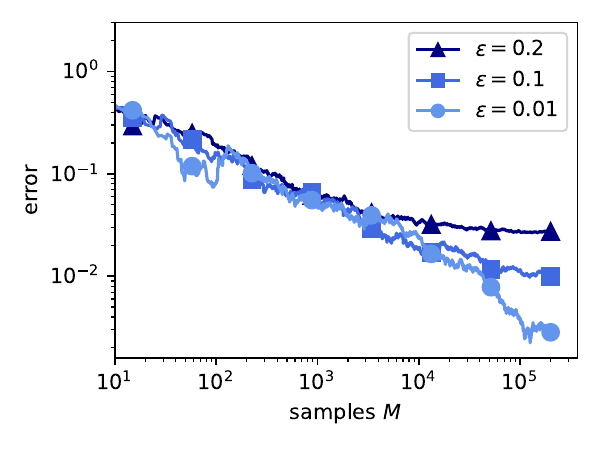}
    \caption[Convergence of the QGT estimator]{Convergence of the estimator $\hat g_M$. As discussed in Section~\ref{sec:spsa}, the SPSA estimator for the Hessian has a $\mathcal{O}(\epsilon)$ bias, which causes the error convergence to plateau.
    }
    \label{fig:qgt_sample_convergence}
\end{figure}

Individual samples $\hat g$ have rank $\leq 2$ and are a poor approximation of the full matrix $g$. In addition, the QNG requires the inversion of $g$ to compute $g^{-1}(\vec\theta)\vec\nabla\mathcal{L}$ and using a single sample $\hat g$ is numerically unstable.
The estimator could be improved by combining individually drawn samples as
\begin{equation}\label{eq:qnspsa_average}
    \hat g_M = \frac{1}{M} \sum_{m=1}^{M} \hat g_{(m)}.
\end{equation}
which converges to $g$ with an error of $\mathcal{O}(M^{-0.5})$, as shown in an example in Fig.~\ref{fig:qgt_sample_convergence}.
However, since the learning rate $\eta_k$ is typically small and the QGT does not change significantly in between QNG iterations, we can also combine the sample $\hat g^{(k)}$ at current parameters $\vec\theta^{(k)}$ with samples of previous iterations~\cite{spall_accelerated_1997}. 
The averaged estimator over all samples in the iteration is
\begin{equation}\label{eq:qnspsa_global_average}
    \bar g^{(k)} = \frac{1}{k+1} \sum_{j=0}^{k} \hat g^{(j)} = \frac{1}{k+1} \hat g^{(k)} + \frac{k}{k + 1} \bar g^{(k-1)}.
\end{equation}
As initial value we use the identity, $\hat g^{(0)} = \bar g^{(0)} = \mathbb{I}$, which equals to starting from standard, first-order SPSA.
To further improve the estimator, we next impose properties of the matrix $g$, namely that it is symmetric and PSD.
The samples are already symmetric by construction and the PSD-ness can, for example, be ensured by replacing the estimator with $\sqrt{\bar g^{(k)} \bar g^{(k)}}$,
which amounts to replacing the eigenvalues of $\bar g^{(k)}$ with their absolute values~\cite{spall_accelerated_1997}.

Finally, the QN-SPSA update step is 
\begin{equation}
    \vec\theta^{(k+1)} = \vec\theta^{(k)} - \eta_k \big(\bar g^{(k)}\big)^{-1} \widehat{\vec\nabla\mathcal{L}}^{(k)},
\end{equation}
where the PSD-version of $\bar g^{(k)}$ is used. In practice, it is crucial to compute the inverse of the QGT estimator in a numerically stable manner, e.g., by solving a linear system of equations instead and by using an appropriate regularization, such as an L-curve method~\cite{cultrera_lcurve_2020}. 
Solving for the update step is discussed in more detail in Chapter~\ref{chap:saqite}, where an accurate solution at minimal regularization is even more crucial as we are not only interested in the final, minimal energy state, but in the precise state evolution at every timestep.

\subsection{Measuring state fidelities}\label{sec:fidelities}

Computing a sample $\hat g$ requires the evaluation of the fidelity of the ansatz $\ket{\phi(\vec\theta)}$ for two different sets of parameters, $\vec\theta$ and $\vec\theta'$.
Evaluating such state overlaps is an ubiquitous task, used, among other fields, in machine learning~\cite{havlicek_supervised_2019} or chemistry~\cite{higgott_vqd_2019} applications, and there a variety of techniques available to compute the fidelity on a quantum computer.
In the following paragraphs we outline different methods to then select the optimal version for near-term quantum processors.
A comparison of resource requirements in shown in Table~\ref{tab:fidelities}.

\begin{table}[htp]
    \centering
    \begin{tabular}{|l||c|c|c|c|c|}
        \hline
         & unitary required & complex parts & depth & width \\
        \hline
         Hadamard test~\cite{cleve_algorithms_1998} & yes & yes & $2D^{(*)}$ & $n + 1$ \\ 
         swap test~\cite{barenco_swap_1997} & no & no & $D + \mathcal{O}(n)$  & $2n + 1$ \\ 
         swap in Bell basis~\cite{garcia_swap_2013} & no & no & $D + 2^{(\dagger)}$ & $2n$ \\ 
         unitary overlap~\cite{havlicek_supervised_2019} & yes & no & $2D$ & $n$ \\
         \makecell[l]{randomized\\ measurements$^{(\ddagger)}$~\cite{elben_overlap_2019}} & no & no & $D$ & $n$ \\
         \hline
    \end{tabular}
    \caption[Comparison of fidelity methods]{Overview of methods for fidelity evaluation. ${}^{(*)}$ Can be reduced to $D$ if the circuits are prepared by the same unitary. ${}^{(\dagger)}$  If pairwise connection between qubit pairs of the two states do not require additional swaps. ${}^{(\ddagger)}$ Requires $\mathcal{O}(2^n)$ measurements.}
    \label{tab:fidelities}
\end{table}

\subsubsection{Swap test}

\begin{figure}[htp]
    \centering
    \includegraphics[width=0.9\textwidth]{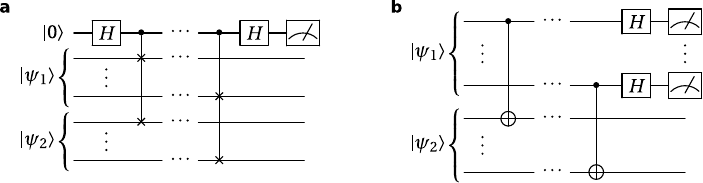}
    \caption[Swap test implementations]{(a) Swap test using controlled Swap gates. (b) Swap test variation based on CX gates and classical post-processing.}
    \label{fig:swap_test}
\end{figure}

The Swap test~\cite{barenco_swap_1997} is a well-known algorithm that does not require explicit knowledge of the state-preparing unitaries. 
For two input states $\ket{\psi_1}$ and $\ket{\psi_2}$ and the circuit shown in Fig.~\ref{fig:swap_test}(a), the probability to measure 0 on the auxiliary qubit is
\begin{equation}
    p_0 = \frac{1 + |\braket{\psi_1|\psi_2}|^2}{2},
\end{equation}
which allows to determine the fidelity.
An imminent drawback of the Swap test is the complexity of implementing the series of controlled-Swap gates. While there is a direct implementation of this gate in quantum optics~\cite{patel_cswap_2016}, other platforms require a, typically expensive, decomposition in terms of basis gates.

By including additional, classical post-processing, the resource requirements of the Swap test can be reduced to the circuit shown in Fig.~\ref{fig:swap_test}(b)~\cite{garcia_swap_2013, cincio_learning_2018}. This method, however, still requires the implementation of CX gates across the state registers, which be challenging to map onto near-term quantum computers efficiently due to the typically limited connectivity between qubits.

\subsubsection{Randomized measurements}

The randomized measurements~\cite{elben_overlap_2019} approach for fidelity calculation requires the least complex circuits, as it only applies local unitaries to the states $\ket{\psi_1}$ and $\ket{\psi_2}$ and measures them separately.
The fidelity for $n$-qubit states is calculated as 
\begin{equation}
    |\braket{\psi_1|\psi_2}|^2 = 2^n \sum_{\vec s, \vec s'} \left(-\frac{1}{2}\right)^{d_H(\vec s, \vec s')} \frac{1}{|\mathcal{U}|} \sum_{U \in \mathcal{U}} p_{\vec s}(U\ket{\psi_1}) p_{\vec s'}(U\ket{\psi_2}),
\end{equation}
where $\vec s, \vec s' \in \{0, 1\}^n$ iterate over all computational basis states, $p_{\vec s}(\ket{\psi})$ is the probability to measure the outcome $\vec s$ on state $\ket{\psi}$, $\mathcal{U}$ contains a number of random local unitaries and $d_H$ is the Hamming distance defined as 
\begin{equation}
    d_H(\vec s, \vec s') = \sum_{j=1}^{n} |s_j - s'_j|.
\end{equation}
The accuracy of this estimation depends number of measurements and number of random unitaries. Due to the iteration over all basis states the randomized measurements approach requires an exponential number of measurements, which makes it unsuitable to scale up to a large number of qubits.

\subsubsection{Hadamard test}

The fidelity can be evaluated with a linear combination of unitaries technique, which we previously applied to estimate quantum circuit gradients in Section~\ref{sec:lcu}. For the application to the overlap of quantum states this approach is commonly referred to as the Hadamard test~\cite{cleve_algorithms_1998}.
To compute the overlap the Hadamard test requires knowledge of the state preparing unitaries, $\ket{\psi_{1,2}} = U_{1,2}\ket{0}$. It then uses the LCU circuit (see Fig.~\ref{fig:innerprod_lcu_circuit}) to evaluate 
\begin{equation}
    |\braket{\psi_1|\psi_2}|^2 =
    |\braket{0|U_1^\dagger U_2|0}|^2 = 
    \mathrm{Re}\left(\braket{0|U_1^\dagger U_2|0}\right)^2 
    + \mathrm{Im}\left(\braket{0|U_1^\dagger U_2|0}\right)^2,
\end{equation}
where real and imaginary parts are evaluated separately.

\begin{figure}[htp]
    \[
    \begin{array}{c}
    \Qcircuit @C=1em @R=.7em {
        \lstick{\ket{0}} & \gate{H} & \gate{P(\alpha)} & \ctrl{1} & \qw & \cdots & & \qw & \ctrl{1} & \gate{H} & \multimeasureD{1}{Z \otimes I^{\otimes n}} \\
        \lstick{\ket{0}^{\otimes n}} & \gate{V_0} & \gate{U_1(\theta_1)} & \gate{U_1(\delta\theta_1)} & \qw & \cdots & & \gate{U_d(\theta_d)} & \gate{U_d(\delta\theta_d)} & \gate{V_d} &  \ghost{Z \otimes I^{\otimes n}}
    }
    \end{array}
    \]
    \caption[Hadamard test implementation]{The Hadamard test for a parameterized state with two different sets of parameters, $\ket{\phi(\vec\theta)}$ and $\ket{\phi(\vec\theta')}$, where $\vec{\delta\theta} = \vec\theta' - \vec\theta$. For $\alpha=0$ the real part of the overlap is computed and for $\alpha=\pi/2$ the imaginary part.}
    \label{fig:hadamard_test_efficient}
\end{figure}
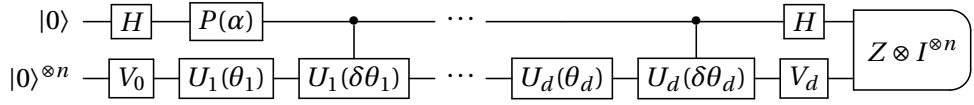

If the states are prepared by the same parameterized unitary with different sets or parameters, the circuit for the Hadamard test can be significantly simplified. Instead of the open and closed controls on the complete unitary, it is sufficient to only control the difference of the two states. For example, if $\ket{\psi_1} = \ket{\phi(\vec\theta)}$ and $\ket{\psi_2} = \ket{\phi(\vec\theta')}$ with
\begin{equation}
    \ket{\phi(\vec\theta)} = V_d U_d(\theta_d) \cdots V_1 U_1(\theta_1) V_0 \ket{0},
\end{equation}
the circuit simplifies to Fig.~\ref{fig:hadamard_test_efficient}.
While this simplification reduces the number of control and total number of gates to be implemented, it still requires $d$ controlled, generally non-local operations, which makes it challenging to implement on near-term hardware.

\subsubsection{Unitary overlap}

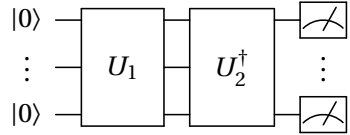
\begin{figure}[htp]
    \centering
    \[
    \begin{array}{c}
    \Qcircuit @C=1em @R=.7em {
        \lstick{\ket{0}} & \multigate{2}{~~U_1~~} & \multigate{2}{~~U_2^\dagger~~} & \meter \\
        \lstick{\raisebox{.5em}{\ensuremath{\vdots~~}}} & \ghost{~~U_1~~} & \ghost{~~U_2^\dagger~~} & \raisebox{.5em}{\ensuremath{\vdots}} \\
        \lstick{\ket{0}} & \ghost{~~U_1~~} & \ghost{~~U_2^\dagger~~} & \meter
    }
    \end{array}
    \]
    \caption[Unitary overlap implementation]{The circuit for unitary overlap. The fidelity is given by the probability to measure 0 on every qubit.}
    \label{fig:unitary_overlap}
\end{figure}

A possibly more near-term friendly method, that does not require qubit connections beyond already present in the state preparation is the unitary-overlap technique, also referred to as compute-uncompute~\cite{havlicek_supervised_2019}.
The fidelity can be written as measurement of the all-zero projector $P_0 = (\ket{0}\bra{0})^{\otimes n}$, as
\begin{equation}
    |\braket{\psi_1|\psi_2}|^2 
    = |\braket{0|U_1^\dagger U_2|0}|^2
    = \braket{0|U_1^\dagger U_2|0}\bra{0}\underbrace{U_2^\dagger U_1\ket{0}}_{=\ket{\psi_{12}}}
    = \braket{\psi_{12}|P_0|\psi_{12}}.
\end{equation}
Thus, measuring the fidelity amounts to preparing the state shown in Fig.~\ref{fig:unitary_overlap} and measuring the probability of all qubits being in state 0. 

The drawback of this approach is still that it requires applying both unitaries sequentially. However, since no additional qubits, non-local connections, or exponential numbers of measurements are required, this is generally the most near-term compatible approach to measuring the fidelity. 

\section{Numerical experiments}

Two important metrics for the efficiency of an optimization algorithm are the convergence speed and stability. 
How fast an algorithm converges can be measured, for example, in the number of iterations, the wall-time or the consumed resources. The stability includes, e.g., the sensitivity of the hyperparameters (does a slight change in the settings have a large impact on the convergence?) or dependence of the convergence to the initial point of the optimization.
In this section we investigate and benchmark the behavior of QN-SPSA in numerical simulations with regard to these metrics.
We compare the performance of QN-SPSA to plain SPSA, which requires less resources per step but does not take into account information geometry, and the non-stochastic versions of GD: standard, or ``vanilla'', GD and QNG.

\subsection{Convergence speed}\label{sec:qnspsa_convergence}

\begin{figure}[thpb]
    \centering
    \[
    \begin{array}{c}
    \Qcircuit @C=1em @R=.3em {
        \lstick{\ket{0}} & \gate{R_Y\left(\frac{\pi}{4}\right)} & \gate{R_{P_1}(\theta_1)} & \ctrl{1} & \qw & \qw & \cdots  \\
        \lstick{\ket{0}} & \gate{R_Y\left(\frac{\pi}{4}\right)} & \gate{R_{P_2}(\theta_2)} & \control\qw & \ctrl{1} & \qw & \cdots \\
        \lstick{\ket{0}} & \gate{R_Y\left(\frac{\pi}{4}\right)} & \gate{R_{P_3}(\theta_3)} & \ctrl{1} & \control\qw & \qw & \cdots \\
        \lstick{\ket{0}} & \gate{R_Y\left(\frac{\pi}{4}\right)} & \gate{R_{P_4}(\theta_4)} & \control\qw & \ctrl{1} & \qw & \cdots \\
        \lstick{\ket{0}} & \gate{R_Y\left(\frac{\pi}{4}\right)} & \gate{R_{P_5}(\theta_5)} & \qw & \control\qw & \qw & \cdots
        \gategroup{1}{3}{5}{5}{1em}{--} \\
    }
    \end{array}
    \]
    \caption[Pauli two-design circuit]{The Pauli two-design circuit shown here for $n=5$ qubits. The dashed box is repeated in each layer and a final layer of rotation gate $R_{P_j}$ is added. The rotation axis $P_j$ is chosen uniformly at random from $\{X, Y, Z\}$.}
    \label{fig:pauli2design_circuit}
\end{figure}
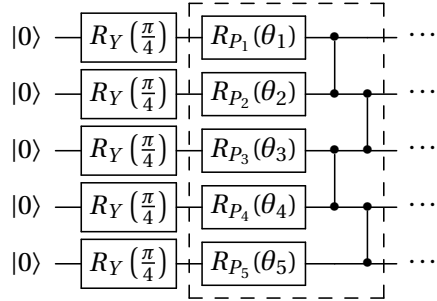

\noindent
In the first experiment we compare the speed of convergence as the value of the loss function against the number of iterations and against the number of circuit executions. 
The loss function is the expectation value of Hamiltonian $H$,
\begin{equation}
    \mathcal{L}(\vec\theta) = \braket{\phi(\vec\theta)|H|\phi(\vec\theta)},
\end{equation}
where $H$ is a local Pauli-$ZZ$ operator and we use a Pauli two-design circuit as ansatz $\ket{\phi(\vec\theta)}$, as in Ref.~\cite{stokes_qng_2020}.
The circuit, shown in Fig.~\ref{fig:pauli2design_circuit}, is initialized with a layer of $R_Y(\pi/4)$ gates followed by a series of rotation and entanglement layers. In the rotation layer single-qubit Pauli rotations are applied, where the rotation axis is randomly selected from $\{X, Y, Z\}$.
The entanglement layer consists of pairwise controlled-$Z$ gates between neighboring qubits.

In this experiment, we use $n=11$ qubits with the $ZZ$ operator acting on qubits 5 and 6, that is $H = Z_5 Z_6$.
We repeat the rotation and entanglement layers 3 times, such that all qubits contribute to the expectation value, and a add a final rotation layer, which leads to a total of $d=44$ parameters. 
All optimizers use a learning rate of $\eta=10^{-2}$ and the SPSA methods use perturbation of  $\epsilon=10^{-2}$.
SPSA and QN-SPSA use a single sample of the gradient (and QGT) per step. For numerical stability in solving the QN-SPSA linear system, we add a regularization of $\beta=10^{-3}$ to the diagonal of the QGT estimate.
GD and QNG are run once, while for the SPSA methods we show mean and one standard deviation over 25 independent experiments.
All optimizers start from the same initial point and each circuit execution is simulated with 8192 noise-free measurements in Qiskit~\cite{Qiskit}.

\begin{figure}[thbp]
    \centering
    \includegraphics[width=0.9\textwidth]{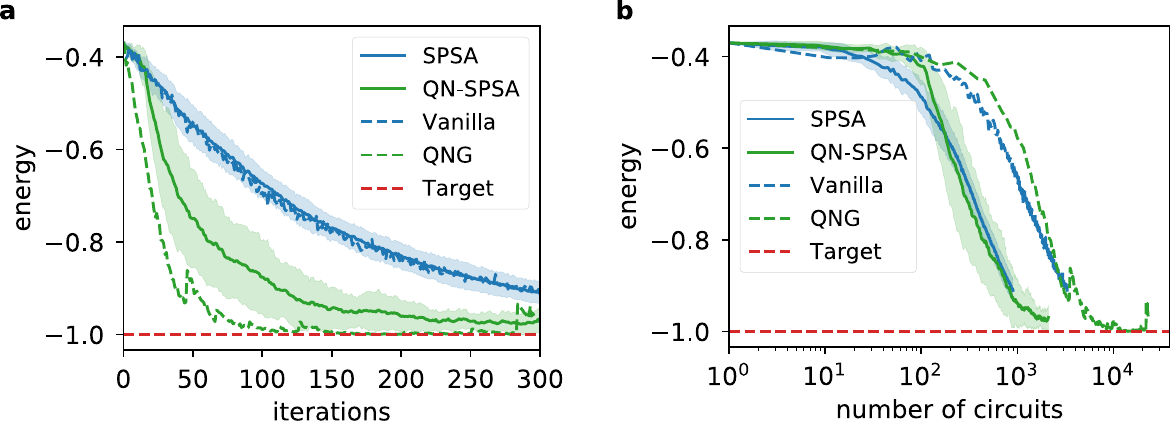}
    \caption[QN-SPSA convergence of a local cost function]{Investigation of the loss for the Pauli two-design circuit with 3 layers for the local observable $H = Z_5 Z_6$ on $n=11$ qubits.
    (a) The loss as function of the number of iterations, and
    (b) as number of executed circuits.
   }
    \label{fig:pauli2design}
\end{figure}

In Fig.~\ref{fig:pauli2design}(a) we show the convergence as function of the iterations for each method.
As previously observed in Ref.~\cite{stokes_qng_2020}, QNG requires less iterations to converge compared to GD.
The jumps in the QNG convergence are caused by numerical instabilities in solving the linear system.
The mean of SPSA closely matches GD, which we expect as the SPSA gradient are unbiased estimates of the true gradient.
Importantly, we see that QN-SPSA outperforms the first-order gradient optimizers, however, it does not match the QNG loss. This is due to the additional regularization, which biases the parameter dynamics towards GD, as shown in Section~\ref{sec:regularization_influence}. Also, QN-SPSA exhibits a larger standard deviation than SPSA, that is likely caused by the preconditioning of the gradient with a noisy, potentially low-rank matrix.

On near-term quantum computers, however, convergence in terms of numbers of iteration is not necessarily the most important metric. Instead, the overall runtime is determined by the number of circuit executions---at least until these can be massively parallelized.
From this point of view, Fig.~\ref{fig:pauli2design}(b) shows that both SPSA techniques clearly exceed the performance of QNG and GD. Even though QN-SPSA performed worse in the per-iteration comparison, it requires only 2100 circuits for the optimization whereas QNG evaluated 679'200, under the favorable assumption that the LCU technique can be used to evaluate the QGT.
Note, that even for QN-SPSA, the number of measurements exceeds the dimension of the Hilbert space and the ground state of the diagonal Hamiltonian could be found by enumeration of all basis states. 
However, the goal of this experiment is to demonstrate the relation of the different optimization algorithms. To outperform enumeration, we can consider larger systems, optimize the algorithms' hyperparameters or consider non-diagonal Hamiltonians.

In this simple example, SPSA and QN-SPSA require roughly the same number of circuits to converge. However for loss functions that have different sensitivities to different parameters, the natural gradient characteristic of QN-SPSA comes more into play.
The following, more challenging example displays the advantage of taking into account the information geometry.

\subsubsection{Challenging landscape}

As a more difficult optimization landscape, we consider a Max-Cut problem on a random graph with random integer weights on $n=5$ nodes.
The Hamiltonian for the specific instance we investigate is 
\begin{equation}
   H = Z_4 Z_5 + 2.5 Z_3 Z_5 + 2.5 Z_3 Z_4 - 0.5 Z_2 Z_5 - 0.5 Z_2 Z_3 - 4.5 Z_1 Z_5 + 3.5 Z_1 Z_3.
\end{equation}
To find the ground state of this Hamiltonian we use a QAOA ansatz~\cite{farhi_quantum_2014} with a mixer Hamiltonian $H_M = \sum_{j=1}^{n} X_j / 20$. The circuit is shown in Fig.~\ref{fig:qnspsa_qaoa_circuit}.
The loss landscape of this optimization problem has various local minima embedded in steep, narrow canyons that are surrounded by flats.
The gradients vary by several orders of magnitude in different parameter dimensions and at different points, which makes this a challenging problem for geometry-agnostic optimizers.

\begin{figure}[thbp]
    \[
    \begin{array}{c}
    \Qcircuit @C=1em @R=.7em {
        \lstick{\ket{0}} & \gate{H} & \gate{P(2\theta_1)} & \gate{P(-9\theta_1)} & \qw & \gate{P(-\theta_1)} & \qw & \gate{P(5\theta_1)} & \gate{R_X(\theta_2/10)} & \qw \\
        \lstick{\ket{0}} & \gate{H} & \ctrl{-1} & \qw & \qw & \qw & \gate{P(5\theta_1)} & \qw & \gate{R_X(\theta_2/10)} & \qw \\
        \lstick{\ket{0}} & \gate{H} & \gate{P(7 \theta_1)} & \qw & \gate{P(-\theta_1)} & \qw & \ctrl{-1} & \ctrl{-2} & \gate{R_X(\theta_2/10)} & \qw \\
        \lstick{\ket{0}} & \gate{H} & \qw & \qw & \ctrl{-1} & \ctrl{-3} & \qw & \qw & \gate{R_X(\theta_2/10)} & \qw \\
        \lstick{\ket{0}} & \gate{H} & \ctrl{-2} & \ctrl{-4} & \qw & \qw & \qw & \qw & \gate{R_X(\theta_2/10)} & \qw
    }
    \end{array}
    \]
    \caption[QAOA circuit for a challenging Max-Cut landscape]{The QAOA ansatz for the specified Max-Cut problem.}
    \label{fig:qnspsa_qaoa_circuit}
\end{figure}

\begin{figure}[thbp]
    \centering
    \includegraphics[width=\textwidth]{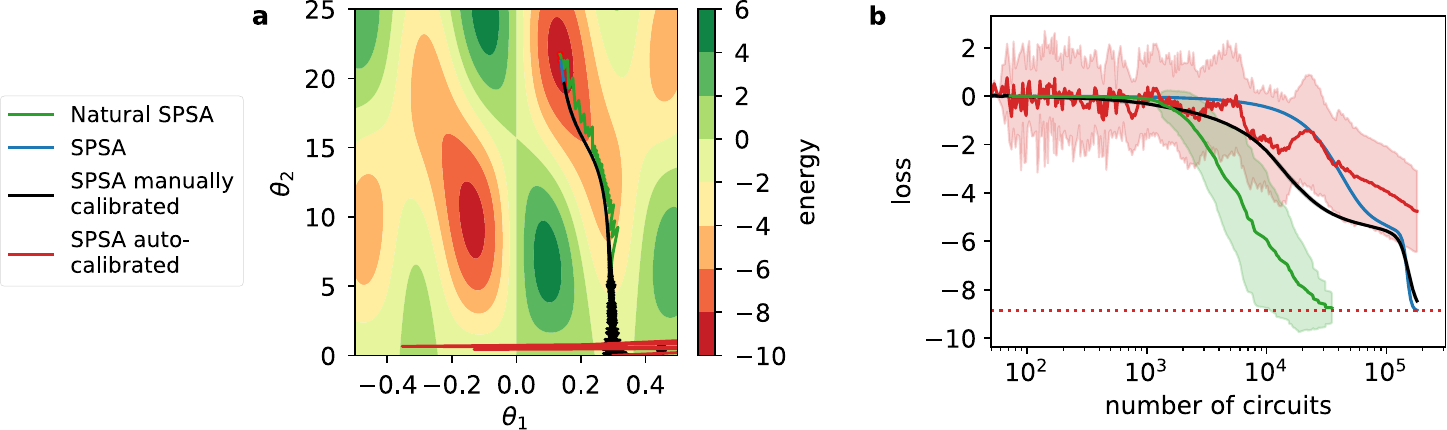}
    \caption[Challenging QAOA landscape]{(a) The QAOA loss landscape with the parameter trajectories of SPSA and QN-SPSA. (b) The convergence as a function of number of circuit evaluations.}
    \label{fig:qnspsa_qaoa}
\end{figure}

We run QN-SPSA for $\eta=10^{-2}$ and $\epsilon=10^{-2}$ and compare the parameter trajectory to SPSA in Fig.~\ref{fig:qnspsa_qaoa}(a).
For SPSA we use three different sets of hyperparameters:
\begin{itemize}
    \item The same as QN-SPSA, $\eta=10^{-2}$ and $\epsilon=10^{-2}$.
    \item Exponential decays, $\eta_k = A k^{-0.602}$ and $\epsilon_k = C k^{-0.101}$, where $C=0.2$ and $A$ is chosen such that the first update step has an approximate magnitude of $\delta=\pi/5$ per parameter dimension, as suggested in Ref.~\cite{kandala_vqe_2017}. Explicitly, this means
    \begin{equation}\label{eq:spsacal}
        A = \frac{\delta}{\frac{\left\|\mathbb E\big[\widehat{\vec\Delta\mathcal{L}}\big(\vec\theta^{(0)}\big)\big]\right\|_1}{d}} 
        \approx \delta \cdot d \cdot \left\|\frac{1}{M} \sum_{j=1}^{M} \frac{\mathcal{L}(\vec\theta^{(0)} + C\vec\Delta_{(j)}) - \mathcal{L}(\vec\theta^{(0)} - C\vec\Delta_{(j)})}{2C} \vec\Delta_{(j)}^{-1} \right\|_1,
    \end{equation}
    where $\vec\Delta_{(j)} \in \mathbb{R}^d$ are i.i.d. sampled perturbation directions and the inverse $\vec\Delta^{-1}_{(j)}$ is element-wise, $\|\cdot\|_1$ is the 1-norm and we used $M=25$.
    \item Exponential decays with a manually tuned update step magnitude of $\delta=0.1$.
\end{itemize}
Only the SPSA settings with sufficiently small learning rate find the target minimum. The calibration from Ref.~\cite{kandala_vqe_2017} starts from a much too large learning rate and oscillates around the loss landscape. 
The advantage of QN-SPSA is clearly shown in Fig.~\ref{fig:qnspsa_qaoa}(b), where we show the convergence as function of number of evaluated circuits.

This experiment emphasises an additional advantage of natural gradients over standard gradient descent: both GD and SPSA need careful, model-dependent hyperparameter tuning.
In addition, in some cases an efficient set of hyperparameters might not exist, since the learning rate acts the same on all parameters but the model has different sensitivities in different dimensions.
Natural gradient approaches, the learning rate directly controls changes in the model instead of the parameters and can, thus, often be set independent of the specific model at hand.
In all our experiments the value of $\eta=10^{-2}$ performed well.

\subsection{Convergence region}

While a faster convergence is a key advantage of natural gradients, it is not the only one.
Since QNG approximates VarQITE, as previously discussed in Section~\ref{sec:ite}, it comes with a convergence guarantee under idealized conditions, including that the initial state has a sufficient overlap with the ground state, and that the ansatz and hyperparameter settings allow to track the parameter dynamics sufficiently close.
But, even if not all these criteria are met, QNG and QN-SPSA benefit from superior convergence properties compared first-order gradient techniques.

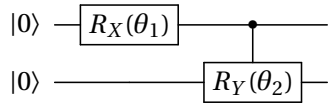
\begin{figure}[htp]
    \centering
    \[
    \begin{array}{c}
    \Qcircuit @C=1em @R=.7em {
        \lstick{\ket{0}} & \gate{R_X(\theta_1)} & \ctrl{1} & \qw \\
        \lstick{\ket{0}} & \qw & \gate{R_Y(\theta_2)} & \qw
    }
    \end{array}
    \]
    \caption[Simple circuit model for QN-SPSA convergence experiments]{The ansatz to minimize the energy of the diagonal Hamiltonian. Note that compared to Ref.~\cite{mcardle_varqite_2019} no explicit global phase parameter is needed as the QGT already accounts for it.}
    \label{fig:mcardle_ansatz}
\end{figure}

To demonstrate this, we investigate a simple two-qubit problem from Ref.~\cite{mcardle_varqite_2019}, which minimizes the energy of the diagonal Hamiltonian 
\begin{equation*}
    H = \begin{pmatrix}
    1 & 0 & 0 & 0 \\
    0 & 2 & 0 & 0 \\
    0 & 0 & 3 & 0 \\
    0 & 0 & 0 & 0
    \end{pmatrix},
\end{equation*}
using the ansatz shown in Fig.~\ref{fig:mcardle_ansatz}.
We investigate the convergence of GD, QNG, SPSA and QN-SPSA for a grid of initial points in $[-\pi, \pi]^2$ without shot noise.
The hyperparameters use the same settings as in Ref.~\cite{mcardle_varqite_2019}, which are $\eta=0.886$ for GD (and SPSA) and $\eta=0.225$ for QNG (and QN-SPSA).
An optimization is marked as converged if the final value is within a distance of $10^{-4}$ to the optimal value.
Since SPSA and QN-SPSA are of stochastic nature, we repeat 10 runs per initial point and consider the experiment as successful if at least one run converged.

\begin{figure}
    \centering
    \includegraphics[width=0.9\textwidth]{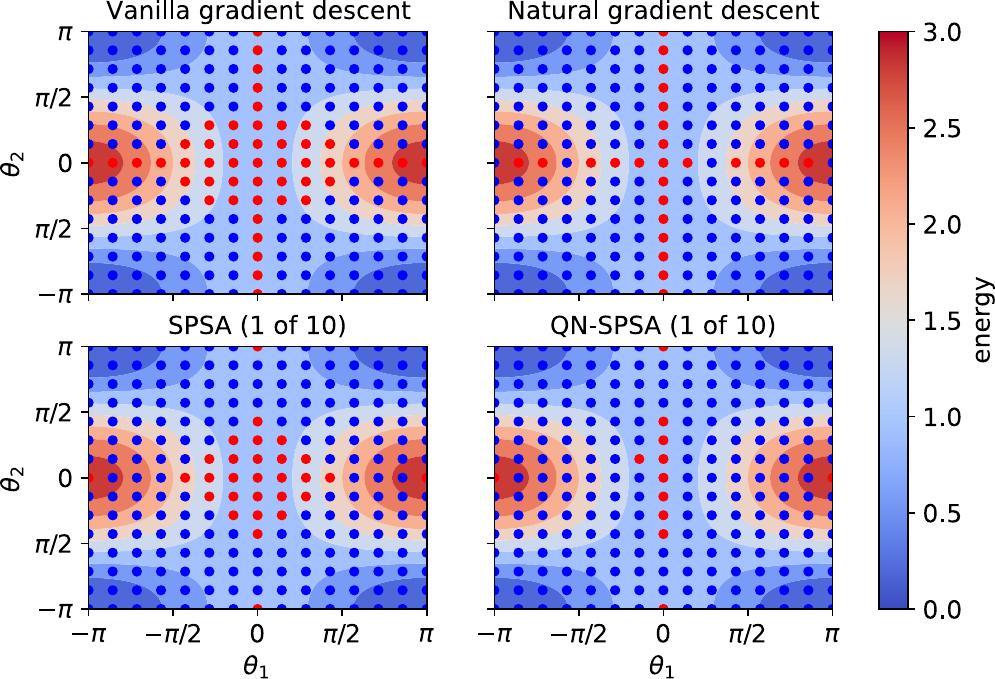}
    \caption[Comparison of convergence regions]{Convergence regions of GD, QNG, SPSA and QN-SPSA. Initial points are labelled blue, if the optimizer converged from it, and red otherwise.}
    \label{fig:convergence_region}
\end{figure}

In Fig.~\ref{fig:convergence_region} we show the loss landscape where the initial points are labelled blue, if the optimization converged from this point, and red otherwise.
The results of GD and QNG match the observations of Ref.~\cite{mcardle_varqite_2019}.
The QNG optimization is successful for all initial points except when one initial parameter is exactly zero, in which case the minimization cannot escape the center lines as there is no orthogonal gradient contribution.
In addition to these points, GD fails to find the optimal solution in a diamond-shaped region around the center saddle point at $(0, 0)$.
SPSA and QN-SPSA largely match the behavior or GD and QNG, but suffer less from the missing orthogonal gradient component. 
This is due to the fact that all gradient components contribute to the magnitude of the randomly selected update direction, which allows to move into directions where the gradient is zero.
In conclusion, we find that QN-SPSA has the largest convergence region of all the compared methods.

\subsection{Influence of regularization}\label{sec:regularization_influence}

The QGT matrix be easily singular, which causes the linear system, that needs to be solved for the parameter derivative, to be ill-conditioned.
For example, assume the simple Pauli two-design circuit in Fig.~\ref{fig:pauli2design} with only $R_Y$ rotations on 2 qubits with $r=1$ repetition at the parameters $\vec\theta = \vec 0$.
The real part of the QGT for this circuit is
\begin{equation}
    g(\vec 0) = \frac{1}{4}
    \begin{pmatrix}
        1 & 0 & 1/\sqrt{2} & 0 \\
        0 & 1 & 0 & 1/\sqrt{2} \\
        1/\sqrt{2} & 0 & 1 & 1/2 \\
        0 & 1/\sqrt{2} & 1/2 & 1
    \end{pmatrix},
\end{equation}
which is a singular matrix with null-space $K = \{x/\sqrt{2}, -x/\sqrt{2}, -x, x ~|~x\in\mathbb{R}\}$.

In addition to the fact that the QGT can be singular, the QN-SPSA estimates $\hat g$ are at most rank-2 matrices, as they are proportional to the outer product $(\vec\Delta(\vec\Delta')^\top)^{-1} + (\vec\Delta'\vec\Delta^\top)^{-1}$.
To ensure the linear system can be solved in a stable manner, we add a regularization constant $\beta > 0$ onto the diagonal of the QGT estimate.
This diagonal shift, however, has an influence on the parameter dynamics.

\begin{figure}[tbhp]
    \centering
    \includegraphics[width=0.5\textwidth]{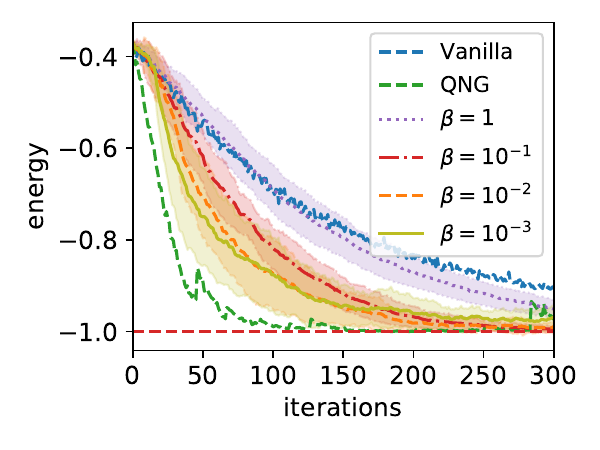}
    \caption[Impact of the regularization on QN-SPSA]{Convergence of QN-SPSA for the Pauli two-design experiment for different regularization constants $\beta$.}
    \label{fig:qnspsa_regularization}
\end{figure}

We assume a normalized diagonal shift as regularization, that is $(g + \beta\mathbb{1})/(1 + \beta)$, such that for $\beta=0$ there is no regularization and for $\beta\rightarrow\infty$ the estimate is the identity $\mathbb{I}$. 
Without the normalizing factor, increasing $\beta$ would decrease the step size.
From the limits we can see that $\beta$ interpolates between the QNG and GD, since using the identity as pre-conditioner simply yields the standard gradient.
This is reflected in the numerical experiment in Fig.~\ref{fig:qnspsa_regularization}, where we repeat the Pauli two-design experiment of Section~\ref{sec:qnspsa_convergence} for different regularization constants. Using a small regularization, which provides a faithful approximation of the QGT is prone to numerical instability.
A large regularization, on the other hand, neglects the information geometry and is closer to a standard gradient descent update.

\section{Applications}

In the following sections we investigate QN-SPSA in the context of molecular ground state search, generative learning, and black box optimization.

\subsection{Molecular ground states}

Finding ground states is a key subroutine in quantum chemistry, which we demonstrate in this section for the lithium-hydride (LiH) molecule using QN-SPSA and the \texttt{ibmq\_montreal} device~\cite{ibm_quantum}.

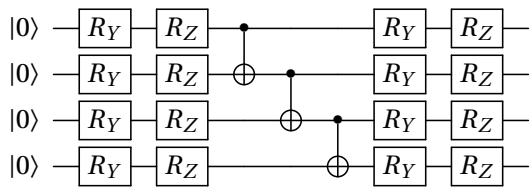
\begin{figure}[thbp]
    \centering
    \[
    \begin{array}{c}
    \Qcircuit @C=1em @R=.3em {
        \lstick{\ket{0}} & \gate{R_Y} & \gate{R_Z} & \ctrl{1} & \qw & \qw & \gate{R_Y} & \gate{R_Z} & \qw \\
        \lstick{\ket{0}} & \gate{R_Y} & \gate{R_Z} & \targ & \ctrl{1} & \qw & \gate{R_Y} & \gate{R_Z} & \qw \\
        \lstick{\ket{0}} & \gate{R_Y} & \gate{R_Z} & \qw & \targ & \ctrl{1} & \gate{R_Y} & \gate{R_Z} & \qw \\
        \lstick{\ket{0}} & \gate{R_Y} & \gate{R_Z} & \qw & \qw & \targ & \gate{R_Y} & \gate{R_Z} & \qw 
    }
    \end{array}
    \]
    \caption[Ansatz for LiH ground-state calculation]{The ansatz used for the LiH calculation. Each Pauli rotation gate is parameterized with an individual parameter.}
    \label{fig:efficientsu2_lih}
\end{figure}

We describe the LiH molecule, at a bond distance of 2.5\AA{}, with an STO3G basis set, which results in six molecular orbitals. 
The one and two body integrals for the system are then obtained using PySCF~\cite{pyscf_2020} to perform a Restricted Hartree-Fock calculation.
To reduce the number of qubits, we further restrict the simulated active space to three molecular orbitals and map the excitation and annihilation operators to qubits using the parity mapping~\cite{bravyi_tapering_2017}, which leads to a 6-qubit Hamiltonian. 
Leveraging the properties of the parity mapping, we can remove two additional qubits and finally obtain a 4-qubit description of the system.
As ansatz to model the wave function we use a hardware-efficient circuit shown in Fig.~\ref{fig:efficientsu2_lih}, which can be mapped to the quantum chip without additional Swap gates.

\begin{figure}[thbp]
    \centering
    \includegraphics[width=0.95\textwidth]{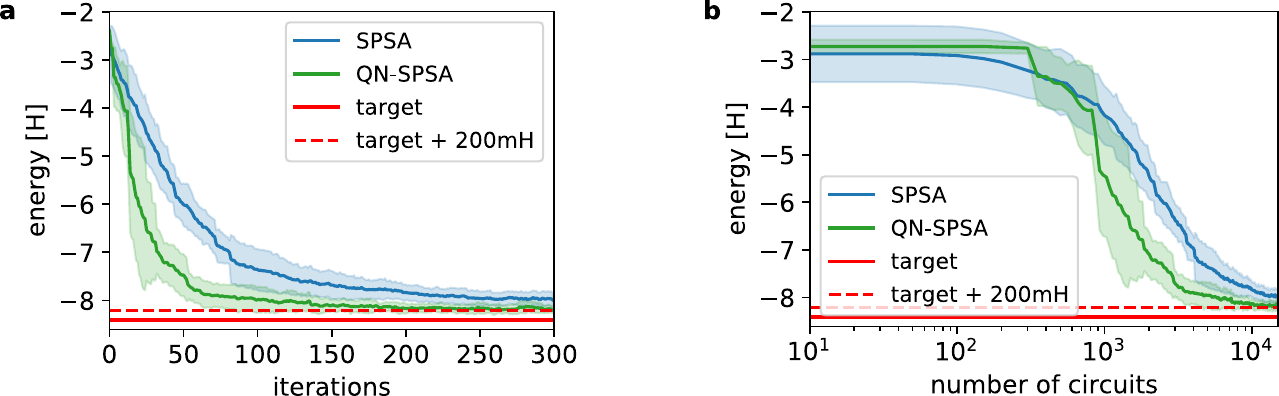}
    \caption[Ground-state approximation of LiH]{Ground-state approximation of LiH, measured in Hartree (H) as function of (a) the number of iterations and (b) the number of evaluated circuits.} 
    \label{fig:lih}
\end{figure}

In Fig.~\ref{fig:lih} we compare the performance of SPSA and QN-SPSA with learning rates $\eta=10^{-2}$, perturbation of $\epsilon=10^{-1}$ where each circuit is sampled with 1024 measurements. QN-SPSA uses a regularization of $\beta=10^{-3}$ and starts with 100 QGT samples in the first two iterations, for a good initial estimation, followed by two samples in the remaining optimization.
We use the same random initial point for both optimizers and show mean and standard deviation over five independent runs, and observe that QN-SPSA converges faster both in number of iterations and required circuit evaluations.
The measured energies at final parameters deviate from the exact value by 200mH, due to the hardware noise. This difference can likely be overcome with error mitigation techniques, such as readout error mitigation and zero-noise extrapolation.

Molecular Hamiltonians typically have a large number of Pauli terms, which affects the number of circuit evaluations required to measure the energy gradient. The QGT, on the other hand, only depends on the ansatz and each QGT sample uses only four fidelity evaluations independent of the structure of the Hamiltonian.
Thus, the overhead to perform a QN-SPSA sample on top of an SPSA gradient is
\begin{equation}
    \frac{\text{QN-SPSA}}{\text{SPSA}} = \frac{4 + 2K}{2K} = 
    \begin{cases}
        3, \text{ diagonal Hamiltonian}, \\
        1.08, \text{ LiH in restricted space}, \\
        1.01, \text{ full LiH with 6-31G orbital basis},
    \end{cases}
\end{equation}
where $K$ is the number of basis that the circuit has to be measured in to evaluate the energy.
The more complex the Hamiltonian, the smaller the overhead of QN-SPSA and for $K\gg 1$ the natural gradient information can be included at a negligible overhead.

\subsection{Quantum Boltzmann machines}

A Boltzmann machines (BM) is a machine learning model, that can be expressed as stochastic Ising model defined by
\begin{equation}
    H = -\sum_{\braket{jk}} W_{jk} Z_j Z_k - \sum_{j} h_j Z_j,
\end{equation}
where $W_{jk}$ are the model weights and $h_j$ the node biases.
From a physics perspective, $W_{jk}$ describes the interaction strength of spins $j$ and $k$ and $h_j$ is the magnetic field strength at site $j$.
BMs can be fully connected, in which case $\braket{jk}$ iterates over all pairs, or restricted to certain connections among the nodes.

BMs assign a probability distribution to a binary state $\vec z \in \{-1, 1\}^n$ given by the Boltzmann distribution
\begin{equation}
   p_{\vec z}(W, \vec h) = \frac{e^{-\beta E_{\vec z}(W, \vec h)}}{Z(\beta, W, \vec h)},
\end{equation}
where $E_{\vec z}$ is the energy,
\begin{equation}
    E_{\vec z}(W, \vec h) = \braket{\vec z|H(W, \vec h)|\vec z} = -\sum_{\braket{jk}} W_{jk} z_j z_k - \sum_{j} h_j z_j,
\end{equation}
and $Z(\beta, W, \vec h)$ is the partition function at inverse temperature $\beta$.
For a sufficient number of sites, BMs are universal function approximators and can represent any classical probability distribution.
They have been used for a wide range of classical applications~\cite{hinton_contrastive_2002, hinton_reducing_2006, ruslan_rbm_2007}, but also to model quantum systems~\cite{melko_rbm_2019}.
Challenges of implementing a BM, however, include the evaluation of the partition function and the evaluation of gradients, in particular for unrestricted BMs.
Since this machine learning model is inherently connected to a quantum mechanical system, quantum Boltzmann machines (QBMs) have recently been suggested~\cite{amin_qbm_2018, zoufal_qbm_2021} as a solution to these bottlenecks.

The probability distribution of a BM can be written in terms of a Gibbs state $\rho_G$, that is
\begin{equation}\label{eq:bm_probability}
    p_{\vec z}(W, \vec h) = \mathrm{Tr}\left(\rho_G(W, \vec h) \ket{\vec z}\bra{\vec z}\right),
\end{equation}
with
\begin{equation}
    \rho_G(W, \vec h) = \frac{e^{-\beta H(W, \vec h)}}{Z(\beta, W, \vec h)}.
\end{equation}
The spin state $\ket{\vec z}$ is identified with a computational basis state $\ket{x}, x \in \mathbb{N}$ by
\begin{equation}
    \ket{\vec z} = \ket{z_n \cdots z_1} \equiv \Ket{\frac{1 - z_n}{2} \cdots \frac{1 - z_1}{2}} = \ket{x_n \cdots x_1} = \ket{x},
\end{equation}
where $z_j \in \{-1, 1\}$ and where $x_j \in \{0, 1\}$ describes the binary representation of the integer $x$.

The Gibbs state can, for example, be prepared via quantum phase estimation~\cite{yung_quantumquantum_2012, bilgin_gibbs_2010} or minimizing the free energy of the system~\cite{wang_gibbsprep_2021}. Here, we instead use the approach outlined in Fig.~\ref{fig:gibbs-state-prep}, which relies on quantum imaginary time evolution of a maximally mixed state.
Since the QNG dynamics coincide with variational imaginary time evolution, we approximate the imaginary time evolution with our QN-SPSA algorithm to obtain an efficient Gibbs-state preparation.

We use the variational QBM (VarQBM) in an illustrative example to learn the distribution of a Bell-state, $(\ket{00} + \ket{11})/\sqrt{2}$, defined by
\begin{equation}
    \vec p^* = \begin{pmatrix}
        \frac{1}{2} \\ 
        0 \\
        0 \\
        \frac{1}{2}
    \end{pmatrix}.
\end{equation}
The VarQBM is trained by optimizing the weights $W$ and biases $\vec h$, such that the Boltzmann distribution at inverse temperature $\beta=1$ matches the target distribution. 
To measure the difference to the target distribution we use the cross entropy, such that the loss function is
\begin{equation}
    \mathcal{L}(W, \vec h) = -\sum_{x=0}^{2^n - 1} p^*_x\log\left(p_{x}(W, \vec h)\right),
\end{equation}
with the Boltzmann probability defined in Eq.~\eqref{eq:bm_probability} and $n=2$.
Each loss function evaluation thus requires the preparation of a Gibbs state.

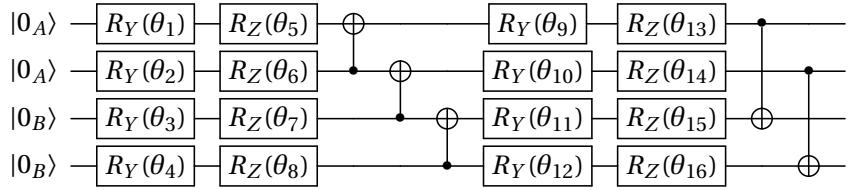
\begin{figure}[htp]
    \centering
    \[
    \begin{array}{c}
    \Qcircuit @C=1em @R=.3em {
        \lstick{\ket{0_A}} & \gate{R_Y(\theta_1)} & \gate{R_Z(\theta_5)} & \targ & \qw & \qw       & \gate{R_Y(\theta_9)} & \gate{R_Z(\theta_{13})} & \ctrl{2} & \qw & \qw \\
        \lstick{\ket{0_A}} & \gate{R_Y(\theta_2)} & \gate{R_Z(\theta_6)} & \ctrl{-1} & \targ & \qw & \gate{R_Y(\theta_{10})} & \gate{R_Z(\theta_{14})} & \qw & \ctrl{2} & \qw \\
        \lstick{\ket{0_B}} & \gate{R_Y(\theta_3)} & \gate{R_Z(\theta_7)} & \qw & \ctrl{-1} & \targ & \gate{R_Y(\theta_{11})} & \gate{R_Z(\theta_{15})} & \targ & \qw & \qw \\
        \lstick{\ket{0_B}} & \gate{R_Y(\theta_4)} & \gate{R_Z(\theta_8)} & \qw & \qw & \ctrl{-1}   & \gate{R_Y(\theta_{12})} & \gate{R_Z(\theta_{16})} & \qw &\targ & \qw
    }
    \end{array}
    \]
    \caption[Ansatz for Gibbs-state preparation]{The parameterized circuit encoding the Gibbs state.}
    \label{fig:gibbs_ansatz}
\end{figure}

To prepare these Gibbs states, we use the ansatz $\ket{\phi(\vec\theta)}$ shown in Fig.~\ref{fig:gibbs_ansatz} with the initial parameters
\begin{equation}
    \theta^{(0)}_j = 
    \begin{cases}
        \frac{\pi}{2}, \text{ if } j \in \{9, 10\}, \\
        0, \text{ otherwise},
    \end{cases}
\end{equation}
such that tracing out subsystem $B$ yields the maximally mixed state on subsystem $A$.. 
We then perform imaginary time evolution up to time $\tau = \beta/2 = 0.5$ by using QN-SPSA with the adapted update rule,
\begin{equation}
    \vec\theta(\tau + \Delta_\tau) = \vec\theta(\tau) - \Delta_t \bar{g}^{-1}(\tau) \frac{\widehat{\vec\nabla E}(\tau)}{2},
\end{equation}
with $E(\tau) = \braket{\phi(\vec\theta(\tau))| H(W, \vec h) |\phi(\vec\theta(\tau))}$.
In the experiments we use 10 integration steps with $\Delta_\tau = 0.05$, a perturbation of $10^{-2}$ and a regularization constant of 0.1.
Since we are interested accurately reproducing the dynamics, we re-sample the QGT and energy gradient 10 times in each iteration, to increase the accuracy.
This approach is further developed in Chapter~\ref{chap:saqite}.

With the Gibbs preparation at hand, we optimize the Hamiltonian parameters $W$ and $\vec h$ with 100 SPSA iterations with a learning rate and perturbation of 0.1.
The initial parameters are chosen uniformly at random from $[-2, 2]$.

\begin{figure}[htp]
    \centering
    \includegraphics[width=\textwidth]{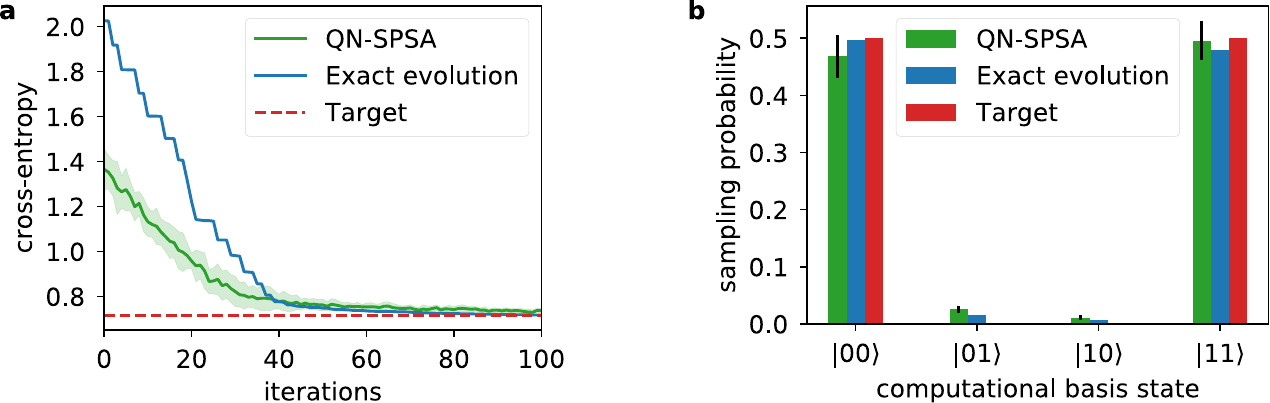}
    \caption[Generative learning with VarQBMs]{Generative learning with VarQBMs. (a) Comparison of the training loss using two different subroutines for the Gibbs-state preparation: QN-SPSA and exact matrix exponentiation. As target value we show the final value of the exact evolution method. For the QN-SPSA training we show mean and standard deviation over 10 independent runs. (c) The final probabilities of the trained Gibbs states. For each of the 10 optimization runs we prepare the 
    final state 10 times to approximate the standard deviation on the final sampling statistics.}
    \label{fig:generative_learning}
\end{figure}

In Fig.~\ref{fig:generative_learning}(a) we show the convergence of the cross entropy of the VarQBM, where we compare performing the imaginary time evolution with QN-SPSA against an exact, matrix-based evolution. 
Even though the preparation via QN-SPSA is subject to sampling noise, the VarQBM reliably converges to the target loss, and the final sampling statistics closely match the target distribution, as seen in Fig.~\ref{fig:generative_learning}(b).

To further improve the results, we could improve the imaginary time evolution, by using more QGT re-samplings and reducing the timestep $\Delta_\tau$, or improve the QBM model. This can be achieved by including additional hidden layers, i.e. additional spins that are not representing the final state.

\subsection{Black box optimization}\label{sec:blackbox}

\paper{This subsection is based on the co-authored article "Variational quantum algorithm for unconstrained black box binary optimization: Application to feature selection" by Zoufal et al., published in Quantum \textbf{7} 909 (2023).}

In binary black box optimization we consider an objective function $f: \{0, 1\}^n \rightarrow \mathbb{R}$, which can be  evaluated efficiently on a classical computer, but beyond the evaluation no information about the internal structure or any gradient information is available.
The task is then to find the optimal solution $\vec x^*$, defined as
\begin{equation}
    \vec x^* = \argmin_{\vec x \in \{0,1\}^n} f(\vec x).
\end{equation}
Without knowledge on the behavior of $f$ or access to its gradient and an exponentially large state space, it is challenging to optimize the objective function.

Here we propose a workflow to implement black box optimization problems on a quantum computer and solve them using QNG, which has advantageous convergence properties over classical optimization routines. 
As discussed before, these properties are derived from VarQITE, which, for a sufficiently expressive ansatz $\ket{\phi(\vec\theta)}$, a small enough integration timestep and an initial overlap with the ground state, guarantees a convergence to the lowest energy of the system.
Since the optimal solution is a computational basis state, an overlap with the solution is easily prepared by initializing the optimization in the equal superposition state,
\begin{equation}
    \ket{\phi_0} = \ket{+}^{\otimes n} = \frac{1}{\sqrt{2^n}} \sum_{x=0}^{2^n - 1} \ket{x}.
\end{equation}

The objective function can be formulated as expectation value of a diagonal Hamiltonian with the each possible binary vector evaluated on the diagonal, that is
\begin{equation}
    H = 
    \begin{pmatrix}
        f(0\cdots 00) & & & \\
        & f(0\cdots 01) & & \\
        & & \ddots & \\
        & & & f(1\cdots 11)
    \end{pmatrix}.
\end{equation}
Imaginary-time evolution starting from the initial state $\ket{\phi_0}$ under this diagonal Hamiltonian leads to the ground state and the solution of the optimization problem.
Constructing the Hamiltonian, however, would require evaluating the objective for the entire $2^n$ possible bitstrings, which is an infeasible, exponential number and would already reveal the solution through brute-force search.

To circumvent this issue, we can formulate the expectation value of the Hamiltonian using only on-demand evaluations of a fixed number of bitstrings. 
Suppose an ansatz $\ket{\phi(\vec\theta)}$ where we seek the parameters $\vec\theta \in \mathbb{R}^d$ to minimize the energy of the above diagonal Hamiltonian, which equals the loss function
\begin{equation}
    \begin{aligned}
        \mathcal{L}(\vec\theta) &= \braket{\phi(\vec\theta)|H|\phi(\vec\theta)}  \\
                                &= \sum_{x=0}^{2^n - 1} p_x(\vec\theta) f(\text{bin}(x)), \\
    \end{aligned}
\end{equation}
where $p_x(\vec\theta) = \braket{\phi(\vec\theta)\ket{x}\bra{x}\phi(\vec\theta)}$ is the probability to measure the qubits in state $x$, $\text{bin}(x)$ returns the binary representation of the integer $x$.
Instead of evaluating all exponential bitstrings, we sample the ansatz $N$ times and approximate the loss function as
\begin{equation}
    \mathcal{L}(\vec\theta) \approx \frac{1}{N} \sum_{j=1}^{N} f(\text{bin}(x_{(j)})),
\end{equation}
where $x_{(j)}$ are integer-valued samples of the quantum state and in practice the samples are first accumulated to avoid repetitive evaluations of the objective function for the same input.
With an efficient routine to evaluate the expectation value of the Hamiltonian at hand, we can now use imaginary-time evolution or QNG to find the ground-state of the system. 

In Ref.~\cite{zoufal_blackbox_2023} we use our proposed workflow to solve a feature-selection problem on up to 59 features on equally many qubits. Since a computing the QGT for this system size is prohibitively expensive, we successfully use QN-SPSA to optimize the loss function.
We refer to the paper for more details and the results, which include a 20-qubit experiment on hardware and a 59-qubit matrix-product-state simulation on a reduced ansatz to demonstrate the scalability of our algorithm.

\section{Conclusion}

In this chapter, we derived a stochastic estimator for the quantum geometric tensor (QGT) using techniques from simultaneous perturbation, stochastic approximation (SPSA) gradients.
Instead of calculating the QGT at a cost of $\mathcal{O}(d^2)$ circuit evaluations for a parameterized circuit with $d$ parameters, samples of this estimator can be obtained at a constant cost.
With this sampling approach we proposed a stochastic version of quantum natural gradients, which we called QN-SPSA, which only requires $\mathcal{O}(1)$ circuit evaluations per iteration.
In detailed numerical experiments we find that this optimization routine inherits the favorable convergence properties and robustness of natural gradients at the computational efficiency of SPSA.
As applications, we considered molecular ground-state preparation and generative learning using variational quantum Boltzmann machines.
In particular for molecular ground states we find a significant improvement in the number of required circuit measurements for QN-SPSA. 
Our algorithm also contributed to the reported 120x speedup in the calculation of the LiH dissociation curve using Qiskit Runtime~\cite{ibm_120x_2021}.
The performance is likely amplified by the fact that for Hamiltonian with a large number of terms, as is typically the case in molecular systems, the overhead to implement QN-SPSA over SPSA is negligible.

The QN-SPSA algorithms allows the incorporation of a variety of improvements and hyperparameter tuning.
These include improving the stability of the QGT estimator in the initial iterations or discarding updates that increase the loss function due to numerical instability.
Further investigations into the optimal choice of these values, as well as the remaining hyper-parameters, are interesting open questions.
In addition, more in-depth comparisons to budget-friendly optimization techniques, such as iCANS~\cite{kubler_icans_2020} or BayesMGD~\cite{stanisic_observing_2022}, could provide insights into best practices for optimization of variational quantum algorithms.
A caveat of QN-SPSA could be the evaluation of the fidelity for the computation of a QGT sample, which can require doubling either width or depth of the ansatz circuit.

To conclude, we provide a general procedure to approximate the QGT.
In an optimization context, this allows us to converge with less measurements and scale problems up to larger sizes the previously possible. 
The introduced technique, however, is more widely applicable and could provide a benefit for algorithms where the QGT evaluation is a bottleneck, such as in variational quantum time evolution which we consider in in Chapter~\ref{chap:saqite}.

%% file: main/ch5_saqite.tex
\chapter[Stochastic approximation of quantum time evolution]{Stochastic approximation\\ of quantum time evolution}\label{chap:saqite}

\summary{This chapter is based on the article "Stochastic Approximation of Variational Quantum Imaginary Time Evolution" by Gacon et al., in the proceedings of the 2023 IEEE International Conference on Quantum Computing and Engineering (QCE).
Building on the previous chapter, we leverage the stochastic approximation of the quantum geometric tensor to reduce the cost of variational quantum time evolution. 
Compared to QN-SPSA, which solely aims to minimize the system energy, the here introduced modifications focus on providing an accurate estimation at each timestep.
As application, we perform the imaginary-time evolution of a hardware-native Ising model on a complete 27 qubit chip.
}

\noindent
As we have discussed earlier, quantum computers are a promising platform to efficiently perform quantum time evolution (QTE)---a task that can be challenging for classical processors due to the exponential scaling of the system's wave function.
Since near-term quantum computers are noisy and have limited connectivities, we introduced variational quantum time evolution (VarQTE)~\cite{yuan_varqte_2019}, which maps the state's time evolution to the evolution of parameters in a parameterized ansatz. Crucially, the ansatz can be restricted to circuit within the device's capabilities. We refer to Section~\ref{sec:varqte} for the theory background used in this chapter.

Computing the parameter dynamics for VarQTE is based on quantum geometric tensor (QGT) and evolution gradient. The QGT describes the sensitivity of the ansatz with respect to its parameters and the evolution gradient encodes the direction of the parameter dynamics on the variational manifold.
For $d \in \mathbb{N}$ variational parameters in the ansatz, calculating the QGT and evolution gradient require evaluating a number of circuits scaling as $\mathcal{O}(d^2)$ and $\mathcal{O}(d)$, respectively. 
In the previous chapter on quantum natural gradients (QNG), see Fig.~\ref{fig:qng_scaling}, we have seen that these evaluations quickly becomes a bottleneck for even moderately sized circuits with a few 100 parameters. The same considerations hold for VarQTE as well, as the QNG update step coincides with VarQTE for imaginary-time evolution, and the overhead for real-time evolution is even greater.
Due to this unfavorable scaling, a range of different approaches to variational time evolution have been proposed, which we already reviewed in Section~\ref{sec:varqte}.

The stochastic approach presented in this chapter allows to perform both real- and imaginary quantum time evolution by extending the simultaneous perturbation, stochastic approximation techniques previously used for QNG~\cite{spall_spsa_1988, gacon_qnspsa_2021}.
Instead of evaluating the QGT and evolution gradients in each timestep, we use accurate initial values and iteratively update the estimators with unbiased samples.
Each sample can be obtained at a constant cost independent of the number of parameters, i.e., $\mathcal{O}(1)$.
In numerical benchmarks we provide evidence that this stochastic approach requires fewer measurements than VarQTE to achieve the same accuracy.
This reduction in computational cost allows us to perform an imaginary-time evolution on 27 qubits on an IBM Quantum processor~\cite{ibm_quantum}.

\section{Stochastic estimation of the parameter dynamics}

As a solution to the prohibitive scaling of the computational costs involved in evaluating the QGT for a large number of parameters, we propose to replace $g$ with a stochastic estimate $\hat g$ that can be computed efficiently~\cite{gacon_qnspsa_2021}.
This estimator, introduced in Chapter~\ref{chap:qnspsa}, leverages the fact that the QGT is the Hessian of the Fubini-Study metric, which allows to construct a sample $\hat g$ at point $\vec\theta$ as
\begin{equation}
    \hat g = -\frac{\delta F}{8 \epsilon^2}\left(\frac{1}{2 \vec\Delta (\vec\Delta')^\top} + \frac{1}{2\vec\Delta' \vec\Delta^\top}\right),
\end{equation}
where $\vec\Delta, \vec\Delta' \in \mathbb{R}^d$ are random perturbation directions whose inverse is understood element-wise, $\epsilon > 0$ is a small perturbation, and $\delta F$ is measure for the curvature defined as 
\begin{equation}
    \delta F = F(\vec\theta, \vec\theta + \epsilon(\vec\Delta + \vec\Delta')) 
                - F(\vec\theta, \vec\theta + \epsilon(\vec\Delta - \vec\Delta'))
                - F(\vec\theta, \vec\theta + \epsilon(\vec\Delta' - \vec\Delta)) 
                + F(\vec\theta, \vec\theta - \epsilon(\vec\Delta + \vec\Delta')).
\end{equation}
The fidelity $F(\vec\theta, \vec\theta') = |\braket{\phi(\vec\theta)|\phi(\vec\theta')}|^2$ can be evaluated with a range of techniques explained in Section~\ref{sec:fidelities}.
The perturbation directions must obey certain properties described in Section~\ref{sec:spsa}, such as having zero mean and existing inverse moments. A commonly used distribution fulfilling all requirements is the Bernoulli distribution with equal probabilities for sampling $1$ or $-1$ in each dimension, i.e., $\vec\Delta, \vec\Delta' \sim \mathcal{U}(\{1, -1\}^d)$.

In variational imaginary-time evolution, $\vec b$ is the gradient of the energy, which also can be sampled from with only a single perturbation direction, that is
\begin{equation}\label{eq:b}
   \hat{\vec b} = -\frac{1}{2}\frac{E(\vec\theta + \epsilon\vec\Delta) - E(\vec\theta - \epsilon\vec\Delta)}{2\epsilon \vec\Delta}.
\end{equation}
Note that for real-time evolution, it is not known how to write the evolution gradient as the gradient of a scalar function. Therfore it cannot be sampled from but has to be evaluated with a cost of $\mathcal{O}(d)$.

As in both estimators, $\hat g$ and $\hat{\vec b}$, all parameter dimensions are perturbed simultaneously, the evaluation cost does not depend on the number of parameters $d$ in the model. Instead of $\mathcal{O}(d^2)$ circuits for $g$ and $\mathcal{O}(d)$ circuits for $\vec b$, only a constant number of circuits, $\mathcal{O}(1)$, have to be evaluated per sample.

\subsection{Improving estimator accuracy}

In the previous chapter on QN-SPSA we have seen that individual samples $\hat g$ can have a very low accuracy to approximate the QGT.
To reduce the error, we therefore averaged over a set of $M$ samples and combined them into a global average with all previous samples (remember Eqs.~\eqref{eq:qnspsa_average} and~\eqref{eq:qnspsa_global_average}).
The task of time evolution, however, requires a higher accuracy than ground-state search as we are not only interested in the final state, but aim to track the dynamics as closely as possible throughout every timestep.

We therefore propose the following adaption of QN-SPSA to make it suitable for time evolution. 
In addition to re-sampling the QGT, we also draw multiple samples of the evolution gradient, such that the samples in every step are
\begin{equation}
    \begin{aligned}
    \hat g_M &= \frac{1}{M} \sum_{m=1}^M \hat g_{(m)}, \\
    \hat{\vec b}_M &= \frac{1}{M} \sum_{m=1}^M \hat{\vec b}_{(m)}.
    \end{aligned}
\end{equation}
Further, a global average over all samples cannot correctly capture the dynamics of $g$ and $\vec b$.
Instead, we suggest to re-combine the samples via a momenta $\tau_1, \tau_2 \in [0, 1]$, such that the estimators in timestep $k$ are given by
\begin{equation}
    \begin{aligned}
    \bar{g}^{(k)} &= \tau_1 \bar{g}^{(k - 1)} + (1 - \tau_1) \hat{g}_M^{(k)}, \\
    \bar{\vec{b}}^{(k)} &= \tau_2 \bar{\vec{b}}^{(k - 1)} + (1 - \tau_2) \hat{\vec{b}}_M^{(k)}.
    \end{aligned}
\end{equation}

Especially in the first few timesteps, the estimators are biased towards the initial values $\bar g^{(0)}$ and $\bar{\vec b}^{(0)}$, thus, 
it is important to find high-accuracy initial estimates.
These initial values could be computed using a large number of samples $M$ or even by paying the $\mathcal{O}(d^2)$ price to use the analytic formulae for $g$ and $\vec b$.
However, in many cases the circuit implementing the variational ansatz $\ket{\phi(\vec\theta)}$ and the initial point have a simple structure that allow to efficiently evaluate the QGT and evolution gradient classically.

Circuits, for example, that are constructed from Pauli rotations and Clifford gates at initial points that are integer multiples of $\pi/2$ require only the evaluation of Clifford circuits to compute $g$ and $\vec b$.
That is because in this case the QGT and evolution gradient can be evaluated with the parameter-shift rules with a shift of $\pi/2$ and at angles $k\pi/2,~k\in\mathbb{Z}$, the Pauli rotations themselves become Clifford, as they can be expressed as 
\begin{equation}
    \begin{aligned}
    R_X\left(\frac{\pi}{2}\right) &= S^\dagger H S^\dagger, \\
    R_Y\left(\frac{\pi}{2}\right) &= X H, \\
    R_Z\left(\frac{\pi}{2}\right) &= H R_X\left(\frac{\pi}{2}\right) H,
    \end{aligned}
\end{equation}
and $R_{X, Y, Z}(k\pi/2) = R^k_{X, Y, Z}(\pi/2)$.

\begin{figure}[t]
    \centering
    \[
    \begin{array}{c}
    \Qcircuit @C=1em @R=.3em {
        \lstick{\ket{0}} & \gate{R_Y(0)} & \ctrl{1} & \qw & \qw   & \gate{R_Y\left(\frac{\pi}{2}\right)} & \qw & \ket{+} \\
        \lstick{\ket{0}} & \gate{R_Y(0)} & \targ & \ctrl{1} & \qw & \gate{R_Y\left(\frac{\pi}{2}\right)} & \qw & \ket{+} \\
        \lstick{\ket{0}} & \gate{R_Y(0)} & \qw & \targ & \ctrl{1} & \gate{R_Y\left(\frac{\pi}{2}\right)} & \qw & \ket{+} \\
        \lstick{\ket{0}} & \gate{R_Y(0)} & \qw & \qw & \targ      & \gate{R_Y\left(\frac{\pi}{2}\right)} & \qw & \ket{+}     }
    \end{array}
        ~~~~
        \stackrel{\partial_2}{\rightarrow}
        ~~~~~~~~
    \begin{array}{c}
    \Qcircuit @C=1em @R=.3em {
        \lstick{\ket{0}} & \qw & \qw & \ctrl{1} & \qw & \qw                   & \gate{H} & \gate{X} & \qw  \\
        \lstick{\ket{0}} & \gate{H} & \gate{X} & \targ & \ctrl{1} & \qw & \gate{H} & \gate{X} & \qw \\
        \lstick{\ket{0}} & \qw & \qw & \qw & \targ & \ctrl{1}                 & \gate{H} & \gate{X} & \qw \\
        \lstick{\ket{0}} & \qw & \qw & \qw & \qw & \targ                      & \gate{H} & \gate{X} & \qw 
    }
    \end{array}
    \]
    \caption[Clifford evaluation of a circuit gradient]{A hardware-efficient circuit used in a feature selection problem in Ref.~\cite{zoufal_blackbox_2023} with the initial circuit on the left, preparing the $\ket{+}$ state on all qubits, and the circuit to evaluate $b_2(\vec\theta) \propto \partial_2 E(\vec\theta)$ using the parameter-shift rule on the right, where all gates are written in their Clifford representation. The parameter $\theta_2$ is the angle of the first $R_Y$ on the second wire.}
    \label{fig:hw_efficient_clifford_angles}
\end{figure}
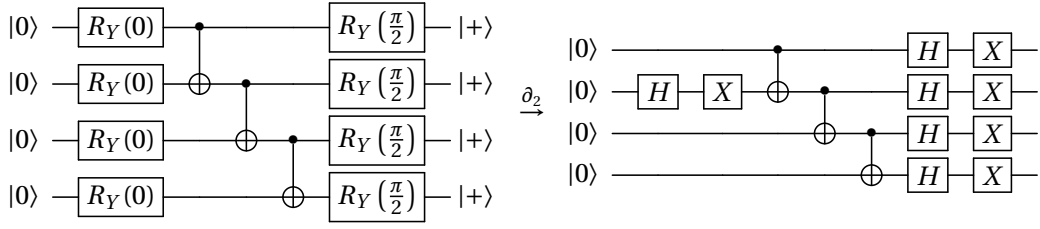

While this might seem restrictive, it includes a commonly used class of circuits used e.g. in ground-state search for molecules, where the variational ansatz first prepares the Hartree-Fock solution, followed by partial swaps~\cite{barkoutsos_ph_2018}, or in combinatorial optimization, which starts in the equal superposition state $\ket{+}$ on all qubits by e.g. using a QAOA ansatz where all parameters are initially set to zero~\cite{farhi_quantum_2014} or a hardware-efficient ansatz with $\pi/2$ angles~\cite{zoufal_blackbox_2023}.
An example of the latter is shown in Fig.~\ref{fig:hw_efficient_clifford_angles}.

\subsection{Stabilizing the linear system}

Solving the linear system $g(\vec\theta) \dot{\vec\theta} = \vec b(\vec\theta)$ for the parameter derivative is only numerically stable for the exact values of the QGT and the evolution gradient~\cite{hackl_varqte_2020}.
The estimates $\bar{g}^{(k)}$ and $\bar{\vec b}^{(k)}$ we have access to, however, are subject to sampling errors due to a finite number of SPSA-based samples, a finite number of measurements in the evaluation of the fidelities and expectation values, and an error due to hardware noise.
Finding a stable solution of the noisy linear system
\begin{equation}\label{eq:noisy_lse}
    \bar g^{(k)}\dot{\vec\theta} = \bar{\vec b}^{(k)},
\end{equation}
therefore requires careful regularization.

A straightforward regularization approach is to add a diagonal shift to the QGT, that is
\begin{equation}\label{eq:diagshift}
    (\bar g^{(k)} + \delta \mathbb{I})\dot{\vec\theta} = \bar{\vec b}^{(k)},
\end{equation}
where $\delta > 0$ is the magnitude of the shift and $\mathbb I \in \mathbb{R}^{d\times d}$ is the identity matrix.
In practice, it is often useful to re-write the linear system as convex, quadratic program, i.e.,
\begin{equation*}
    \dot{\vec\theta} = \argmin_{\dot{\vec\theta}} \frac{\dot{\vec\theta}^\top (\bar g^{(k)} + \delta\mathbb I) \dot{\vec\theta}}{2} - \dot{\vec\theta}^\top \bar{\vec b}^{(k)},
\end{equation*}
and solve it with a minimization subroutine (in this work we use COBYLA~\cite{powell_cobyla_1994}).
This diagonal shift is equivalent to adding an offset $\delta$ to the eigenvalues of the QGT estimate, which decreases the condition number and results in a more stable linear system.
As shown in Chapter~\ref{chap:qnspsa} a diagonal shift, however, has an impact on the parameter dynamics and can therefore be problematic for time evolution.

Another option to stabilize the linear system is to solve it only the stable subspace where the eigenvalues of the QGT estimator are above a given threshold.
Instead of damping the magnitude of the update step, as under a diagonal shift, this regularization only takes parameter dimensions into account that have sufficient contribution to the dynamics. 
Since the QGT estimate is a real, symmetric matrix it can be diagonalized, and the linear system becomes 
\begin{equation*}
    \bar g^{(k)}\dot{\vec\theta} = B\Lambda B^\top \dot{\vec\theta},
\end{equation*}
where $B$ is an orthonormal matrix and $\Lambda = \mathrm{diag}(\lambda_1, \lambda_2, ..., \lambda_d)$ is a diagonal matrix containing the real eigenvalues $\{\lambda_j\}_{j=1}^d$ of $\bar{g}^{(k)}$.
By defining the basis-transformed vectors $\dot{\vec\theta}^B = B^\top \dot{\vec\theta}$ and $\vec b^B = B^\top\bar{\vec b}^{(k)}$, we can write the linear system for the parameter update in a diagonal form, given by
\begin{equation}
    \Lambda\dot{\vec\theta}^B = \vec{b}^B.
\end{equation}
We now solve this system but only consider the well-conditioned components, that is
\begin{equation}\label{eq:stable_subspace}
    \dot\theta^B_j = \begin{cases}
    b^B_j / \lambda_j, \text{ if } \lambda_j \geq \delta, \\
    0, \text{ otherwise},
    \end{cases}
\end{equation}
and then transform the solution back to the original basis, as $\dot{\vec\theta} = B\dot{\vec\theta}^B$.
The threshold $\delta > 0$ should be chosen large enough to cut off noisy signals due to finite sampling and measurement errors, but as small as possible to not ignore significant contributions to the parameter dynamics.

\section{Numerical experiments}

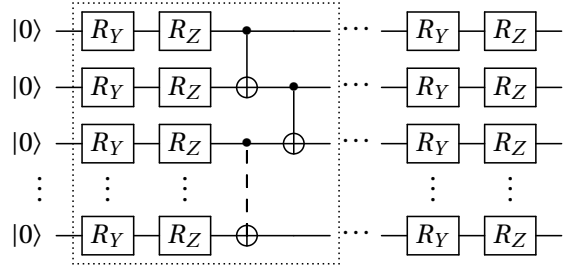
\begin{figure}[htbp]
    \centering
    \[
    \begin{array}{c}
    \Qcircuit @C=1em @R=.7em {
        \lstick{\ket{0}} & \gate{R_Y} & \gate{R_Z} & \ctrl{1} & \qw & \qw & \cdots & & \gate{R_Y} & \gate{R_Z} & \qw\\
        \lstick{\ket{0}} & \gate{R_Y} & \gate{R_Z} & \targ & \ctrl{1} & \qw & \cdots & & \gate{R_Y} & \gate{R_Z} & \qw \\
        \lstick{\ket{0}} & \gate{R_Y} & \gate{R_Z} & \control\qw & \targ & \qw & \cdots & & \gate{R_Y} & \gate{R_Z} & \qw\\
        \lstick{\vdots} & \vdots & \vdots & & &  & &  & \vdots & \vdots &  \\
        & & & & & & & & & & \\
        \lstick{\ket{0}} & \gate{R_Y} & \gate{R_Z} & \targ & \qw & \qw & \cdots & & \gate{R_Y} & \gate{R_Z} & \qw 
                \gategroup{1}{2}{6}{6}{.7em}{.}
                \gategroup{3}{4}{6}{4}{-.8em}{--}
    }
    \end{array}
    \]
    \caption[Ansatz for SA-QITE experiments]{The circuit preparing $\ket{\phi(\vec\theta)}$ in the following numerical experiments. The dashed box is repeated $r = \lceil\log(n)\rceil$ times.}
    \label{fig:efficientsu2}
\end{figure}

In this section we benchmark the performance of SA-QITE compared to standard VarQITE on a transverse field Ising model on $n$ spin-$1/2$ particles, given by
\begin{equation}\label{eq:ising}
    H = J \sum_{j=1}^{n - 1} Z_j Z_{j + 1} + h\sum_{j=1}^{n} X_i,
\end{equation}
for $J=1/2$ and $h=-1$.
We select a variational ansatz $\ket{\phi(\vec\theta)}$ that is prepared by a circuit reflecting the nearest-neighbor interactions of the system. The circuit, shown in Fig.~\ref{fig:efficientsu2}, consists of $r$ repetitions of rotation layers, implemented by individually parameterized Pauli-$Y$ and -$Z$ rotations, and entanglement layers made up of pairwise CX gates.
This circuit has a total of $d = 2n(r+1)$ parameters.
Before turning to the accuracy benchmarks, we investigate which regularization method provides the most accurate approximation to the parameter dynamics to improve the results of SA-QITE.

\subsection{Solving for the parameter update}

\begin{figure}[hbtp]
    \centering
    \includegraphics[width=0.9\textwidth]{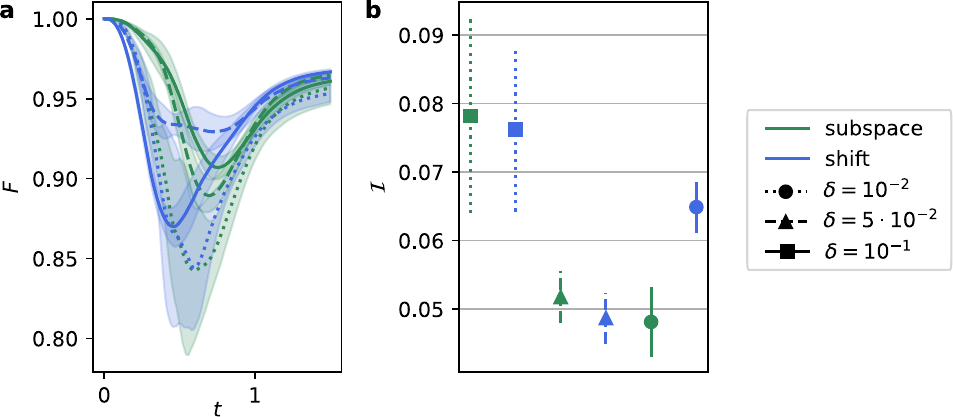}
    \caption[Regularization techniques for SA-QITE]{Comparison of techniques to solve for the parameter derivative. (a) Fidelity $F$ compared to the exact time-evolved state at each time $t$. (b) Integrated infidelity $\mathcal{I}$ for the methods.}
    \label{fig:solvers}
\end{figure}

We conduct a numerical experiment to benchmark both regularization approaches and identify which provides the more accurate time evolution
on the Ising Hamiltonian of Eq.~\eqref{eq:ising} with $n=8$ spins.
In Figure~\ref{fig:solvers} we perform the imaginary time evolution up to time $T=1.5$ for different thresholds $\delta$ and compare the diagonal shift and stable subspace regularizations. 
We measure the accuracy of a specific setting by measuring the fidelity $F$ and average integrated infidelity $\mathcal{I}$, which is defined as
\begin{equation}
    \mathcal{I}(T) = \frac{1}{T} \int_0^T \left(1 - |\braket{\phi(\vec\theta(t))|\Psi(t)}|^2 \right) \mathrm{d}t,
\end{equation}
where $\ket{\Psi(t)}$ is the exact time-evolved state and $\ket{\phi(\vec\theta(t))}$ is the SA-QITE solution.
We find that solving the linear system in the stable subspace at a threshold of $\delta=10^{-2}$ provides the best result and is less sensitive to changes, than using a diagonal shift of $\delta$.

\subsection{Resource requirements}

\begin{figure}[t]
    \centering
    \includegraphics[width=\linewidth]{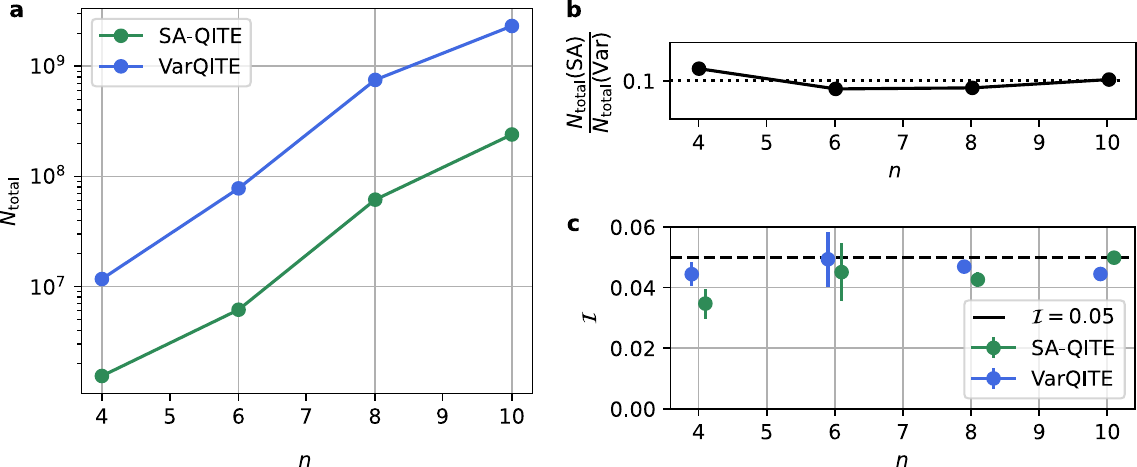}
    \caption[Resource comparison of SA-QITE and VarQITE]{Total number of measurements, $N_\mathrm{total}$, required to achieve the target accuracy for SA-QITE and VarQITE, along with the fraction of both resource counts. SA-QITE requires $\approx 10\%$ of the number of measurements compared to VarQITE.}
    \label{fig:resources}
\end{figure}

To measure the performance of SA-QITE and VarQITE we implement the imaginary time evolution up to $T=1.5$ and count the total number of circuit measurements $N_\text{total}$ required to achieve an average integrated fidelity of $\mathcal{I} = 0.05$.
We repeat the experiment for $n=4$ to $10$ qubits, using $r = \lceil\log{n}\rceil$ repetitions of rotation and entangling layers.
For each system size we optimize the hyperparameters of both algorithms to use the least possible resources while still maintaining the target accuracy.
In Fig.~\ref{fig:resources}(a) we present the resource counts for both algorithms, along with the fraction of measurements taken by SA-QITE and VarQITE.
We find that the proposed SA-QITE approximately requires an order of magnitude less measurements than VarQITE, as shown in Fig~\ref{fig:resources}(b), but shows the same asymptotic scaling.
In a setting, where not all parameters contribute to the dynamics, however, we expect that SA-QITE needs less samples to converge, while VarQITE requires $\mathcal{O}(d^2)$ measurements regardless. 
In Table~\ref{tab:saqite_settings} we detail the settings and visualize the achieved accuracies in Fig.~\ref{fig:resources}(c).

\begin{table}[htp]
    \centering
    \subfloat[][]{
    \centering{
    \begin{tabular}{c|cc}
        $n$ & $N$ & $\delta$ \\ \hline
        4 & 128 & 0.05 \\
        6 & 400 & 0.05 \\
        8 & 1024 & 0.01 \\
        10 & 2048 & 0.05 
    \end{tabular}
    }
    }
    \subfloat[][]{
    \centering{
    \begin{tabular}{c|ccccc}
        $n$ & $N$ & $M$ & $\tau_1$ & $\tau_2$ & $\delta$ \\ \hline
        4 & 128 & 10 & 0.99 & 0.7 & 0.05 \\
        6 & 256 & 20 & 0.99 & 0.9 & 0.05  \\
        8 & 512 & 75 & 0.99 & 0.7 & 0.05  \\
        10 & 800 & 250 & 0.99 & 0.7 & 0.05  \\
    \end{tabular}
    }
    }
    \caption[Algorithm settings for the SA-QITE resource estimation]{Algorithm settings for the resource estimations for (a) VarQITE and (b) SA-QITE, including the number of qubits $n$, the number of measurements per basis $N$, the number of samples for the QGT and energy gradient $M$, the momenta $\tau_1$ and $\tau_2$, and the cutoff $\delta$ in the stable subspace solver.}
    \label{tab:saqite_settings}
\end{table}

\section{Applications}

\subsection{27-qubit imaginary-time evolution}

\begin{figure}[thbp]
    \centering
    \includegraphics[width=0.7\linewidth]{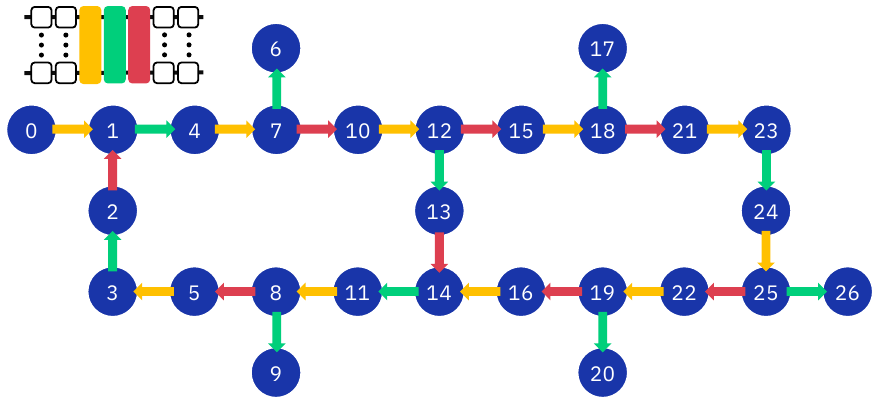}
    \caption[Hardware-efficient heavy-hex circuit]{The coupling map of IBM's 27-qubit Falcon processor. The connections are colored depending on the CX layer they are executed in, and the arrows are pointing from control to target qubit.}
    \label{fig:falcon_chip}
\end{figure}

Since SA-QITE promises a resource reduction compared to VarQITE, we now scale the Ising model up to a larger system size of $n=27$ qubits and perform the imaginary-time evolution on \texttt{ibm\_auckland}, which is one of the IBM Quantum Falcon processors \cite{ibm_quantum}.
In this experiment we consider spin interactions implemented by the coupling map of the quantum processor, which is shown in Fig.~\ref{fig:falcon_chip}, an interaction strength of $J=0.1$ and transversal field strength of $h=-1$.
This choice enables us to efficiently implement an ansatz circuit that reflects the spin interactions without introducing additional Swap gates upon compiling the circuit.
Here, we use the same structure as in the previous experiments with $r=1$, see Fig.~\ref{fig:efficientsu2}, but the pairwise CX interactions are between all connections of the device's coupling map. This circuit can be implemented with a minimal CX depth of three.

\begin{figure}[thbp]
    \centering
    \includegraphics[width=0.6\linewidth]{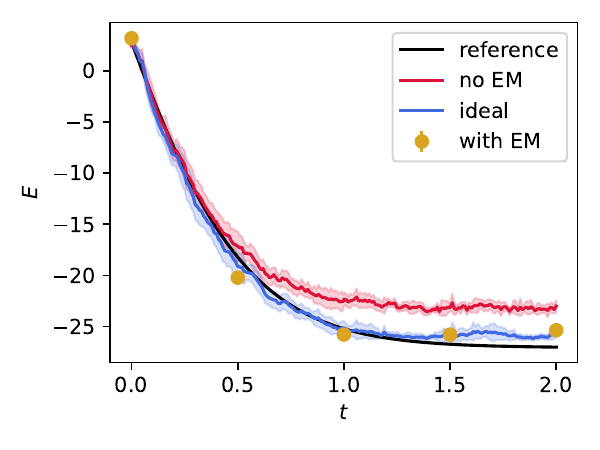}
    \caption[Hardware results for 27-qubit SA-QITE]{The energy $E$ of the imaginary-time evolution as a function of time $t$. 
        The black reference line is obtained by a classical Taylor expansion, the red line shows mean and standard deviation
        over five SA-QITE runs without any error mitigation, and the blue line are mean and standard deviation of the noisy parameters
        with ideal, statevector-based evaluation. 
        The golden circles show the error-mitigated results including standard deviation over five experiments.
        }
    \label{fig:27q_energies}
\end{figure}

We perform SA-QITE from the initial state $\ket{0}^{\otimes n}$, which is prepared by the initial parameters $\vec\theta_0 = \vec 0$, and integrate up to time $T=2$ with a timestep of $\Delta_t = 10^{-2}$. In each timestep we average over $M=10$ samples of the QGT and evolution gradient and re-combine them with previous samples using the momenta terms $\tau_1 = 0.99$ and $\tau_2 = 0$. Each circuit is measured with $N=1024$ shots.
In Fig.~\ref{fig:27q_energies} we show the evolution of the energy measured on the hardware and compare to a classical reference calculation, based on a Taylor expansion of the imaginary-time evolution operator, see Appendix~\ref{app:taylor_qite}.

Even without any error mitigation (EM), the energies match the reference up to $t \approx 0.5$, but then proceed to converge to a plateau with constant offset to the final energy.
To understand whether this offset is due to a faulty imaginary-time evolution or due to noise in the energy evaluation, we evaluate the noisy parameters obtained by the experiment in an exact, statevector-based calculation. We find that the energies closely follow the reference calculation, which suggests that, despite the hardware noise, SA-QITE correctly tracked the parameter dynamics and that with sufficient error-mitigation we can obtain an accurate estimation of the correct energies.

\begin{figure}[thbp]
    \centering
    \includegraphics[width=0.8\linewidth]{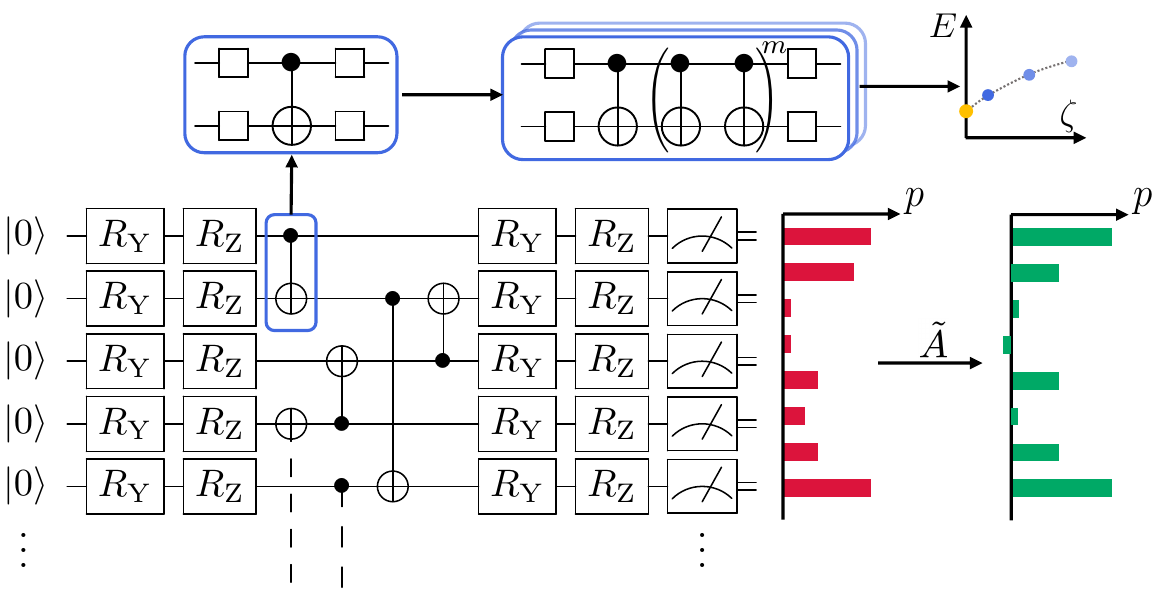}
    \caption[Error mitigation scheme in practice]{The error mitigation scheme used on \texttt{ibm\_peekskill}. The CX gates are first twirled, such that the total action remains unchanged, and then folded with $m \in \{0, 1, 2\}$ to increase the noise. Each energy evaluation is further measurement mitigated, before we extrapolate to the final error-mitigated result.
    }
    \label{fig:em}
\end{figure}

To evaluate the energy to a higher accuracy at specific times we use a second 27-qubit device, \texttt{ibm\_peekskill}, and use a set of error mitigation techniques summarized in Fig.~\ref{fig:em}.
To mitigate measurement errors, which are especially dominant in shallow circuits as in this experiment, we use the M3 measurement mitigation~\cite{nation_m3_2021} calibrated on 1000 measurements. As explained in detail in Section~\ref{sec:mem}, this method is scalable to a large number of qubits as the dimension of the transfer matrix $\tilde{A}$ is limited by the number of taken measurements. 
We then use zero-noise extrapolation (ZNE)~\cite{giurgica_zne_2020} to first increase the hardware noise in the circuit by repeating all CX gates, as
\begin{equation}
    \mathrm{CX} \rightarrow \mathrm{CX}^{2m + 1},
\end{equation}
where we use $m \in \{0, 1, 2\}$, and then extrapolate to the zero-noise limit. For this so-called gate-folding we expect an exponential increase in the error and therefore use an exponential fit to extrapolate the errors, that is
\begin{equation}
    E(\zeta) = a + be^{c\zeta},
\end{equation}
where $\zeta \in \{1, 3, 5\}$ is the number of CX gates and is extrapolated to $E(\zeta=0)$.
In practice, ZNE performs best if combined with Pauli twirling~\cite{kim_scalable_2023}, which is a technique to transform coherent noise into stochastic noise. After repeating the CX gates, we therefore twirl the CX block with random, but unitary-preserving Pauli gates and average over 25 repetitions. See Section~\ref{sec:zne} for more detail.
After applying the EM techniques, we find that indeed we are able to closely match the reference calculations. 
This shows that SA-QITE can be executed with no, or minimal, EM to obtain the variational parameters, followed by EM for the points of interest.

\subsection{Ground-state search}

Imaginary-time evolution is guaranteed to converge to the ground state of a system, if the initial state has sufficient overlap with it.
Variational methods have additional conditions to fulfill for this guarantee, including that the variational manifold is expressive enough to correctly track the dynamics and that the time integration is sufficiently precise. 
Even without these properties however, which are generally hard to verify, there is a strong motivation to use variational imaginary time evolution for ground-state preparation. As previously discussed in Chapter~\ref{chap:qnspsa}, these properties transfer to Quantum Natural Gradients.

\begin{figure}[!t]
    \centering
    \includegraphics[width=0.4\linewidth]{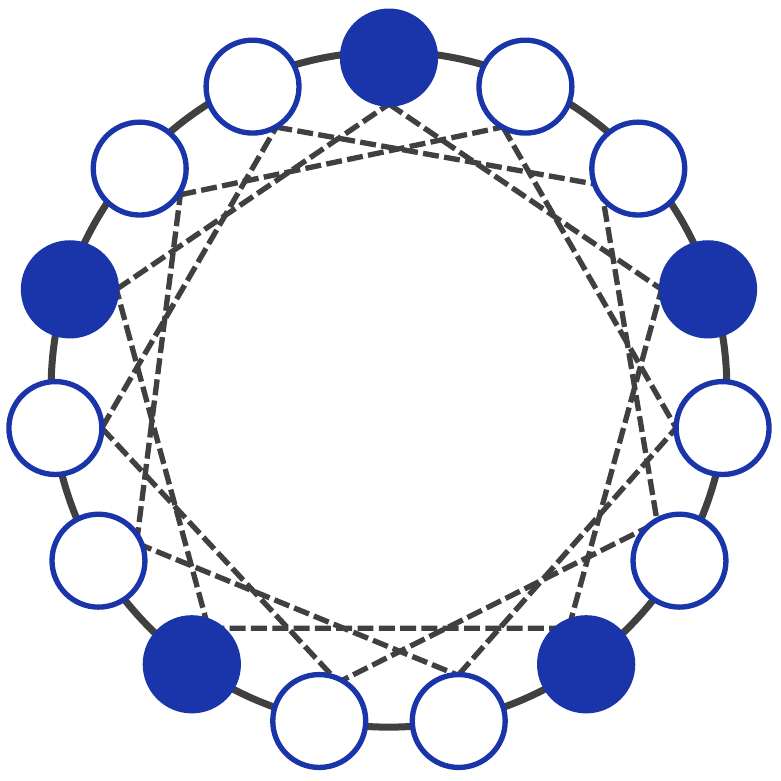}
    \caption[Circular graph for Max-Cut]{
        The circular graph in the Max-Cut application. The edge weights are $w_1 = 20$ on solid lines and $w_2 = -20$ on dashed lines.
        The node colors show one of the six optimal configurations. The other optimal solutions can be obtained by rotating the coloring, which gives two additional configurations, and by inverting every solution, which gives additional three solutions.
    }
    \label{fig:circle}
\end{figure}

In a ground-state search application the exact trajectory is not required, but we are only interested in the final state. In this case, we can relax the number of samples $M$ in SA-QITE to trade-off the accuracy during the time evolution and resources.
As an example, we solve the Max-Cut problem on a circular graph with $n=15$ nodes, shown in Fig.~\ref{fig:circle}, using the Quantum Approximate Optimization Algorithm (QAOA)~\cite{farhi_quantum_2014}.

The goal of Max-Cut is to group the nodes of a weighted graph into two groups, such that the sum of the edge weights across the groups is maximized.
This problem can be stated as maximizing the following cost function
\begin{equation}
    C(\vec x) = \sum_{\braket{jk}} \frac{w_{jk}}{2} \left(x_j (1 - x_k) + x_k (1 - x_j) \right),
\end{equation}
where $x_j \in \{0, 1\}$ is a binary variable indicating the class of the node, $w_{jk}$ is the edge weight between nodes $j$ and $k$ and $\braket{jk}$ iterates over the connections of the graph. 
We can cast this problem to an Ising Hamiltonian by replacing the binary variables $x_j$ with spin variables $z_j \in \{-1, 1\}$, as
\begin{equation}
    x_j \mapsto z_j = \frac{1 - x_j}{2},
\end{equation}
and then promoting the spin variable $z_j$ to a Pauli-$Z$ operator acting on qubit $j$.
Dropping all constant terms, as they do not affect the solution, the cost Hamiltonian for the circular graph is given by
\begin{equation*}
    H_C = w_1 \sum_{j=1}^{n} Z_j Z_{(j+1) \text{ mod } n} + w_2 \sum_{j=1}^{n} Z_j Z_{(j + 3) \text{ mod } n},
\end{equation*}
with $w_1 = -w_2 = 20$. 
The ground-state of this Hamiltonian encodes the solution of the Max-Cut problem.

QAOA is a variational algorithm that minimizes the energy of $H_C$ by optimizing the parameters of a specific variational ansatz, defined as 
\begin{equation*}
    \ket{\phi(\vec\gamma, \vec\beta)} = \left( \prod_{p=r}^{1} e^{-i\beta_p H_M} e^{-i \gamma_p H_C}\right) \ket{+}^{\otimes n},
\end{equation*}
with the mixer $H_M = -\sum_{i=1}^n X_i$, parameters $\vec\gamma, \vec\beta \in \mathbb{R}^{r}$, and we choose $r=2$. 
This form is motivated by simulated annealing from $H_M$ to $H_C$, where the ground-state of $H_M$ should be easy to prepare, see also Section~\ref{sec:ansatz}. 
In this experiment, the mixer's ground state is $\ket{+}^{\otimes n}$ and can simply be constructed by Hadamard gates on all qubits and setting all variational parameters to $0$.
The ansatz can be directly implemented on a gate-based quantum computer, as $\exp(-\beta_p H_M)$ equals a layer of $R_X$ gates, and $\exp(-\gamma_p H_C)$ can be implemented using $R_{ZZ}$ gates.

\begin{figure}[!t]
    \centering
    \includegraphics[width=0.495\linewidth]{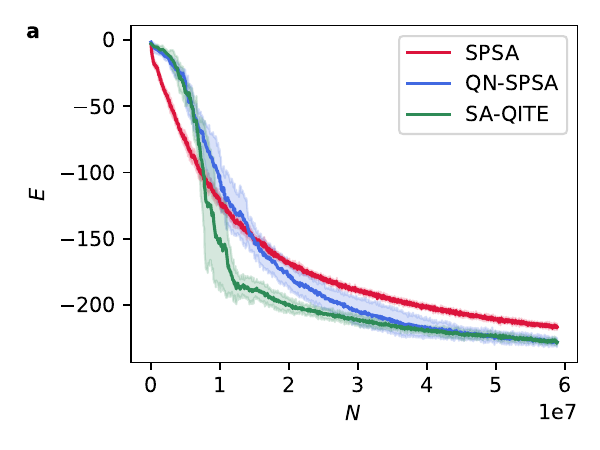}
    \includegraphics[width=0.495\linewidth]{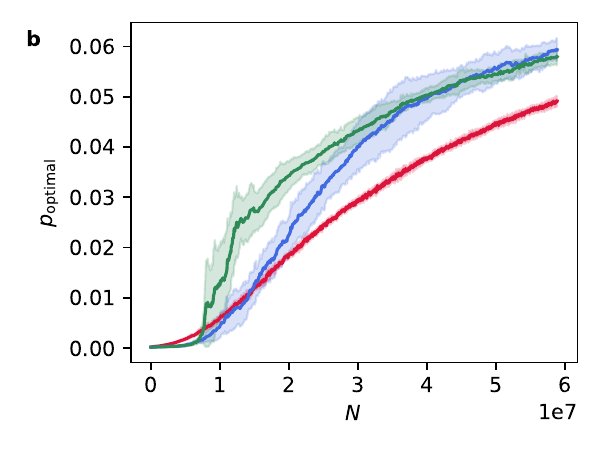}
    \caption[Max-Cut loss and optimal sample probabilities]{
    (a) The energies $E$ of the cost Hamiltonian as a function of total number of measurements $N$ for
    SPSA and the natural gradient adaptations.
    (b) The probability $p_\text{optimal}$ to sample one of the optimal states as a function of $N$.
    }
    \label{fig:qaoa_energies}
\end{figure}

In the first iteration of SA-QITE we use $M_0 = 10$ samples, which we then reduce to $M_k = \max\{1, \lfloor (0.9)^{k} M_0 \rfloor\}$ in the $k$th iteration.
We use a QGT momentum of $\tau_1 = 0.99$, but do not use momentum on the evolution gradient, to avoid any bias once the optimization has converged.
At the initial values $(\vec \beta, \vec\gamma) = (\vec 0, \vec 0)$ the circuit is Clifford for any parameter-shift of $\pi/2$, which allows to evaluate the QGT and evolution gradients exactly on a classical computer.
To stabilize the linear system solution, we use a stable subspace regularization with a threshold of $\delta=10^2$, which is approximately $0.5\%$ of the largest eigenvalue of the initial QGT. 
As perturbation for the finite-difference approximations we use $\epsilon=10^{-2}$ at $8 \cdot 10^3$ measurements per circuit evaluation and the perturbation directions are sampled from a Bernoulli distribution over $\{-1, 1\}$, i.e. $\vec\Delta, \vec\Delta' \in \mathcal{U}(\{-1, 1\}^{2r})$. We use a timestep of $\Delta_t = 10^{-3}$ for the forward Euler integration.

We compare the performance of SA-QITE against two resource-efficient optimizers; SPSA~\cite{spall_spsa_1988} and QN-SPSA~\cite{gacon_qnspsa_2021}, which have previously been introduced in Section~\ref{sec:spsa} and Chapter~\ref{chap:qnspsa}, respectively. 
Note that QN-SPSA equals SA-QITE without improvements of this chapter, i.e., it uses a global average over all $\hat g$ samples and initializes the estimator with the identity, $\bar g^{(0)} = \mathbb{I}$.
Both these optimizers use the same perturbation and number of shots as SA-QITE and QN-SPSA uses the learning rate equivalent to SA-QITE's timestep. We calibrate SPSA's learning rate to the largest, still converging possible value of $\eta = 5 \cdot 10^{-7}$. 
Due to the symmetry of the QAOA loss landscape, we start the optimization from a small perturbation at $\vec\gamma^{(0)} = (10^{-3}, 10^{-3})$ and $\vec\beta^{(0)} = (10^{-2}, 10^{-2})$, to ensure the optimizers converge to the same minimum and can be compared fairly.

The energy as function of the circuit evaluations is shown in Fig.~\ref{fig:qaoa_energies}(a). Similar to the Max-Cut experiment in Section~\ref{sec:qnspsa_convergence}, SPSA requires a small learning rate to reliably converge in the challenging loss landscape. Both QN-SPSA and SA-QITE do not suffer from this issue as they take into account the model sensitivity. Importantly, we further observe that SA-QITE initially outperforms QN-SPSA due to the exact initialization and improved estimators. Towards the end of the optimization, they show similar results.
While the ground state encodes the solution of the optimization problem, in combinatorial optimization it is usually sufficient to measuring an optimal solution, even if the optimization is not yet fully converged.
We therefore measure the probability $p_\text{optimal}$ to sample one of the six optimal bitstrings and show the results in Fig.~\ref{fig:qaoa_energies}(b).
The improved performance of SA-QITE, especially in the beginning of the optimization, amplifies the likelihood to measure an optimal solution significantly faster than the other methods. 
To achieve a $1\%$ probability, amplified from $2^{-15}$, SA-QITE only requires approximately $64\%$ of the number of measurements of SPSA or QN-SPSA.

\section{Conclusion}

In this chapter we leveraged sampling access to the quantum geometric tensor (QGT), previously introduced in Chapter~\ref{chap:qnspsa}, to formulate a stochastic approach to variational quantum imaginary time evolution (VarQITE), which we called SA-QITE.
Instead of $\mathcal{O}(d^2)$ circuit evaluations per timestep, the number of circuit evaluations is $\mathcal{O}(M)$, where $M$ is the number of samples per iteration.

Using an exact initialization and momentum terms for the estimator of the QGT and energy gradients, we find an empirical order of magnitude reduction in the number of measurements of SA-QITE compared to the direct approach of VarQITE.
We expect this advantage to further increase for overparameterized circuits, where the QGT does not have a complex structure and can be efficiently sampled instead of explicitly evaluated.
To demonstrate the cost reduction of SA-QITE enables scaling up to larger systems, we performed a hardware experiment on 27-qubit imaginary time evolution, which provided accurate results in combination with error mitigation.
Due to the connection of imaginary time evolution and natural gradients, reducing the number of samples connects SA-QITE to QN-SPSA of Chapter~\ref{chap:qnspsa}.
We show that the improvements due to initialization and momentum introduced in this chapter also have the potential to speed up ground-state preparations.

While we assumed fix schedules for the number of samples during the time evolution and a fixed time step, the performance of SA-QITE could be further improved by adaptive schemes. 
For example, during times where the system and the QGT do not change significantly, less samples are sufficient.
As with QN-SPSA, the dependence on evaluating the fidelity is a drawback of our approach, especially on near-term quantum computers. 
Without adaptive approaches, the correct setting of hyper-parameters, such as the number of samples, can also be a caveat of our algorithm.

In conclusion, we extended the prominent approach of simultaneous perturbations to variational quantum time evolution. This paves the way to scaling experiments up to larger sizes and probe the limit of current devices.

%% file: main/ch6_dualqte.tex
\chapter{Dual quantum time evolution}\label{chap:dual}

\summary{This chapter is based on the article "Variational Quantum Time Evolution without the Quantum Geometric Tensor" by Gacon et al., published in Phys. Rev. Research \textbf{6} 013143 (2023).
Instead of reducing the cost of variational quantum time evolution by leveraging resource-efficient approximations of the quantum geometric tensor (QGT), we here introduce a dual formulation that completely avoids it.
We show in numerical experiments and a sample complexity analysis that this new approach promises an asymptotic reduction of the number of measurements required for variational time evolution to converge.
As application, we implement the quantum minimally entangled typical thermal states (QMETTS) algorithm to compute thermal averages, which uses imaginary time evolution as subroutine.
}

\noindent
In the previous chapters, we reduced the resource requirements for quantum time evolution and quantum natural gradients (QNG) by approximating the quantum geometric tensor (QGT) with stochastic samples. 
Stochastic approximations have the advantage that they only require a constant number of circuits per sample, independent of the number of variational parameters, and they have proven stable in presence of hardware noise.
In this chapter, we pursue a different approach: instead of finding efficient approximations of the QGT, we use a \emph{dual} formulation that does not require the QGT in the first place.
This formulation solves an optimization problem based on the fidelity and is applicable to real- and imaginary-time evolution, and to the QNG.

\begin{figure}[t]
    \centering
    \includegraphics[width=0.6\linewidth]{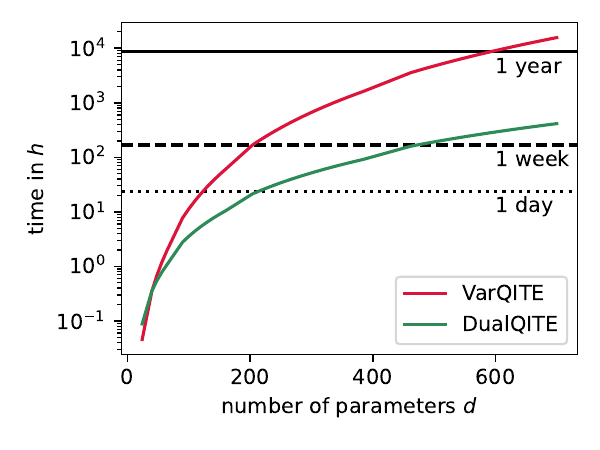}
    \caption[Runtime estimation for VarQITE]{Runtime prognosis of VarQITE and the proposed dual approach (DualQITE) on current superconducting hardware as a function of the number of parameters $d$ in the variational circuit model.
    }
    \label{fig:varqte_scaling}
\end{figure}

We show that this new algorithm significantly reduces the number of measurements required for these applications and, thereby, the expected runtimes on near-term quantum processors.
The estimated speedup is shown in Fig.~\ref{fig:varqte_scaling}, which compares the runtime prognosis of variational quantum imaginary-time evolution (VarQITE) to our proposed dual algorithm (DualQITE), assuming superconducting qubit processor specifications detailed in Appendix~\ref{app:dual_prognosis}.
For already 200 parameters in the circuit model we estimate DualQITE to reduce the runtime from a week computation time to only a single day---a difference that increases with system size.

\section{Dual formulation of variational quantum time evolution}

Under McLachlan's variational principle, variational quantum time evolution (VarQTE) computes the parameter dynamics $\dot{\vec\theta} \in \mathbb{R}^d$ of the variational ansatz $\ket{\phi(\vec\theta)}$ as
\begin{equation}\label{eq:dual_lse}
    g(\vec\theta)\dot{\vec\theta} = \vec b(\vec\theta),
\end{equation}
with the real part of the quantum geometric tensor (QGT),
\begin{equation}
    g_{jk}(\vec\theta) = \mathrm{Re}\left\{\braket{\partial_j \phi(\vec\theta)|\partial_k \phi(\vec\theta)} - \braket{\partial_j \phi(\vec\theta)|\phi(\vec\theta)}\braket{\phi(\vec\theta)|\partial_k \phi(\vec\theta)}\right\},
\end{equation}
and the evolution gradient $\vec b$, which for real-time evolution (VarQRTE) is given by
\begin{equation}\label{eq:b_real}
    b^{(R)}_j(\vec\theta) = \mathrm{Im}\big(\braket{\partial_j\phi(\vec\theta)|H|\phi(\vec\theta)} -\braket{\partial_j\phi(\vec\theta)|\phi(\vec\theta)} E(\vec\theta) \big), 
\end{equation}
and for imaginary-time evolution (VarQITE) by
\begin{equation}\label{eq:b_imag}
    b^{(I)}_j(\vec\theta) = -\mathrm{Re}\big(\braket{\partial_j\phi(\vec\theta)|H|\phi(\vec\theta)}\big) = -\frac{\partial_j E(\vec\theta)}{2},
\end{equation}
with the energy $E(\vec\theta) = \braket{\phi(\vec\theta)|H|\phi(\vec\theta)}$.
See also Section~\ref{sec:varqte} for more details.
In the following we drop the superscript for the evolution gradient, as the techniques apply to both. 

The idea of this chapter is based on solving the dual formulation of the linear system from Eq.~\eqref{eq:dual_lse}, defined as
\begin{equation}\label{eq:argmin_qgt}
   \dot{\vec\theta} = \argmin_{\dot{\vec\theta}} \frac{\dot{\vec\theta}^\top g(\vec\theta) \dot{\vec\theta}}{2} - \dot{\vec\theta}^\top \vec{b}(\vec\theta).
\end{equation}
The term $\dot{\vec\theta}^\top g(\vec\theta) \dot{\vec\theta}$ is a metric describing the magnitude of the derivative $\dot{\vec\theta}$ by how much change it induces in the ansatz $\ket{\phi(\vec\theta)}$ and is derived from the Fubini-Study metric.
For infinitesimally small displacements $\vec{\delta\theta}$, the metric can be written as~\cite{stokes_qng_2020}
\begin{equation}\label{eq:qgt_approximation}
    \begin{aligned}
        ||\vec{\delta\theta}||^2_{g(\vec\theta)} &= \vec{\delta\theta}^\top g(\vec\theta) \vec{\delta\theta} \\
                                       &= 1 - F(\vec\theta,\vec\theta + \vec{\delta\theta}) + \mathcal{O}(\|\vec{\delta\theta}\|_2^3),
    \end{aligned}
\end{equation}
with the fidelity $F(\vec\theta, \vec\theta') = |\braket{\phi(\vec\theta)|\phi(\vec\theta')}|^2$.

We can, thus, reformulate the quadratic optimization problems in terms of the fidelity by explicitly writing the derivative as difference quotient, i.e. $\dot{\vec\theta} = \vec{\delta\theta}/\delta\tau$, for a time perturbation $\delta\tau > 0$. 
We obtain the dual quantum time evolution (DualQTE) update rule
\begin{equation}\label{eq:argmin_dual}
   \vec{\delta\theta}
   \approx \argmin_{\vec{\delta\theta}} \frac{1 - F(\vec\theta, \vec\theta+\vec{\delta\theta})}{2(\delta\tau)^2} - \frac{\vec{\delta\theta}^\top \vec b(\vec\theta)}{\delta\tau} 
   = \argmin_{\vec{\delta\theta}} \frac{\mathcal{L}(\vec{\delta\theta})}{(\delta\tau)^2},
\end{equation}
where we introduce the loss function 
\begin{equation}\label{eq:dual_cost}
\mathcal{L}(\vec{\delta\theta}) = \frac{1 - F(\vec\theta, \vec\theta+\vec{\delta\theta})}{2} - \delta\tau \cdot \vec{\delta\theta}^\top \vec b(\vec\theta).
\end{equation}
The loss function can be optimized without the factor $(\delta\tau)^{-2}$, which, in practice, improves the numerical stability as the quadratic term of the loss is not influenced by the time perturbation.
This dual formulation can also be derived from the QNG~\cite{stokes_qng_2020, sbahi_mirror_2022}, as shown in Appendix~\ref{app:dual_qng}.

Where VarQTE evaluates $\mathcal{O}(d^2)$ circuits per timestep to compute the QGT, the proposed DualQTE algorithm optimizes a loss function based on the fidelity, which only requires the evaluating of a single circuit.
The efficiency of DualQTE thus depends on the number of optimization steps required to minimize the loss function: if the optimal displacement $\delta{\vec\theta}$ can be found with less than $\mathcal{O}(d^2)$ circuit evaluation it improves over VarQTE.
In the following sections we show in numerical experiments and sampling complexity proof that this is indeed the case and DualQTE requires asymptotically less samples to achieve the same accuracy as VarQTE.

\subsubsection{Illustrative example}

\begin{figure}[t]
    \centering
    \includegraphics[width=0.49\textwidth]{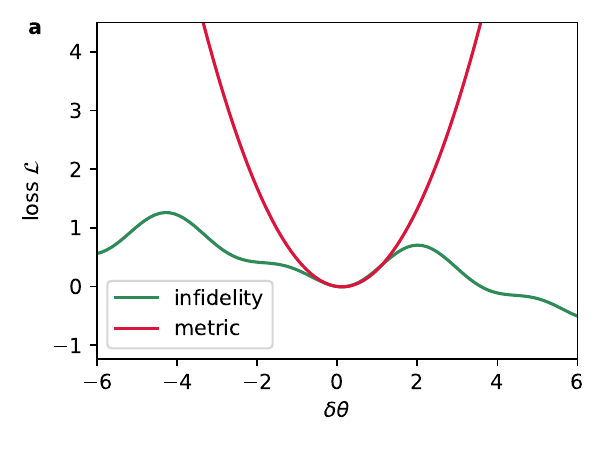}
    \includegraphics[width=0.49\textwidth]{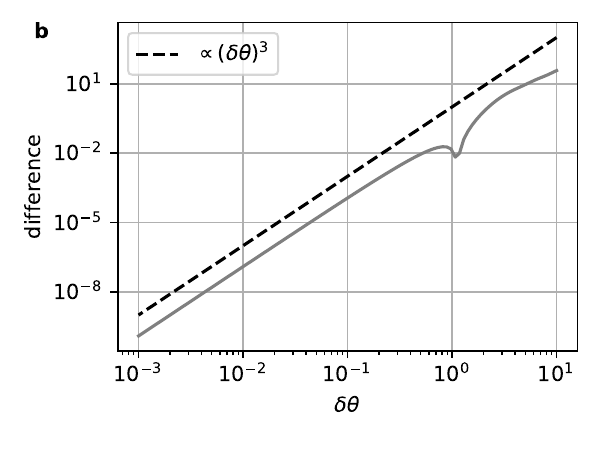}
    \caption[QGT vs. fidelity-based loss function]{(a) Values of the loss function $\mathcal{L}$ for evaluation with the QGT metric in Eq.~\eqref{eq:argmin_qgt}, and with introduced infidelity approximation in Eq.~\eqref{eq:argmin_dual}
    (b) Difference of the QGT metric and infidelity as function of the perturbation $\delta\theta$.}
    \label{fig:illustrative}
\end{figure}

To build an intuition for the relation of the QGT norm and the infidelity, 
we provide an illustrative example using the ansatz $\ket{\phi(\theta)} = R_\mathrm{Z}(\theta) R_\mathrm{Y}(\theta)\ket{0}$ with a single parameter, and the Hamiltonian $H = Z$.
In Fig.~\ref{fig:illustrative}(a) we visually compare the true loss functions, based on the QGT, with the DualQTE loss function, based on the infidelity, for imaginary-time evolution at the parameter value $\theta = \pi / 4$.
The difference between the two function scales as $\delta\theta^3$, as shown in Fig.~\ref{fig:illustrative}(b) and at $\delta\theta = 0$ both function coincide, since $0 g(\theta) 0 = 1 - F(\theta, \theta) = 0$.

Since the infidelity is bounded in $[0, 1]$ but the linear term $\vec b^\top(\vec\theta) \vec{\delta\theta}$ is not, the DualQTE loss function has an unbounded global minimum.
To find the update step $\vec{\delta\theta}$, however, we must find the local minimum that matches the minimum of the QGT-based loss function. 
By using a local optimization routine, such as gradient descent (GD), we can ensure to find the correct minimum.

\subsection{Solving the dual objective}

The DualQTE loss function $\mathcal{L}$ is locally convex in the vicinity of $\vec{\delta\theta} = \vec{0}$ because its Hessian $\vec\nabla\vec\nabla^\top \mathcal{L}(0) = g/2$ and $g$ is positive semi-definite. 
We take advantage of this property and employ gradient descent as local optimization routine.
This also allows the use of analytic gradient formulas, which have demonstrated better stability in the presence of shot noise than a direct optimization based on the loss function values or gradient approximations based on finite differences~\cite{piskor_gradients_2022}.
This is due the fact that, for small timesteps, the fidelity is close to $1$ and the readout errors due finite sampling statistics or device noise can easily mask changes in the cost function. 
Analytic gradient formulas are less prone to this issue, since the evaluate the 
parameter-shift gradients suffer less from this problem, as they allow to evaluate the cost function over larger perturbations, and do not amplify the noise by dividing by a small constant.

The gradient of $\mathcal{L}$ must be calculated with respect to the parameter perturbation $\vec{\delta\theta}$ and is given by
\begin{equation}
    \vec\nabla_{\vec{\delta\theta}} \mathcal{L}(\vec{\delta\theta}) = -\frac{\vec\nabla_{\vec{\delta\theta}} F(\vec\theta, \vec\theta + \vec{\delta\theta})}{2} - \delta\tau \cdot \vec{b}(\vec\theta).
\end{equation}
The fidelity gradient can, for instance, be evaluated using a parameter-shift rule
\begin{equation}
    \frac{\partial F}{\partial (\delta\theta)_j} = \frac{\lambda}{2} \left(F(\vec\theta, \vec\theta + \vec{\delta\theta} + \vec e_j s) - F(\vec\theta, \vec\theta + \vec{\delta\theta} - \vec e_j s)\right),
\end{equation}
where we assume that the gate of the $j$th parameter is implemented as $\exp(-i\theta_j G_j)$ for a Hermitian generator $G_j$ with eigenvalues $\pm\lambda$, $s=\pi/(4\lambda)$ is the parameter shift and $\vec e_j$ is the $j$th unit vector.
The gradient descent update step is then
\begin{equation}
    \vec{\delta\theta}^{(k+1)} = \vec{\delta\theta}^{(k)} - \eta_k \vec\nabla \mathcal{L}\left(\vec{\delta\theta}^{(k)}\right),
\end{equation}
with the learning rate $\eta_k > 0$.
Per timestep we perform gradient descent update until a maximum number of iterations or a convergence criterion, such as a threshold of the gradient norm or a minimum change of the loss function, is met.

A natural initial point for the gradient descent iteration is $\vec{\delta\theta}^{(0)} = \vec{0}$, i.e., no adjustment in the parameters. 
A more efficient alternative might be to warmstart the optimization by initializing the iteration with the final update step from the preceding timestep. 
This approach is motivated by the fact that, particularly for brief timesteps, the parameter derivatives $\dot{\vec\theta}$ usually do not vary substantially.
This approach would be particularly efficient if the first optimization can be omitted by solving for the parameter derivative with a classically efficient technique, possible, for example, for circuits that require only of Clifford gates for the evolution gradient and QGT at the initial parameter values~\cite{gacon_saqite_2023}.

Another important hyperparameter in the DualQTE loss function is the time perturbation $\delta\tau$, whose selection requires a trade-off: The QGT approximation error scales with $(\delta\tau)^3$ but, on the other hand, reducing the perturbation amplifies any errors in measuring the loss function since the parameter derivative is computed as $\dot{\vec\theta} = \vec{\delta\theta} / \delta\tau$.
This trade-off is discussed in detail in Appendix~\ref{app:dualqte_time_perturbation} through an illustrative example.

\subsection{Trainability}\label{sec:trainable}

Recent studies have shown that, under specific conditions, the gradient of the loss function for variational algorithms decreases exponentially to zero. As a result, efficient evaluation becomes unfeasible since it would require an exponentially large number of measurements. This phenomenon of flat regions in the optimization landscape, so-called \emph{barren plateaus} can arise from various scenarios~\cite{ragone_unified_2023}. For instance, they can occur when the loss function involves measuring a global observable~\cite{cerezo_cost-induced_2021}, if the quantum circuit preparing the parameterized ansatz state is too expressive~\cite{mcclean_barren_2018, cerezo_cost-induced_2021} or generates too much entanglement~\cite{ortiz_entanglement-induced_2021}, or if the measurements are too noisy~\cite{wang_noise-induced_2021}.

Variational quantum dynamics is steered by the evolution gradient defined Eqs.~\eqref{eq:b_real} and~\eqref{eq:b_imag} which can be affected by barren plateaus, which can lead to a failure in tracking the evolution of the quantum state. 
A key condition for barren plateaus is that the gradients only decay exponentially in system size on average when initialized randomly.
In quantum time evolution, however, the initial quantum state, and therefore the initial variational parameters, are specifically selected.
Moreover, the physical systems usually have local interactions, which leads to Hamiltonians without global terms.
By selecting a circuit depth that scales logarithmically with system size, the problem of barren plateaus can then be avoided. 
An alternative approach is to use an application-tailored ansatz with a limited number of variational parameters. This can help combat the challenges of barren plateaus. Examples include circuits based on Hamiltonian evolutions~\cite{ollitrault_uvcc_2020, park_hva_2023}.

Besides the evolution gradient, the DualQTE loss function depends on the fidelity of the variational state evaluated at two parameter points $\vec\theta$ and $\vec\theta'$.
The fidelity can be written as
\begin{equation}
    F(\vec\theta, \vec\theta') = |\braket{0|U^\dagger(\vec\theta) U(\vec\theta')|0}|^2 = \braket{\lambda|P_0|\lambda},
\end{equation}
where $U$ is the unitary preparing the ansatz and $P_0 = \ket{0}\bra{0}^{\otimes n}$ is the \emph{global} projector on the all-zero state.
The fidelity is, thus, a global observable and evaluating its gradient for random parameters will, on average, yield exponentially small values.
In DualQTE, the optimization begins with zero perturbations, that is, $\vec\theta = \vec\theta'$. At this point, the unitary preparing the state we measure for the fidelity evaluation is simply the identity, represented as $\ket{\lambda} = U^\dagger(\vec\theta)U(\vec\theta)\ket{0} = \mathbb{I}\ket{0}$. Notably, this type of initialization has been demonstrated to circumvent barren plateaus, even when dealing with global cost functions~\cite{grant_bp-initialization_2019}. 

In addition to a successful initialization, the loss function is locally convex and the distance to the optimum can be controlled by the time perturbation $\delta\tau$. 
While this does not prove that the loss function can be efficiently optimized, it does provide a compelling motivation, which we support with numerical evidence in Appendix~\ref{app:vanishing_gradients}. There, we show that for a local Hamiltonian and a circuit with logarithmic depth, neither the evolution gradient or the fidelity gradients decay exponentially with system size.

\subsection{Sample complexity}\label{sec:sampling}

When implementing VarQTE on quantum hardware, several error sources appear:
\begin{itemize}
    \item The model $\ket{\phi(\vec\theta)}$ might not be expressive enough to accurately represent the dynamics,
    \item both the QGT and the evolution gradient can suffer from sampling errors due to a limited number of measurements, and 
    \item errors can arise from the time integration scheme,
    \item each circuit operation is subject to hardware noise.
\end{itemize}
We distinguish into the idea VarQTE parameters $\vec\theta(t)$, without errors due to a finite number of samples or faulty operations, and the noisy parameters $\tilde{\vec\theta}(t)$. 
The error $\varepsilon(t)$ at time $t$ can, then, be split 
\begin{equation}
    \begin{aligned}
    \varepsilon(t) &= D_B(\phi(\tilde{\vec\theta}(t)), \Psi(t)) \\
                   &\leq \varepsilon_M(t) + \varepsilon_S(t),
    \end{aligned}
\end{equation}
where $\ket{\Psi(t)}$ is the exact time-evolved state and the error is measured in Bures distance
\begin{equation}
    D_B(\psi, \phi) = \sqrt{2(1 - |\braket{\psi|\phi}|)}.
\end{equation}
Here, $\varepsilon_M(t) = D_B(\phi(\vec\theta(t)), \Psi(t))$ describes the model error, due to missing expressitivity of the variational ansatz and time integration error, and $\varepsilon_S(t) =  D_B(\phi(\tilde{\vec\theta}(t)), \phi(\vec\theta(t)))$ is the error due to a noisy VarQTE solution.
While the introduced DualQTE algorithm promises a decrease in the measurement expense of VarQTE, this chapter does not address issues related to the choice of ansatz or hardware noise. Thus, our primary focus is to study the scaling behavior of the sampling error. The model error, denoted as $\varepsilon_M$, can be constrained using a-posteriori error bounds~\cite{zoufal_errorbounds_2021}, which we further investigate in Sec.~\ref{sec:real}.

To determine a specific bound that relies on attributes like the system energy or number of variational parameters, we must make certain assumptions about the circuit structure. 
In this chapter, our analysis assumes a circuit where parameterized gates are exclusively Pauli rotations $R_P(\theta_j)$, $P\in\{X,Y,Z\}$, $\theta_j \in \mathbb{R}$.  
In this setup, each parameter $\theta_j$ is unique and does not have any coefficients.
These assumptions can be relaxed and the presented bounds be modified to accommodate different circuit structures and parameterization methods. 
Furthermore, we introduce an assumption of a minimum value cutoff, $\delta_c > 0$, for the smallest eigenvalue of $g$. This is a consequence of the regularization implemented in the linear system.

Under these assumptions, we can formulate the following upper bound on the number of samples $N$ needed to achieve a sampling error of $\varepsilon_S$,
\begin{equation}\label{eq:vq_upper_bound}
    N \leq \mathcal{O}\left(\frac{d^3 E_\text{max}^2 \Delta_t^2}{\delta_c^4 \varepsilon_S^2}\right),
\end{equation}
where $E_\text{max}$ is the largest eigenvalue of the Hamiltonian.
In comparison to a similar work in Ref.~\cite{endo_varqte_2020} which expresses an upper bound in terms of the QGT and evolution gradient, we derive the bound in concrete terms, such as the number of variational parameters and the system energy, which can be readily evaluated.
We also improve the existing result and provide a tighter bound by employing Latala's theorem, a result from random matrix theory, to find an upper bound on the sampling error in the QGT~\cite{latala_bound_2005}.

Since our proposed algorithm avoids the explicit construction of the QGT and only uses the evolution gradient, we expect a reduction in sampling complexity by a factor of $d$, at the additional cost of $K$ optimization steps per timestep.
In line with this, we can demonstrate the following upper bound on the number of samples,
\begin{equation}\label{eq:dual_upper_bound}
    N \leq \mathcal{O}\left(\frac{d^2 K^2 \Delta_t^2}{\varepsilon_S^2}\left(\frac{1}{\delta\tau} + E_\text{max}\right)^2\right).
\end{equation}
Importantly, we empirically find that the number of iteration $K$ can likely be chosen constant, if the optimizations are warmstarted, for the systems we investigated, see Appendix~\ref{app:warmstarting}.
However, $K$ will generally depend on the system at hand and the optimization's hyperparameters.

The bounds for VarQTE and DualQTE are derived in Appendix~\ref{app:sampling_error}.
Though it is possible to design circuits for which different parts of the bounds are tight, Sec.~\ref{sec:imag} illustrates that, in practice, the actual number of samples needed is often less than the theoretical upper limit.

\section{Imaginary time evolution}\label{sec:imag}

This section benchmarks DualQTE and VarQTE for imaginary-time evolution and compares the required resources to achieve a target accuracy. We then use our proposed algorithm to generate quantum minimally entangled typical thermal states (QMETTS) of the Heisenberg model and compute thermodynamic observables.

\subsection{Heisenberg model}\label{sec:imag_heisen}

As quantum mechanical system we consider the Heisenberg model on a $n$-qubit circular topology with nearest-neighbor interactions and a transversal field. The Hamiltonian of this system is given by
\begin{equation}\label{eq:heisencomb}
    H = J \sum_{\braket{jk}} \left( X_j X_k + Y_j Y_k + Z_j Z_k \right) + h \sum_{j=1}^n Z_j,
\end{equation}
where $J$ is the interaction strength, $\braket{jk}$ iterates over neighboring pairs of qubits and $h$ is the transversal field strength. Throughout this chapter, we use $J=1/4$ and $h=-1$.
We consider a hardware-efficient variational ansatz composed of rotation layers with Pauli-$Y$ and -$Z$ gates that alternate with entangling layers using pairwise CX gates. A representation of the circuit is shown in Fig.~\ref{fig:efficientsu2}.
The initial state for the evolution is an equal superposition of all qubits, $\ket{+}^{\otimes n}$, which is achieved by setting the rotation angles of the final Pauli-$Y$ layer to $\pi/2$, while keeping the angles of the remaining layers at $0$.

In DualQITE, the optimization problems in each timestep are solved using gradient descent with a constant learning rate of $\eta = 0.1$ and a time perturbation set at $\delta\tau = 0.01$. The first iteration performs $K_0=100$ update steps and the subsequent iterations use $K_{>0}=10$ iterations, as they benefit from a warm start.
The values are heuristic and are chosen according to a calibration described in  Appendix~\ref{app:warmstarting}. We use fixed values for the iterations in the simulations with shot noise, as defining a termination criterion is difficult with only access to a noisy evaluations of loss functions and gradients.
To integrate the parameters, we use an explicit Euler scheme with a timestep of  $\Delta_t = 0.01$, that is
\begin{equation}
    \vec\theta(t + \Delta_t) = \vec\theta(t) + \Delta_t\dot{\vec\theta}(t) = \vec\theta(t) + \Delta_t\frac{\vec{\delta\theta}}{\delta\tau}.
\end{equation}
We point out that the integration timestep $\Delta_t$, which affects the number of time steps and the accuracy in the time integration, does not have to be the same as the time perturbation $\delta\tau$, which determines the QGT approximation error.

\begin{figure}[t]
    \centering
    \includegraphics[width=\textwidth]{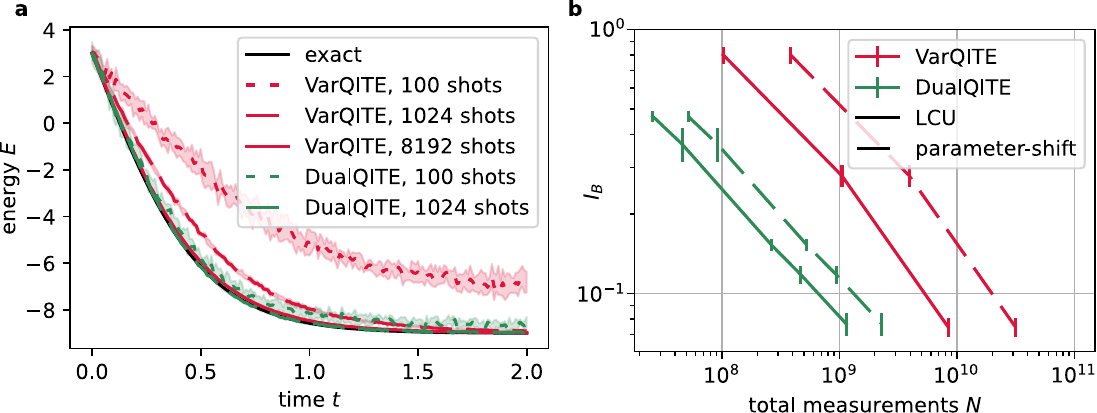}
    \caption[Convergence of DualQITE]{
    (a) Mean and standard deviation, as average over 5 independent experiments, for DualQITE and VarQITE for different number of shots.
    (b) The accuracy of DualQTE and VarQTE, averaged over 5 independent experiments, measured in integrated Bures distance $I_B$ (see Eq.~\eqref{eq:integrated_bures}) as a function of total number of measurements $N$ throughout the entire evolution.
    The resources are shown for two different methods of evaluating the gradient and QGT: the parameter-shift rule (dashed) and a linear combination of unitaries (sold lines).
    }
    \label{fig:heisen}
\end{figure}

The performance of DualQITE is compared against VarQITE, which is implemented with the same integration scheme. To ensure a stable solution of the linear system, we use an L-curve regularization~\cite{cultrera_lcurve_2020}.
In Fig.~\ref{fig:heisen}(a) we show the energy throughout the time evolution for $n=12$ qubits and different numbers of shots, and compare them to the energies obtained with exact diagonalization.
Even with only $100$ measurements per circuit evaluation, DualQITE can implement the imaginary-time evolution and, until time $t \approx 1$, is more accurate than VarQITE using $1024$ shots. Increasing the measurements of DualQITE to the same $1024$ shots enables our dual approach to faithfully follow the exact evolution towards the ground state, with a higher accuracy than VarQITE with $8192$ shots.

\subsection{Convergence}

In the previous experiment, DualQITE requires fewer circuit evaluations to reach the same energy accuracy as VarQITE.
For a more detailed performance comparison, we now investigate the convergence of both algorithms measured in terms of the fidelity.
Because the goal is tracking the imaginary-time dynamics accurately at each timestep, instead of only finding the final state, we measure the error of DualQITE as the integrated Bures distance to the exact time-evolved state $\ket{\Psi(t)}$, i.e.
\begin{equation}\label{eq:integrated_bures}
    I_B(T) = \frac{1}{T}\int_0^T D_B(\phi(\vec\theta(t)), \Psi(t)) \mathrm{d}t.
\end{equation}
The state fidelity in $D_B$ is evaluated exactly by computing the state vectors of $\ket{\phi(\vec\theta)}$ and $\ket{\Psi(t)}$ and taking the inner product.

In Fig.~\ref{fig:heisen}(b) we present the results for an final time of $T=2$. 
DualQITE consistently achieves the same accuracy using about one order of magnitude less measurements.
We expect this difference to scale with the number of parameters in the ansatz, as DualQITE only only requires the computation of small corrections in each timestep whereas VarQITE has to compute the full QGT. This is also what the bounds on the sampling complexity in Eq.~\eqref{eq:dual_upper_bound} suggests.

In DualQITE, the resources are balanced between the number of measurements used to evaluated the gradients and the number of optimization steps, whereas in VarQITE they are only determined by the number of measurements. The settings of both algorithms for the above experiments are listed in Appendix~\ref{app:dual_resources}.
An additional factor for the resource count is the selected gradient method. We here show number of measurements for both parameter-shift (PSR) and linear combination of unitaries (LCU) approaches, which differ in the structure and number of circuits executed. 
If the Hamiltonian contains $P$ Pauli terms, the total number of  circuits $C$ per timestep is
\begin{equation}
    C^\text{VarQITE}_\text{LCU} = \frac{d(d+5)}{2} + Pd,~~  
    C^\text{VarQITE}_\text{PSR} = 2d(d + P + 1),
\end{equation}
i.e., LCU uses less circuits, but requires an auxiliary qubit and non-local operations, which PSR does not need but instead runs more circuits.
For DualQITE, these numbers
\begin{equation}
    C^\text{DualQITE}_\text{LCU} = Pd + Kd = \frac{C^\text{DualQITE}_\text{PSR}}{2},
\end{equation}
where $K$ is the number of gradient descent iterations per timestep.
The total number of measurements $N$ is obtained by multiplying $C$ by the number of shots per circuit evaluation.

\subsection{Sample complexity}

Next, we examine how the resource requirements scale with system size. We repeat the experiments from before for a varying number of qubits from $n=4$ to $12$ and adjust the number of circuit layer repetitions to $r=\lceil\log_2(n)\rceil$.
The settings of VarQITE and DualQITE are tuned to achieve an average accuracy of $I_B \leq 0.1$ across 5 runs. This threshold equals a per-timestep fidelity of $0.995$.

\begin{figure}[t]
    \centering
    \includegraphics[width=\linewidth]{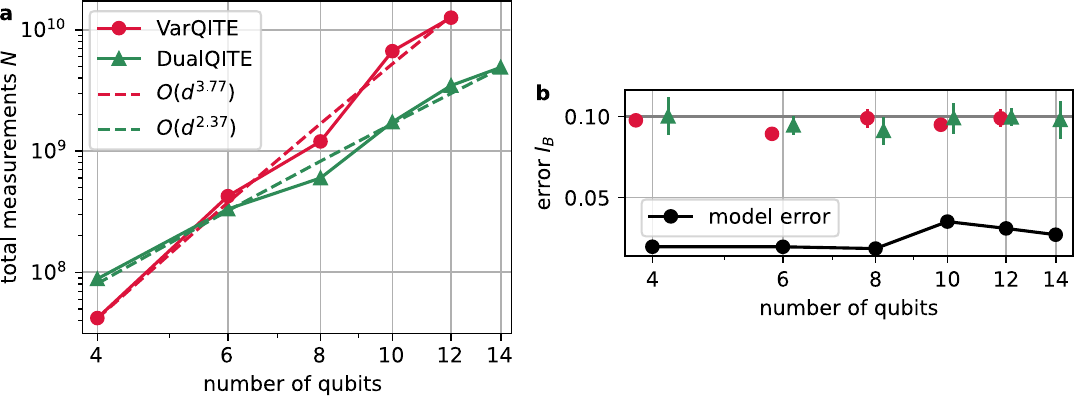}
    \caption[Resource scaling for DualQITE and VarQITE]{(a) The total number of measurements necessary for an accuracy of $I_B \leq 0.1$ over an average of 5 independent runs.
    Polynomials fits are shown as dashed lines, which show bumps as the number of parameters $d$ depends on $\lceil\log_2(n)\rceil$, which is discontinuous.
    For VarQITE the $n=14$ point is not evaluated as it requires too many measurements.
    (b) The mean accuracy including standard deviation of each experiment, along with the model error $\varepsilon_M$. 
    The grey line shows the target accuracy, $I_B = 0.1$.}
    \label{fig:sizescaling}
\end{figure}

In Fig.~\ref{fig:sizescaling}, we present the number of measurements required by both algorithms as a function of system size. 
For small system sizes, which only have a fewer parameters and computing the QGT is still cheap, VarQITE performs better than DualQITE.
However, this quickly changes as system size grows and from $n=8$ qubits on the quadratic scaling of VarQITE becomes a bottleneck, and DualQITE asymptotically performs better.

With this data we can verify the derived upper bounds on the number of measurements of Sec.~\ref{sec:sampling}.
Since the model error $\varepsilon_M$, shown in Fig.~\ref{fig:sizescaling}(b), is negligible, we have $\varepsilon_S \approx I_B \approx 0.1$.
To evaluate the bound we require the maximal energy of the system, which for the Heisenberg model on a chain scales with the number of qubits. As bound on the energy we can therefore use $E_\text{max} = \mathcal{O}(n) \leq \mathcal{O}(n\log(n)) = \mathcal{O}(d)$. 
In total we obtain expected scalings of $\mathcal{O}(d^5)$ for VarQTE and $\mathcal{O}(d^4 K)$ for DualQTE.
The experimental data, which is matched remarkably well by polynomial fits, is considerably below the expected scaling at 
$d^{3.77}$ for VarQTE and $d^{2.37}$ for DualQTE.
As further discussed in Appendix~\ref{app:sampling_error}, this experiment suggests that the bounds are valid but not yet tight.

\subsection{Calculating thermodynamics observables}

We use imaginary-time evolution to measure thermodynamic properties with the quantum minimally entangled thermal states method (QMETTS) \cite{stoudenmire_minimally_2010, motta_determining_2020}. The METTS approach is originally based on matrix product states (MPS), since it only computes imaginary-time evolution of initially low-entangled states. However, classical methods struggle if a lot of entanglement occurs during the evolution, as can be the case, e.g., in low-temperature, 2D systems. 
Implementing the evolutions on a quantum computer is, thus, a promising application.

The QMETTS algorithm computes the ensemble average of an observable $A$ at inverse temperature $\beta$ by averaging over samples $\{A_m\}_m$, as
\begin{equation}
    \braket{A}_\text{ens} = \frac{\tr(e^{-\beta H} A)}{\tr(e^{-\beta H})} \approx \frac{1}{M} \sum_{m=1}^{M} A_m.
\end{equation}
The samples $A_m$ are constructed using the following Markov chain protocol:
\begin{enumerate}
    \item Start from a product state $\ket{\phi_m(t=0)}$.
    \item Evolve up to imaginary time $t = \beta/2$
    \begin{equation*}
        \ket{\phi_m(\beta/2)} \propto e^{-\beta H/2 }\ket{\phi_m(0)}.
    \end{equation*}
    \item Evaluate the observable to obtain the sample
    \begin{equation*}
        A_m = \braket{\phi_m(\beta/2) | A |\phi_m(\beta/2)}.
    \end{equation*}
    \item Measure $\ket{\phi_m(\beta/2)}$ in some basis to obtain
    the next random product state $\ket{\phi_{m+1}(0)}$.
\end{enumerate}

We compute the energy per site, $\braket{H}/n$, of the Heisenberg model on a chain of $n=6$ spins, with the system values $J=1/4$ and $h=-1$.
Switching the measurement basis in the Markov chain can reduce the auto-correlation length of the samples and speed up the convergence to the ensemble average.
Here, we flip between the $X$ and $Y$ basis in between every sample.
We avoid using the $Z$ basis since the Heisenberg Hamiltonian preserves the number of qubit excitations.
Evolving a $Z$-basis state would only allow for a narrow exploration of the available states which increases the variance of the Markov chain.

The imaginary-time evolution is implemented with DualQITE using an ansatz similar to Fig.~\ref{fig:efficientsu2}, whose entangling CX gates reflect the system interactions.
To evolve states prepared in the $X$ basis, i.e. $\ket{\pm}$, we use rotation layers with Pauli-$Y$ and -$Z$ rotations, whereas for the $Y$ basis states $\ket{\pm i}$ we use Pauli-$X$ and -$Z$ gates.
We prepare the initial product states $\ket{\phi_m(0)}$ by tuning the parameter values in the final rotation layers to the following values, while leaving the remaining angles at 0,
\begin{equation}
    \begin{aligned}
        \ket{\pm} &= R_\mathrm{Y}\left(\frac{\pm \pi}{2}\right) R_\mathrm{Z}(0)\ket{0} \\
        \ket{+i} &= R_\mathrm{X}\left(\frac{\pi}{2}\right) R_\mathrm{Z}(\pi)\ket{0}, \\
        \ket{-i} &= R_\mathrm{X}\left(\frac{\pi}{2}\right) R_\mathrm{Z}(0)\ket{0}.
    \end{aligned}
\end{equation}
The DualQITE optimization problems are solved with GD with $K_0=100$ iterations in the first timestep and $K_{>0}=10$ in the following, each using a learning rate of $\eta=0.1$ and a time perturbation of $\delta\tau=0.01$.
The parameters are integrated with a forward Euler method and a timestep of $\Delta_t = 0.01$ and the circuits are evaluated with $1024$ measurements.

\begin{figure}[t]
    \centering
    \includegraphics[width=0.6\linewidth]{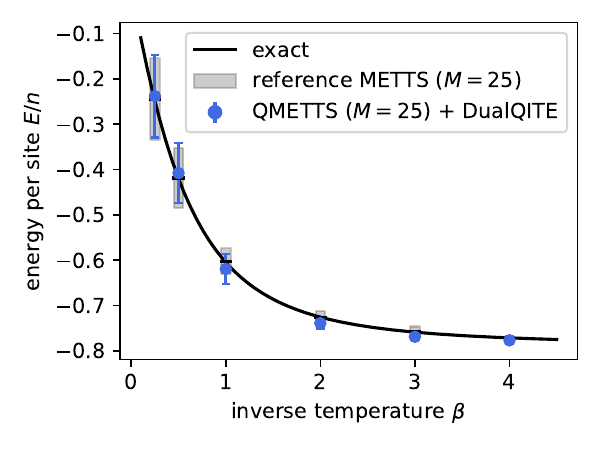}
    \caption[QMETTS for the Heisenberg model]{Energy per site for the Heisenberg model on a 6-spin chain, comparing mean and standard deviation of  QMETTS with DualQITE (blue circle and errorbars) with a reference METTS implementation (black line and grey shade).}
    \label{fig:energy_per_site}
\end{figure}

In Fig.~\ref{fig:energy_per_site} we show the estimated energy per site, including standard deviation, for a range of inverse temperatures $\beta$.
With alternating $X-Y$ bases, the Markov chain converges quickly and only $M=25$ samples are enough to obtain an accurate estimate.
We compare QMETTS with DualQITE as subroutine to a reference calculation where the imaginary-time evolutions are implemented with exact diagonalization.
Since the results obtained with DualQITE match the reference closely, we conclude that the proposed algorithm can accurately reproduce the required dynamics.

\section{Real time evolution}\label{sec:real}

The proposed DualQTE algorithm is of particular interest for the application to imaginary-time evolution, due to a lack of efficiently implementable variational algorithms.
Our algorithm can equally be applied to real-time evolution, however previously suggested algorithms on projecting variational quantum dynamics have a similar structure.
The p-VQD~\cite{barison_pvqd_2021} algorithm, for example, projects a single Suzuki-Trotter evolution step onto the variational model by solving
\begin{equation}
    \vec\theta(t + \Delta_t) = \mathrm{argmax}_{\vec\theta'} \big|\braket{\phi(\vec\theta') | e^{-iH\Delta_t} | \phi(\vec\theta(t))}\big|^2,
\end{equation}
which, like DualQRTE, is a fidelity-based optimization.
Yet, there are key differences between the approaches.

DualQRTE does not require the implementation of a Suzuki-Trotter step which, or Hamiltonians with many Pauli terms or long-range interactions, such as those arising in molecular dynamics, can already lead to complex, non-local circuits. 
Furthermore, a-posteriori error bounds can be evaluated at almost no overhead in DualQRTE.
A drawback of our algorithm is that it requires the LCU method to compute the imaginary part of the energy and state gradients, using an auxiliary qubit and an additional entangling gate.

\subsection{Heisenberg model}

\begin{figure}[htp]
    \centering
    \includegraphics[width=0.5\textwidth]{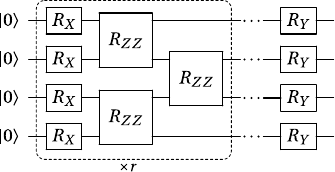}
    \caption[Brickwall ansatz for a real-time evolution]{The variational ansatz used for the real-time evolution experiments. The dashed part is repeated $r=3$ times. Note that the $R_{ZZ}$ gates commute and can be aligned in a compact, pairwise fashion.}
    \label{fig:heisen_real_circuit}
\end{figure}

We investigate the real-time evolution under the same Heisenberg Hamiltonian as for the imaginary-time before, but on a linear chain with $n=4$ spins. 
The variational circuit is motivated from the Suzuki-Trotter decomposition and consists of Pauli-$X$ rotation layers, alternating with Pauli-$ZZ$ entangling gates that reflect the linear interactions. Finally, we apply a layer of Pauli-$Y$ rotations.
The circuit is shown in Fig.~\ref{fig:heisen_real_circuit} and in our experiments we use $r=3$ repetitions of the described circuit block.
The initial state of the evolution is $\ket{+}^{\otimes n}$, which can be prepared by setting the angles of the final Pauli-$Y$ rotations to $\pi/2$ and the remaining parameters to $0$.

We compare DualQRTE, p-VQD and VarQRTE by measuring the magnetization in $X$ and $Z$ directions during the evolution, which are given by
\begin{equation}
    \braket{X} = \frac{1}{n} \sum_{j=1}^n \braket{X_j}, ~~
    \braket{Z} = \frac{1}{n} \sum_{j=1}^n \braket{Z_j}.
\end{equation}
The Heisenberg Hamiltonian preserves the qubit excitations, therefore the $Z$ expectation value should remain at the initial value of $0$ throughout the evolution.
All algorithms evolve up to $T=2$ using an explicit Euler scheme with 100 equidistant timesteps.

\begin{figure}[t]
    \centering
    \includegraphics[width=\linewidth]{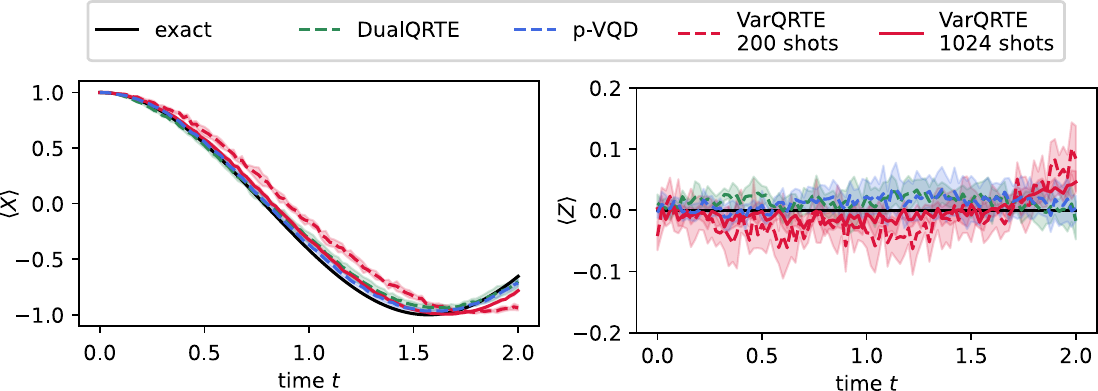}
    \caption[Magnetization during a time evolution]{Mean $X$- and $Z$-magnetization during the real-time evolution for DualQRTE, p-VQD and VarQRTE.}
    \label{fig:magnetization}
\end{figure}

In Fig.~\ref{fig:magnetization} we present the results. We observe a similar performance of DualQRTE and p-VQD, which show accurate results for only $200$ shots per circuit. To achieve a similar accuracy, VarQRTE requires $1024$ shots.

\subsection{Error bounds}

The derivative of the model error $\dot{\varepsilon}_M$, which describes the error due to projection onto the variational manifold, can, for real-time evolution, be expressed as
\begin{equation}
    \begin{aligned}
        \dot\varepsilon_M &:= \left\|\sum_{k=1}^d \dot\theta_k \ket{\partial_k \phi(\vec\theta)} + i H\ket{\phi(\vec\theta)}\right\|_2^2  \\
                          &= \mathrm{Var}(E) + \dot{\vec\theta}^\top g(\vec\theta) \dot{\vec\theta} - 2 \dot{\vec\theta}^\top \vec{b}(\vec\theta),
    \end{aligned}
\end{equation}
where $\mathrm{Var}(E) = \braket{\phi(\vec\theta)|H^2|\phi(\vec\theta)} - E^2(\vec\theta)$~\cite{zoufal_errorbounds_2021}.
By integrating the error rate, we obtain an expression for the integrated Bures distance, as
\begin{equation}
    D_B(\phi(\vec\theta(T)), \Psi(T)) \leq \int_0^T \dot\varepsilon_M(t) \mathrm{d} t, 
\end{equation}
with the exact time-evolved state $\ket{\Psi(t)}$ and the time-dependence of $\varepsilon_M$ comes from the fact that the parameters $\vec\theta = \vec\theta(t)$ are time dependent.

The model error rate matches the DualQRTE loss function, with exception of the additional variance factor $\mathrm{Var}(E)$, since
\begin{equation}
    \begin{aligned}
    \dot\varepsilon_M &= \mathrm{Var}(E) + \frac{1 - F(\vec\theta, \vec\theta + \vec{\delta\theta})}{(\delta\tau)^2} - \frac{2 \vec{\delta\theta}^\top \vec{b}(\vec\theta)}{\delta\tau} + \mathcal{O}(\delta\tau) \\
    &= \mathrm{Var}(E) + \frac{2 \mathcal{L}(\vec{\delta\theta})}{(\delta\tau)^2} + \mathcal{O}(\delta\tau).
    \end{aligned}
\end{equation}
This allows us to evaluate $\dot\varepsilon_M$ as a by-product of the DualQRTE evolution, with the only overhead of evaluating the variance of the energy.
A forward Euler integration of the model error with a timestep of $\Delta_t$ has an error scaling with $\mathcal{O}(\Delta_t + T\delta t)$. For a meaningful estimate we therefore have to choose $\delta t \ll T^{-1} = \Delta_t$, which highlights the difference of the two factors.

\begin{figure}[t]
    \centering
    \includegraphics[width=0.9\textwidth]{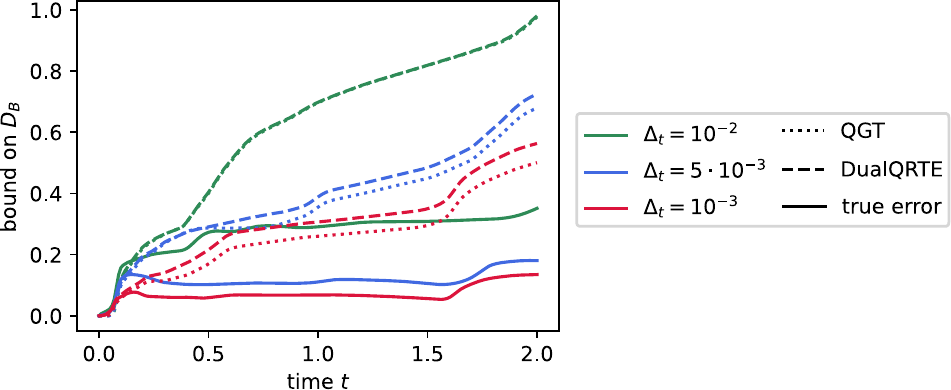}
    \caption[Error of a real-time evolution]{Error of the real-time evolution in Bures distance, plus error bounds obtained with VarQRTE and DualQRTE.}
    \label{fig:errorbounds}
\end{figure}

We compute the error bounds using with VarQRTE and DualQRTE for varying timesteps $\Delta_t$ in exact, state vector simulations. DualQRTE uses a fixed time perturbation of $\delta\tau = 10^{-3}$.
The results, presented in Fig.~\ref{fig:errorbounds}, show that the bounds generally hold and are converge to the true error for a decreasing timestep.
We observe that the larger the difference of $\delta\tau$ and $\Delta_t$, the closer the approximate DualQRTE bound is to the VarQRTE result. This is because for larger $\Delta_t$, the integration error in $\mathcal{O}(\Delta_t + T\delta\tau)$ dominates, but the smaller the timestep is the more the dual approximation error falls into weight.

\section{Conclusion}

In this chapter we use the intrinsic properties of the quantum geometric tensor (QGT) to reformulate variational quantum time evolution in terms of the fidelity. 
In contrast to previous chapters, which followed a stochastic approach to reduce the cost of computing the QGT, this dual formulation circumvents the explicit QGT entirely and instead solves an optimization problem in each timestep.
In numerical experiments we show that the resulting algorithm, DualQTE, requires less measurements than the direct variational approach, VarQTE, especially as the dimension of the problem and the number of parameters in the ansatz increase.
This observation is supported by bounds on the sample complexity.

As applications, we consider the imaginary-time evolution of the Heisenberg model, which we then use as subroutine to compute thermal averages via the quantum minimally entangled typical thermal states (QMETTS) algorithm.
We also perform an illustrative real-time evolution and compare DualQTE to another resource-efficient algorithm based on a variational projection of the Suzuki-Trotter step.
A key difference of the dual approach is that the dynamics are induced by the evolution gradient and no Suzuki-Trotter step has to implemented, which could become a bottleneck for Hamiltonians with a large number of Paulis or non-local interactions.

While, in this chapter, we considered a gradient descent scheme with fixed learning rate to solve the optimization problem in each timestep, we expect that more elaborate optimizers could further improve DualQTE.
In addition, it could be combined with exact initialization of the QGT and evolution gradient introduced in Chapter~\ref{chap:saqite} to skip the first timestep where warm starting is not available.
A potential drawback of DualQTE is that it relies on an accurate evaluation of the fidelity, which can be costly to evaluate on near-term devices.
Since we are not interested in minimizing the fidelity itself, but rather the values of the fidelity close to the minimum using local approximations of the fidelity is not an option.
A possible resolution could be the development of fidelity-specific error mitigation techniques or building surrogate models, e.g., based on quasi-probability decompositions.

To conclude, we introduced a novel perspective to variational quantum time evolution which promises a favorable scaling in the number of parameters compared to direct approaches. 
This cost reduction allows scaling variational time evolution up to larger problem sizes and enables the investigation of a range of interesting tasks beyond pure time evolution, such as mixed time evolution or the preparation of thermal states.
Such resource reductions are an important step to performing experiments on the full sizes of current and upcoming quantum computers, and to advance to the demonstration of practically relevant applications.

%% file: main/ch7_conclusion.tex
\chapter{Conclusion}\label{ch:conclusion}

\section{Main results}

In this thesis, we investigate scalable variational quantum algorithms on noisy quantum computers.
We focus on the simulation of physical systems with the tasks of ground-state preparation, and real- and imaginary-time evolution.
In this context, the evaluation of the quantum geometric tensor (QGT) becomes computationally prohibitively expensive and hinders scaling up algorithms to practically relevant system sizes.
Our main results are the development of two novel techniques to reduce the resource costs.
They are based on a stochastic approximation of the QGT and on a dual formulation of variational quantum time evolution.

In the stochastic approach, we introduce an unbiased estimator of the QGT which allows to construct samples at a constant cost instead of scaling quadratically with the number of parameters in the variational ansatz circuit.
This allows to formulate novel, stochastic versions of quantum natural gradient descent, called QN-SPSA, and of quantum time evolution, called SA-QTE.
Extensive numerical benchmarks show that these algorithms require significantly less measurements than their non-stochastic counterparts, for both ground-state preparation and time evolution. 
Due to this resource reduction, we are able to scale an imaginary-time evolution up to a 27-qubit system on current quantum hardware, including error-mitigation techniques.

In the dual formulation, we exploit the intrinsic connection of the QGT and the fidelity to derive a QGT-free formulation of variational quantum time evolution, called DualQTE. 
This algorithm relies on solving a fidelity-based optimization problem in each timestep and is equally applicable to real- and imaginary-time evolution.
In combination with a suitable initialization, we observe that this approach requires asymptotically less measurements in numerical benchmarks than the QGT-based approach to time evolution.
This result is further supported by bounds on the sampling complexity of both algorithms.

While the focus on time evolution and ground state preparation is motivated by the simulation of physical systems, our algorithms find applications in other fields such as, for example, optimization or machine learning. 
This is demonstrated by several experiments in this thesis, such as solving Max-Cut problems, black-box optimizations or generative learning tasks.

\section{Outlook}

In theory, quantum computers are able to store and propagate a quantum system's wave function efficiently.
Yet, in practice, a range of open problems currently limit the scale of demonstrations either to hardware-tailored model-systems~\cite{kim_utility_2023, ebadi_256_2021} or small sizes~\cite{stanisic_observing_2022}.
Scaling experiments up to practically relevant system sizes requires developments and synergies across the quantum computing stack.
Commencing at the lowest level, the main tunable layers include the following.
\begin{itemize}
    \item Improvements in quality (and scale) of quantum hardware: smaller gate errors and longer coherence times allow for more complex circuits, and better connectivities reduce the depth of the compiled circuit.
    \item Optimized circuit compilation: circuit results can be improved by minimizing the number of circuit operations, selecting the optimal qubits, and finding efficient mappings from the circuit to the hardware. In addition, the wall-time to compile the circuit must achieve reasonable runtimes for variational workflows.
    \item Algorithmic developments (this thesis): by finding hardware-friendly and resource-efficient algorithms, larger problems can be tackled without relying on device improvements.
    \item Problem-specific knowledge: with insights into the individual problem at hand, specific ansatz circuits, initial parameter values, and error-mitigation techniques can impact the trainability and solution quality.
\end{itemize}
In addition to algorithmic developments, we expect the last point to be especially important for current variational quantum algorithms. The chosen ansatz influences not only the ability to represent the solution and how challenging the algorithm is to simulate classically, but also directly affects the runtime. Obeying the symmetries of a physical system, for instance, could provide an expressive model with a limited number of tunable parameters.
Such a model could further enable problem-specific error mitigation by, for example, discarding samples that break the symmetry~\cite{barron_cvar_2023}.

A practical use of quantum computers for the simulation of physical systems can also include synergies with classical methods. 
Embedding techniques, for instance, use a classical algorithm in a regime where it finds a high-quality solution and uses a quantum computer in a small, specifically selected subspace~\cite{barison_embedding_2023}.
Another example is to solve bottlenecks of previously classical methods, such as in Markov Chain Monte Carlo~\cite{layden_qmcmc_2023}.

Finally, as the quality and scale of current quantum computers increase, understanding their true potential and limitations requires performing experimental demonstrations. 
While a rivalry to classical methods helps to push the limits of both computational paradigms, the final goal remains a joint one: solving challenging tasks to uncover the laws of physics.

%% file: tail/appendix.tex
\appendix

\chapter{Quantum computing basics}

\section{(Quantum) logic gates}\label{app:gates}

In this section, we list the matrix forms of commonly used gates. Some gates represent classical logic gates; $X$ (NOT), CX (reversible XOR) and Toffoli (reversible AND), which are used in Fig.~\ref{fig:classical_program}.
Gates typically have a circuit symbol that labels their action, e.g., the Hadamard gate $H$ is represented as 
\[
\begin{array}{c}
    \Qcircuit @R=.7em @C=1em {
        & \gate{H} & \qw
    }
\end{array},
\]
however, if a gate has a special symbol it is denoted below.

\subsubsection{Single-qubit operations}

\begin{itemize}
    \item The Pauli operations are
\begin{equation}
    X \equiv \begin{pmatrix} 0 & 1 \\ 1 & 0 \end{pmatrix},~~
    Y \equiv \begin{pmatrix} 0 & -i \\ i & 0 \end{pmatrix},~~
    Z \equiv \begin{pmatrix} 1 & 0 \\ 0 & -1 \end{pmatrix}.
\end{equation}
The $X$ gate equals the classical NOT gate and is denoted with the symbol
\[
\begin{array}{c}
    \Qcircuit @R=.7em @C=1em {
        & \targ & \qw
    }
\end{array}.
\]
\item The explicit forms of Pauli rotations are
\begin{equation}\label{eq:pauli_rotations}
    R_X(\theta) \equiv \begin{pmatrix}
        \cos\left(\frac{\theta}{2}\right) & \!\!\!\!\!\!\! -i\sin\left(\frac{\theta}{2}\right) \\
        -i\sin\left(\frac{\theta}{2}\right) & \!\!\!\!\!\!\! \cos\left(\frac{\theta}{2}\right) 
    \end{pmatrix},
    ~~ 
    R_Y(\theta) \equiv \begin{pmatrix}
        \cos\left(\frac{\theta}{2}\right) & \!\!\!\!\!-\sin\left(\frac{\theta}{2}\right) \\
        \sin\left(\frac{\theta}{2}\right) & \!\!\!\!\!\cos\left(\frac{\theta}{2}\right) 
    \end{pmatrix},
    ~~ 
    R_Z(\theta) \equiv \begin{pmatrix}
        e^{-i\frac{\theta}{2}} & \!\!\!\!\!\! 0 \\ 0 & \!\!\!\!\!\!e^{i\frac{\theta}{2}} 
    \end{pmatrix}.
\end{equation}
\item The Hadamard gate $H$, the phase gate $S$, and the $\sqrt{X}$ gate are defined as 
\begin{equation}
    H \equiv \frac{1}{\sqrt{2}} \begin{pmatrix}
        1 & 1 \\ 1 & - 1
    \end{pmatrix}, 
    ~~
    S \equiv \begin{pmatrix}
        1 & 0 \\ 0 & i
    \end{pmatrix},
    ~~
    \sqrt{X} \equiv \frac{1}{2} 
    \begin{pmatrix}
        1 + i & 1 - i \\
        1 - i & 1 + i
    \end{pmatrix}.
\end{equation}
\item The matrix form of the general single-qubit gate is 
\begin{equation}
    U(\theta, \phi, \lambda) \equiv 
    \begin{pmatrix}
        \cos(\theta/2) & -ie^{i\lambda} \sin(\theta/2) \\
        -ie^{i\phi}\sin(\theta/2) & e^{i(\phi + \lambda)}\cos(\theta/2)
    \end{pmatrix}.
\end{equation}
\end{itemize}

\subsubsection{Two-qubit gates}

\begin{itemize}
    \item The controlled-$X$ (CX) gate also corresponds to a reversible XOR gate. Its gate symbols and matrix form are
    \begin{equation}
        \begin{array}{c}
        \Qcircuit @R=2em @C=.7em {
            & \ctrl{1} & \qw \\
            & \targ & \qw
        }
        \end{array}:~~
        \text{CX} \equiv \begin{pmatrix}
            1 & 0 & 0 & 0 \\
            0 & 1 & 0 & 0 \\
            0 & 0 & 0 & 1 \\
            0 & 0 & 1 & 0
        \end{pmatrix}. 
    \end{equation}
    \item The controlled-$Z$ (CX) gate is invariant under exchange of control and target, which is reflected by the circuits symbol. It implements the unitary given by
    \begin{equation}
        \begin{array}{c}
        \Qcircuit @R=2em @C=.7em {
            & \ctrl{1} & \qw \\
            & \control\qw & \qw
        }
        \end{array}:~~
        \text{CZ} \equiv 
        {\small
        \begin{pmatrix}
            1 & 0 & 0 & 0 \\
            0 & 1 & 0 & 0 \\
            0 & 0 & 1 & 0 \\
            0 & 0 & 0 & -1
        \end{pmatrix}
        }.
    \end{equation}
    \item The Swap gate is 
    \begin{equation}
        \begin{array}{c}
        \Qcircuit @R=2em @C=.7em {
            & \qswap & \qw \\
            & \qswap\qwx & \qw
        }
        \end{array}:~~
        \text{Swap} \equiv \begin{pmatrix}
            1 & 0 & 0 & 0 \\
            0 & 0 & 1 & 0 \\
            0 & 1 & 0 & 0 \\
            0 & 0 & 0 & 1
        \end{pmatrix}.
    \end{equation}
\end{itemize}

\subsection{Three-qubit gates}

The Toffoli gate is a reversible AND gate, as it maintains the two inputs. 
\begin{equation}
    \begin{array}{c}
    \Qcircuit @R=2em @C=.7em {
        & \ctrl{1} & \qw \\
        & \ctrl{1} & \qw \\
        & \targ & \qw
    }
    \end{array}:~~
    \text{CZ} \equiv 
    {\small
    \begin{pmatrix}
        1 & 0 & 0 & 0 & 0 & 0 & 0 & 0 \\
        0 & 1 & 0 & 0 & 0 & 0 & 0 & 0 \\
        0 & 0 & 1 & 0 & 0 & 0 & 0 & 0 \\
        0 & 0 & 0 & 1 & 0 & 0 & 0 & 0 \\
        0 & 0 & 0 & 0 & 1 & 0 & 0 & 0 \\
        0 & 0 & 0 & 0 & 0 & 1 & 0 & 0 \\
        0 & 0 & 0 & 0 & 0 & 0 & 0 & 1 \\
        0 & 0 & 0 & 0 & 0 & 0 & 1 & 0 \\
    \end{pmatrix}
    }
    .
\end{equation}

\section{Tensor product arithmetic}\label{app:tensoryoga}

States of multiple qubits are described using the tensor product of single qubit states. 
We usually drop the explicit tensor product for brevity and denote
\begin{equation}
    \ket{q_n q_{n-1} \cdots q_1} = \ket{q_n} \otimes \ket{q_{n-1}} \otimes \cdots \otimes \ket{q_1}.
\end{equation}
There exists a natural isomorphism of the tensor vector space $(\mathbb{C}^2)^{\otimes n}$ to the complex vector space $\mathbb{C}^{2^n}$, which is implemented with the Kronecker product. 
This allows to identify $n$-qubit states as $2^n$-dimensional complex vectors and, analogously, operators on $n$ qubits as $2^n \times 2^n$-dimensional complex matrices. 
See Appendix~\ref{app:tensoryoga} for an explicit example.

As an example for working with tensor products, we consider the expectation value of the state $\ket{+} \otimes \ket{1}$ for the operator $Z \otimes Z$.
Expanding both using the Kronecker product, we obtain
\begin{equation}
    \ket{+} \otimes \ket{1} = 
    \frac{1}{\sqrt{2}}
    \begin{pmatrix}
        1 \\ 
        1
    \end{pmatrix}
    \otimes
    \begin{pmatrix}
        0 \\ 
        1 
    \end{pmatrix}
    \iso 
    \frac{1}{\sqrt{2}}
    \begin{pmatrix}
    1 \cdot \begin{pmatrix}
        0 \\ 1
    \end{pmatrix} \\ 
    \\
    1 \cdot \begin{pmatrix}
        0 \\ 1
        \end{pmatrix}
    \end{pmatrix}
    =
    \frac{1}{\sqrt{2}}
    \begin{pmatrix}
        0 \\ 1 \\ 0 \\ 1
    \end{pmatrix},
\end{equation}
and
\begin{equation}
    Z \otimes Z = 
    \begin{pmatrix} 1 & 0 \\ 0 & -1 \end{pmatrix}
    \otimes
    \begin{pmatrix} 1 & 0 \\ 0 & -1 \end{pmatrix}
    \iso 
    \begin{pmatrix} 
    1 \cdot  \begin{pmatrix} 1 & 0 \\ 0 & -1 \end{pmatrix}
    &
    0 \cdot \begin{pmatrix} 1 & 0 \\ 0 & -1 \end{pmatrix}
    \\ \\
    0 \cdot \begin{pmatrix} 1 & 0 \\ 0 & -1 \end{pmatrix}
    & 
    -1 \cdot \begin{pmatrix} 1 & 0 \\ 0 & -1 \end{pmatrix}
    \end{pmatrix}
    = 
    \begin{pmatrix}
        1 & 0 & 0 & 0 \\
        0 & -1 & 0 & 0 \\
        0 & 0 & -1 & 0 \\
        0 & 0 & 0 & 1
    \end{pmatrix}.
\end{equation}
Performing the matrix multiplications we obtain the expectation value
\begin{equation}
    \big(\bra{+} \otimes \bra{1}\big)\big(Z \otimes Z\big)\big(\ket{+} \otimes \ket{1}\big) 
    =
    \frac{1}{2}
    \begin{pmatrix}
        0 & 1 & 0 & 1
    \end{pmatrix}
    \begin{pmatrix}
        1 & 0 & 0 & 0 \\
        0 & -1 & 0 & 0 \\
        0 & 0 & -1 & 0 \\
        0 & 0 & 0 & 1
    \end{pmatrix}
    \begin{pmatrix}
        0 \\ 1 \\ 0 \\ 1
    \end{pmatrix} 
    = 0.
\end{equation}
If readers do not have a penchant for large matrices, a more concise way to evaluate expectation values is typically to directly work with the tensor products, as 
\begin{equation}
    \begin{aligned}
    \big(\bra{+} \otimes \bra{1}\big)\big(Z \otimes Z\big)\big(\ket{+} \otimes \ket{1}\big) 
    &= \braket{+|Z|+} \cdot \braket{1|Z|1}  \\
    &= \frac{\braket{0|Z|0} + \braket{1|Z|0} + \braket{0|Z|1} + \braket{1|Z|1}}{2} \cdot \braket{1|Z|1} \\
    &= \frac{1 + 0 + 0 - 1}{2}\cdot 1  \\
    & = 0.
    \end{aligned}
\end{equation}

\chapter{Quantum computing stack}

\section{Superconducting qubits}\label{app:sc}

\subsection{From quantum harmonic oscillator to charge qubit}

\textit{In this section we use a different convention as in the rest of this thesis: quantum mechanical operators are denoted with a hat and $\hbar$ is explicitly stated. These exceptions are made to distinguish between operators and variables with the same symbol and to match the common nomenclature in the field.}

\noindent
A quantum harmonic oscillator, as shown in Fig.~\ref{fig:qho}, is described by the Hamiltonian
\begin{equation}
    \hat H = \frac{\hat{Q}^2}{2C} + \frac{\hat{\Phi}^2}{2L},
\end{equation}
with inductance $L$, capacitance $C$, magnetic flux $\Phi$ through the inductance and charge $Q$ on the capacitance\cite{krantz_sc_2019}.
To compare this system to the anharmonic oscillator that will be introduced later on, it is instructive to rewrite the Hamiltonian in terms of the reduced charge $\hat n = \hat Q/(2e)$ and reduced flux $\hat\varphi = 2e\hat\Phi/\hbar$.
The operators $\hat \varphi$ and $\hat n$ are conjugate variables, i.e., they satisfy $[\hat\varphi, \hat n] = i$.
Since at superconducting temperatures electrons occur only as Cooper pairs, $n$ describes the excess number of Cooper pairs on one side of the capacitance, a.k.a. the ``island''.
In this picture, the Hamiltonian is
\begin{equation}
    \hat H = \frac{E_C}{2} \hat{n}^2 + \frac{E_L}{2} \hat{\varphi}^2,
\end{equation}
where $E_C = (2e)^2/C$ is the energy required for a Cooper pair to hop on the island and $E_L = (2e)^{-2} / L$ is the inductive energy~\cite{krantz_sc_2019}.

This Hamiltonian can be diagonalized by introducing operators that create or annihilate excitations in the resonator circuit,
\begin{equation}
    \hat{a} = \frac{1}{2}\left(\frac{\hat\varphi}{\varphi_0} - i\frac{\hat n}{n_0}\right),
\end{equation}
with the zero-point fluctuations $n_0 = (E_L / (4E_C))^{1/4}$ and $\varphi_0 = (E_C / (4E_L))^{1/4}$.
Under this transformation, the Hamiltonian takes on the form
\begin{equation}
    \hat H = \hbar\omega_c \left(\hat{a}^\dagger \hat a + \frac{1}{2}\right),
\end{equation}
and we can read off the energy levels according to the number of excitations $k$ as 
\begin{equation}
    E_k = \hbar\omega_c\left(k + \frac{1}{2}\right).
\end{equation}

The Hamiltonian in the anharmonic oscillator case, where we replace the inductance with a Josephson junction is
\begin{equation}
    \hat H = \frac{E_C}{2} (\hat n - n_g)^2 - E_J \cos(\hat \varphi),
\end{equation}
where $n_g$ is called the gate charge and describes the number of Cooper pairs on the island due to the applied voltage $V$~\cite{koch_transmon_2007, egger_sc_2014}.
This offset charge is not considered in the harmonic case, as it does not have an impact on the differences of the energy levels.
It is possible to solve for the eigenvalues of the Hamiltonian analytically by representing the problem in the phase basis and solving the Schrödinger equation.
This approach yields solutions in terms of the Mathieu characteristic values~\cite{cottet_scqubit_2002, devoret_sc_2004}.

Here, we follow a different approach by reformulating the Hamiltonian in the charge basis $\{\ket{n} | n \in \mathbb{Z}\}$ and numerically diagonalizing the system. 
The contribution due to Cooper pairs on the island is already in the right basis and the cosine term can be evaluated as
\begin{equation}
    \begin{aligned}
    \bra{m}\cos(\hat \varphi)\ket{n} &= \frac{\bra{m}\left(e^{i\hat \varphi} + e^{-i\hat \varphi}\right)\ket{n}}{2} \\
                             &= \frac{\braket{m|n+1} + \braket{m|n-1}}{2},
    \end{aligned}
\end{equation}
where we used the relation $\exp(i\ell \hat \varphi)\ket{n} = \ket{n + \ell}$ for $\ell \in \mathbb{Z}$, which is due to the operators $\hat n$ and $\hat \varphi$ being conjugate~\cite{cottet_scqubit_2002}.
The resulting representation has a tridiagonal shape and is defined by
\begin{equation}
    \hat H = \sum_{n=-\infty}^\infty \left(\frac{E_C}{2}(n - n_g)^2 \ket{n}\bra{n} - \frac{E_J}{2}\left(\ket{n}\bra{n+1} + \ket{n+1}\bra{n}\right)\right).
\end{equation}
To numerically solve this Hamiltonian, we only sum over $n \in [-n_\text{max}, n_\text{max}]$ for a cutoff value $n_\text{max}$.
This approximation works well as we are only interested in the lowest energy eigenvalues, which will be concentrated around $n = 0$, i.e. few Cooper pairs on the island.

\begin{figure}[t]
    \centering
    \includegraphics[width=\textwidth]{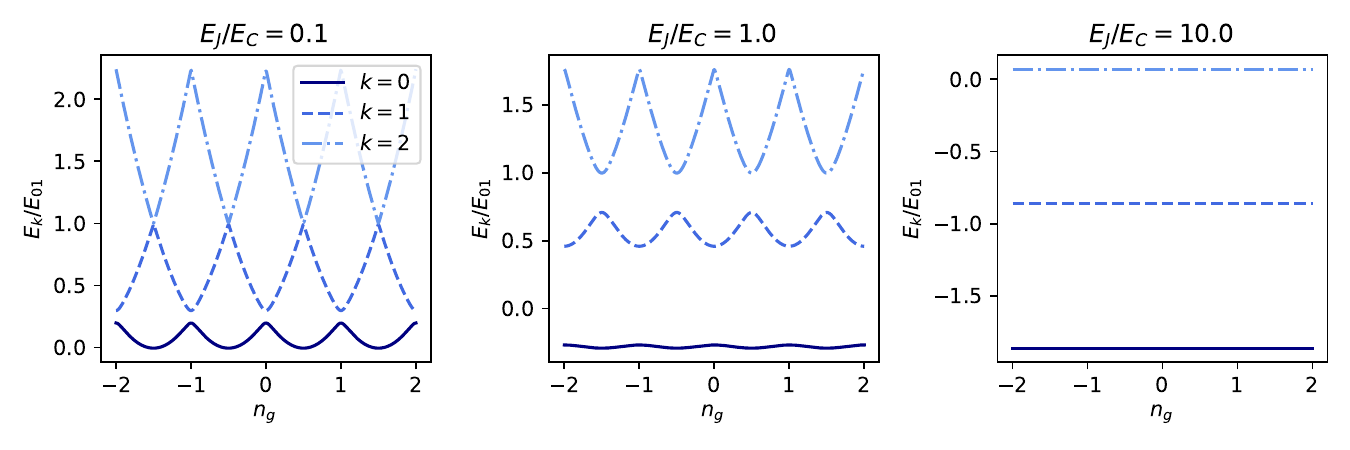}
    \caption[Energy levels of a Cooper pair box]{The lowest 3 energy levels of the Cooper pair box for different fractions $E_J/E_C$ and gate charges $n_g$. The energies are normalized by the transition energy from the $k=0$ to $1$ excitation at $n_g=1/2$.}
    \label{fig:cpb_energies}
\end{figure}

In Fig.~\ref{fig:cpb_energies} we show the energy levels of the Cooper pair box for different $n_g$ and fractions of $E_J/E_C$.
An ideal qubit has clearly distinct transition energies between the energy levels, which can be achieved with a moderate fraction of $E_J/E_C$ at a specific gate charge $n_g$, or $E_J/E_C \gg 1$.
Either approaches comes with other trade-offs.
A system with large $E_J/E_C$ has less energy fluctuations due to noise in gate charge and greater coherence times, but the anharmonicity in the energy level transitions $\Delta E_k = E_{k+1} - E_k$ decreases. This can lead to reduced fidelities in qubit operations or leakage~\cite{blais_cqed_2021}.

\subsection{Virtual Z rotation}\label{app:virtualz}

We use the identity~\cite{mckay_virtualz_2017} 
\begin{equation}
    R_Z(-\phi) R_X(\theta) R_Z(\phi) = e^{i\frac{\phi}{2} Z} e^{-i\frac{\theta}{2} X} e^{-i\frac{\phi}{2} Z} 
    = e^{-i\frac{\theta}{2} \left(\cos(\phi) X + \sin(\phi) Y\right)} 
    = R_{\phi}(\theta),
\end{equation}
to show Eq.~\eqref{eq:virtualz}.
Assume a $Z$ rotation with angle $\phi$ before two rotations with angles $\theta_j$ and rotation axis defined by $\phi_j$.
Using the above identity we have
\begin{equation}
    \begin{aligned}
        &R_{\phi_2}(\theta_2) R_{\phi_1}(\theta_1) R_Z(\phi)  \\
        &\equiv R_Z(-\phi) R_{\phi_2}(\theta_2) R_Z(\phi) R_Z(-\phi)R_{\phi_1}(\theta_1) R_Z(\phi) \\
        &= R_Z(-\phi) R_Z(-\phi_2) R_Z(\theta_2) R_Z(\phi_2) R_Z(\phi) R_Z(-\phi)  R_Z(-\phi_1)R_{X}(\theta_1) R_Z(\phi_1) R_Z(\phi) \\
        &= R_Z(-\phi-\phi_2) R_Z(\theta_2) R_Z(\phi_2+\phi) R_Z(-\phi-\phi_1)R_{X}(\theta_1) R_Z(\phi_1+\phi) \\
        &= R_{\phi_2+\phi}(\theta_2) R_{\phi_1+\phi}(\theta_1).
    \end{aligned}
\end{equation}
For any rotation following $R_Z(\phi)$ the rotation axis obtains a $\phi$-shift. Since the preceding gates are unaffected, this shows Eq.~\eqref{eq:virtualz}.

\section{Compiling a Trotter circuit}\label{app:compile_trotter}

In this section we show the concrete steps performed to compile the circuit of Fig.~\ref{fig:trotter_logical}.
Remember that we implement a single step of the first-order Trotter expansion of
\begin{equation}
    H = Z_3 Z_4 + Z_1 Z_3 + Z_1 Z_2 + X_2 X_3 + Y_1 Y_3,
\end{equation}
to evolve a pair of Bell states, $\ket{\psi_0} = (\ket{00} + \ket{11})^{\otimes 2}/2$.
The basis gate set we consider consists of $\{\sqrt{X}, R_Z, \text{CX}\}$ on linear topology,
i.e. CX gates on neighboring qubits are supported. Each step of the compilation is illustrated in Fig.~\ref{fig:compile_example}.

\begin{landscape}
\begin{figure}
    \centering
    \includegraphics[width=0.99\linewidth]{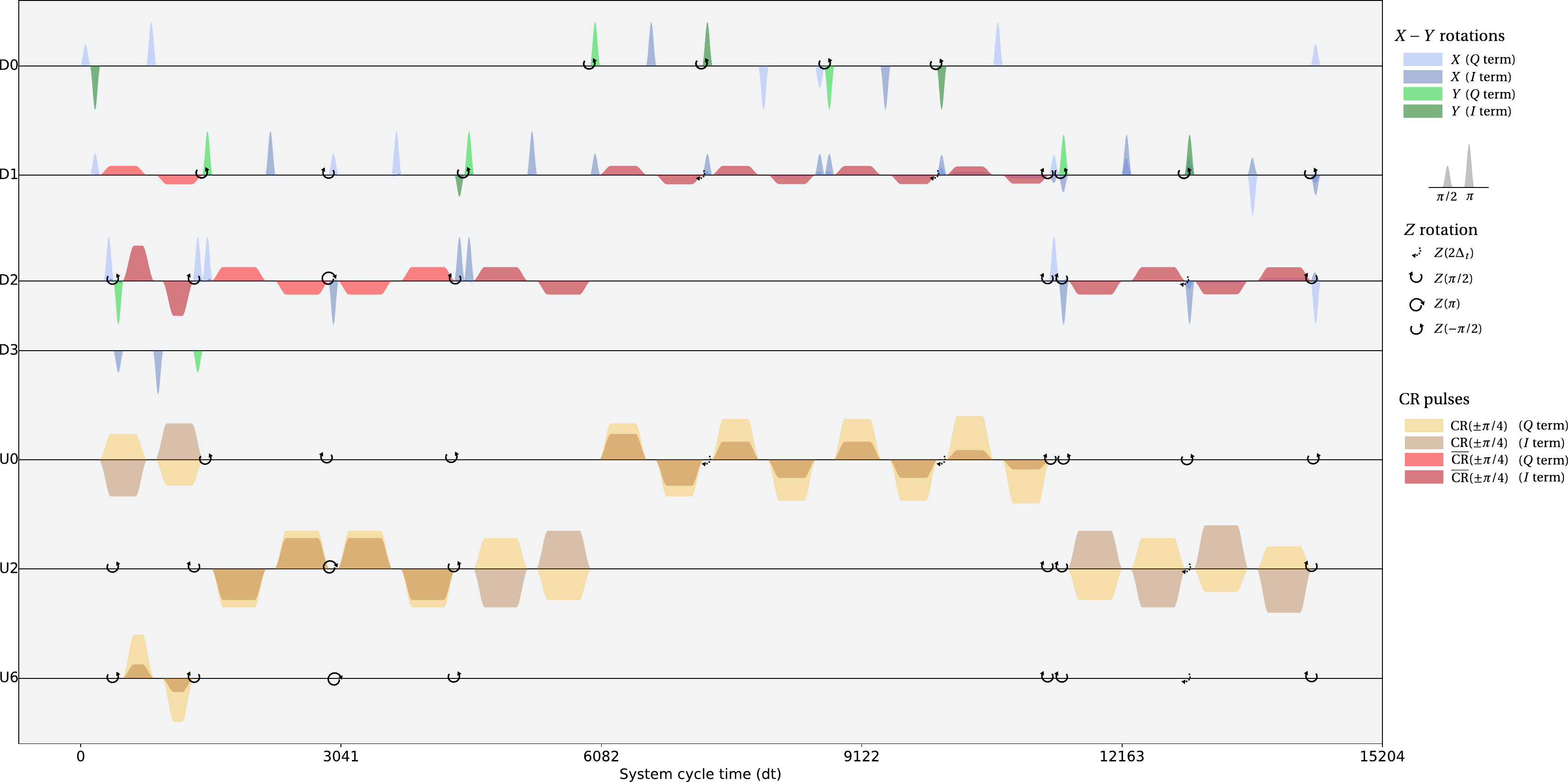}
    \caption[Pulse schedule of a Suzuki-Trotter expansion]{The pulse schedule for the physical Trotter circuit of Fig.~\ref{fig:trotter_logical}(b) generated for $\texttt{ibm\_cairo}$, which is an IBM Quantum Falcon chip~\cite{ibm_quantum}.
    Each qubit has a single drive line, however the schedules are drawn separately for single qubit rotations (D0, D1, D2, D3) and CR pulses (U0, U2, U6) for better readability.
    The Gaussian envelopes driving $X$ and $Y$ rotations are shown in blue and green and virtual $Z$ rotations are indicated as circular arrows.
    During the Gaussian square CR pulse on the control qubit, a cancellation tone $\overline{\text{CR}}$ is applied to the target qubit to suppress undesired dynamics.
    Each pulse shows the shape of the $I$ and $Q$ parts of the envelope given by $s(t) = \mathrm{Re}\left((I + iQ)e^{i\omega_d t}\right)$, where $\omega_d$ is the drive frequency.
    With some few exceptions, the $\pi/2$ pulses have half the amplitude of the $\pi$ pulse and we omit to indicate the signs of the rotations for visual clarity.
    }
    \label{fig:trotter_pulse}
\end{figure}
\end{landscape}

\begin{enumerate}
    \item \textit{High-level optimization}~ High-level descriptions of the gate operations and algorithmic knowledge allows to perform optimizations.
        \begin{itemize}
            \item The $R_{ZZ}$ gates commute, which allows to pull the third $R_{ZZ}$ gate to the front to reduce the circuit depth.
            \item Knowing that we perform a first-order Trotter expansion, the error scaling is not affected by switching the $R_{XX}$ and $R_{YY}$ gates. Grouping gates that act on the same qubit connections allows for easier routing later on.
        \end{itemize}
    \item \textit{Routing}~ The problem of mapping the logical qubits to physical qubits and routing the circuit does not have a unique solution. In this particular case, we find an optimal solution with only a single Swap gate and trivial routing: the $j$th logical qubit is mapped to the $j$th physical qubit.
    \item \textit{Basis translation}~ We replace each non-basis gate with a decomposition in the device's basis.
    \item \textit{Low-level optimization}~ The low-level circuit allows for further optimizations which, in this instance, are
        \begin{itemize}
            \item Adjoint $CX$ gates cancel.
            \item Re-synthesize the single qubit chain 
            \begin{equation*}
                R_Z(\pi/2) \sqrt{X} R_Z(\pi/2) \sqrt{X},
            \end{equation*}
            into
            \begin{equation*}
                e^{i 5\pi/4} R_Z(-\pi) \sqrt{X} R_Z(\pi/2),
            \end{equation*}
            and drop the global phase.
            \item Optimize the equal superposition construction from $R_Z(\pi/2) \sqrt{X} R_Z(\pi/2) \ket{0}$ to $R_Z(\pi/2) \sqrt{X} \ket{0}$.
            \item Use that diagonal gates commute with control-operations.
            \item Remove $R_Z$ gates acting on $\ket{0}$, as they only apply a global phase.
            \item Remove diagonal gates (i.e. $R_Z$) before measurements, as they do not affect the readout.
        \end{itemize}
    \item \textit{Pulse translation} Finally, the basis gates are translated to a pulse schedule according to the available device calibrations. This is shown in Fig.~\ref{fig:trotter_pulse}.
\end{enumerate}

\begin{figure}
    \centering
    \flushleft{\textit{Initial circuit}}
    \[
        \begin{array}{c}
            \Qcircuit @C=1em @R=.7em {
                \lstick{\ket{0}} & \gate{H} & \ctrl{1} & \qw & \multigate{2}{R_{ZZ}} & \multigate{1}{R_{ZZ}} & \qw  & \multigate{2}{R_{YY}} & \qw\\
                \lstick{\ket{0}} & \qw & \targ & \qw &\ghost{R_{ZZ}} & \ghost{R_{ZZ}} & \multigate{1}{R_{XX}} & \ghost{R_{YY}} & \qw  \\
                \lstick{\ket{0}} & \gate{H} & \ctrl{1} & \multigate{1}{R_{ZZ}} & \ghost{R_{ZZ}}  & \qw & \ghost{R_{XX}} & \ghost{R_{YY}} & \qw \\
                \lstick{\ket{0}} & \qw & \targ & \ghost{R_{ZZ}} & \qw & \qw & \qw & \qw & \qw
            }
        \end{array}
    \]
    \flushleft{\textit{High-level optimizations}}
    \[
        \begin{array}{c}
            \Qcircuit @C=1em @R=.7em {
                \lstick{\ket{0}} & \gate{H} & \ctrl{1} & \multigate{1}{R_{ZZ}} & \multigate{2}{R_{ZZ}} & \multigate{2}{R_{YY}} & \qw & \qw\\
                \lstick{\ket{0}} & \qw & \targ & \ghost{R_{ZZ}} & \ghost{R_{ZZ}} & \ghost{R_{YY}} & \multigate{1}{R_{XX}} & \qw  \\
                \lstick{\ket{0}} &\gate{H} & \ctrl{1} & \multigate{1}{R_{ZZ}} & \ghost{R_{ZZ}}  & \ghost{R_{YY}} & \ghost{R_{XX}}  & \qw \\
                \lstick{\ket{0}} & \qw & \targ & \ghost{R_{ZZ}} & \qw & \qw & \qw & \qw
            }
        \end{array}
    \]
    \flushleft{\textit{Routing}}
    \[
        \begin{array}{c}
            \Qcircuit @C=1em @R=.7em {
                \lstick{q_0: \ket{0}} & \gate{H} & \ctrl{1} & \multigate{1}{R_{ZZ}} & \qw & \multigate{1}{R_{ZZ}} & \multigate{1}{R_{YY}} & \qw & \qw\\
                \lstick{q_1: \ket{0}} & \qw & \targ & \ghost{R_{ZZ}} & \qswap & \ghost{R_{ZZ}} & \ghost{R_{YY}} & \multigate{1}{R_{XX}} & \qw  \\
                \lstick{q_2: \ket{0}} & \gate{H} & \ctrl{1} & \multigate{1}{R_{ZZ}} & \qswap\qwx & \qw & \qw & \ghost{R_{XX}} & \qw \\
                \lstick{q_3: \ket{0}} & \qw & \targ & \ghost{R_{ZZ}} & \qw & \qw & \qw & \qw & \qw 
            }
        \end{array}
    \]
    \flushleft{\textit{Basis translation}}
    \[
        {\footnotesize
        \begin{array}{c}
            \Qcircuit @C=.1em @R=.7em {
                \lstick{q_0: \ket{0}} & \gate{R_Z^{\pihalf}} & \gate{\sqrt{X}} & \gate{R_Z^{\pihalf}} & \ctrl{1} & \ctrl{1} & \qw & \ctrl{1} & \qw & \qw & \qw & \ctrl{1} & \qw & \ctrl{1} & \gate{\sqrt{X}} & \ctrl{1} & \qw & \ctrl{1} & \gate{\sqrt{X}} & \qw & \qw & \qw & \qw & \qw & \qw & \qw & \qw & \qw & \qw\\
                \lstick{q_1: \ket{0}} & \qw & \qw & \qw & \targ & \targ & \gate{R_Z} & \targ & \ctrl{1} & \targ & \ctrl{1} & \targ & \gate{R_Z} & \targ & \gate{\sqrt{X}} & \targ & \gate{R_Z} & \targ & \gate{\sqrt{X}} & \gate{R_Z^{\pihalf}} & \gate{\sqrt{X}} & \gate{R_Z^{\pihalf}} & \ctrl{1} & \qw & \ctrl{1} & \gate{R_Z^{\pihalf}} & \gate{\sqrt{X}} & \gate{R_Z^{\pihalf}} & \qw  \\
                \lstick{q_2: \ket{0}} & \gate{R_Z^{\pihalf}} & \gate{\sqrt{X}} & \gate{R_Z^{\pihalf}} & \ctrl{1} & \ctrl{1} & \qw & \ctrl{1} & \targ & \ctrl{-1} & \targ & \qw & \qw & \qw & \qw & \qw & \qw & \qw & \qw & \gate{R_Z^{\pihalf}} & \gate{\sqrt{X}} & \gate{R_Z^{\pihalf}} & \targ & \gate{R_Z} & \targ & \gate{R_Z^{\pihalf}} & \gate{\sqrt{X}} & \gate{R_Z^{\pihalf}} & \qw \\
                \lstick{q_3: \ket{0}} & \qw & \qw & \qw & \targ & \targ & \gate{R_Z} & \targ & \qw & \qw & \qw & \qw & \qw & \qw & \qw & \qw & \qw & \qw & \qw & \qw & \qw & \qw & \qw & \qw & \qw & \qw & \qw & \qw & \qw & \qw
            }
        \end{array}
        }
    \]
    \flushleft{\textit{Low-level optimizations}}
    \[
        \begin{array}{c}
            \Qcircuit @C=.1em @R=.7em {
                \lstick{q_0: \ket{0}} & \gate{\sqrt{X}} & \gate{R_Z^{\pi/2}} & \ctrl{1} & \qw & \qw & \qw & \ctrl{1} & \qw & \ctrl{1} & \gate{\sqrt{X}} & \ctrl{1} & \qw & \ctrl{1} & \gate{\sqrt{X}} & \qw & \qw & \qw & \qw & \qw & \qw & \qw\\
                \lstick{q_1: \ket{0}} & \qw & \qw & \targ & \ctrl{1} & \targ & \ctrl{1} & \targ & \gate{R_Z} & \targ & \gate{\sqrt{X}} & \targ & \gate{R_Z} & \targ & \gate{R_Z^{\pi/2}} & \gate{\sqrt{X}} & \ctrl{1} & \qw & \ctrl{1} & \gate{R_Z^{-\pi/2}} & \gate{\sqrt{X}} & \qw  \\
                \lstick{q_2: \ket{0}} & \gate{\sqrt{X}} & \gate{R_Z^{\pi/2}} & \ctrl{1} & \targ & \ctrl{-1} & \targ & \qw & \qw & \qw & \qw & \qw & \qw & \gate{R_Z^{\pi/2}} & \gate{\sqrt{X}} & \gate{R_Z^{\pi/2}} & \targ & \gate{R_Z} & \targ & \gate{R_Z^{\pi/2}} & \gate{\sqrt{X}} & \qw \\
                \lstick{q_3: \ket{0}} & \qw & \qw & \targ & \qw & \qw & \qw & \qw & \qw & \qw & \qw & \qw & \qw & \qw & \qw & \qw & \qw & \qw & \qw & \qw & \qw & \qw & \qw 
            }
        \end{array}
    \]
    \caption[Detailed compilation workflow]{Compilation workflow for the first-order Suzuki-Trotter example.}
    \label{fig:compile_example}
\end{figure}

\section{Measurement error mitigation}\label{app:mem}

We are interested in computing the expectation value of an observable $O$ with respect to a given state $\ket{\psi}$. For simplicity, we assume $O$ is diagonal in the computational basis $\{\ket{x}\}_x$---if it is not, it can be decomposed into a sum of terms diagonalizable with single qubit transformations.
The expectation value can be written as
\begin{equation}
    \braket{\psi|O|\psi} = \sum_{x = 0}^{2^n - 1} c^*_x c_x \braket{x|O|x} = \sum_{x=0} p_x O(x),
\end{equation}
where we expanded $\ket{\psi} = \sum_x c_x \ket{x}$, introduced the probability to sample the state $\ket{x}$ as $p_x = |c_x|^2$ and used the notation $O(x) = \braket{x|O|x}$.
To compute the error mitigated expectation value, we evaluate the expectation value for the probabilities $A^{-1}\tilde{\vec p}$, i.e.
\begin{equation}
    \braket{\psi|O|\psi} = \sum_{x=0}^{2^n - 1} \left(A^{-1} \tilde{\vec p}\right)_x O(x).
\end{equation}

If we calculate the expectation value in an experiment, we only know the probabilities $p_x$ by sampling the state $N$ times and measuring the outcome bitstrings $\{s_j\}_j$ as
\begin{equation}
    \tilde p_x = \frac{1}{N} \sum_{j=1}^N |\braket{s_j|x}|^2.
\end{equation}
Plugging this definition into the error-mitigated expectation value we obtain
\begin{equation}
    \begin{aligned}
        \braket{\psi|O|\psi} 
        &= \sum_{x=0}^{2^n-1} \left(\sum_{y=0}^{2^n -1} \tilde p_y A^{-1} \ket{y}\right)_x O(x) \\
        &= \sum_{x=0}^{2^n-1} \sum_{y=0}^{2^n-1} \tilde p_y \bra{x} A^{-1} \ket{y} O(x) \\
        &= \frac{1}{N} \sum_{j=1}^{N} \sum_{x=0}^{2^n-1} \sum_{y=0}^{2^n-1} |\braket{s_j|y}|^2 \bra{x} A^{-1} \ket{y} O(x) \\
        &= \frac{1}{N} \sum_{j=1}^{N} \sum_{x=0}^{2^n-1} \bra{x} A^{-1} \ket{s_j} O(x),
    \end{aligned}
\end{equation}
where we used $\tilde{\vec p} = \sum_x \tilde p_x \ket{x}$ and $\braket{s_j|x} = \delta_{s_j, x}$.

If we now express $A = A_n \otimes \cdots \otimes A_1$ and $O = O_n \otimes \cdots \otimes O_1$, then we can rewrite the expectation value in a form using only a linear number of operations in $n$.
Before exploiting the tensor-product form, we rewrite the expectation value in a more convenient form, that is
\begin{equation}
    \begin{aligned}
        \braket{\psi|O|\psi} &= \frac{1}{N} \sum_{j=1}^{N} \sum_{x=0}^{2^n-1} \bra{x} O A^{-1} \ket{s_j} \\
        &= \frac{\sqrt{2^n}}{N} \sum_{j=1}^{N} \bra{+}^{\otimes n} O A^{-1} \ket{s_j}, \\
    \end{aligned}
\end{equation}
where we exploited that $O$ is diagonal, i.e. $\bra{x} O = \bra{x} O(x)$ and $\ket{+} = (\ket{0} + \ket{1})/\sqrt{2}$.
Leveraging that all states and operators are now product of single-qubit operations, we can write
\begin{equation}
    \braket{\psi|O|\psi} = \frac{\sqrt{2^n}}{N} \sum_{j=1}^{N} \prod_{k=1}^n \bra{+} O_k A_k^{-1} \ket{s_j^k},
\end{equation}
which only requires iterating over $Nn$ terms and can be done efficiently.

\section{Pauli twirling on \texttt{ibm\_cairo}}\label{app:pt}

In the experiment of Fig.~\ref{fig:twirling}, we perform a Pauli process tomography on the circuit
\begin{equation}
    \mathcal{U} = \left(R_{ZZ}\left(\frac{\pi}{2}\right) R_{ZZ}\left(-\frac{\pi}{2}\right)\right)^r,
\end{equation}
for $r=10$ repetitions, whose noise-free Pauli process tomography should result in the identity matrix.
We twirl each $R_{ZZ}$ gate individually with a random two-qubit Pauli operation $T = P_1 \otimes P_2$ as 
\begin{equation}
    R_{ZZ}(\theta) \rightarrow T R_{ZZ}((-1)^{s_T} \theta) T,
\end{equation}
where the sign of the rotation must be flipped if an odd number of Paulis does not commute with $R_{ZZ}$.
More formally,
\begin{equation}
    s_T = \prod_{P \in \{P_1, P_2\}} 
    \begin{cases}
        1, \text{ if } [P, Z] = 0, \\
        -1, \text{ otherwise}.
    \end{cases}
\end{equation}
The sign $(-1)^{s_T}$ must be adjusted to ensure the operation is twirled in a unitary-preserving manner, e.g.
\begin{equation}
    \begin{aligned}
        R_{ZZ}(\theta) &= YY R_{ZZ}(\theta) YY, \\
        R_{ZZ}(\theta) &= XZ R_{ZZ}(-\theta) XZ.
    \end{aligned}
\end{equation}

The Pauli process tomography is performed by preparing any possible combination of the single-qubit initial states
\begin{equation}
    \{\ket{0}, \ket{1}, \ket{+}, \ket{i}\}^{\otimes 2},
\end{equation}
where $\ket{+} = (\ket{0} + \ket{1})/\sqrt{2}$ and $\ket{i} = (\ket{0} + i\ket{1})/\sqrt{2}$, and we measure in every possible Pauli basis $\{I, X, Y, Z\}^{\otimes 2}$.
For the twirled instance, the process tomography is repeated 10 times, each with individual twirls, and then averaged.

\chapter{Variational quantum algorithms}

\section{Variational principles}\label{app:vp}

Table~\ref{tab:vp} compares formulations and drawbacks of McLachlan's VP, the TDVP, and Dirac-Frenkel's VP. They are defined in terms of the QGT, 
\begin{equation}
    G_{jk}(\vec\theta) = \braket{\partial_j\phi(\vec\theta) | \partial_k\phi(\vec\theta)} - \braket{\partial_j \phi(\vec\theta)|\phi(\vec\theta)}\braket{\phi(\vec\theta)|\partial_k\phi(\vec\theta)},
\end{equation}
and the evolution gradient $\vec B$,
\begin{equation}
     B_k(\vec\theta) = \braket{\partial_k\phi(\vec\theta) | H | \phi(\vec\theta)} - \braket{\partial_k \phi(\vec\theta)|\phi(\vec\theta)}.
\end{equation}
McLachlan's VP for imaginary-time evolution has a different evolution gradient $\vec B^{(I)}$, given by
\begin{equation}
    B^{(I)}_k(\vec\theta) = \braket{\partial_k\phi(\vec\theta) | H | \phi(\vec\theta)}.
\end{equation}
The TDVP for real parameters in not applicable for imaginary-time evolution, as the update rule would yield purely imaginary parameters.

\begin{table}[htbp]
    \centering
    \begin{tabular}{|c|c|c|c|}
        \hline
        & McLachlan & TDVP & Dirac-Frenkel \\
        \hline\hline
        real & $\mathrm{Re}(G)\dot{\vec\theta} = \mathrm{Im}(\vec B)$ & $\mathrm{Im}(G)\dot{\vec\theta} = -\mathrm{Re}(\vec B)$ & $G\dot{\vec\theta} = -i\vec B$ \\
        imaginary & $\mathrm{Re}(G)\dot{\vec\theta} = -\mathrm{Re}(\vec B^{(I)})$ & N.A. & $G\dot{\vec\theta} = -\vec B$ \\
        drawbacks & $-$ & unstable & $\dot{\vec\theta}$ is complex \\
             &     & misses dynamics & more costly \\
        \hline
    \end{tabular}
    \caption[Comparison of variational principles]{Variational principles update rules and drawbacks.}
    \label{tab:vp}
\end{table}

Under certain conditions described in Appendix~\ref{app:compare_vp}, they are equivalent. However, for real parameters simple cases can be constructed where they are not, as shown in Appendix~\ref{app:tdvp_fails}.

\subsection{Equivalence of variational principles}\label{app:compare_vp}

This thesis focuses on McLachlan's VP, due to practical drawbacks in implementing the TDVP or Dirac-Frenkel VP. 
In this section we briefly outline the derivations of these two VPs and set all three into context.

The TDVP uses the Lagrangian formulation of the Schrödinger equation,
\begin{equation}
    L(\psi,\dot\psi) = \Braket{\psi(t) \left| \left(\frac{\mathrm{d}}{\mathrm{d} t} + iH\right)\right|\psi(t)}
\end{equation}
and seeks the stationary point of the action functional,
\begin{equation}
    S[L] = \int_{t_1}^{t_2} L(\psi(t), \Dot\psi(t)) \mathrm{d}t.
\end{equation}
Solving the minimization for $\psi(t) = \ket{\phi(\vec\theta(t))}$ leads to the equation
\begin{equation}
    \mathrm{Im}(G(\vec\theta))\dot{\vec\theta} = \mathrm{Re}(\vec B(\vec\theta)).
\end{equation}

The Dirac-Frenkel VP requires the residual $\ket{r(\vec\theta)}$ to be orthogonal to the tangent space of the variational manifold at point $\vec\theta$~\cite{broeckhove_equivalence_1988}, 
\begin{equation}
    \braket{\delta\phi(\vec\theta) | r(\vec\theta)} = 0.
\end{equation}
This is equivalent to directly projecting the dynamics given by $-iH\ket{\phi(\vec\theta)}$ onto the space of the derivative $\partial/\partial t \ket{\phi(\vec\theta)}$~\cite{yuan_varqte_2019} and leads to
\begin{equation}
    G(\vec\theta)\dot{\vec\theta} = -i\vec B(\vec\theta).
\end{equation}
For a model with real parameters, the formulation due to Dirac and Frenkel requires truncation of possibly complex parameters and additional resources to compute both real- and imaginary parts of $G$ and $\vec B$.

If the parameters $\vec\theta$ are complex, all three variational principles yield the same dynamics.
However, in quantum circuits the parameters are typically real, as they determine rotation angles, in which case the variational principles may differ.
McLachlan's and the Dirac-Frenkel VP coincide if the solution of the latter is real.
The TDVP equals both if~\cite{broeckhove_equivalence_1988}
\begin{equation}
    \ket{\partial_j \phi(\vec\theta)} = i \ket{\partial_k \phi(\vec\theta)},
\end{equation}
which is typically not satisfied in quantum circuit models~\cite{yuan_varqte_2019}.

\subsection{Failure of the TDVP}\label{app:tdvp_fails}

The TDVP often produces an unstable linear system for real parameters as the diagonal of the QGT is real. Solving for the parameter derivative, thus, requires a regularization which comes at the cost of introducing a bias.
In addition, the TDVP may fail to exhibit dynamics at all, even for simple settings where the variational model allows to follow the dynamics exactly. 
For example the Hamiltonian $H = X$ with model
\begin{equation}
    \ket{\phi(\theta)} = R_X(\theta)\ket{0},
\end{equation}
at point $\theta = \pi/2$ has the QGT $G = 1/4$ and evolution gradient $B = i/2$. The TDVP attempts to solve the equation $0\dot{\vec\theta} = \vec 0$.
On a gate-based quantum computer, McLachlan's VP is therefore usually the most suitable method to implement variational time evolution. 

\subsection{Derivation of McLachlan's variational principle}\label{app:mclachlan_proof}

This section derives the parameter dynamics under real-time evolution using McLachlan's variational principle.

We consider a variational state with explicit global phase dependency,
\begin{equation}
    \ket{\Phi(\theta_0, \vec\theta)} = e^{i\theta_0} \ket{\phi(\vec\theta)},
\end{equation}
since the global phase affects the state derivative, and thus the dynamics, even though it cannot be measured on the state itself.
The time derivative of this state is
\begin{equation}
    \ket{\dot\Phi} := \frac{\partial}{\partial t}\ket{\Phi} = i\dot\theta_0 e^{i\theta_0}\ket{\phi} + e^{i\theta_0}\sum_{k} \dot\theta_k \ket{\partial_k \phi},
\end{equation}
where we dropped explicit parameter dependencies of $\ket{\Phi}$ and $\ket{\phi}$ for convenience.

The norm of the residual $\ket{r}$ is minimized by variation over the parameter derivatives $\dot{\vec\theta}$ and $\dot\theta_0$.
Using the notation of variational calculus we write optimization as
\begin{equation}
    0 = \delta \|\ket{r}\|_2
    = \delta \left\| \left(\frac{\partial}{\partial t} + iH\right)\ket{\Phi}\right\|_2 
    = \delta \left\| \ket{\dot\Phi} + iH\ket{\Phi}\right\|_2,
\end{equation}
where $\delta$ denotes the variation over the parameter derivatives. Note that, since we only vary the parameter derivatives, the variation of $\ket{\Phi}$ is 0, i.e. $\ket{\delta\Phi} = 0$.
Since $\|\ket{r}\|^2 = 0$ implies $\|\ket{r}\| = 0$, we minimize the squared residual and obtain
\begin{equation}\label{eq:mclachlan_imag_projection}
    \begin{aligned}
      0 &= \delta \braket{r|r} \\
        &= \delta \left(\braket{\dot\Phi|\dot\Phi} + i\braket{\dot\Phi|H|\Phi} - i\braket{\Phi|H|\dot\Phi} + \braket{\Phi|H^2|\Phi}\right) \\
        &= \braket{\delta\dot\Phi|\dot\Phi} + \braket{\dot\Phi|\delta\dot\Phi} + i\braket{\delta\dot\Phi|H|\Phi} - i\braket{\Phi|H|\delta\dot\Phi} \\
        &= -i\braket{\delta\dot\Phi|i\dot\Phi} + i\braket{i\dot\Phi|\delta\dot\Phi} + i\braket{\delta\dot\Phi|H|\Phi} - i\braket{\Phi|H|\delta\dot\Phi} \\
        &= -i\braket{\delta\dot\Phi|i\dot\Phi - H\Phi} + i\braket{i\dot\Phi - H\Phi|\delta\dot\Phi} \\
        &= 2\mathrm{Im}\left(\braket{\delta\dot\Phi|i\dot\Phi - H\Phi}\right).
    \end{aligned}
\end{equation}

As remark, we would like to point out that this representation of McLachlan's variational principle allows an intuitive comparison to the TDVP, which can be written as 
\begin{equation}
    0 = \mathrm{Re}\left(\braket{\delta\dot\Phi|i\dot\Phi - H\Phi}\right),
\end{equation}
and DF, 
\begin{equation}
    0 = \left(\braket{\delta\dot\Phi|i\dot\Phi - H\Phi}\right).
\end{equation}
McLachlan's principle and TDVP thus require different parts of the projection to vanish, and if the DF is satisfied it implies that both McLachlan and TDVP are also obeyed~\cite{broeckhove_equivalence_1988}.

The variation of the time-derivative is
\begin{equation}
    \ket{\delta\dot\Phi} = i\delta\dot\theta_0 e^{i\theta_0}\ket{\phi} + e^{i\theta_0}\sum_{k} \delta\dot\theta_k \ket{\partial_k \phi}.
\end{equation}
Plugging the definitions of $\ket{\dot\Phi}$ and $\ket{\delta\dot\Phi}$ into Eq.~\eqref{eq:mclachlan_imag_projection} we obtain
\begin{equation}
    \begin{aligned}
    0 &= \mathrm{Im}\Bigg(
    i\dot\theta_0 \delta\dot\theta_0 
    + \delta\dot\theta_0\sum_k \dot\theta_k \braket{\phi|\partial_k\phi}
    + i\delta\dot\theta_0 E
    - \dot\theta_0 \sum_k \delta\dot\theta_k \braket{\partial_k\phi|\phi}  \\
    &~~~~~~~~~~~~ + i \sum_{j, k} \delta\dot\theta_k \dot\theta_j \braket{\partial_k\phi|\partial_j\phi}
    - \sum_k \delta\dot\theta_k \braket{\partial_k\phi|H|\phi}\Bigg) \\
    &= \delta\dot\theta_0 \mathrm{Im}\Bigg(\dot\theta_0 + \sum_k \dot\theta_k\braket{\phi|\partial_k\phi} + E \Bigg) \\
    &~~~~ + \sum_k \delta\dot\theta_k \mathrm{Im}\Bigg( 
        -\dot\theta_0 \braket{\partial_k\phi|\phi}
        + i\sum_j \dot\theta_j \braket{\partial_k\phi|\partial_j\phi} 
        - \braket{\partial_k \phi | H |\phi}
    \Bigg).
    \end{aligned}
\end{equation}
where we used $\braket{\phi|\phi} = 1$, $E = \braket{\phi|H|\phi}$ and that the parameter derivatives and it's variations are real.
Since this expression must be satisfied for any variation, we set the $\delta\dot\theta_0$ and $\delta\dot\theta_k$ terms to 0.

The variation of $\delta\dot\theta_0$ vanishing implies
\begin{equation}\label{eq:theta0_variation_imag}
    \dot\theta_0 = -\sum_{k} \dot\theta_k \mathrm{Im}\left(\braket{\phi|\partial_k\phi}\right) - E,
\end{equation}
where we again used $\dot\theta_k \in \mathbb{R}$.
Note that the derivative $\braket{\phi|\partial_k\phi}$ has no real part due to the normalization of the state, as shown previously in Eq.~\eqref{eq:zero_real_gradient}. This allows to re-write the above as
\begin{equation}\label{eq:theta0_variation_complex}
    \dot\theta_0 = i\sum_k \dot\theta_k \braket{\phi|\partial_k \phi} - E.
\end{equation}
Depending on how $\dot\theta_0$ is defined we obtain two different, but equivalent, descriptions of $\dot{\vec\theta}$. 
We first show the result using the first definition, which results in a description that requires less circuit evaluations, and then using the second one, which gives the form commonly used in literature.

Inserting Eq.~\eqref{eq:theta0_variation_imag} into the variations of $\delta\dot\theta_k$ yields
\begin{equation}\label{eq:thetak_variation}
    \begin{aligned}
    0 &= \mathrm{Im}\left(-\dot\theta_0 \braket{\partial_k\phi|\phi}\right)
        + \mathrm{Re}\left(\sum_j \dot\theta_j \braket{\partial_k\phi|\partial_j\phi}\right)
        - \mathrm{Im}\left(\braket{\partial_k\phi | H |\phi}\right) \\
    &= \mathrm{Im}\left(\sum_{j} \dot\theta_j \mathrm{Im}\left(\braket{\phi|\partial_j\phi}\right) \braket{\partial_k\phi|\phi}\right)
        + E \mathrm{Im}\left(\braket{\partial_k\phi|\phi}\right)
        + \sum_j \dot\theta_j \mathrm{Re}\left(\braket{\partial_k\phi|\partial_j\phi} \right)
        - \mathrm{Im}\left(\braket{\partial_k\phi | H |\phi}\right) \\
    &= \sum_j \dot\theta_j \Bigg(\mathrm{Re}\left(\braket{\partial_k\phi|\partial_j\phi}\right) + \mathrm{Im}\left(\braket{\phi|\partial_j\phi}\right) \mathrm{Im}\left(\braket{\partial_k\phi|\phi}\right) \Bigg)
    + \mathrm{Im}\left(\braket{\partial_k\phi|\phi}E - \braket{\partial_k \phi | H |\phi}\right),
    \end{aligned}
\end{equation}
where we reformulated $\mathrm{Im}(iz) = \mathrm{Re}(z)$ for any $z \in \mathbb{C}$.

This defines a linear system of equations for $\dot{\vec\theta}$,
\begin{equation}\label{eq:varqte_lse_efficient}
    A\dot{\vec\theta} = \vec b,
\end{equation}
with 
\begin{equation}
    A_{jk} = \mathrm{Re}\left(\braket{\partial_k\phi|\partial_j\phi}\right) + \mathrm{Im}\left(\braket{\phi|\partial_j\phi}\right) \mathrm{Im}\left(\braket{\partial_k\phi|\phi}\right),
\end{equation}
and 
\begin{equation}
    b_k = \mathrm{Im}\left(\braket{\partial_k | H |\phi} - \braket{\partial_k\phi|\phi}E\right).
\end{equation}

As pointed out before, this is not the same formulation as Eq.~\eqref{eq:varqte_lse}, since the definition of $A$ does not exactly match the real part of the QGT, $g$.
This formulation is obtained by inserting Eq.~\eqref{eq:theta0_variation_complex} into Eq.~\ref{eq:thetak_variation}, which gives
\begin{equation}
\begin{aligned}
    0 &= \mathrm{Im}\left(-i\sum_j \dot\theta_j \braket{\phi|\partial_j\phi} \braket{\partial_k\phi|\phi}\right)
        + E \mathrm{Im}\left(\braket{\partial_k\phi|\phi}\right)
        + \sum_j \dot\theta_j \mathrm{Re}\left(\braket{\partial_k\phi|\partial_j\phi} \right)
        - \mathrm{Im}\left(\braket{\partial_k | H |\phi}\right) \\
    &= \sum_j \dot\theta_j \Bigg(\mathrm{Re}\left(\braket{\partial_k\phi|\partial_j\phi}\right) 
    - \mathrm{Re}\left({\partial_k\phi|\phi}\braket{\phi|\partial_j\phi}\right) \Bigg)
    + \mathrm{Im}\left(\braket{\partial_k\phi|\phi}E - \braket{\partial_k | H |\phi}\right).
\end{aligned}
\end{equation}
This equals the expected linear system 
\begin{equation}\label{eq:varqte_lse_inefficient}
    g\dot{\vec\theta} = \vec b.
\end{equation}

We emphasize that both formulations, via $A$ and $g$, are equivalent and provide the same dynamics $\dot{\vec\theta}$.
The formulation via the QGT exposes the connection of VarQTE and QNG and allows to view the all three variational principles as components of the same equation. In practice however, using Eq.~\eqref{eq:varqte_lse_efficient} requires implementing less circuits, as only the imaginary parts of $\braket{\phi|\partial_k\phi}$ are required, whereas Eq.~\eqref{eq:varqte_lse_inefficient} requires both real and imaginary parts to correctly evaluate the product.
If the gradients are evaluated with a LCU technique, both equations use $d(d+1)/2$ circuits for the $\mathrm{Re}(\braket{\partial_k\phi|\partial_j\phi})$ parts and Eq.~\eqref{eq:varqte_lse_efficient} uses $d$ for the phase-fix part where Eq.~\eqref{eq:varqte_lse_inefficient} requires $2d$.
$\qed$

\section{Imaginary-time evolution}\label{app:ite}

The imaginary-time evolved state is
\begin{equation}
    \ket{\psi(\tau)} = \frac{\exp(-\tau H)\ket{\psi_0}}{Z(\tau)},
\end{equation}
where $Z(\tau) = \sqrt{\braket{\psi_0|\exp(-2\tau H)|\psi_0}}$.
Expanding in the eigenbasis $\{\lambda_n\}_{n \geq 0}$ of the Hamiltonian with $H\ket{\lambda_n} = E_n\ket{\lambda_n}$, we obtain
\begin{equation}
    \begin{aligned}
    \ket{\psi(\tau)} &= \frac{e^{-H\tau}}{Z(\tau)} \sum_{n\geq 0} c_n \ket{\lambda_n} \\
                     &= \frac{1}{Z(\tau)} \sum_{n\geq 0} c_n e^{-E_n \tau}\ket{\lambda_n} \\
                     &= \frac{e^{-E_0\tau}}{Z(\tau)} \left(c_0 \ket{\lambda_0} + \sum_{n > 0} c_n e^{-(E_n - E_0) \tau}\ket{\lambda_n}\right),
    \end{aligned}
\end{equation}
where $c_n = \braket{\lambda_n|\psi_0}$
Similarly, the normalization can be written as
\begin{equation}
    \begin{aligned}
        Z(\tau) &= \left(\sum_{n \geq 0} |c_n|^2 e^{-2E_n \tau}\right)^{1/2} \\
                &= e^{-E_0 \tau} \left(|c_0|^2 + \sum_{n \geq 0} |c_n|^2 e^{-2(E_n - E_0) \tau}\right)^{1/2}.
    \end{aligned}
\end{equation}
The time-evolved state is then
\begin{equation}
    \ket{\psi(\tau)} = \frac{c_0 \ket{\lambda_0} + \sum_{n > 0} c_n e^{-(E_n - E_0) \tau}\ket{\lambda_n}}{\sqrt{|c_0|^2 + \sum_{n \geq 0} |c_n|^2 e^{-2(E_n - E_0) \tau}}},
\end{equation}
which is stated in Eq.~\eqref{eq:itevolved}.
Taking the long-time limit and assuming $c_0 \neq 0$, ordered energy eigenstates, and a non-degenerate ground state $E_0 < E_1$, we obtain
\begin{equation}
    \lim_{\tau\rightarrow\infty} \ket{\psi(\tau)} = \frac{c_0}{|c_0|} \ket{\lambda_0},
\end{equation}
which converges at a rate $E_1 - E_0$.

More quantitatively, the fidelity of the ground state with the imaginary-time evolved state is
\begin{equation}
        |\langle \lambda_0 | \psi(\tau)\rangle|^2
        = \frac{|c_0|^2}{|c_0|^2 + \sum_{n \geq 0} |c_n|^2 e^{-2(E_n - E_0) \tau}}.
\end{equation}
This expression can be lower bounded by assuming that, except for the overlap with the ground state, the overlap with the first excited state is maximal. Since this term decays the slowest, this maximizes the denominator and lower bounds the fidelity. We obtain
\begin{equation}
    \begin{aligned}
        |\langle \lambda_0 | \psi(\tau)\rangle|^2
        &\geq \frac{|c_0|^2}{|c_0|^2 + |c_1|^2 e^{-2(E_1 - E_0) \tau}} \\
        &\geq \left( 1 + \frac{ e^{-2(E_1 - E_0) \tau}}{|c_0|^2} \right)^{-1},
    \end{aligned}
\end{equation}
where we further bounded the expression by setting $|c_1|^2 = 1$. 
The overlap is therefore lower bounded depending on an exponential decay with a rate of the spectral gap, relative to the initial overlap.
polynomial time as long as the spectral gap is not exponentially small.

\section{Gibbs state preparation}\label{app:gibbsprep_proof}

Here, we show the circuit designed in Fig.~\ref{fig:gibbs-state-prep} indeed prepares a Gibbs state.
We consider two $n$-qubit subsystems $A$ and $B$ and prepare Bell state pairs across the subsystems. This constructs the state
\begin{equation}
    \begin{aligned}
    \ket{\psi_1} &= \frac{1}{\sqrt{2^n}} \left(\ket{0_A 0_B} + \ket{1_A 1_B}\right)^{\otimes n} \\
    &= \frac{1}{\sqrt{2^n}} \left(\ket{0_A 0_B~~0_A 0_B~\cdots~0_A 0_B} + \ket{0_A 0_B~~0_A 0_B~\cdots~1_A 1_B} + \cdots + \ket{1_A 1_B~~1_A 1_B~\cdots~1_A 1_B}  \right) \\
    &= \frac{1}{\sqrt{2^n}} \left( \ket{00\cdots 0_A~~00\cdots 0_B} + \ket{00\cdots 1_A~~00\cdots 1_B} + \cdots + \ket{11\cdots 1_A + 11\cdots 1_B}\right) \\
    &= \frac{1}{\sqrt{2^n}} \sum_{x=0}^{2^n - 1} \ket{x_A x_B},
    \end{aligned}
\end{equation}
where $\ket{x} := \ket{\mathrm{bin}(x)}$ is the binary representation of the state $\mathrm{bin}(x)$.

Onto this state we act with the operation
\begin{equation}
    \begin{aligned}
        e^{-\frac{\beta}{2} H_A \otimes \mathbb{1}_B} &= \sum_k \left(-\frac{\beta}{2}\right)^k \frac{(H_A \otimes \mathbb{1}_B)^k}{k!} \\
        &= \sum_k \left(-\frac{\beta}{2}\right)^k \frac{H_A^k \otimes \mathbb{1}_B^k}{k!} \\
        &= \left(\sum_k \left(-\frac{\beta}{2}\right)^k \frac{H_A^k}{k!} \right) \otimes \mathbb{1}_B \\
        &= e^{-\frac{\beta}{2} H_A} \otimes \mathbb{1}_B,
    \end{aligned}
\end{equation}
such that we obtain
\begin{equation}
    \ket{\psi_2} = \frac{1}{\sqrt{2^n}} \sum_{x=0}^{2^n - 1} \left(e^{-\frac{\beta}{2} H_A}\ket{x_A}\right) \otimes \ket{x_B}.
\end{equation}
Tracing out system $B$ yields
\begin{equation}
    \begin{aligned}
    \rho_3 &= \mathrm{Tr}_B(\ket{\psi_2}\bra{\psi_2}) \\
    &= \frac{1}{2^n} \sum_{m=0}^{2^n - 1}\sum_{x,y=0}^{2^n - 1} \left(e^{-\frac{\beta}{2} H_A}\ket{x_A}\right)\left(\bra{y_A}e^{-\frac{\beta}{2} H_A}\right) \braket{m_B | x_B}\braket{y_B | m_B} \\
    &= \frac{1}{2^n} \sum_{m=0}^{2^n - 1} e^{-\frac{\beta}{2} H_A}\ket{m_A}\bra{m_A}e^{-\frac{\beta}{2} H_A} \\
    &= \frac{1}{2^n} e^{-\frac{\beta}{2} H_A}\left(\sum_{m=0}^{2^n - 1}\ket{m_A}\bra{m_A}\right)e^{-\frac{\beta}{2} H_A} \\
    &= \frac{1}{2^n} e^{-\beta H_A},
    \end{aligned}
\end{equation}
using that $\sum_m \ket{m}\bra{m} = \mathbb{1}$ for any complete basis set $\{\ket{m}\}_m$.
The output state is normalized, thus, we finally have
\begin{equation}
    \frac{\rho_3}{\mathrm{Tr}(\rho_3)} = \frac{e^{-\beta H_A}}{Z(\beta)},
\end{equation}
with $Z(\beta) = \mathrm{Tr}(\exp(-\beta H_A))$, which is the Gibbs state $\rho_G$ on subsystem $A$. $\qed$

\section{Unitary coupled cluster ansatz}\label{app:uccsd}

In this section, we introduce the unitary coupled cluster (UCC) ansatz, which is a symmetry-preserving ansatz used primarily in molecular ground-state preparation.
In a Born-Oppenheimer approximation, the Hamiltonian in second quantization describes the number of electrons in an orbital and is given by
\begin{equation}
    H = \sum_{jk} h_{jk} a^\dagger_j a_k + \sum_{jk\ell m} h_{jk\ell m} a^\dagger_j a^\dagger_k a_\ell a_m,
\end{equation}
where $h_{jk}, h_{jk\ell m} \in \mathbb{R}$ describe overlaps of the orbital basis functions and $a_j$ is the annihilation operator of an electron is orbital $j$.
Since the total number of electrons in the system is conserved, the UCC ansatz leverages the cluster operators $T$ which describe the exchange of electrons in the orbitals and is defined as
\begin{equation}
    \begin{aligned}
        T(\vec\theta) &= \sum_{M} T^{(M)}(\vec\theta), \\
        T^{(M)}(\vec\theta) &= \sum_{\substack{j_1 \cdots j_M\\k_1 \cdots k_M}} \theta_{j_1 \cdots j_M}^{k_1 \cdots k_M} a^\dagger_{j_1} \cdots a^\dagger_{j_M} a_{k_M} \cdots a_{k_1},
    \end{aligned}
\end{equation}
i.e., $T^{(M)}$ describes the simultaneous exchange of $M$ electrons in between different orbitals.
To ensure the ansatz is unitary, the ansatz generator is taken to be $T - T^\dagger$, that is
\begin{equation}
    \ket{\phi(\vec\theta)} = e^{T(\vec\theta) - T^\dagger(\vec\theta)}.
\end{equation}
The initial state should have the correct number of electrons, which can be obtained classically efficiently with the Hartree-Fock solution~\cite{hartree_fock_1935}, which provides a single-configuration approximation to the system's ground state.

Commonly only the first few excitations are used, e.g. $T^{(1)}$ and $T^{(2)}$, which provides a sufficiently good approximation at a reduced computational complexity~\cite{omalley_uccsd_2016, romero_uccsd_2019}.
Since the circuit are still prohibitively expensive as the molecule increases in size, adaptive techniques recently gained a lot of interest, where excitations are added individually to the circuit based on which provides the largest energy gradient at the current point in the optimization~\cite{grimsley_adaptvqe_2019, tang_adaptvqe_2021}.

\section{Derivation of gradient rules}

\subsection{Parameter-shift rule}\label{app:paramshift_proof}

This section proves Eq.~\eqref{eq:paramshift} for the gradient calculation of quantum circuit expectation values.

The $j$th derivative of the expectation value of the state $\ket{\phi(\vec\theta)}$ of an observable $O$ is
\begin{equation}
    \begin{aligned}
    \partial_j \ell(\vec\theta) &= \partial_j \braket{\phi(\vec\theta)| O | \phi(\vec\theta)} \\
    &= \braket{\partial_j \phi(\vec\theta)| O | \phi(\vec\theta)} + \text{c.c.} \\
    &= \braket{0| V_0^\dagger U_1^\dagger(\theta_1) \cdots i G_j U_j^\dagger(\theta_j) \cdots U_d^\dagger(\theta_d) V_d^\dagger| O | \phi(\vec\theta)} + \text{c.c.} \\
    &= \braket{\phi^{(j)} | iG_j O^{(j)} \mathbb{1} | \phi^{(j)}} + \text{c.c.},
    \end{aligned}
\end{equation}
where we introduced
\begin{equation}
    \begin{aligned}
        &\ket{\phi^{(j)}} = V_{j-1} U_{j-1}(\theta_{j-1}) \cdots U_1(\theta_1) V_0 \ket{0}, \\
        &O^{(j)} = U_j^\dagger(\theta_j) V_j^\dagger \cdots U_d^\dagger(\theta_d) V_d^\dagger O V_d U_d(\theta_d) \cdots V_j U_{j}(\theta_{j}).
    \end{aligned}
\end{equation}
Next, we use the following identity, which holds for any two operators $A$ and $B$, and any state $\ket{\phi}$~\cite{schuld_evaluating_2019}
\begin{equation}\label{eq:innerprod_sum_decompose}
   \braket{\phi|A^\dagger O B|\phi} + \text{c.c.} = \frac{1}{2}\left(\braket{\phi|(A + B)^\dagger O (A + B)|\phi} - \braket{\phi|(A - B)^\dagger O (A - B)|\phi}
   \right),
\end{equation}
and introduce a factor $\lambda$, whose role becomes clear later, to write
\begin{equation}
    \begin{aligned}
        \partial_j \ell(\vec\theta) &= \lambda \Braket{\phi^{(j)} | \frac{i G_j}{\lambda} O^{(j)} \mathbb{1} |\phi^{(j)}} + \text{c.c.} \\
        &= \frac{\lambda}{2}\left(\Braket{\phi^{(j)}|\left(\mathbb{1} - \frac{iG_j}{\lambda}\right)^\dagger O^{(j)} \left(\mathbb{1} - \frac{iG_j}{\lambda}\right)|\phi^{(j)}} 
        - \Braket{\phi^{(j)}|\left(\mathbb{1} + \frac{iG_j}{\lambda}\right)^\dagger O^{(j)} \left(\mathbb{1} + \frac{iG_j}{\lambda}\right)|\phi^{(j)}}\right).
    \end{aligned}
\end{equation}
We now assume that $G_j$ only has two distinct eigenvalues, which w.l.o.g. can be shifted to $\pm \lambda, \lambda \in \mathbb{R}$ with a global phase on the expectation value. 
Then the above expectations can be evaluated for certain parameter values, which can be seen using $G^2 = \lambda^2 \mathbb{1}$ and writing
\begin{equation}
    \begin{aligned}
        U_j(\theta_j) &= e^{-i\theta_j G_j} \\
        &= \sum_{k=0}^\infty \frac{(-i\theta_j G_j)^k}{k!} \\
        &= \sum_{k=0}^\infty \frac{(-i\theta_j G_j)^{2k}}{(2k)!} + 
           \sum_{k=0}^\infty \frac{(-i\theta_j G_j)^{2k + 1}}{(2k + 1)!} \\
        &= \mathbb{1} \sum_{k=0}^\infty \frac{(\theta_j \lambda)^{2k}}{(2k)!} 
           -\frac{i G_j}{\lambda} \sum_{k=0}^\infty \frac{(-1)^k (\theta_j \lambda)^{2k + 1}}{(2k + 1)!}  \\
        &= \mathbb{1} \cos(\theta_j \lambda) - \frac{iG_j}{\lambda}\sin(\theta_j \lambda).
    \end{aligned}
\end{equation}
For the parameter value $s = \pi/(4\lambda)$ we obtain 
\begin{equation}
    U_j(\pm s) = \frac{1}{\sqrt{2}} \left(\mathbb{1} \mp \frac{i G_j}{\lambda}\right).
\end{equation}
Hence the gradient is 
\begin{equation}
    \begin{aligned}
    \partial_j \ell(\vec\theta)
        &= \frac{\lambda}{2}
        \left(\Braket{\phi^{(j)}|U_j(s)^\dagger O^{(j)} U_j(s)|\phi^{(j)}}  -
        \Braket{\phi^{(j)}|U_j(-s)^\dagger O^{(j)} U_j(-s)|\phi^{(j)}} \right) \\
        &= \frac{\lambda}{2} 
        \left(\braket{\phi(\vec\theta + s\vec e_j)|O|\phi(\vec\theta + s\vec e_j)} 
        - \braket{\phi(\vec\theta - s\vec e_j) | O | \phi(\vec\theta - s\vec e_j)}\right),
        \end{aligned}
\end{equation}
where we used $U_j(\theta_j)U_j(\pm s) = U_j(\theta_j \pm s)$. $\qed$

\subsection{Linear combination of unitaries}\label{app:lcu_proof}

This section shows the derivation of the LCU circuits in Fig.~\ref{fig:lcu_circuit} for the gradient calculation of quantum circuit expectation values.

Using a phase gate $P(\alpha)$, the auxiliary qubit is prepared in the state
\begin{equation}
    \begin{aligned}
        \frac{\ket{0} + e^{i\alpha}\ket{1}}{\sqrt{2}}.
    \end{aligned}
\end{equation}
Applying the unitary gates and the controlled generator $G$ we obtain the state
\begin{equation}
    \begin{aligned}
        &\frac{1}{\sqrt{2}}\ket{0} \big(V_d U_d(\theta_d) \cdots V_1 U_1(\theta_1) \ket{0}\big)
        + \frac{e^{i\alpha}}{\sqrt{2}} \ket{1} \big(V_d U_d(\theta_d) \cdots V_j U_j(\theta_j) G_j \cdots V_1 U_1(\theta_1) \ket{0}\big) \\
        &= \frac{1}{\sqrt{2}}\ket{0}\ket{\phi(\vec\theta)} + \frac{i e^{i\alpha}}{\sqrt{2}} \ket{1}\ket{\partial_j \phi(\vec\theta)},
    \end{aligned}
\end{equation}
where the additional complex phase appears since $U_j(\theta_j) G_j = i \partial U_j(\theta_j)$.
By computing the expectation value of the Pauli-$Z$ operator on the auxiliary qubit this state can be used to evaluate the gradient of a target observable $O$ on the state register. 
After the Hadamard gate on the auxiliary qubit the expectation value of the LCU circuit is
\begin{equation}
    \begin{aligned}
        & \frac{1}{2} \big(\bra{+}\bra{\phi(\vec\theta)} - i e^{-i\alpha} \bra{-}\ket{\partial_j \phi(\vec\theta)}\big) \big(Z \otimes O\big)
        \big(\ket{+}\ket{\phi(\vec\theta)} + i e^{i\alpha} \ket{-}\ket{\partial_j \phi(\vec\theta)}\big) \\
        &= \frac{i}{2} \left(e^{i\alpha} \braket{\phi(\vec\theta) |O|\partial_j  \phi(\vec\theta)} - e^{-i\alpha} \braket{\phi(\vec\theta) |O|\partial_j  \phi(\vec\theta)} \right).
    \end{aligned}
\end{equation}
Depending on the value of $\alpha$, this represents the real- or imaginary part of the expectation value gradient, as
\begin{equation}\label{eq:lcu}
    \begin{aligned}
        \alpha = 0 &\rightarrow \frac{i}{2} \left(\braket{\phi(\vec\theta) |O| \partial_j \phi(\vec\theta)} - \braket{\partial_j \phi(\vec\theta) |O| \phi(\vec\theta)} \right) = \mathrm{Im}(\braket{\partial_j \phi(\vec\theta) |O| \phi(\vec\theta)}), \\
        \alpha = \frac{\pi}{2} &\rightarrow -\frac{1}{2} \left(\braket{\partial_j \phi(\vec\theta) |O| \phi(\vec\theta)} + \braket{\partial_j \phi(\vec\theta) |O| \phi(\vec\theta)} \right) = -\mathrm{Re}(\braket{\partial_j \phi(\vec\theta) |O| \phi(\vec\theta)}).
    \end{aligned}
\end{equation}
$\qed$

\subsubsection{Quantum geometric tensor}

The derivation for the QGT circuit in Fig.~\ref{fig:lcu_qgt_circuit} is analogous to the gradient circuit, with one additional open control to apply the derivative with respect to another parameter on the state left of the observable.

Before the final Hadamard on the auxiliary qubit, the circuit prepares the state
\begin{equation}
    \begin{aligned}
        \frac{i}{\sqrt{2}}\ket{0} \ket{\partial_j\phi(\vec\theta)} + \frac{ie^{i\alpha}}{\sqrt{2}} \ket{1} \ket{\partial_k\phi(\vec\theta)}.
    \end{aligned}
\end{equation}
Applying the Hadamard and computing the expectation value with Pauli-$Z$ on the auxiliary qubit and the identity on the state register, we obtain
\begin{equation}
    \begin{aligned}
        \frac{1}{2}\left(e^{i\alpha}\braket{\partial_j\phi(\vec\theta)|\partial_k\phi(\vec\theta)} + e^{-i\alpha}\braket{\partial_j\phi(\vec\theta) | \partial_k\phi(\vec\theta)}\right),
    \end{aligned}
\end{equation}
which for $\alpha = 0$ yields $\mathrm{Re}(\braket{\partial_j \phi(\vec\theta)|\partial_k\phi(\vec\theta)})$ and for 
for $\alpha = \pi/2$ yields $-\mathrm{Im}(\braket{\partial_j \phi(\vec\theta)|\partial_k\phi(\vec\theta)})$.

The LCU circuits for the QGT evaluations can be further simplified.
Since the observable measured on the state register is the identity, the gates after the last differentiated gates cancels,
\begin{equation}
    \begin{aligned}
        \braket{\partial_j\phi(\vec\theta) | \phi(\vec\theta)} &= i\braket{0|V_0^\dagger \cdots G_j V_{j-1} U_{j-1}(\theta_{j - 1}) \cdots V_0 |0} \\
        \braket{\partial_j\phi(\vec\theta) | \partial_k \phi(\vec\theta)} &= \braket{0|V_0^\dagger \cdots G_j \cdots U^\dagger_{k-1}(\theta_{j - 1}) V_{k-1}^\dagger G_k V_{k-1} U_{k-1} (\theta_{k-1}) \cdots V_0 |0}.~(\text{w.l.o.g.} j \leq k)\\
    \end{aligned}
\end{equation}
$\qed$

\subsection{Unbiased gradient samples}\label{app:spsa_unbiased}

This section shows the SPSA gradient is an unbiased estimator up to $\mathcal{O}(\epsilon^2)$.
By Taylor expansion of the perturbation $\epsilon\vec\Delta$ the expectation value of the gradient estimator is
\begin{equation}
    \begin{aligned}
        \mathbb{E}\left[\widehat{\vec\nabla\ell}\right] 
        &= \mathbb{E}\left[\frac{\ell(\vec\theta + \epsilon\vec\Delta) - \ell(\vec\theta - \epsilon\vec\Delta)}{2\epsilon} \vec\Delta^{-1}\right] \\
        &= \mathbb{E}\left[\frac{\ell(\vec\theta) + \epsilon\vec\Delta^\top\vec\nabla\ell(\vec\theta) + \frac{\epsilon^2}{2}\vec\Delta^\top H(\vec\theta) \vec\Delta - \ell(\vec\theta) + \epsilon\vec\Delta^\top\vec\nabla\ell(\vec\theta) - \frac{\epsilon^2}{2}\vec\Delta^\top H(\vec\theta) \vec\Delta  + \mathcal{O}(\epsilon^3)}{2\epsilon}\vec\Delta^{-1}\right] \\
        &= \mathbb{E}\left[\vec\Delta^\top\vec\nabla\ell(\vec\theta)\vec\Delta^{-1}\right] + \mathcal{O}(\epsilon^2),
    \end{aligned}
\end{equation}
where $H(\vec\theta) = \vec\nabla \vec\nabla^\top \ell(\vec\theta)$ is the Hessian of $\ell$.
Dropping the explicit parameter dependence of $\vec\nabla\ell$, the $j$th entry of the expectation value is
\begin{equation}
    \begin{aligned}
        \mathbb{E}\left[\left(\widehat{\vec\nabla\ell}\right)_j\right] &=
        \mathbb{E}\left[\vec\nabla\ell_j + \Delta_j^{-1}\sum_{k\neq j} \Delta_k \vec\nabla\ell_k \right] + \mathcal{O}(\epsilon^2) \\
        &= \vec\nabla\ell_j + \mathbb{E}\left[\Delta_j^{-1}\right]\sum_{k\neq j} \mathbb{E}[\Delta_k] \vec\nabla\ell_k + \mathcal{O}(\epsilon^2) \\
        &= \vec\nabla\ell_j + \mathcal{O}(\epsilon^2),
    \end{aligned}
\end{equation}
where we used that all $\Delta_j$ are i.i.d., that $\mathbb{E}[\Delta_j] = 0$ and that $\mathbb{E}[\Delta_j^{-1}] \leq \mathbb{E}[|\Delta_j^{-1}|] < \infty$. $\qed$

\subsection{Unbiased Hessian samples}\label{app:2spsa_unbiased}

Analogous to the gradient samples, we can show the Hessian samples are unbiased up to leading order in $\epsilon$.

We remember the Hessian sample of a function $f$ at parameters $\vec\theta$ is defined as 
\begin{equation}
    \hat H = \frac{\delta f}{4\epsilon^2} \left(\frac{1}{2\vec\Delta(\vec\Delta')^\top} + \frac{1}{2 \vec\Delta'\vec\Delta^\top}\right),
\end{equation}
where the matrix in the denominator is understood as element-wise inverse, and with 
\begin{equation}
    \delta f = \sum_{s_1, s_1 \in \{1, -1\}} s_1 s_2 \cdot f\left(\vec\theta + \epsilon(s_1\vec\Delta + s_2\vec\Delta')\right).
\end{equation}

We expand each term in the curvature $\delta f$ as second-order Taylor series, that is
\begin{equation}
    \begin{aligned}
    f(\vec\theta + \epsilon(s_1\vec\Delta + s_2\vec\Delta')) &= 
        f(\vec\theta) + \epsilon(s_1\vec\Delta + s_2\vec\Delta') \vec\nabla f(\vec\theta) \\
        &+ \frac{\epsilon^2}{2} (s_1\vec\Delta + s_2\vec\Delta')^\top H(\vec\theta) (s_1\vec\Delta + s_2\vec\Delta') + \mathcal{O}(\epsilon^3),
    \end{aligned}
\end{equation}
where $H(\vec\theta) = \vec\nabla\vec\nabla^\top f(\vec\theta)$ is the Hessian of $f$.
Summing over the different perturbations, we find that the first-order gradient contributions cancel and we obtain
\begin{equation}
    \delta f = 2\epsilon^2\left(\vec\Delta^\top H(\vec\theta) \vec\Delta' + (\vec\Delta')^\top H(\vec\theta) \vec\Delta\right) + \mathcal{O}(\epsilon^3).
\end{equation}

The expectation value of the Hessian sample is then
\begin{equation}
    \begin{aligned}
       \mathbb{E}\left[\hat H\right] &= \frac{2\epsilon^2}{4 \epsilon^2} \mathbb{E}\left[\left(\vec\Delta^\top H(\vec\theta) \vec\Delta' + (\vec\Delta')^\top H(\vec\theta) \vec\Delta\right) \left(\frac{1}{2\vec\Delta(\vec\Delta')^\top} + \frac{1}{2 \vec\Delta'\vec\Delta^\top}\right) \right] + \mathcal{O}(\epsilon) \\
       &= \frac{1}{4} \left(
        \mathbb{E}\left[\frac{\vec\Delta^\top H(\vec\theta) \vec\Delta'}{\vec\Delta(\vec\Delta')^\top}\right]
        + \mathbb{E}\left[\frac{\vec\Delta^\top H(\vec\theta) \vec\Delta'}{\vec\Delta'\vec\Delta^\top}\right]
        + \mathbb{E}\left[\frac{(\vec\Delta')^\top H(\vec\theta) \vec\Delta}{\vec\Delta(\vec\Delta')^\top}\right]
        + \mathbb{E}\left[\frac{(\vec\Delta')^\top H(\vec\theta) \vec\Delta}{\vec\Delta'\vec\Delta^\top}\right]
       \right) + \mathcal{O}(\epsilon).
    \end{aligned}
\end{equation}
We investigate the summands individually and write the $(a,b)$-th element of the first summand as
\begin{equation}
    \begin{aligned}
    &\mathbb{E}\left[\frac{1}{\Delta_a \Delta'_b}\sum_{jk} \Delta_j \Delta'_k H_{jk}(\vec\theta)\right]  \\
    &=\mathbb{E}\Bigg[H_{ab}(\vec\theta) + \frac{1}{\Delta_a \Delta'_b}\sum_{\substack{j\neq a \\ k\neq b}} \Delta_j \Delta'_k H_{jk}(\vec\theta)\Bigg] \\
    &=H_{ab}(\vec\theta) + \mathbb{E}[\Delta_a^{-1}]\mathbb{E}[(\Delta'_b)^{-1}]\sum_{\substack{j\neq a \\ k\neq b}} \mathbb{E}[\Delta_j]\mathbb{E}[\Delta'_k] H_{jk}(\vec\theta) \\
    &=H_{ab}(\vec\theta),
    \end{aligned}
\end{equation}
where, as in Appendix~\ref{app:spsa_unbiased}, we used that all elements of $\vec\Delta$ and $\vec\Delta'$ are i.i.d. and that each element has zero mean and the inverse expectation exists.
The other three terms can be simplified analogously, such that we finally obtain
\begin{equation}
    \mathbb{E}\left[\hat H\right] = H(\vec\theta) + \mathcal{O}(\epsilon).
\end{equation}
$\qed$

\section{Derivation of the quantum natural gradient}\label{app:qng_derivation}

This section proves the QNG formula in two steps. First, we show that the Fubini-Study metric can be locally approximated by the QGT, and then we show how the direct update rule is derived from the minimization formulation.

\subsubsection{Approximation of the Fubini-Study metric}

By Taylor expansion we have
\begin{equation}
    \arccos^2(\sqrt{x}) = 1 - x + \mathcal{O}\left((1 - x)^2\right),
\end{equation}
where we identify $x = |\braket{\phi(\vec\theta)|\phi(\vec\theta + \vec{\delta\theta})}|^2 =: F(\vec\theta, \vec\theta + \vec{\delta\theta})$ as the fidelity.
To understand the quartic error term we expand the fidelity as Taylor series, that is
\begin{equation}\label{eq:fidelity_taylor}
    \begin{aligned}
    F(\vec\theta, \vec\theta + \vec{\delta\theta}) 
    &= F(\vec\theta, \vec\theta) + \vec{\delta\theta}^\top \vec\nabla F(\vec\theta', \vec\theta)\Big|_{\vec\theta' = \vec\theta} + \frac{1}{2} \vec{\delta\theta}^\top \left(\vec\nabla\vec\nabla^\top F(\vec\theta', \vec\theta)\Big|_{\vec\theta' = \vec\theta}\right)\vec{\delta\theta} +  \mathcal{O}(\|\vec{\delta\theta}\|_2^3) \\
    &= 1 + \frac{1}{2}\vec{\delta\theta} H_F(\vec\theta) \vec{\delta\theta} + \mathcal{O}(\|\vec{\delta\theta}\|_2^3),
    \end{aligned}
\end{equation}
since the fidelity is maximal at $F(\vec\theta, \vec\theta) = 1$, and $H_F$ denotes the Hessian of the fidelity. The vertical line with subtext $\vec\theta' = \vec\theta$ means that we first take the derivative with respect to $\vec\theta$, treating $\vec\theta'$ as independent variable, and inserting $\vec\theta' = \vec\theta$ after the differentiation.

Including only the constant term and ignoring the second-order Hessian term at an error of $\mathcal{O}(\|\vec{\delta\theta}\|_2^2)$ yields the first equality, as 
\begin{equation}
    \begin{aligned}
    \arccos^2\left(\sqrt{F(\vec\theta, \vec\theta + \vec{\delta\theta})}\right) &= 1 - F(\vec\theta, \vec\theta + \vec{\delta\theta}) + \mathcal{O}\left(\left(1 - (1 + \mathcal{O}(\|\vec{\delta\theta}\|_2^2)\right)^2\right) \\
    &= 1 - F(\vec\theta, \vec\theta + \vec{\delta\theta}) + \mathcal{O}\left(\|\vec{\delta\theta}\|_2^4)\right). \\
    \end{aligned}
\end{equation}
Note that only the fidelity in the error term is expanded to obtain an error in terms of the perturbation $\vec{\delta\theta}$.

The second equality of Eq.~\eqref{eq:fs_approximation} with the QGT is now follows by replacing the fidelity with it's second-order Taylor expansion and showing that $H_F = -2g$.

The Hessian elements are
\begin{equation}
    \begin{aligned}
        \big(H_F(\vec\theta)\big)_{jk} 
        &= \partial_j \partial_k \left|\braket{\phi(\vec\theta')|\phi(\vec\theta)}\right|^2\Big|_{\vec\theta'=\vec\theta} \\
        &= \partial_j \partial_k \left(\braket{\phi(\vec\theta)|\phi(\vec\theta')}\braket{\phi(\vec\theta')|\phi(\vec\theta)}\right)\Big|_{\vec\theta'=\vec\theta} \\
        &= 2 \partial_j \mathrm{Re} \left(\braket{\phi(\vec\theta)|\phi(\vec\theta')}\braket{\phi(\vec\theta')|\partial_k\phi(\vec\theta)}\right)\Big|_{\vec\theta'=\vec\theta} \\
        &= 2 \mathrm{Re} \left(\braket{\partial_j\phi(\vec\theta)|\phi(\vec\theta)}\braket{\phi(\vec\theta)|\partial_k\phi(\vec\theta)} + \braket{\phi(\vec\theta)|\partial_j \partial_k\phi(\vec\theta)}\right)
    \end{aligned}
\end{equation}
The second-order derivative can be rewritten by using that the quantum state is normalized, $\braket{\phi(\vec\theta)|\phi(\vec\theta)} = 1$.
Differentiating both sides of this equation we obtain
\begin{equation}\label{eq:zero_real_gradient}
    0 = \partial_k \braket{\phi(\vec\theta)|\phi(\vec\theta)} = 2 \mathrm{Re}\left(\braket{\phi(\vec\theta)|\partial_k \phi(\vec\theta)}\right).
\end{equation}
A second differentiation of $0 = \mathrm{Re}(\braket{\phi|\partial_k\phi})$ yields
\begin{equation}
    0 = \mathrm{Re}\left(\braket{\partial_j\phi(\vec\theta)|\partial_k\phi(\vec\theta)} +  \braket{\phi(\vec\theta)|\partial_j\partial_k\phi(\vec\theta)}\right),
\end{equation}
therefore
\begin{equation}
    \mathrm{Re}\left(\braket{\phi(\vec\theta)|\partial_j\partial_k\phi(\vec\theta)}\right) 
    = -\mathrm{Re}\left(\braket{\partial_j\phi(\vec\theta)|\partial_k\phi(\vec\theta)}\right).
\end{equation}
Inserting this into the formula for the Hessian element we obtain
\begin{equation}
    \begin{aligned} 
        \big(H_F(\vec\theta)\big)_{jk} 
        &= 2 \mathrm{Re} \left(\braket{\partial_j\phi(\vec\theta)|\phi(\vec\theta)}\braket{\phi(\vec\theta)|\partial_k\phi(\vec\theta)} - \braket{\partial_j\phi(\vec\theta)|\partial_k\phi(\vec\theta)}\right) \\
        &= -2 \big(g(\vec\theta)\big)_{jk}.
    \end{aligned} 
\end{equation}

Plugging the equality $H_F = -2g$ into the Taylor expansion, we have shown that
\begin{equation}
    \begin{aligned}
    d_{FS}^2(\vec\theta, \vec\theta + \vec{\delta\theta}) &= \arccos^2\left(\sqrt{F(\vec\theta, \vec\theta + \vec{\delta\theta})}\right) \\
    &= 1 - F(\vec\theta, \vec\theta + \vec{\delta\theta}) + \mathcal{O}\left(\|\vec{\delta\theta}\|_2^4)\right) \\
    &= 1 - \left(1 + \frac{\vec{\delta\theta}^\top H_F(\vec\theta) \vec{\delta\theta}}{2} + \mathcal{O}\left(\|\vec{\delta\theta}\|_2^3\right)\right) + \mathcal{O}\left(\|\vec{\delta\theta}\|_2^4)\right) \\
    &= \vec{\delta\theta}^\top g(\vec\theta) \vec{\delta\theta} + \mathcal{O}\left(\|\vec{\delta\theta}\|_2^3\right). 
    \end{aligned}
\end{equation}
$\qed$

\subsubsection{Direct QNG update rule}

For completeness, we show that the direct update rule, Eq.~\eqref{eq:qng}, is obtained by solving Eq.~\eqref{eq:qng_argmin}. 
The extremal condition for the solution $\vec\theta^*$ of the minimization is that the gradient vanishes, i.e.,
\begin{equation}
    \begin{aligned}
    \vec 0 &= \vec\nabla \left(\Braket{\vec\theta^* - \vec\theta^{(k)}, \vec\nabla\mathcal{L}\big(\vec\theta^{(k)}\big)} + \frac{1}{2\eta_k} d^2_{FS}\left(\vec\theta^*, \vec\theta^{(k)}\right)\right) \\
    &\approx \vec\nabla \left(\Braket{\vec\theta^* - \vec\theta^{(k)}, \vec\nabla\mathcal{L}\big(\vec\theta^{(k)}\big)} + \frac{1}{2\eta_k} (\vec\theta^* - \vec\theta^{(k)})^\top g\big(\vec\theta^{(k)}\big) (\vec\theta^* - \vec\theta^{(k)})^\top\right) \\
    &= \vec\nabla\mathcal{L}\big(\vec\theta^{(k)}\big) + \frac{1}{2\eta_k} 2 g\big(\vec\theta^{(k)}\big) (\vec\theta^* - \vec\theta^{(k)}),
    \end{aligned}
\end{equation}
where on the second line we used Eq.~\eqref{eq:fs_approximation} with $\vec{\delta\theta} = \vec\theta - \vec\theta^{(k)}$. 
Reshuffling the terms we obtain the solution of the minimization as
\begin{equation}
    \vec\theta^* = \vec\theta^{(k+1)} = \vec\theta^{(k)} - \eta_k g^{-1}\big(\vec\theta^{(k)}\big)\vec\nabla\mathcal{L}\big(\vec\theta^{(k)}\big).
\end{equation}
$\qed$

\section{Automatic differentiation}\label{app:ad}

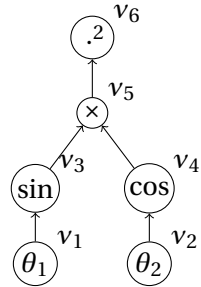
\begin{figure}[thbp]
    \centering
    \begin{tikzpicture}[
        level distance=10mm,
        every node/.style={circle,draw,inner sep=1pt},
        every child/.style={edge from parent/.style={draw, <-}},
    ]
    \node[label=25:$\nu_6$] {$~\cdot^2$}
        child {node[label=25:$\nu_5$] {$\times$}
            child {node[label=25:$\nu_3$] {$\sin$}
                child {node[label=25:$\nu_1$] {$\theta_1$}}
            }
            child {node[label=25:$\nu_4$] {$\cos$}
                child {node[label=25:$\nu_2$] {$\theta_2$}}
            }
        };
    \end{tikzpicture}
    \caption[Computational graph for automatic differentiation]{Computational graph for $f(\theta_1, \theta_2) = \sin^2(\theta_1)\cos^2(\theta_2)$.
             The node $\nu_5$ describes the operation $\times: x, y \mapsto xy$ (order irrelevant as the inputs are scalars) 
             and the node $\nu_6$ applies the mapping $\cdot^2: x \mapsto x^2$.}
    \label{fig:autodiff_simple_graph}
\end{figure}

Automatic differentiation (AD) describes a powerful strategy to evaluate derivatives of functions that are described by computational graphs.
A function $f: \mathbb{R}^n \rightarrow \mathbb{R}^m$ thus has $n$ input and $m$ output nodes, and each node $\nu_i$ of a graph denotes an intermediate function value after the application of a fundamental arithmetic operation.
For example, the graph of the function
\begin{equation}
    f(\theta_1, \theta_2) = \sin^2(\theta_1)\cos^2(\theta_2),
\end{equation}
is shown in Fig.~\ref{fig:autodiff_simple_graph}.

The derivatives of $f$ can then be computed by means of the chain rule and correctly chaining the Jacobians of the nodes in the graph.
By traversing the computational graph and employing the chain rule on each node, the derivative $\dot\nu_k$ of each node is computed as 
\begin{equation}\label{eq:ad_forward}
    \dot \nu_k = \sum_{j \in \text{pred}(k)} \frac{\partial \nu_k}{\partial \nu_j} \dot \nu_j,
\end{equation}
where $\partial \nu_k / \partial \nu_j$ denotes the Jacobian of node $\nu_k$ with respect to the input node $\nu_j$.
If node $\nu_k$ has $n_k$ inputs and $m_k$ outputs, the Jacobian is a map from $\mathbb{R}^{n_k} \rightarrow \mathbb{R}^{m_k}$, which depends on the values of the input nodes.

In this, so-called \emph{forward mode} of AD, the gradient to any input variable $\partial f / \partial \theta_j$ can be evaluated by setting $\dot\theta_j = \dot\nu_j = 1$ and the other inputs to 0.
A single forward graph traversal (or ``sweep'') allows to compute the gradient with respect to a single input variable. To compute the full gradient for $d$ parameters, we therefore have to repeat the procedure $d$ times.
The \emph{reverse mode} of AD circumvents this overhead.

Instead of evaluating the chain rule starting from the input nodes, the reverse mode starts from the output node. 
It recursively evaluates the adjoint variables $\bar\nu_k = \partial f / \partial \nu_k$ as
\begin{equation}\label{eq:ad_reverse}
    \bar\nu_k = \sum_{j \in \text{desc}(k)} \left(\frac{\partial \nu_j}{\partial \nu_k}\right)^\dagger \bar\nu_j,
\end{equation}
where $\text{desc}(k)$ is the set of all descendants of node $\nu_k$ and $(\partial \nu_j / \partial \nu_k)^\dagger$ is the adjoint of the Jacobian. As initial value the adjoint of the output node is set to 1.

The advantage of the reverse mode is that a single sweep allows to compute the gradient with respect to any input variable. 
However, this comes at a cost: The evaluation of the adjoint variables $\bar\nu_k$ generally requires the values of the nodes $\nu_k$.
Since the values have to be computed starting from the inputs, a reverse sweep requires a prior forward traversal of the graph and to store all node values at an $\mathcal{O}(M)$ memory cost, where $M$ is the number of nodes.
The forward AD mode does not have this limitation, since the value $\nu_k$ can be computed in the same sweep as the derivatives $\nu_k$.
Which mode is more efficient therefore depends on the function. 
Generally, for a multivariate function $f: \mathbb{R}^n \rightarrow \mathbb{R}^m$, the forward mode is more efficient if $m \geq n$, that is if the number of outputs is smaller or about the number of inputs.
Due to the additional memory overhead, reverse mode is typically only more efficient if $n \gg m$~\cite{margossian_autodiff_2019}.

\subsubsection{Illustrative example}

As an example we explicitly evaluate the gradients of the function in Fig.~\ref{fig:autodiff_simple_graph} both with the forward and reverse mode of AD.
The graph describes the operations
\begin{equation}
    \begin{aligned}
        \nu_1 &= \theta_1, \\
        \nu_2 &= \theta_2, \\
        \nu_3 &= \sin(\nu_1), \\
        \nu_4 &= \cos(\nu_2), \\
        \nu_5 &= \nu_3\nu_4, \\
        \nu_6 &= \nu_5^2,
    \end{aligned}
\end{equation}
such the function value equals the final node, $\nu_6 = f(\theta_1, \theta_2) = \sin^2(\theta_1)\cos^2(\theta_2)$.

The gradient $\partial f / \partial \theta_1$ using forward mode is obtained by setting $\dot\nu_1 = 1,~\dot\nu_2=0$ and traversing the graph using Eq.~\eqref{eq:ad_forward} until the gradient of the final node, $\dot\nu_6$ is reached. Explicitly,
\begin{equation}
    \begin{aligned}
        \dot\nu_3 &= \frac{\partial\nu_3}{\partial \nu_1} \dot\nu_1 = \cos(\nu_1) \cdot 1 = \cos(\nu_1), \\
        \dot\nu_4 &= \frac{\partial\nu_4}{\partial \nu_2} \dot\nu_2 = -\sin(\nu_2) \cdot 0 = 0, \\
        \dot\nu_5 &= \frac{\partial\nu_5}{\partial\nu_3}\dot\nu_3 + \frac{\partial\nu_5}{\partial\nu_4}\dot\nu_4 = \nu_4\dot\nu_3 + \nu_3\dot\nu_4 = \cos(\nu_2) \cos(\nu_1), \\
        \dot\nu_6 &= \frac{\partial \nu_6}{\partial \nu_5} \dot\nu_5 = 2\nu_5\dot\nu_5 = 2\sin(\nu_1)\cos(\nu_2)\cos(\nu_2)\cos(\nu_1),
    \end{aligned}
\end{equation}
and plugging in the parameter values the derivative is $\dot\nu_6 = 2\sin(\theta_1)\cos(\theta_1)\cos^2(\theta_2)$.

For the reverse mode we require the adjoint of the Jacobians, which in the scalar case is simply the derivative. 
Note that the adjoint might be more complex for vector- or matrix-valued functions~\cite{giles_autodiffmat_2008}.
We start by seeding the sweep with $\bar\nu_6 = 1$ and traverse the graph top to bottom, computing the adjoints of each node via Eq.~\eqref{eq:ad_reverse}, as
\begin{equation}
    \begin{aligned}
        \bar\nu_5 &= \frac{\partial\nu_6}{\partial\nu_5} \bar\nu_6 = \nu_5 \cdot 1, \\
        \bar\nu_4 &= \frac{\partial\nu_5}{\partial\nu_4} \bar\nu_5 = 2\nu_5 \nu_3, \\
        \bar\nu_3 &= \frac{\partial\nu_5}{\partial\nu_3} \bar\nu_5 = 2\nu_5 \nu_4, \\
        \bar\nu_2 &= \frac{\partial\nu_4}{\partial\nu_2} \bar\nu_4 = -2\nu_5\nu_3\sin(\nu_2), \\
        \bar\nu_1 &= \frac{\partial\nu_3}{\partial\nu_1} \bar\nu_3 = 2\nu_5\nu_4\cos(\nu_1),
    \end{aligned}
\end{equation}
where we used that in this case $(\partial\nu_k / \partial\nu_j)^\dagger = \partial\nu_k / \partial \nu_j$.
By plugging in all values we obtain $\partial f / \partial \theta_1 = \bar \nu_1 = 2\sin(\theta_1)\cos^2(\theta_2)\cos(\theta_1)$.
In contrast to a forward mode sweep, we already have all information available to also compute the derivative $\partial f / \partial \theta_2 = \bar\nu_2$, however we require storing the values of the nodes $\nu_k$ prior to the evaluation.

\subsubsection{AD for quantum circuit simulation}

For a loss function based on the expectation value of a parameterized quantum circuits there is only one output dimension,
which favors the reverse mode. An exemplary computational graph is shown in Fig.~\ref{fig:autodiff_quantum_circuit}.
However, storing the intermediate node values would require storing the full state-vector of the circuit, which makes the standard 
reverse AD prohibitively expensive.

\begin{figure}[htp]
    \centering
    \begin{tikzpicture}[
        level distance=10mm,
        every node/.style={circle,draw,inner sep=1pt},
        every child/.style={edge from parent/.style={draw, <-}},
        default/.style={edge from parent/.style={solid, thin, draw, <-}},
        emph/.style={edge from parent/.style={dotted, draw, <-}},
    ]
    \node {$\braket{O}$}
        child[default] {node[label=55:$\nu_d$] at (0, 0) {$\times$}
            child[emph] {node[label=55:$\nu_3$] at (-1, 0.3) {$\times$}
                child[default] {node[label=55:$\nu_2$, black, solid] at (-0.5, 0.5) {$\times$}
                    child {node[label=55:$\nu_1$] at (-0.5, 0.5) {$\times$}
                        child {node {$\ket{0}$}}
                        child {node {$U_1$} child {node {$\theta_1$}}}
                    }
                    child[default] {node[solid, thin] {$U_2$} child {node {$\theta_2$}}}
                }
                child[default] {node[solid, thin] {$U_{3}$} child {node {$\theta_{3}$}}}
            }
            child {node {$U_d$} child {node {$\theta_d$}}}
        };
    \end{tikzpicture}
    \caption[Quantum computational graph of an expectation value]{Computational graph for a quantum circuit expectation value $\braket{\phi(\vec\theta)|O|\phi(\vec\theta)}$
        for a parameterized circuit $\ket{\phi(\vec\theta)} = U_d(\theta_d) \cdots U_1(\theta_1)\ket{0}$.
        In comparison to the state of Eq.~\eqref{eq:ansatz_state}, the unparameterized unitaries $V_i$ have
        been absorbed into the $U_i$ to simplify the graph.
        Nodes with $\times$ denote multiplications of form $\times: L, R \mapsto RL$ where $R$ is the right input node and $L$ the left.
        The final node describes the operation $\braket{O}: \ket{\psi} \mapsto \braket{\psi|O|\psi}$.
        }
    \label{fig:autodiff_quantum_circuit}
\end{figure}

Fortunately, this memory overhead can be significantly reduced by leveraging the reversibility of the quantum circuit graph.
If the operations $U_i$ are unitary, i.e. $U_k^\dagger U_k = \mathbb{1}$ then the value of node $\nu_k$ can be computed from the \emph{following} node $\nu_{k+1}$, as 
\begin{equation}
    \nu_k = U_k \cdots U_1 \ket{0} = U^\dagger_{k+1} \nu_{k+1}.
\end{equation}
A reverse AD sweep can therefore compute the each node value within the same reverse graph traversal without requiring storing $\mathcal{O}(M)$ state-vectors.
This technique is leveraged by quantum computing simulators~\cite{luo_yaojl_2020}, and is has also been used in reversible neural networks~\cite{gomez_reversible_2017}.

\section{Quantum time evolution by Taylor expansion}\label{app:taylor_qite}

If exact matrix exponentiation is not feasible anymore, such as in the 27-qubit experiments of Chapter~\ref{chap:saqite}, we can instead approximate the quantum time evolution with a Taylor expansion.

For real-time evolution, we can express the Taylor expansion as unitary operation~\cite{forcrand_cqp_2009} by writing
\begin{equation}
    \begin{aligned}
    \ket{\psi(t + \Delta_t)} &= e^{-i\Delta_t H} \ket{\psi(t)} \\
                             &= \left(e^{i\frac{\Delta_t}{2} H}\right)^{-1}e^{-i\frac{\Delta_t}{2} H} \ket{\psi(t)} \\
                             &= \left(\mathbb{I} + \frac{i\Delta_t}{2} H\right)^{-1} \left(\mathbb{I} - \frac{i\Delta_t}{2} H\right) \ket{\psi(t)} + \mathcal{O}(\Delta_t^3),
    \end{aligned}
\end{equation}
where $\mathbb{I}$ is the identity matrix of appropriate dimension $2^n \times 2^n$.
The unitary property is straightforward to verify by the condition that if $UU^\dagger = \mathbb{I}$ the matrix $U$ is unitary. We have
\begin{equation}
    \begin{aligned}
     &\left(\mathbb{I} + \frac{i\Delta_t}{2} H\right)^{-1} \left(\mathbb{I} - \frac{i\Delta_t}{2} H\right)
     \left[
     \left(\mathbb{I} + \frac{i\Delta_t}{2} H\right)^{-1} \left(\mathbb{I} - \frac{i\Delta_t}{2} H\right)
     \right]^\dagger \\
     &= \left(\mathbb{I} + \frac{i\Delta_t}{2} H\right)^{-1} \left(\mathbb{I} - \frac{i\Delta_t}{2} H\right)
      \left(\mathbb{I} + \frac{i\Delta_t}{2} H\right)\left(\mathbb{I} - \frac{i\Delta_t}{2} H\right)^{-1} \\
     &= \left(\mathbb{I} + \frac{i\Delta_t}{2} H\right)^{-1} \left(\mathbb{I} + \frac{i\Delta_t}{2} H\right)
      \left(\mathbb{I} - \frac{i\Delta_t}{2} H\right)\left(\mathbb{I} - \frac{i\Delta_t}{2} H\right)^{-1} \\
    &= \mathbb{I},
     \end{aligned}
\end{equation}
where we have used that the inverse and complex conjugate are interchangeable and that for any scalar $c$ the terms $(\mathbb{I} + cH)$ and $(\mathbb{I} - cH)$ commute. $\qed$

Instead of evaluating the inverse, in practice the equation is solved as implicit scheme, that is
\begin{equation}
    \left(\mathbb{I} + \frac{i\Delta_t}{2} H\right)\ket{\psi(t + \Delta_t)} = \left(\mathbb{I} - \frac{i\Delta_t}{2} H\right) \ket{\psi(t)}.
\end{equation}

For imaginary-time evolution, which is not unitary to begin with, we can simply use a first order Taylor expansion and update the state as
\begin{equation*}
    \begin{aligned}
        \ket{\tilde\psi(t + \Delta_t)} &= (\mathbb{I} - \Delta_t H)\ket{\psi(t)} \\
        \ket{\psi(t + \Delta_t)} &= \frac{\ket{\tilde\psi(t + \Delta_t)}}{\|\ket{\tilde\psi(t + \Delta_t)} \|_2}.
    \end{aligned}
\end{equation*}
To ensure the timestep is chosen small enough, we run the full imaginary time evolution for decreasing timesteps $\Delta_t$ until a smaller timestep does no longer significantly change the evolution of the energy.
In the experiment of Chapter~\ref{chap:saqite}, for example, the calculation converged for $\Delta_t = 10^{-3}$.

\chapter{Quantum Natural SPSA}

\section{Runtime estimation on superconducting qubits}\label{app:qng_scaling}

To estimate the runtimes of QNG and VarQITE on superconducting hardware
we translate the circuit to a typical basis gate set on superconducting qubits, consisting of $\sqrt{X}$, $X$, $R_\mathrm{Z}$, and CX gates, commonly used e.g. by IBM Quantum devices.
This gate set allows to compile any sequence of single qubit gates into 2 $\sqrt{X}$ gates, interleaved with 3 virtual $R_Z$ rotations.
The time for a single measurement of the circuit in Fig.~\ref{fig:qng_scaling}(b) with $r$ repetitions is then
\begin{equation}
    t_\text{shot} = 2r t_\text{CX} + 2(r + 1) t_{\sqrt{X}} + t_\text{meas} + t_\text{reset},
\end{equation}
with the durations $t_\text{CX}$, $t_{\sqrt{X}}$, $t_\text{meas}$ and $t_\text{reset}$ of a CX gate, a $\sqrt{X}$ gate, a measurement and a qubit reset, respectively. 
In this qubit architecture, $R_Z$ rotations can be implemented as rotations of the reference frame and do not contribute to the runtime, but only affect the following pulses. For $N$ shots, the overall runtime is then $N t_\text{shot}$.

To calculate the number of circuits we assume a linear-combination of unitaries (LCU) method can be used for both the QGT and energy gradient. This is an idealistic assumption, since most superconducting devices have a sparse topology and using LCU typically requires additional Swap gates which we omit here.
For a Hamiltonian with $P$ Pauli terms, the number of circuits $C$ per timestep is
\begin{equation}
    C_\text{LCU} = \frac{d(d+5)}{2} + Pd,
\end{equation}
where $d$ is the number of parameters in the circuit and we assume $P=2$, which covers the class of Ising Hamiltonians.

The exact gate durations and fidelities depend on the qubit implementation and even varies within the family of superconducting qubits~\cite{kjaergaard_superconducting_2020}.
In our estimation we assume gate times reported by \texttt{ibm\_peekskill} (v2.6.5), which is an IBM Quantum Falcon processors~\cite{ibm_quantum}, of $t_{CX} = 451ns$, $t_{\sqrt{X}} = 36ns$ and $t_\text{meas} = 860ns$.
The qubit reset can be achieved in different fashions, such as natural relaxation of active resets~\cite{tornow_restless_2022}. Here, we assume $t_\text{reset} = 2\mu s$, which is possible to achieve using higher excited qubit states~\cite{egger_reset_2018, magnard_reset_2018}.

\section{QGT via the fidelity}\label{app:qgt_formulas}

Eq.~\eqref{eq:qgt_sample} shows a different representation of the QGT than introduced in Eq.~\eqref{eq:qgt_direct}. 
Here we show their equivalence.
A single element of the Fubini-Study metric tensor according to Eq.~\eqref{eq:qgt_sample} is
\begin{equation*}
    \begin{aligned}
        & -\frac{\partial_j \partial_k}{2} |\braket{\phi(\vec\theta^\prime) | \phi(\vec\theta)}|^2 \bigg\vert_{\vec\theta^\prime = \vec\theta} \\
        &= -\frac{\partial_j \partial_k}{2}  \braket{\phi(\vec\theta) | \phi(\vec\theta^\prime)} \braket{\phi(\vec\theta^\prime) | \phi(\vec\theta)} \bigg\vert_{\vec\theta^\prime = \vec\theta} \\
        &= -\partial_j \mathrm{Re}\left\{\braket{\phi(\vec\theta) | \phi(\vec\theta^\prime)} \braket{\phi(\vec\theta^\prime) | \partial_k\phi(\vec\theta)} \right\}\bigg\vert_{\vec\theta^\prime = \vec\theta}\\
        &= -\mathrm{Re}\left\{\braket{\phi(\vec\theta) | \phi(\vec\theta^\prime)} \braket{\phi(\vec\theta^\prime) | \partial_j\partial_k \phi(\vec\theta)} + \braket{\partial_j\phi(\vec\theta) | \phi(\vec\theta^\prime)} \braket{\phi(\vec\theta^\prime) | \partial_k \phi(\vec\theta)} \right\} \bigg\vert_{\vec\theta^\prime = \vec\theta}\\
        &= \mathrm{Re}\left\{-\braket{\phi(\vec\theta) | \partial_j\partial_k\phi(\vec\theta)} - \braket{\partial_j\phi(\vec\theta) | \phi(\vec\theta)}\braket{\phi(\vec\theta) | \partial_k(\vec\theta)}\right\}.
    \end{aligned}
\end{equation*}
We can rewrite the first summand using the identity we obtain from differentiating both sides of the equation $1 = \braket{\phi(\vec\theta)|\phi(\vec\theta)}$ with 
respect to $\vec\theta$,
\begin{equation*}
    \begin{aligned}
    0 &= \partial_j \partial_k \braket{\phi(\vec\theta)|\phi(\vec\theta)} \\
      &= 2 \mathrm{Re}\left\{\partial_j \braket{\partial_k\phi(\vec\theta) | \phi(\vec\theta)}\right\}  \\
      &= 2 \mathrm{Re}\left\{\braket{\partial_j \partial_k \phi(\vec\theta) | \phi(\vec\theta)} + \braket{\partial_j\phi(\vec\theta) | \partial_k \phi(\vec\theta)} \right\}\\
   \Leftrightarrow -\mathrm{Re}\left\{\braket{\partial_j \partial_k \phi(\vec\theta) | \phi(\vec\theta)} \right\} &= \mathrm{Re}\left\{\braket{\partial_j \phi(\vec\theta) | \partial_k \phi(\vec\theta)}\right\}.
    \end{aligned}
\end{equation*}
Replacing the second derivative of $\ket{\phi(\vec\theta)}$ with the two first order derivatives we obtain
\begin{equation*}
    g_{jk}(\vec\theta) = \mathrm{Re}\left\{\braket{\partial_j \phi(\vec\theta) | \partial_k \phi(\vec\theta)} - \braket{ \partial_j \phi(\vec\theta) | \phi(\vec\theta) } \braket{\phi(\vec\theta) | \partial_k \phi(\vec\theta)}\right\},
\end{equation*}
which is the same as Eq.~\eqref{eq:qgt_direct}.

\chapter{Dual quantum time evolution}

\section{Runtime on superconducting hardware}\label{app:dual_prognosis}

We assume the same processor specification as described in Appendix~\ref{app:qng_derivation}. The key difference to the QNG resource estimation in Fig.~\ref{fig:qng_scaling} is that we now determine the number of shots based on the required accuracy in the time evolution.

To estimate the total number of required measurements for VarQITE and DualQITE, we extrapolate the Heisenberg benchmark of Fig.~\ref{fig:sizescaling} to 50 qubits using the indicated scalings of $d^{3.77}$ and $d^{2.37}$, respectively.

\section{Derivation via quantum natural gradients}\label{app:dual_qng}

The QNG update rule is
\begin{equation}\label{eq:app_qng}
    \vec\theta^{(k+1)} = \vec\theta^{(k)} - \eta_k g^{-1}(\vec\theta^{(k)}) \vec\nabla\mathcal{L}(\vec\theta^{(k)}),
\end{equation}
which can be expressed in a dual formulation as
\begin{equation}\label{eq:app_qng_dual}
    \vec\theta^{(k+1)} = \argmin_{\vec\theta} \braket{\vec\nabla\mathcal{L}(\vec\theta^{(k)}, \vec\theta - \vec\theta^{(k)})} + \frac{1}{2 \eta} d^2(\vec\theta, \vec\theta^{(t)}),
\end{equation}
where $\mathcal{L}$ is the loss function, $\eta_k$ the learning rate and $d$ is the Fubini-Study distance metric, defined as 
\begin{equation}
    d(\vec\theta, \vec\theta') = \arccos|\braket{\phi(\vec\theta)|\phi(\vec\theta')}|.
\end{equation}

We can approximate the Fubini-Study distance metric locally for a small perturbation $\vec{\delta\theta}$ as
\begin{equation}\label{eq:app_fs_approximation}
    \begin{aligned}
    d^2(\phi(\vec\theta), \phi(\vec\theta + \vec{\delta\theta})) &= \arccos^2|\braket{\phi(\vec\theta)|\phi(\vec\theta + \vec{\delta\theta})}| \\
    &= 1 - F(\vec\theta, \vec\theta + \vec{\delta\theta})  + \mathcal{O}(||\vec{\delta\theta}||_2^4) \\
    &= \braket{\vec{\delta\theta}, g(\vec\theta)\vec{\delta\theta})} + \mathcal{O}(||\vec{\delta\theta}||_2^3),
    \end{aligned}
\end{equation}
with $F(\vec\theta, \vec\theta') = |\braket{\phi(\vec\theta)|\phi(\vec\theta + \vec{\delta\theta})}|^2$.
The standard form of the QNG, as in Eq.~\eqref{eq:app_qng}, is obtained by inserting the approximation in terms of the QGT into the Eq.~\eqref{eq:app_qng_dual} and solving the optimization problem explicitly.
The dual formulation, analogous to Eq.~\eqref{eq:argmin_dual}, appears by instead using the second line of the approximation, which uses the infidelity.

\section{Selecting the time perturbation}\label{app:dualqte_time_perturbation}

Since the error in approximating the QGT inner product with the infidelity scales with the norm of $\vec{\delta\theta}$, see e.g.~\eqref{eq:app_fs_approximation}, and $\dot{\vec\theta} = \vec{\delta\theta} / \delta\tau$ a smaller time perturbation $\delta\tau$ reduces the approximation error.
Because in practice, however, the loss function is subject to measurement noise we have errors in the solution $\vec{\delta\theta}$, which are amplified by $1/\delta\tau$.
This creates a trade-off between an approximation error scaling as $\mathcal{O}((\delta\tau)^3)$ and an error amplification scaling as $\mathcal{O}((\delta\tau)^{-1})$.

\begin{figure}[htp]
    \centering
    \includegraphics[width=\textwidth]{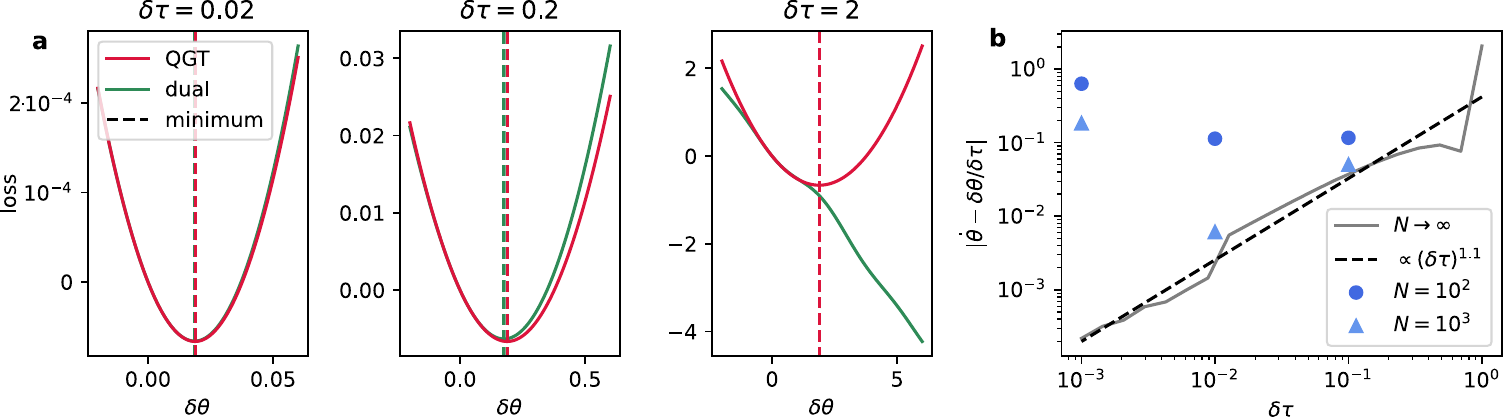}
    \caption[Impact of $\delta\tau$ on DualQTE loss function]{(a) Loss landscapes and optimal solutions of the original, QGT-based loss function and the dual loss function for different $\delta\tau$. (b) Error in calculating the parameter derivative $\dot\theta$ depending on $\delta\tau$ and the number of measurements $N$.}
    \label{fig:dtau_scaling}
\end{figure}

In Fig.~\ref{fig:dtau_scaling}(a), the effect of $\delta\tau$ on the loss landscape is illustrated. For small time perturbations, the landscapes of the QGT-based and DualQTE loss are almost identical. However, if $\delta\tau$ is too large, they increasingly differ---up to the point wehere the DualQTE loss function does not have a local minimum and the algorithm breaks down. 
Fig.~\ref{fig:dtau_scaling}(b) shows that, if the loss function is evaluated exactly with $N\rightarrow\infty$ measurements and no device noise, using the smallest possible $\delta\tau$ provides the smallest error.
However, if only a finite number of measurements is used, which leads to an error in $\delta\tau$, there is an optimal $\delta\tau$ that balances approximation error and error amplification.

To guarantee that DualQTE converges we have to ensure the local minimum exists, unlike for the situation in Fig.~\ref{fig:dtau_scaling} for $\delta\tau = 2$.
We can derive an upper bound on the time perturbation by using the necessary condition that $\vec\nabla\mathcal{L} = \vec{0}$ at the minimum, which can be written as the condition
\begin{equation}
    \forall j \in \{1, \dots, d\}: \frac{1}{2}\frac{\partial}{\partial(\delta\theta)_j} F(\vec\theta, \vec\theta + \vec{\delta\theta}) = -\delta\tau \cdot b_j(\vec\theta).
\end{equation}
Since the fidelity gradient is bounded but the linear part $-\delta\tau \cdot b_j^(\vec\theta)$ is not, there is a maximum feasible range for the value of $\delta\tau$. 
The bound depends on the maximum achievable value of the fidelity gradient. 
For a circuit with unique parameters and only Pauli rotations gates, as used in Chapter~\ref{chap:dual} for example, the gradient can be bounded to be in $[-1/2, 1/2]$ parameter-shift rule (see also Appendix~\ref{app:sampling_error}). A necessary condition for $\delta\tau$ is, thus,
\begin{equation}
    \forall j \in \{1, \dots, d\}: \delta\tau \in \left[\frac{-1}{4|b_j(\vec\theta)|}, \frac{1}{4|b_j(\vec\theta)|}\right],
\end{equation}
which can be generalized other circuits.
It is important to highlight that this is only a necessary condition for the presence of a minimum, not a sufficient one. Depending on the circuit's architecture, the fidelity gradient might not span the complete range $[-1/2, 1/2]$.

\section{Sampling error}\label{app:sampling_error}

Here, we derive the upper bound on the sample complexities of VarQTE and DualQTE to achieve a target error given by
\begin{equation}
    \varepsilon_S = \frac{1}{T}\int_0^T \sqrt{2(1 - |\braket{\phi(\vec\theta) | \phi(\tilde{\vec\theta})}|)}\mathrm{d} t.
\end{equation}
We assume the parameters are integrated with a forward Euler integration, which allows to write error as a function of the QGT, as
\begin{equation}\label{eq:bures_as_l2}
    \begin{aligned}
    \varepsilon_S &= \frac{1}{T}\int_0^T \sqrt{2 \left(1 - \sqrt{1 - \Delta_t^2 \Delta\dot{\vec\theta}^\top g(\vec\theta) \Delta\dot{\vec\theta} }\right)} \mathrm{d} t\\
    &= \frac{1}{T}\int_0^T \Delta_t \sqrt{\Delta\dot{\vec\theta} g(\vec\theta)\Delta\dot{\vec\theta}} \mathrm{d}t\\
    &\leq \frac{1}{T} \int_0^T \Delta_t \|g(\vec\theta)\|_2 \|\Delta\dot{\vec\theta}\|_2 \mathrm{d}t \\
    &\leq \Delta_t \sqrt{\lambda_\text{max}} \|\Delta\dot{\vec\theta}_\mathrm{max}\|_2,
    \end{aligned}
\end{equation}
with $\Delta\dot{\vec\theta} = \tilde{\dot{\vec\theta}} - \dot{\vec\theta}$, $\lambda_\text{max} \geq 0$ describing the largest eigenvalue of $g$ and, similarly, $\|\Delta\dot{\vec\theta}_\mathrm{max}\|_2$ being an upper bound on the norm $\Delta\dot{\vec\theta}$.
In the first line, we omit terms scaling with $\Delta_t^3$,
in the second line we approximate the inner square root with a first order Taylor-expansion and in the third line bound the inner product with $g$ using the definition of the operator.

\subsection{VarQTE}

Every VarQTE iterations comptues the parameter derivative as solution of a LSE, where both the system matrix, given by the real part of the QGT, and the right-hand side, given by the evolution gradient, are affected by sampling error.
We write these noisy quantities as
$\tilde g(\vec\theta) = g(\vec\theta) + \Delta g(\vec\theta)$ and $\tilde{\vec b}(\vec\theta) = \vec{b}(\vec\theta) + \Delta \vec{b}(\vec\theta)$ and, then, must solve the noisy LSE
\begin{equation}
    \tilde g(\vec\theta) \tilde{\dot{\vec\theta}} = \tilde{\vec b}(\vec\theta),
\end{equation}
where $\tilde{\dot{\vec\theta}} = \dot{\vec\theta} + \Delta\dot{\vec\theta}$ is the noisy solution.
As discussed in the main part of the thesis, the LSE is stabilized using a diagonal shift $\delta_c > 0$ on $g$.

The norm of the error in the LSE solution is
\begin{equation}\label{eq:deltatheta_bound}
    \begin{aligned}
    \|\Delta\dot{\vec\theta}\|_2 &= \|(g + \Delta g)^{-1}(\vec b + \Delta \vec b) - g^{-1}\vec b\|_2  \\
    &\approx \|(g^{-1} - g^{-1}\Delta g g^{-1})(\vec b + \Delta\vec b) - g^{-1} \vec b \|_2 \\
    &= \|g^{-1}\Delta \vec b - g^{-1} \Delta g g^{-1} \vec b - g^{-1} \Delta g^{-1} g^{-1}\Delta \vec b \|_2  \\
    &\approx \|g^{-1}\Delta \vec b - g^{-1}\Delta g \dot{\vec\theta} \|_2 \\
    &\leq \|g^{-1}\|_2 \left(\|\Delta\vec b\|_2 + \|\Delta g\|_2 \|\dot{\vec\theta}\|_2\right),
    \end{aligned}
\end{equation}
where we omit to explicitly state the parameter dependency to improve legibility.
In the second line we approximate $(g + \Delta g)^{-1} = g^{-1} - g^{-1}\Delta g g^{-1} + \mathcal{O}(\|\Delta g\|_2^2 \|g^{-1}\|_2^3)$, which is the von Neumann series, and on the fourth line we dropped quadratic error terms.

The following paragraphs derive bounds for the individual components of the above equation to finally obtain an error bound in available quantities, such as the system energy or the number of parameters.

\subsubsection{QGT spectrum}

The elements of the real part of the QGT can be computed as~\cite{gacon_qnspsa_2021}
\begin{equation}
    g_{jk}(\vec\theta) = -\frac{1}{2} \partial_j\partial_k F(\vec\theta', \vec\theta) \Bigg\vert_{\vec\theta' = \vec\theta}.
\end{equation}
Under the assumption of a circuit with unique parameters in Pauli rotation gates and no coefficients, we can bound the entries using the parameter-shift rule by
\begin{equation}
    g_{jk}(\vec\theta) = -\frac{1}{2} \frac{F^{(++)}_{jk} - F^{(+-)}_{jk} - F^{(-+)}_{jk} + F^{(--)}_{jk}}{4} \in \left[-\frac{1}{4}, \frac{1}{4}\right],
\end{equation}
where we used the notation $F^{(\pm\pm)}_{jk} = F(\vec\theta, \vec\theta \pm \vec{e}_j \pi/2 \pm \vec{e}_k \pi/2)$ with unit vectors $\vec{e}_j$ and $\vec{e}_k$, and we used that the fidelity is in $[0, 1]$.
As each entry of $g$ is bounded, we can leverage Gershgorin's circle~\cite{gershgorin_circle_1931} theorem to bound the maximal eigenvalue $\lambda_\text{max}$ by the maximal sum over columns or rows of the matrix, that is
\begin{equation}
    \lambda_\text{max} \leq \sum_{i=1}^{d} \frac{1}{4} = \frac{d}{4}.
\end{equation}

Note that this bound can easily be generalized to circuits with different parameter structures or gates by using chain and product rules. 
If, for example, we allow parameters to be repeated $m$ times, the bound on $\lambda_\text{max}$ becomes $md/4$.

\subsubsection{Norm of the update step}

The norm of the parameter derivative can be bounded from above as $\|\dot{\vec\theta}\|_2 \leq \|g^{-1}\|_2 \|\vec b\|_2$, where $\|g^{-1}\|_2 \leq \delta_c^{-1}$ and $\|\vec b\|_2$ can 
be bounded via the parameter-shift rule, as
\begin{equation}
    \begin{aligned}
    |b_k| = \frac{|E^{(+)}_k - E^{(-)}_k|}{2} \leq \frac{|E^{(+)}_k| + |E^{(-)}_k|}{2} \leq E_\text{max},
    \end{aligned}
\end{equation}
where $E^{(\pm)}_k = E(\vec\theta \pm \vec{e}_k \pi/2)$ and $E_\text{max}$ is the maximum absolute value of the system energy.
For the entire vector we have $\|\vec b\|_2 \leq \sqrt{d} E_\text{max}$, which leads to
\begin{equation}\label{eq:dottheta_bound}
    \|\dot{\vec\theta}\|_2 \leq \frac{\sqrt{d} E_\text{max}}{\delta_c},
\end{equation}
for any $\vec\theta \in \mathbb{R}^d$.

\subsubsection{Sampling errors in $g$ and $\vec b$}

The error due to a finite number of samples is unbiased, which means that $\Delta g = \tilde g - g$ is a random variable with zero mean and i.i.d. entries.
For such a random variable we can use apply Latala's theorem~\cite{latala_bound_2005} to state
\begin{equation}
    \mathbb{E}[\|\Delta g\|_2] \leq C \left(\max_{j} \sqrt{\sum_{k=1}^d \mathbb{E}[(\Delta g)_{jk}^2]} + \max_{k} \sqrt{\sum_{j=1}^d \mathbb{E}[(\Delta g)_{jk}^2]} + \sqrt[4]{\sum_{j, k=1}^d \mathbb{E}[(\Delta g)_{jk}^4]} \right),
\end{equation}
for a constant $C \in \mathbb{R}$.

In Ref.~\cite{gentinetta_pegasos_2022} the similar problem of sampling the matrix $[F(x_j, x_k)]_{j,k=1}^d$ for parameters $\{x_j\}_{j=1}^d$ is investigated, where the matrix elements are Bernoulli-distributed with probability $F(x_j, x_k)$. 
We can follow an analogous derivation by observing that the entries of $g_{jk}$ follow a Poisson binomial distribution~\cite{wang_number_1993} with probabilities $[F^{(++)}_{jk}, 1 - F^{(+-)}_{jk}, 1-F^{(-+)}_{jk},F^{(--)}_{jk}]$ over a shifted support $[0, 1, 2, 3, 4] \rightarrow [-2, -1, 0, 1, 2]$.
Since the means of this distribution are independent of the number of circuit parameters, we can show that
\begin{equation}
    \mathbb{E}[|(\Delta g)_{jk}|^2] = \mathcal{O}\left(\frac{1}{N}\right) \text{ and }
    \mathbb{E}[|(\Delta g)_{jk}|^4] = \mathcal{O}\left(\frac{1}{N^2}\right),
\end{equation}
leading to 
\begin{equation}
    \mathbb{E}[\|\Delta g\|_2] = \mathcal{O}\left(\sqrt{\frac{d}{N}}\right).
\end{equation}

It suffices to bound $\|\Delta \vec b\|_2$ by the sampling error, as the bound does not need to be tighter than $\|\Delta g\|_2 \|\dot{\vec\theta}\|_2$.
By the parameter-shift rule we have
\begin{equation}
    \begin{aligned}
    |\Delta b_k| &= |\tilde b_k - b_k| = \frac{|\tilde E^{(+)}_k - \tilde E^{(-)}_k - E^{(+)}_k + E^{(-)}_k|}{2}  \\
    &\leq \frac{|\tilde E^{(+)}_k - E^{(+)}_k| + |\tilde E^{(-)}_k - E^{(-)}_k|}{2}  \\
    &= \mathcal{O}\left(\frac{\sqrt{\mathrm{Var}(E^{(+)}_k)} + \sqrt{\mathrm{Var}(E^{(-)}_k)}}{2\sqrt{N}}\right).
    \end{aligned}
\end{equation}
The variance of the energy can further be bounded by 
\begin{equation}
    \mathrm{Var}(E) = \braket{\psi|H^2|\psi} - E^2 \leq \braket{\psi|H^2|\psi} \leq E_\text{max}^2,
\end{equation}
for any state $\ket{\psi}$.
By summing over all elements in the evolution gradient we obtain
\begin{equation}
    \|\Delta\vec b_\mathrm{max}\|_2 = \mathcal{O}\left(\frac{\sqrt{d}E_\text{max}}{\sqrt{N}}\right).
\end{equation}

\subsubsection{Final bound} 

Putting together the results from previous paragraphs into Eq.~\eqref{eq:deltatheta_bound} and Eq.~\eqref{eq:bures_as_l2}, we can finally state 
\begin{equation}
    \varepsilon_S \leq \mathcal{O}\left(\frac{d^{3/2} E_\text{max} \Delta_t}{\delta_c^2 \sqrt{N}}\right).
\end{equation}
The same asymptotic bound can alternatively be derived with a moment expansion on the expectation $\mathbb{E}[\dot{\vec\theta} - \tilde{\dot{\vec\theta}}]$.

\subsubsection{Illustrative example}

To illustrate the above result, we present a simple product-state model. This example shows the tightness of different bounds used in the derivation.

We consider the first timestep under an $n$-qubit Hamiltonian defined as $H = \sum_{j=1}^n Z_j$. The chosen ansatz consists of a single layer of Pauli-$Y$ rotations, with each rotation having its distinct parameter. The initial state for this system is $\ket{+}^{\otimes n}$, which is achieved by setting every parameter to $\pi/2$. For calculating each expectation value, $N = 1000$ measurements are used, and the QGT is regularized with a diagonal shift of $\delta_c = 10^{-2}$.
We keep track of different error contributions for a varying number of qubits from $n = 2$ to $10$. Due to the stochastic nature of $\Delta g$ and $\Delta\vec b$ we average each experiment over 10 executions.

\begin{figure}[htp]
    \centering
    \includegraphics[width=\textwidth]{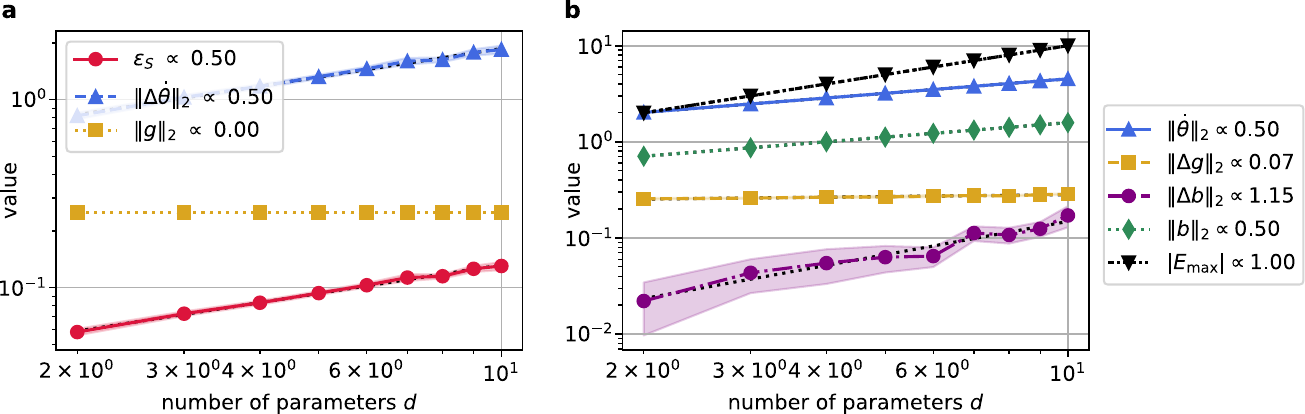}
    \caption[DualQTE error bounds of a product state]{Contributions to the error bound for a simple product state. The labels $\propto \alpha$ indicate a scaling with number of parameters as $d^\alpha$.}
    \label{fig:error_scalings}
\end{figure}

The QGT measures the correlation between parameter derivatives in the model and, as there is no light-cone connecting any two parameterized gates in the product state ansatz, the QGT is diagonal for this example.
The QGT norm is, thus, $\|g\|_2 = 1/4$ independent of system size. With this in mind, we find in Fig.~\ref{fig:error_scalings}(a) that the bound on the Bures metric in Eq.~\eqref{eq:bures_as_l2} is tight as $\varepsilon_S \propto \|\Delta\dot{\vec\theta}\|_2$.
All bounds used in the proof are obeyed, though we observe that the bound on $\|\dot{\vec\theta}\|_2$ in particular is loose: we upper bound the derivative with with $\sqrt{d}E_\text{max} \propto d^{1.5}$, but in this example only measure $d^{0.5}$ scaling. 
This suggests that the bound on the parameter derivative could be further improved. We could, for example, consider that the magnitude of the derivative is connected to the change induced of the evolution operator $\exp(-\Delta_t H)$, which is independent of the number of parameters $d$.

\subsection{DualQTE}

Under the assumption that DualQTE requires $K$ steps to converge, the error in the parameter update $\vec{\delta\theta}$ is 
\begin{equation}
    \begin{aligned}
        \|\Delta(\vec{\delta\theta})\|_2 = \|\Delta(\vec{\delta\theta}^{(K)})\|_2 &= \|\widetilde{\vec{\delta\theta}}^{(K - 1)} - \eta\widetilde{\vec\nabla\mathcal{L}}(\widetilde{\vec{\delta\theta}}^{(K-1)}) - \vec{\delta\theta}^{(K-1)} + \eta\vec\nabla\mathcal{L}(\vec{\delta\theta}^{(K-1)}) \|_2\\
        &\leq \|\Delta(\vec{\delta\theta}^{(K-1)})\|_2 + \eta \|\widetilde{\vec\nabla\mathcal{L}}(\widetilde{\vec{\delta\theta}}^{(K-1)}) - \vec\nabla\mathcal{L}(\vec{\delta\theta}^{(K-1)}) \|_2 \\
        &\leq \|\Delta(\vec{\delta\theta}^{(K-1)})\|_2 + \eta \|\Delta (\vec\nabla\mathcal{L})_\mathrm{max}\|_2 \\
        &\leq \eta K \|\Delta (\vec\nabla\mathcal{L})_\mathrm{max}\|_2,
    \end{aligned}
\end{equation}
where we used that initial error is zero, $\|\Delta(\vec{\delta\theta}^{(0)}\|_2 = 0$.
We write the error in measuring the gradient of the loss function as 
\begin{equation}
    \begin{aligned}
    \|\Delta(\vec\nabla\mathcal{L}(\vec{\delta\theta}))\|_2 &= \left\|\frac{\Delta (\vec\nabla F(\vec\theta, \vec\theta + \vec{\delta\theta}))}{2} + \delta\tau \Delta \vec{b}(\vec\theta)  \right\|_2 \\
    &\leq \frac{\|\Delta (\vec\nabla F(\vec\theta, \vec\theta + \vec{\delta\theta}))\|}{2} + \delta\tau \|\Delta \vec{b}(\vec\theta)\|_2 \\
    &\leq \frac{\|\Delta (\vec\nabla F)_\mathrm{max}\|}{2} + \delta\tau \|\Delta\vec b_\mathrm{max}\|_2,
    \end{aligned}
\end{equation}
where $\Delta (\vec\nabla F(\vec\theta, \vec\theta + \vec{\delta\theta})) = \widetilde{\vec\nabla F}(\vec\theta, \vec\theta + \vec{\delta\theta}) - \vec\nabla F(\vec\theta, \vec\theta + \vec{\delta\theta})$ and $\|\Delta (\vec\nabla F)_\mathrm{max}\|_2$ is an upper bound on the largest error the fidelity gradient can attain.

The asymptotic error in $\vec\nabla F$ can be quantified using the parameter-shift rule, as
\begin{equation}
    \begin{aligned}
    |\Delta \partial_j F| &= \frac{\Delta F^{(+)}_j - \Delta F^{(-)}_j}{2}  \\
    &= \mathcal{O}\left(\sqrt{\frac{\mathrm{Var}(F^{(+)}_j)}{N}} + \sqrt{\frac{\mathrm{Var}(F^{(-)}_j)}{N}}\right) \\
    &= \mathcal{O}\left(\frac{1}{\sqrt{N}}\right),
    \end{aligned} 
\end{equation}
where we used 
\begin{equation}
    \begin{aligned}
    \mathrm{Var}(F) = \braket{\psi|P_0^2|\psi} - \braket{\psi|P_0|\psi}^2 
                    = \braket{\psi|P_0|\psi} - \braket{\psi|P_0|\psi}^2  
                    = F(1 - F) 
                    \leq \frac{1}{4}.
    \end{aligned}
\end{equation}
The error of $\vec\nabla F$ is, then,
\begin{equation}
    \|\Delta(\vec\nabla F)_\mathrm{max}\|_2 = \mathcal{O}\left( \sqrt{\frac{d}{N}} \right).
\end{equation}

Since we already derived a bound $\|\Delta \vec b_\mathrm{max}\|_2$ we can state 
\begin{equation}
    \|\Delta(\vec{\delta\theta})\|_2 = \mathcal{O}\left(\frac{\sqrt{d}K(1 + \delta\tau E_\mathrm{max})}{\sqrt{N}} \right),
\end{equation}
which can be translated to scaling of the integrated Bures distance by the definition $\dot{\vec\theta} = \vec{\delta\theta} / \delta\tau$. We obtain
\begin{equation}
    \varepsilon_S \leq \Delta_t \sqrt{\lambda_\mathrm{max}} \frac{\|\Delta(\vec{\delta\theta})\|_2}{\delta\tau}
    = \mathcal{O}\left( \sqrt{\frac{\lambda_\mathrm{max} d}{N}}\frac{\Delta_t K (1 + \delta\tau E_\mathrm{max})}{\delta\tau} \right).
\end{equation}

\section{Gradient benchmark}\label{app:vanishing_gradients}

In this section, we investigate the norm of the loss function gradient, given by
\begin{equation}
    \vec\nabla_{\vec{\delta\theta}} \mathcal{L}(\vec\theta) = -\frac{\vec\nabla_{\vec{\delta\theta}} F(\vec\theta, \vec\theta+\vec{\delta\theta})}{2} - \delta\tau \cdot \vec{b}(\vec\theta),
\end{equation}
behaves with changing number of qubits in the system.
As discussed in Sec.~\ref{sec:trainable}, we expect neither the fidelity gradients nor the evolution gradient to decay exponentially and we expect the loss function to be measurable efficiently.

\begin{figure}
    \centering
    \includegraphics[width=\textwidth]{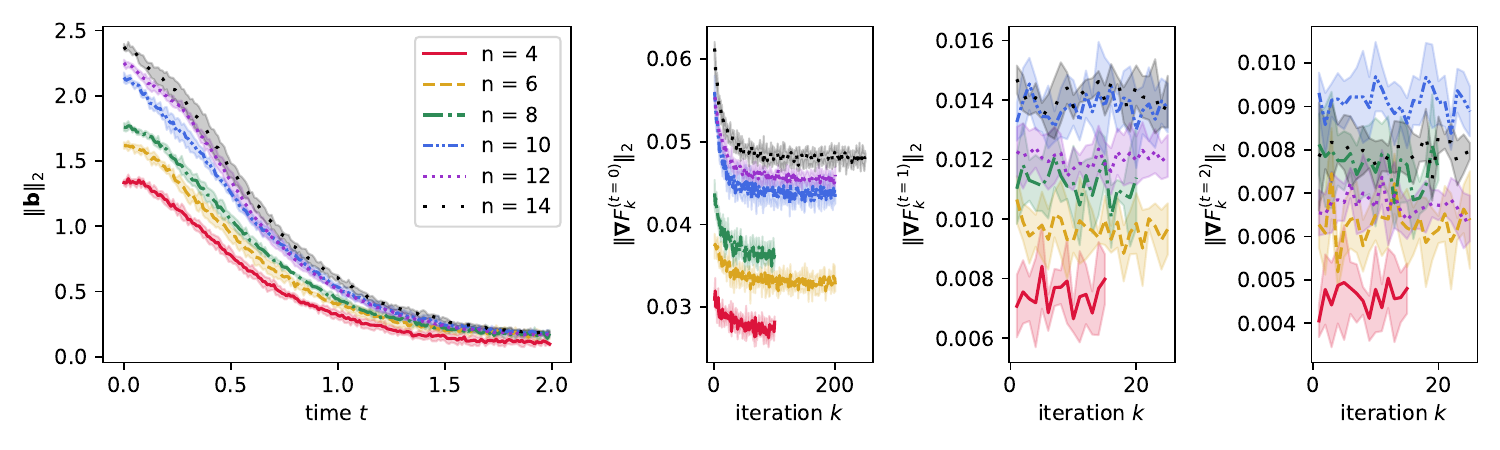}
    \caption[Gradient norms in the DualQTE optimization]{
    The $\ell_2$ norms the evolution gradient and fidelity gradients for varying number of qubits $n$.
    Since the evolution gradient is computed only once per iteration, we can show it for every time. 
    The fidelity gradients are computed per GD step in every timestep, therefore we present the behavior within the optimization iterations $k$ at selected times $t$.
    The number of points for $\vec\nabla F$ differ because the number of GD iterations varies per experiments, see Table~\ref{tab:resources_sizescaling}.}
    \label{fig:gradscaling}
\end{figure}

This is supported by the numerical evidence provided in Fig.~\ref{fig:gradscaling}. 
Instead of decaying, $\|\vec b\|_2$ in fact increases with system size, reflecting the property that the energy in the Heisenberg model is extensive.
The gradients only decay for large evolution times, which is expected as the system converges to the stationary ground state.
We observe a similar behavior for the fidelity gradients, which are shown for three different snapshots at times $t=0$, $1$ and $2$.
If present, barren plateaus would typically already manifest themselves for the system sizes presented here, see e.g. Ref.~\cite{mcclean_barren_2018}.

\section{Termination and warmstarting}\label{app:warmstarting}

Termination criteria for gradient descent algorithms are typically based on two metrics: the difference in the loss function in subsequent iterations or the gradient norm. 
If, however, we only have access to noisy evaluations of the loss function these metrics cannot be reliably verified. 
This is because, even though the algorithm may have converged, the noise exceeds the convergence thresholds. 

A potential workaround could be to track moving averages that takes into account a selection of recent iterations. But, depending on the noise level, the batch size for a reliable estimate could be large and require many iterations.
In the context of DualQTE, where we compute small corrections to the parameters, only a few iterations are required.
This is especially true if small timesteps are used and the parameters are warmstarted, i.e., initialized with the previous solution.

We therefore pursue a heuristic approach using many iterations in the first timestep and a fixed, small number in the remaining evolution. 
To calibrate the required number of steps---and to demonstrate the effectiveness of warmstarting the optimizations---we investigate DualQITE in an ideal, noise-free setting without exact state vector simulations.
We perform two simulations of the Heisenberg model of Sec.~\ref{sec:imag_heisen} for $n=12$ qubits with $r=6$ circuit layer repetitions.
In the first experiment, we always start the optimizations from $\vec{\delta\theta}^{(0)} = \vec 0$ (0-start), whereas in the second the initial guess is set to the solution of the previous iteration (warmstart).
In both experiments we perform GD iterations with $\eta=0.1$ until the change in the loss function drops below a set threshold of $10^{-6}$.
 
\begin{figure}[th]
    \centering
    \includegraphics[width=0.49\textwidth]{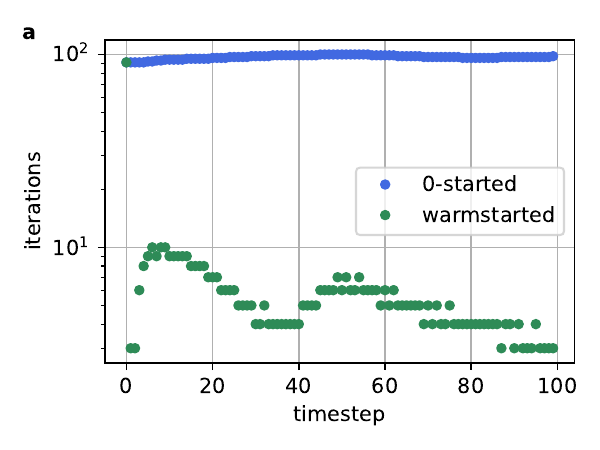}
    \includegraphics[width=0.49\textwidth]{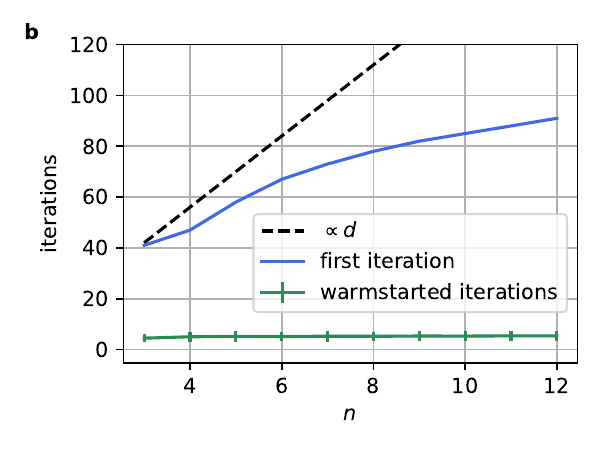}
    \caption[Warmstarting DualQTE]{(a) The number of iterations per timestep needed to reach the convergence criterion.
    (b) The number of iterations required in the warmstarting case, for both the initial timestep and the following timesteps. 
    }
    \label{fig:circuit_requirements}
\end{figure}

The number of iterations are for both approaches are shown in Fig.~\ref{fig:circuit_requirements}(a). 
We find that warmstarting significantly reduces the number of iterations needed to achieve the convergence criterion.
In Fig.~\ref{fig:circuit_requirements}(b) the experiment is repeated for different system sizes and we observe that the number of iterations in the warmstarted iterations seems to be constant independent of system size.

\subsection{Algorithm settings for resource benchmarks}\label{app:dual_resources}

In this section we list the settings of VarQITE and DualQITE used in the experiments of Sec.~\ref{sec:imag}. 
Table~\ref{tab:resources} corresponds to Fig.~\ref{fig:heisen}(b)
and Table~\ref{tab:resources_sizescaling} to Fig.~\ref{fig:sizescaling}.

\begin{table}[htp]
    \centering
    \subfloat[][]{
    \centering{
    \begin{tabular}{c|c|c}
        $I_B$ & shots & $N$ \\ \hline
        1.601 & 100 & $\sim 10^8$ \\
        0.558 & 1024 & $\sim 10^9$ \\
        0.149 & 8192 & $\sim 8 \cdot 10^9$
    \end{tabular}
    }}
    \hfill
    \subfloat[][]{
    \centering{
    \begin{tabular}{c|c|c|c|c}
        $I_B$ & shots & $K_0$ & $K_{>0}$ & $N$ \\ \hline
        0.937 & 100 & 100 & 10 & $\sim 2.5 \cdot 10^7$ \\
        0.735 & 100 & 200 & 20 & $\sim 5 \cdot 10^7$ \\
        0.305 & 1024 & 100 & 10 & $\sim 2.5 \cdot 10^8$ \\
        0.236 & 1024 & 200 & 20 & $\sim 5 \cdot 10^8$ \\
        0.153 & 2048 & 250 & 25 & $\sim 10^9$
    \end{tabular}
    }}
    \caption[VarQITE and DualQITE settings for Fig.~\ref{fig:heisen}(b)]{The settings for the accuracy benchmark at fixed number of qubits $n=12$ for (a) VarQITE and (b) DualQITE. For each point we show 
    the achieved integrated Bures distance, $I_B$, the number of shots per circuits, and the total number of measurements throughout the evolution, $N$.
    For DualQITE, we additionally list the number of iterations in the first and subsequent optimizations, $K_0$ and $K_{>0}$.}
    \label{tab:resources}
\end{table}

\begin{table}[htp]
    \centering
    \subfloat[][]{
    \centering{
    \begin{tabular}{c|c|c}
        $n$ & shots & $N$\\ \hline
        4 & 500 & $4.2 \cdot 10^7$ \\
        6 & 1500 & $4.2 \cdot 10^8$ \\
        8 & 2500 & $1.2 \cdot 10^9$ \\
        10 & 6000 & $6.7 \cdot 10^9$ \\
        12 & 8000 & $1.3\cdot 10^{10}$ 
    \end{tabular}
    }}
    \hfill
    \subfloat[][]{
    \centering{
    \begin{tabular}{c|c|c|c|c|c}
        n & shots & $K_0$ & $K_{>0}$ & $\eta$ & $N$\\ \hline
        4 & 500 & 100 & 15 & 0.07 & $8.8 \cdot 10^7$ \\
        6 & 600 & 200 & 25 & 0.07 & $3.3 \cdot 10^8$\\
        8 & 1000 & 100 & 20 & 0.1 & $6 \cdot 10^8$\\
        10 & 1500 & 200 & 25 & 0.12 & $1.7 \cdot 10^9$\\
        12 & 2500 & 200 & 25 & 0.1 & $3.5 \cdot 10^9$\\
        14 & 3000 & 250 & 25 & 0.12 & $4.9 \cdot 10^9$
    \end{tabular}
    }}
    \caption[VarQITE and DualQITE settings for Fig.~\ref{fig:sizescaling}]{The settings for sample complexity experiments of (a) VarQITE and (b) DualQITE, see Table~\ref{tab:resources} for a description of the variables.}
    \label{tab:resources_sizescaling}
\end{table}

%% file: tail/biblio.tex
\cleardoublepage
\phantomsection
\setcounter{biburlnumpenalty}{9000}
\emergencystretch=1em

\printbibliography
\addcontentsline{toc}{chapter}{Bibliography}